\documentclass[11pt,a4paper]{article}
\pdfoutput=1
\usepackage{jheppub}
\usepackage[utf8]{inputenc}
\usepackage{soul,xcolor}
\usepackage{makecell}
\usepackage{diagbox}
\usepackage{titlesec}
        \titleformat{\subsubsection}
           {\bfseries\normalsize}{\thesubsubsection}{1em}{}
\usepackage{ulem}
\usepackage{graphicx}
\usepackage{comment}
\usepackage{bm,array}
\usepackage{graphics}
\usepackage{epsf}
\usepackage{color}
\usepackage{amsmath}
\usepackage{amssymb}
\usepackage{latexsym}
\usepackage{slashed}
\usepackage{float}
\usepackage{cases}
\usepackage{multirow}
\usepackage{framed}
\def\ba{\begin{array}}
\def\ea{\end{array}}
\def\l{\lambda}

\def\alambda{A_\lambda}
\def\akappa{A_\kappa}
\def\mueff{\mu_\mathrm{eff}}

\def\sQ3{\widetilde{Q}_3}
\def\sU3{\widetilde{U}_3}
\def\sD3{\widetilde{D}_3}
\def\stleft{\tilde{t}_L}
\def\stright{\tilde{t}_R}

\def\higgsd{H_d^0}

\def\higgsu{H_u^0}

\def\bino{\widetilde{B}}
\def\wino{\widetilde{W}}

\def\higgsinod{\widetilde{H}^0_d}
\def\higgsinou{\widetilde{H}^0_u}
\def\singlino{\widetilde{S}}
\def\mssq{m_{_S}^2}
\def\mpsq{m_{_P}^2}
\def\mhiggsdsq{m_{_{H_d}}^2}
\def\mhiggsusq{m_{_{H_u}}^2}
\def\mhiggsusqhat{\hat{m}_{_{H_u}}^2}

\def\mhsmsq{m_{_H{_{_{\mathrm{SM}}}}}^2}
\def\msQthree{m_{\widetilde{Q}_3}}
\def\msUthree{m_{\widetilde{U}_3}}
\def\msDthree{m_{\widetilde{D}_3}}
\def\mstleft{m_{_{\tilde{t}_L}}}
\def\mstright{m_{_{\tilde{t}_R}}}

\def\mstone{m_{\tilde{t}_1}}
\def\msttwo{m_{\tilde{t}_2}}

\def\msinglino{m_{_{\widetilde{S}}}}
\def\mntrlone{m_{{_{\chi}}_1^0}}

\def\mhone{m_{h_1}}
\def\maone{m_{a_1}}
\def\vev{{\it vev}}
\def\vevs{{\it vevs}}
\def\vu{v_u}
\def\vd{v_d}
\def\vs{v_{_S}}
\def\sthreet{S_3(T)}

\def\topt{T_\mathrm{opt}}
\def\veva{{\tt Vevacious}}
\newcommand{\cred}[1]{{\color{red}#1}}
\newcommand{\cblue}[1]{{\color{blue}#1}}
\newcommand{\cgreen}[1]{{\color{green}#1}}

\newcommand{\beq}{\begin{equation}}
\newcommand{\eeq}{\end{equation}}
\newcommand{\bea}{\begin{eqnarray}}
\newcommand{\eea}{\end{eqnarray}}
\title{Exploring viable vacua of the $Z_3$-symmetric NMSSM}
\author[a, b, c]{Jyotiranjan Beuria}
\author[d]{Utpal Chattopadhyay}
\author[a, b]{AseshKrishna Datta}
\author[e]{Abhishek Dey}
\affiliation[a]{Harish-Chandra Research Institute, Allahabad 211019, India}
\affiliation[b]{Homi Bhabha National Institute, Training School Complex, \\
                Anushakti Nagar, Mumbai 400085, India} 
\affiliation[c]{Regional Centre for Accelerator-based Particle Physics \\
                Harish-Chandra Research Institute, Allahabad 211019, India} 
\affiliation[d]{Department of Theoretical Physics, Indian Association for the
                Cultivation of Science, \\ 2A \& B Raja S. C. Mullick Road,
                Jadavpur, Kolkata 700 032, India}
\affiliation[e]{Maulana Azad College, Government of West Bengal, \\
                8 Rafi Ahmed Kidwai Road, Kolkata 700013, India}
\emailAdd{jyotiranjan@hri.res.in, tpuc@iacs.res.in, asesh@hri.res.in,
          {\mbox{dey.abhishek111@gmail.com}}} 
\preprint{HRI-P-16-12-001 \\ 
\vspace*{-0.8cm}
\begin{flushright}
RECAPP-HRI-2016-013
\end{flushright}
}
\abstract{
We explore the vacua of the $Z_3$-symmetric Next-to-Minimal Supersymmetric
Standard Model (NMSSM) and their stability by going beyond the simplistic
paradigm that works with a tree-level neutral scalar potential and adheres
to some specific flat directions in the field space. We work in the so-called
phenomenological NMSSM (pNMSSM) scenario. Also, for our purpose, we adhere to
a reasonably `natural' setup by requiring $|\mueff|$ not too large. Key effects
are demonstrated by first studying the profiles of this potential under various
circumstances of physical interest via a semi-analytical approach. The results
thereof are compared to the ones obtained from a dedicated package like \veva 
~which further incorporates the thermal effects to the potential. Regions of the
pNMSSM parameter space that render the desired symmetry breaking (DSB) vacuum
absolutely stable, long- or short-lived (in relation to the age of the Universe)
under quantum/thermal tunneling are delineated. Regions that result in the
appearance of color and charge breaking (CCB) minima are also presented. It is
demonstrated that light singlet scalars along with a light LSP (lightest
supersymmetric particle) having an appreciable singlino admixture are compatible
with a viable DSB vacuum. Their implications for collider experiments are
commented upon.
}
\begin{document} 
\setstcolor{red}

\maketitle
%
\section{Introduction}
\label{sec:intro}
The Standard Model (SM) of particle physics is quite successful in explaining 
electroweak and strong interactions along with the generation of masses for 
fermions and electroweak gauge bosons via the Higgs mechanism. However, there 
are some theoretical and experimental results that cannot be explained while 
staying within the SM. Since the discovery of the SM-like Higgs boson 
($H_{\mathrm{SM}}$) at the Large Hadron Collider (LHC) at about a mass of 125 
GeV \cite{Aad:2012tfa, Chatrchyan:2012xdj}, studies Beyond the Standard Model 
(BSM) are motivated by the quest for new physics that could provide a potential 
solution to the so-called gauge hierarchy problem, massive neutrinos, 
matter-antimatter asymmetry and particle candidates for the dark matter. 
Supersymmetry (SUSY) is one of the most widely explored paradigms of the BSM 
physics.

The simplest SUSY extension of the SM, the minimal supersymmetric standard model 
(MSSM) addresses most of the above issues rather satisfactorily. However, the 
MSSM suffers from an intricate theoretical problem related to `naturalness' 
issues. The so-called `$\mu$-problem' \cite{Kim:1983dt} in the MSSM arises due 
to the presence of the SUSY Higgs/higgsino mass related to the `$\mu$-term', 
where `$\mu$' is a parameter with the dimension of mass. Successful ElectroWeak 
Symmetry Breaking (EWSB) and consistency with experimental results require 
`$\mu$' to be  of the order of the electroweak (EW) scale. However, there is no 
a priori theoretical reason for a superpotential (hence SUSY-conserving) 
parameter to assume a value near the SUSY-breaking soft mass scale. A larger 
value of `$\mu$' might reintroduce an unacceptably large fine-tuning in the 
generation of the SM gauge boson spectrum with experimentally observed masses. 
On the other hand, much smaller `$\mu$' would result in a very light charged 
Higgsino that is already ruled out by experiments.

The Next-to-Minimal Supersymmetric Standard Model (NMSSM) 
\cite{Ellwanger:2009dp, Maniatis:2009re}, 
the simplest extension of the MSSM, is obtained by adding one SM gauge singlet 
superfield $\hat{S}$ to the MSSM superpotential. Imposition of $Z_3$ symmetry 
then forbids a $\mu$-term in the superpotential. Instead, an effective $\mu$-term 
($\mueff$) is generated once the singlet scalar `$S$' acquires a vacuum 
expectation value (\vev). Since this is associated with EWSB, $\mueff$ 
automatically takes values close to the EW scale thus solving the $\mu$-problem. 
Furthermore, in the NMSSM there are added contributions to the mass of the Higgs 
boson already at the tree level. This is in contrast to the MSSM where one 
requires rather heavy top squarks that contribute radiatively to the Higgs mass 
and help it attain the experimentally observed value. However, this comes at the 
cost of a larger fine-tuning in the MSSM. Clearly, one does not need to bank 
heavily on radiative contributions in the NMSSM and hence on heavier top squarks 
thus ameliorating the fine-tuning issue. Phenomenology of such light top squarks
(squarks from the third generation, in general) has recently been discussed in
much detail \cite{Beuria:2015mta, Beuria:2016mur} in the context of the 
NMSSM. On the other hand, the very structure of the NMSSM that provides this 
extra contribution to the Higgs mass, also triggers mixing between the singlet 
scalar and the doublet Higgs states and modifies their couplings with the gauge 
bosons. Also, the fermionic sector now gets extended to include a corresponding
`singlino' state. This renders the sector to be phenomenologically richer with
its eigenstates having potentially nontrivial admixtures of gauginos, higgsinos
and the singlino.

In spite of all these appealing features, there is one theoretical problem, that 
gets aggravated in the NMSSM. Lorentz invariance of the vacuum allows only the 
Lorentz scalars to acquire a non-zero \vev. The only scalar field of the SM, the 
Higgs being $SU(3)$ singlet, color is always conserved by the ground state of 
the Higgs potential. Besides, one may define the unbroken $U(1)$ generator as 
the electric charge due to the presence of physically equivalent continuum of 
degenerate minima in the SM Higgs potential. Being away from the origin about
which the potential has a symmetry, these correspond to EWSB and the Desired 
Symmetry Breaking (DSB) minima with simultaneous conservation of electric charge 
and color symmetry. For the stability of the DSB minimum, this has to be the 
global minimum of the scalar potential. This, however, results in theoretical 
constraints that affect the allowed region of the parameter space of a given 
scenario. There have been rigorous studies on the stability of the SM vacuum 
\cite{Sher:1988mj}. 

The MSSM has two electroweak Higgs doublets with opposite hypercharge as 
compared to one in the SM. This makes the vacuum structure involving the MSSM 
neutral scalar potential significantly different from that of the SM. Besides 
there appear new kinds of vacua where the scalar SUSY partners of the SM quarks 
and leptons acquire non-vanishing \vevs. If such a minimum is deeper (global) 
than the DSB vacuum the latter could undergo quantum tunneling to the former 
(the panic vacuum) thus becoming unstable. This results in Charge Color Breaking 
(CCB) vacua which are phenomenologically unacceptable. Stringent theoretical 
constraints are obtained while ensuring the DSB vacuum, and not such CCB vacua, 
is the global minimum of the potential
\cite{AlvarezGaume:1983gj,Gunion:1987qv,Komatsu:1988mt,Casas:1995pd,Bordner:1995fh,
Strumia:1996pr,Baer:1996jn,Abel:1998cc,Abel:1998wr,Ferreira:2000hg,LeMouel:2001ym,
LeMouel:2001sf,Brhlik:2001ni,Cerdeno:2003yt,Ferreira:2004yg}.
Furthermore, one may  consider a `metastable' DSB vacuum with a large enough
tunneling time (to the deeper minimum) when compared to the age of the universe. 
Such a long-lived DSB vacuum is perfectly viable indicating that the Universe is 
trapped at a `false' vacuum. This consideration relaxes the constraint discussed 
above and allows for a wider region of viable parameter space of the MSSM
\cite{Brandenberger:1984cz,Kusenko:1995jv,Riotto:1995am,Falk:1996zt,Kusenko:1996jn,
Kusenko:1996xt,Kusenko:1996vp,Cohen:2013kna}.
Extensive studies on  occurrence and implications of global CCB vacua had been 
taken up in the past in various SUSY scenarios via analytical and semi-analytical
means
\cite{Datta:2000xy,Datta:2001dc,Datta:2001qa,Gabrielli:2001py,Datta:2004hr,
Kobayashi:2010zx,Hisano:2010re}.
After the discovery of the SM-like Higgs boson, the issue has also been put to 
context in the framework of the MSSM and its constrained version (CMSSM) 
\cite{Carena:2012mw,Camargo-Molina:2013sta,Chowdhury:2013dka,Blinov:2013uda,
Blinov:2013fta,Camargo-Molina:2014pwa,Chattopadhyay:2014gfa,Hollik:2015pra,Hollik:2016dcm}. 
These studies often involved the state-of-the-art treatments both at the 
analytical and numerical levels.

In the presence of a neutral singlet scalar field (`$S$') the vacuum structure 
of the NMSSM gets to be much involved even in the absence of CCB minima. For, 
along with the neutral components of the two Higgs doublets, `$S$' could also 
acquire non-vanishing \vev ~at the ground state. This renders the study of the
NMSSM vacua rather complicated when compared to its MSSM counterpart.
Theoretical analysis is possible only under simplified assumptions over the 
configurations in the multi-dimensional field-space 
\cite{Ellwanger:1996gw, Ellwanger:1999bv, Kanehata:2011ei, Kobayashi:2012xv}. 
Even with three non-vanishing neutral scalar fields, finding all the minima of 
the potential, which is so essential for the purpose, becomes a rather involved 
task. Consideration of the CCB vacua in such a scenario may make the situation 
all the more complicated. On top of that, radiative and finite temperature 
corrections significantly modify the overall structure of the potential. 
Inclusion of temperature (thermal) effects 
\cite{Affleck:1980ac, Linde:1980tt, Linde:1981zj, Masoumi:2015psa}
has been shown to crucial 
\cite{Brignole:1993wv} while analyzing the stability of DSB vacuum against
tunneling to a deeper minimum. In fact, even a DSB vacuum which is found to
be long-lived under quantum tunneling may be rendered unstable when thermal 
effects are considered.

Under the circumstances, it turns out that the problem at hand has got an 
essentially numerical aspect when it comes to its reliable and comprehensive 
resolution. To the best of our knowledge, the present work is the first one to 
address these issues with vacuum stability in the $Z_3$-symmetric NMSSM in an 
exhaustive way. The scope of our study is primarily guided by two issues that 
are of current relevance: (i) the interesting phenomenological possibility of 
the existence of new Higgs-like states which might get lighter than the recently 
discovered SM-like Higgs boson, could have an appreciable singlet admixture and 
are yet to show up in experiments and (ii) an overall setup that ensures the 
scenario to be `natural', i.e., requires less of a finetuning, in the 
conventional sense. While the first one would constrain some key NMSSM 
parameters, the latter is achieved for not so large values of $\mueff$
\cite{Baer:2012up, Baer:2013ava, Mustafayev:2014lqa, Baer:2015rja}. Furthermore, 
such a setup could naturally lead to light neutralinos which can have appreciable 
admixtures of, in particular, higgsinos and the singlino. Note that such a 
composition would have nontrivial implications for the phenomenology at the 
collider and of dark matter alike.

Fortunately, all the complicated aspects discussed above can now be addressed in 
a framework like \veva ~\cite{Camargo-Molina:2013qva} that we adopt at length in 
the present study. Nonetheless, to the extent possible, we strive to have an 
analytical/semi-analytical/graphical understanding of the proceedings. This may 
shed crucial light on the interplay of the basic parameters leading to an 
eventual reconciliation of results obtained from \veva. Close agreements among 
the basic results obtained via these two approaches add credence to such an 
understanding. Note that for a scenario with multiple scalars, it is extremely 
difficult, if not impossible, to develop an insight into the proceedings solely 
out of \veva ~results. In principle, such agreements also underscore the 
applicability of an analytical/semi-analytical approach in more detailed studies 
of vacuum structure. This comes with an added bonus that one could track the 
proceedings in analytical terms unlike what a hard-core numerical approach could 
offer.

Due to the presence of an associated neutral fermionic state `singlino', the 
neutralino sector of the MSSM gets extended. The lightest SUSY particle (LSP) 
which can be a candidate for the dark matter (DM) thus can have a significant 
singlino admixture. Along with singlet scalars, such an LSP could turn out to 
be rather light ($\sim$few tens of GeV) but might have escaped detection at 
various collider experiments because of its suppressed interaction with other 
particles. It is thus not unexpected that the nature of the stability of the 
DSB vacuum might be connected to such light spectra. This, in turn, would have 
definite ramifications for the collider and the DM experiments.

The paper is organized as follows. In section 2 we discuss the theoretical
framework of the $Z_3$-symmetric NMSSM in reference to its scalar and neutralino 
sectors. The \veva ~approach to the analysis of stability of the DSB vacuum is 
also summarized. Section 3 is devoted to the discussion of the vacuum structure 
of the scenario with a particular emphasis on the false (DSB) vacua appearing 
along specific and arbitrary field directions. Stability of such false vacua is 
analyzed in detail via semi-anlaytic and dedicated numerical (using \veva) means. 
Regions of the parameter space yielding viable EWSB vacua are then delineated. 
The spectral pattern of the light scalars and the LSP are discussed along with 
their implications for the ongoing experiments. In section 4 we discuss issues 
pertaining to the appearance of the CCB vacua in such a scenario. Their 
implications in the presence of a minimum of the potential driven by the singlet 
neutral scalar field are also pointed out. We conclude in section 5.
%
\section{The theoretical framework}
\label{sec:framework}
%
The framework of the NMSSM features an extra singlet superfield $\hat{S}$ in 
addition to the MSSM ones. In the popular $Z_3$-symmetric version of the NMSSM 
on which the present work is based, we ignore the linear and bilinear terms in 
$\hat{S}$. Also, the $Z_3$ symmetry prohibits the explicit presence of the usual 
higgsino mass term, i.e., the well-known $\mu$-term of the MSSM, in the NMSSM 
superpotential which is given by
\beq
\mathcal{W}=\mathcal{W_{MSSM}}|_{\mu=0}+ \lambda \hat{S} \hat{H}_u.\hat{H}_d
  + {\kappa \over 3} \hat{S}^3
\label{eq:superpot}
\eeq
with
\beq
\mathcal{W_{MSSM}}|_{\mu=0}= y_d \hat{H}_d\cdot\hat{Q} \hat{D}_R^c 
                           + y_u \hat{Q} \cdot  \hat{H}_u\hat{U}_R^c 
                           + y_e \hat{H}_d \cdot \hat{L} \hat{E}_R^c \quad .
\eeq
In the above expression, ${\cal{W_{MSSM}}}|_{\mu=0}$ stands for the MSSM 
superpotential less the $\mu$-term, $\hat{H}_{u}$ and $\hat{H}_{d}$ are the 
doublet Higgs superfields while $\hat{S}$ is the gauge singlet superfield 
mentioned above. The superfields $\hat{Q}$, $\hat{U}_R$ and $\hat{D}_R$ 
represent the $SU(2)$ quark-doublet, up-type $SU(2)$ singlet quark and 
down-type $SU(2)$ singlet quark superfields. On the other hand, 
$\hat{L}$ and $\hat{E}_R$ denote the $SU(2)$ doublet and singlet lepton 
superfields, respectively. Furthermore, $y_{f=d,u,e}$ stand for the 
corresponding Yukawa couplings. In the subsequent subsections we discuss
the composition of the resulting scalar potential of the scenario which the 
present work crucially depend upon.
%
\subsection{The scalar potential}
%
The scalar potential of the $Z_3$-symmetric NMSSM is comprised of several
components, i.e., the soft SUSY-breaking part and the $F$- and the $D$-term
contributions. The first one is given by \cite{Ellwanger:2009dp}
\bea
V_\mathrm{soft} &=& m_{H_d}^2 | H_d |^2 + m_{H_u}^2 | H_u |^2 + m_{S}^2 |S|^2+m_Q^2|Q^2| + m_U^2|U_R^2| + m_D^2|D_R^2| \nonumber \\
&+&  m_L^2|L^2| +m_E^2|E_R^2| + y_u A_u Q \cdot H_u U_R^c - y_d A_d Q \cdot H_d D_R^c 
- y_{e} A_{e} L \cdot H_d E_R^c \nonumber \\
&+& \lambda A_\lambda H_u \cdot H_d S + \frac{1}{3} \kappa A_\kappa S^3 + h.c. \quad .
\label{eq:V-soft}
\eea
In the above expression, the third term and the last two terms represent the new
soft SUSY breaking terms (beyond what the MSSM scalar potential already has) 
appearing in the $Z_3$-symmetric NMSSM. We use standard notations to indicate 
various fields and the corresponding soft masses \cite{Ellwanger:2009dp}. Along 
with $\lambda$ and `$\kappa$' that appear in equation \ref{eq:superpot}, 
$A_{\lambda}$ and $A_{\kappa}$, appearing in the last two terms of the above 
expression (and having the dimension of mass) are going to be the new input 
parameters of the NMSSM scenario under consideration. $\mssq$, appearing in the 
third term, is the soft SUSY breaking mass-squared term for the scalar field 
`$S$' corresponding to the chiral superfield $\hat{S}$. $y_u$, $y_d$ and $y_e$ 
are the corresponding Yukawa couplings. Eventually, the tree-level scalar 
potential of the $Z_3$-symmetric NMSSM is given by
\beq
V_\mathrm{scalar}  = V_\mathrm{soft} + V_F + V_D ,
\label{eq:scalar-pot}
\eeq
where $V_{F,D}$ are the $F$- and $D$-term contributions \cite{Kanehata:2011ei}
mentioned previously (the sum of the squares of the matter and the gauge 
auxiliary fields, respectively). 

During electroweak symmetry breaking (EWSB) which preserves charge and color
symmetries, only the $CP$-even neutral, scalar degrees of freedom of $H_d$, 
$H_u$ and `$S$' (corresponding to the superfields $\hat{H}_d$, $\hat{H}_u$ and 
$\hat{S}$, respectively) may acquire vacuum expectation values (\vev) $\vd$, 
$\vu$ and $\vs$, respectively. The last one generates an effective $\mu$ term 
($\mueff$) which is given by $\mueff = \lambda \vs$. This provides an elegant 
solution to the so-called ``$\mu$-problem'' \cite{Kim:1983dt} plaguing the MSSM 
and the NMSSM draws its original motivation in this (see \cite{Ellwanger:2009dp} 
and references therein). Situations under which charged and colored states like 
the scalar leptons (sleptons) and scalar quarks (squarks) acquire \vevs ~lead to 
minima that break charge and color symmetries. These are heavily restricted on 
observational grounds and, as we shall see, serve as major constraints on the 
parameter space that could trigger desired EWSB.

While exploring the charge and color breaking (CCB) minima of the scalar 
potential would involve the entire scalar potential $V_\mathrm{scalar}$ given 
in equation \ref{eq:scalar-pot}, studying viable EWSB that preserves charge and 
color concerns only the part of the scalar potential that involves the neutral 
doublet Higgs fields and the neutral singlet scalar field `$S$'. At the minima 
of the Higgs potential, the field values of the neutral scalars are the 
corresponding \vevs ~and the neutral physical Higgs fields are obtained by 
expanding the scalar potential around the real neutral \vevs ~$\vd$, $\vu$ and 
$\vs$ as
\beq
H_d^0 = \vd + \frac{H_{dR} + iH_{dI}}{\sqrt{2}} , \quad
H_u^0 = \vu + \frac{H_{uR} + iH_{uI}}{\sqrt{2}} , \quad
S = \vs + \frac{S_R + iS_I}{\sqrt{2}} ,
\label{eq:vev-expand}
\eeq
where the subscripts `$R$' and `$I$' indicate the real and imaginary parts of the
scalar fields, respectively. As a result, these neutral scalar states are found 
to mix. The resulting Higgs potential, in terms of the \vevs ~of the neutral 
scalars, at the tree level, is given by
\bea
V^{^{\mathrm{neutral}}}_\mathrm{Higgs}|_\mathrm{tree} & = & 
(m_{H_d}^2 + |\lambda S|^2) |\higgsd|^2 +
(m_{H_u}^2 + |\lambda S|^2) |\higgsu|^2 + 
m_{S}^2\, |S|^2 \nonumber \\
&+& |F_S|^2 + V_D^H - \lambda \alambda\, \higgsd \higgsu S +  \frac{1}{3}\kappa \akappa\, S^3 + h.c. \;\; ,
\label{eq:higgs-pot-tree-org}
\eea
where the mass-squared terms directly correspond to the same in equation
\ref{eq:V-soft} and $F_S$ and $V_D^H$ are the $F$- and the $D$-term contributions
to the neutral Higgs potential given by
\bea
F_S &=& -\lambda \higgsd \higgsu + \kappa S^2 \nonumber \\
V_D^H &=& \frac{g_1^2+g_2^2}{8}(|\higgsd|^2 - |\higgsu|^2)^2 .
\label{eq:f-and-d-terms}
\eea

For the potential to have an extremum, one must make sure that its derivatives 
with respect to each of the three \vevs ~vanish simultaneously, i.e.,
\beq
T_i={\partial V^{^{\mathrm{neutral}}}_\mathrm{Higgs}|_\mathrm{tree} \over \partial \phi_i}\biggr\rvert_{\phi_{iR}= 
\phi_{iI}=0} = 0
\label{eq:first-derivative}
\eeq
with $\phi_i \equiv \left\{H_d, H_u, S \right\}$. These lead to the following set 
of three tree-level `tadpole' equations: 
\begin{subequations}
\bea
m_{H_d}^2 + \mueff^2 + \lambda^2 \vu^2 + \frac{g_1^2+g_2^2}{4} (\vd^2-\vu^2)-\frac{\vu}{\vd} \mueff (A_\lambda+\kappa \vs) &=& 0  \\
\label{eq:tadpole-a}
m_{H_u}^2 + \mueff^2 + \lambda^2 \vd^2 + \frac{g_1^2+g_2^2}{4} (\vu^2-\vd^2)-\frac{\vd}{\vu} \mueff (A_\lambda+\kappa \vs) &=& 0 \\
\label{eq:tadpole-b}
m_S^2 + \kappa A_{\kappa} \vs + 2\kappa^2 \vs^2 + \lambda^2 (\vd^2+\vu^2) -2 \lambda \kappa \vu \vd -\lambda \frac{\vu \vd}{\vs} A_{\lambda} &=& 0 \quad .
\label{eq:tadpole-c}
\eea
\label{eq:tadpoles}
\end{subequations}
These form a set of coupled cubic equations and their solutions give the 
locations of the minima of the potential in the field space. Eliminating 
$m_{H_d}^2$, $m_{H_u}^2$ and $m_{S}^2$ in favour of three \vevs ~by using 
these tadpole equations and setting $\mueff = \lambda \vs$, one finds the
value of the neutral scalar potential at the DSB minimum to be
\bea
V^{^{\mathrm{neutral}}}_\mathrm{Higgs}|_{_\mathrm{tree}}  &=& -\kappa^2 \vs^4 -\frac{1}{3} \kappa A_{\kappa} \vs^3 
- \lambda^2 \vs^2 (\vd^2+\vu^2) -\lambda \vs \vd \vu (A_\lambda+2 \lambda \vs) \nonumber \\
& & -\frac{g_1^2+g_2^2}{8} (\vd^2 -\vu^2)^2 -\lambda^2 \vd^2 \vu^2 \quad .
\label{eq:higgs-pot-tree-sub}
\eea
At the lowest order, the singlet-extended Higgs sector of the $Z_3$-invariant 
NMSSM can be described by the following six independent input parameters: 
$\lambda$, $\kappa$, $A_\lambda$, $A_\kappa$, $\tan\beta \, (=\vu / \vd)$ and 
$\mueff$. In the present study we take all of them to be real. 
We also adopt the popular sign-convention \cite{Ellwanger:2009dp, Cerdeno:2004xw} with $\lambda$ and
$\tan\beta$ to be positive while the others can have both signs.

The present study involves the Higgs potential beyond the tree level and would make
use of leading 1-loop correction to the same. This correction is given by
(in the $\overline{DR}$-scheme) \cite{Quiros:1999jp, Martin:2014bca}):
\bea
\Delta V_\mathrm{rad.corr.}  
=  \sum_{i} \frac{n_i}{64 \pi^2} m_i^4 (\ln \frac{m_i^2}{Q^2}-\frac{3}{2}) \;\; ,
\label{eq:radcorr}
\eea
where the sum runs over all real scalars, Weyl fermions and vector degrees of 
freedom that are present in the scenario with
\beq
n_i  =  (-1)^{2s_i} (2s_i+1) Q_i C_i \nonumber
\eeq
and $Q_i=2 (1)$ for charged particles (neutral particles), $C_i$ is the 
color degrees of freedom, $s_i$ is the spin of the particle, $m_i$ is the mass 
of the same and $Q$ is the cut-off scale employed.

As indicated in the Introduction, thermal effects can be important in the study
of the stability of the DSB vacuum. The leading thermal effects to the Higgs
potential formally arise at the 1-loop level. Therefore, in a consistent treatment, 
these are to be added only to the 1-loop (radiatively) corrected Higgs potential obtained 
from equations \ref{eq:higgs-pot-tree-sub} and \ref{eq:radcorr}. The thermal 
correction is given by \cite{Brignole:1993wv}
\beq\label{eq:pot-thermal}
\Delta V_\mathrm{thermal} = \frac{1}{2\pi^2}\sum{ T^4 J_{\pm}\left(m^2/T^2\right)} ,
\eeq
where
\beq
J_{\pm}(r) = \pm \int_{0}^{\infty} dx ~x^2 \ln(1 \mp e^{-\sqrt{x^2+r}})
\label{eq:jpm}
\eeq
and the sum in equation \ref{eq:pot-thermal} runs over all particle degrees of 
freedom (that couple to the scalar fields including the scalar fields themselves),
viz., the bosons as a set of real scalars ($J_{+}$) and the fermions as 
a set of Weyl fermions ($J_{-}$), $m$ is the mass of the particle and `$T$' is the 
temperature. $J_{\pm}(\frac{m^2}{T^2})$ asymptotically approaches $-\infty$ 
(zero) as $\frac{m^2}{T^2}$ approaches zero ($~\infty$). These tell that thermal 
corrections would always lower the potential \cite{Camargo-Molina:2013qva}. 
Quantitatively, this lowering 
depends upon the value of $\frac{m^2}{T^2}$. Thus, thermal corrections play an 
essential role in the hunt for even deeper minima of the potential and the
thermally corrected Higgs potential at one loop is given by
\beq
V^{^{\mathrm{neutral}}}_\mathrm{Higgs} = V^{^{\mathrm{neutral}}}_\mathrm{Higgs}|_{_\mathrm{tree}} 
                 + \Delta V_\mathrm{rad.corr.} 
                 + \Delta V_\mathrm{thermal} \quad .
\label{eq:full-pot}
\eeq
Given a new physics scenario with multiple scalars (such as the NMSSM, which we 
study here), this potential is the single most important quantity in the study 
of the stability of the DSB vacuum and can be computed in a straight-forward 
(albeit tedious) way. The salient theoretical considerations and technical 
aspects involved in such an exercise are discussed in section \ref{subsec:veva} 
in reference to the publicly available package \veva ~\cite{Camargo-Molina:2013qva}.

As would soon become apparent, a detailed study of the stability of the NMSSM 
Higgs potential would invariably find the NMSSM Higgs and the neutralino sectors 
to be in a broader phenomenological reference. This is since some common NMSSM 
input parameters (like $\lambda$, $\kappa$) control these sectors. Hence we 
briefly outline them in the next subsection.
%
\subsection{The Higgs and the neutralino sectors}
\label{subsec:higgs-neutralino-sectors}
%
The Higgs mass matrices at the tree level are obtained by expanding the scalar
potential of equation \ref{eq:scalar-pot} about the real neutral \vevs ~as
indicated in equation \ref{eq:vev-expand} \cite{Ellwanger:2009dp}. After
eliminating $m_{H_d}^2$, $m_{H_u}^2$ and $m_{S}^2$ by using equation 
\ref{eq:tadpoles}, in the basis \{$H_{dR}$, $H_{uR}$, $S_R$\}, the elements of 
the $3 \times 3$ $CP$-even mass-squared matrix $\mathcal{M}_S^2$ are as follows:
\bea
\mathcal{M}_{S,11}^2 &=& g^2 \vd^2 + \mueff (\alambda + \kappa \vs) \tan\beta \nonumber \\
\mathcal{M}_{S,22}^2 &=& g^2 \vu^2 + \mueff (\alambda + \kappa \vs) / \tan\beta \nonumber \\
\mathcal{M}_{S,33}^2 &=& \lambda \alambda \, {{\vd \vu} \over {\vs}}  + \kappa \vs (\akappa + 4 \kappa \vs) \nonumber \\
\mathcal{M}_{S,12}^2 &=& (2 \lambda^2 - g^2) \vd \vu - \mueff (\alambda + \kappa \vs) \nonumber \\
\mathcal{M}_{S,13}^2 &=& \lambda \, [2 \mueff \, \vd - (\alambda + 2 \kappa \vs) \, \vu] \nonumber \\
\mathcal{M}_{S,23}^2 &=& \lambda \, [2 \mueff \, \vu - (\alambda + 2 \kappa \vs) \, \vd]  ,
\label{eq:msq-cp-even}
\eea
where $\vd$, $\vu$ and $\vs$ are the \vevs ~of $\higgsd$, $\higgsu$ and $S$ at
the DSB minimum of the potential.
One of the eigenvalues of the upper left $2 \times 2$ block matrix gives the 
upper bound on the lightest eigenvalue of $\mathcal{M}_S^2$ and would refer to
the SM-like (doublet) Higgs state. In general, the lightest eigenvalue of 
$\mathcal{M}_S^2$ has a smaller value compared to the one discussed above and 
is likely to refer to a singlet-like $CP$-even scalar. It is to be noted that the 
SM-like Higgs state mentioned above may characteristically become the lighter 
or the heavier of the two $CP$-even doublet Higgs eigenstates.
Phenomenological studies of such light Higgs boson(s) that could have escaped
detection at the past and recent experiments have already been taken up
\cite{Belanger:2012tt, Bhattacherjee:2013vga, Bhattacherjee:2015qct}. 

The observed mass of the SM-like Higgs boson ($\sim 125$ GeV) has 
turned out to be an extremely crucial input in the phenomenological studies, 
post Higgs-discovery. In the $Z_3$-symmetric NMSSM, this is given by 
\cite{Ellwanger:2011sk} 
\beq
\label{eq:sm-higgs-mass}
\mhsmsq = m_Z^2 \cos 2\beta + \lambda^2 v^2 \sin^2 2\beta + \Delta_{\rm mix}+ 
\Delta_{\rm rad. corr.} \, ,
\eeq  
where $v = \sqrt{\vd^2+\vu^2} \simeq 174 \, {\rm GeV}$ and 
$\beta=\tan^{-1} {\vu \over \vd}$.
The first term gives the squared mass of the lightest $CP$-even Higgs boson in the 
MSSM at the tree level. Tree-level NMSSM contribution to the same is given by the
second term. The third term arises from the so-called singlet-doublet mixing. For
a weak mixing this is approximated as \cite{Ellwanger:2011sk}
\beq
\Delta_{\rm mix} \approx \frac{4 \lambda^2 v_s^2 v^2 
(\lambda -\kappa \sin 2\beta)^2} {\tilde{m}_h^2-m_{ss}^2} ,
\label{eq:sdmix}
\eeq
where $\tilde{m}_h^2 = \mhsmsq - \Delta_{\rm mix}$ and 
$m_{ss}^2 = \kappa v_s(A_{\kappa}+4 \kappa v_s)$.  
It is observed that if the SM-like Higgs boson is the lighter (heavier) $CP$-even state
this results in a reduction of (increase in) mass \cite{Huang:2014ifa}.
On the other hand, $\Delta_{\rm rad. corr.}$ stands for the radiative contribution
to the mass of the SM-like Higgs boson at the 1-loop level and is given by 
\cite{Carena:1995bx, Carena:1995wu, Haber:1996fp, Djouadi:2005gj}
\beq
\Delta_{\rm rad. corr.} = \frac{3 m_t^4}{2 \pi^2 v^2 \sin^2 \beta} 
\left[2 \log \frac{M_{\mathrm{SUSY}}}{m_t} 
+ \frac{X_t^2}{2 M_{\mathrm{SUSY}}^2} \left(1- \frac{X_t^2}{6 M_{\mathrm{SUSY}}^2} \right) \right].
\label{eq:radcorr-mhiggs}
\eeq
In the above expression, $m_t$ denotes the mass of the top quark,
$M_{\mathrm{SUSY}}=\sqrt{\mstone \msttwo}$, $X_t=A_t-\mu\cot\beta$,
$m_{\tilde{t}_1,\tilde{t}_2}$ are the mass-eigenvalues of the two physical top 
squark states and $A_t$ is the soft trilinear term for the top quark sector appearing 
in the scalar potential.

The elements of the $2 \times 2$ $CP$-odd mass-squared matrix, $\mathcal{M}_P^2$, 
in the basis \{$A, S_I$\}, are given by \cite{Ellwanger:2009dp}
\bea
\mathcal{M}_{P,11}^2 &=& {{2 \mueff \, (\alambda + \kappa \vs)} \over {\sin 2\beta}} \nonumber \\
\mathcal{M}_{P,22}^2 &=&  \lambda (\alambda + 4 \kappa \vs) {{\vd \vu} \over \vs} - 3 \kappa \akappa \vs \nonumber \\
\mathcal{M}_{P,12}^2 &=& \lambda (\alambda - 2 \kappa \vs) \, v ,
\label{eq:msq-cp-odd}
\eea
where $v=\sqrt{\vd^2+\vu^2}$. Basis vector `$A$' is obtained from
the imaginary pair $\{H_{dI}, H_{uI}\}$ after dropping the massless neutral
Goldstone mode. Similarly, the charged Higgs sector also contains a massless
Goldstone mode and the mass of its two physical states is given by \cite{Ellwanger:2009dp}
\bea
\mathcal{M}_{\pm}^2 &=& {{2 \mueff \, (\alambda + \kappa \vs)} \over {\sin 2\beta}} 
                      +   v^2 \left( {g^2 \over 2} -\lambda^2\right).
\label{eq:msq-charged}
\eea
%
%

The $5 \times 5$ symmetric neutralino mass matrix, in the basis 
\{$\bino, \wino, \higgsinod, \higgsinou, \singlino$ \}, is given by
\cite{Ellwanger:2009dp}
\beq\label{eq:mneut}
{\cal M}_0 =
\left( \begin{array}{ccccc}
M_1 & 0 & -\frac{g_1 \vd}{\sqrt{2}} & \frac{g_1 \vu}{\sqrt{2}} & 0 \\
& M_2 & \frac{g_2 \vd}{\sqrt{2}} & -\frac{g_2 \vu}{\sqrt{2}} & 0 \\
& & 0 & -\mueff & -\l \vu \\
& & & 0 & -\l \vd \\
& & & & 2 \kappa \vs
\end{array} \right) \;,
\eeq
where $M_1$ and $M_2$ stand for the soft SUSY-breaking masses of the $U(1)$ 
($\bino$) and the $SU(2)$ ($\wino$) gauginos, respectively and $g_1$ and $g_2$ 
are the corresponding gauge couplings. Note that there is no direct mixing among 
the gauginos ($\bino$ and $\wino$) and the singlino ($\singlino$). However, a
small such mixing is introduced indirectly via the neutral higgsino ($\higgsinod, 
\, \higgsinou$) sector. On the other hand, the higgsinos and the singlino could 
mix directly via the off-diagonal terms of ${\cal M}_0$ that are proportional to
$\lambda$. Hence scenarios with relatively small $\mueff$ ($\lesssim 500$ GeV), 
which have recently attracted much attention 
\cite{Baer:2012up, Baer:2013ava, Mustafayev:2014lqa, Baer:2015rja}
as the agent that could efficiently 
render the same `natural', would result in 
lighter neutralinos with significant admixture of the singlino and 
the higgsinos over interesting regions of the NMSSM parameter space.
We adhere to such a scenario throughout this work.

It would be helpful to note that the broad requirement for the LSP to be 
singlino-dominated is 
$\msinglino (=2\kappa \vs = 2\kappa {\mueff \over \lambda}) < \mueff$, i.e., 
$\kappa < {\lambda \over 2}$. Furthermore, a singlino-like LSP is keenly related
in mass to those of the singlet scalars. This can be gleaned from the expressions
for $\mathcal{M}_{S,33}^2$ ($=\mssq$, in equation \ref{eq:msq-cp-even}) and that 
for $\mathcal{M}_{P,22}^2$ ($=\mpsq$, in equation \ref{eq:msq-cp-odd}). At the 
DSB minimum for which $\vd, \vu << \vs$, these get simply related by
\beq
\msinglino^2 \approx \mssq + \frac{1}{3} \mpsq \quad.
\label{eq:singlet-singlino-mass}
\eeq
%
\subsection{Stability of the DSB vacuum: the \veva ~approach}
\label{subsec:veva}
%
In this section we briefly outline the \veva ~approach elaborated in
references \cite{Camargo-Molina:2013qva, Camargo-Molina:2014pwa, Camargo-Molina:2015qpa}.

Estimating the stability of the DSB vacuum requires an exhaustive 
knowledge of the (deeper) minima of the potential to which it can tunnel.
Finding them all for a potential involving `$N$' scalar fields entails solving 
`$N$' coupled polynomial equations of up to degree three. \veva ~employs a 
dedicated package {\tt HOM4PS2} \cite{hom4ps2} which is based on the so-called 
principle of polyhedral homotopy continuation. This exercise is carried out with 
tree-level scalar potential.  Hence the solutions are the tree-level extrema. 
These extrema serve as the starting points for the so-called gradient-based 
minimization of the 1-loop effective potential using {\tt PyMINUIT}
\cite{pyminuit}, a {\tt Python} wrapper for the minimization routine 
{\tt MINUIT} \cite{minuit}.

Quantitatively, it is sufficient to check the tunneling probability of the false
(DSB) vacuum to the deeper (true) minimum nearest to it in the field space 
\cite{vevacious-code}. This tunneling (decay) finds a thermodynamic analogue in 
nucleation of a bubble of the true vacuum which expands and eventually engulfs 
the DSB vacuum \cite{Coleman:1977py, Callan:1977pt}.
Stability of the DSB vacuum is then given by its decay width per unit volume 
($\frac{\Gamma}{\mathcal{V}}$). At zero temperature, for which the expanding 
bubble exhibits an $O(4)$ spherical symmetry in the Euclidean space, this is given by 
\cite{Coleman:1977py, Callan:1977pt}
\beq
\frac{\Gamma}{\mathcal{V}}= A \exp(-S_4) \, ,
\label{eq:decay-zero}
\eeq
while for finite temperatures the symmetry breaks down to an $O(3)$ cylindrical
one and the expression in equation \ref{eq:decay-zero} takes the form 
\cite{Linde:1980tt, Linde:1981zj}:
\beq
\frac{\Gamma(T)}{\mathcal{V}(T)}= A(T) \exp[-S_3(T)/T] \, ,
\label{eq:decay-thermal}
\eeq
where $S_4$ and $\sthreet$ are the so-called Euclidean ``bounce'' actions. These 
quantify to what extent the false vacuum could be reflected back by the potential 
barrier thus failing to tunnel. Clearly, the higher the ``bounce'', the lower is 
the tunneling probability. $\sthreet$ contains the thermally corrected potential 
of equation \ref{eq:full-pot}. Prefactors $A$ and $A(T)$ are rather difficult to 
compute. Fortunately, since the dominant effect comes from the exponentials, 
assuming $A (A(T)) \approx Q^4 (T^4)$ suffices on dimensional grounds, $Q(T)$ 
being the renormalization scale. 

\veva ~uses {\tt CosmoTransitions} ~\cite{Wainwright:2011kj} to estimate the 
survival probability of the false (DSB) vacuum, i.e., its probability of not 
tunneling to the true vacuum, within the interval when the Universe is at 
temperatures $T_i$ and at $T_f$, with $T_f < T_i$. This is given by
\bea
P(T_i,T_f) 
&=& \exp\left(-\int_{T_i}^{T_f} \frac{dt}{dT} \, \Gamma(T) \, dT\right) \nonumber \\
&=& \exp\left(-\int_{T_i}^{T_f} \frac{dt}{dT} \, \mathcal{V}(T) A(T) \exp[-S_3(T)/T] 
\, dT\right) \, .
\label{eq:survival}
\eea
Assuming (i) that the Universe is radiation-dominated during its evolution from
$T_i$ to $T_f$, (ii) that entropy conservation holds during this period and 
(iii) that tunneling occurs at $T < M_{\mathrm{New \, Physics}}$ such that only the SM 
degrees of freedom are relativistic, the integral under the exponent in equation 
\ref{eq:survival} 
can be reduced to \cite{Camargo-Molina:2013qva} 
\beq
\int_{T_i}^{T_f} {dt \over dT} \, \mathcal{V}(T) A(T) \exp[-S_3(T)/T] \, dT 
\simeq 1.581 \times 10^{106} \, \mathrm{GeV} \int_{T_f}^{T_i} T^{-2} \exp[-S_3(T)/T] \, dT
\quad .
\label{eq:survival-reduced}
\eeq
As $\sthreet$ increases with temperature it is assumed that it does so monotonically. 
It can then be shown that equation \ref{eq:survival-reduced} leads to an upper bound 
of the survival probability given by \cite{Camargo-Molina:2013qva}
\beq
P(T_i=T, \, T_f=0) < \exp\left(-\exp[244.53-S_3(T)/T-\ln(S_3(T)/\mathrm{GeV})]
\right) \quad .
\label{eq:upper-bound-survival}
\eeq
An exclusion strategy based on such an upper bound would right away discard a 
point in the parameter space if, for it, the bound falls below an user-supplied 
threshold (in reference to the age of the Universe). Note that a continuous evaluation 
of $\sthreet$ (as `$T$' varies) would make things prohibitively slow.
If a single evaluation of $\sthreet$ has to suffice, it is
to be done at an optimal temperature ($T_\mathrm{opt}$) which maximizes the 
right-hand side of equation \ref{eq:upper-bound-survival}. 

However, at finite temperatures the optimal path for
tunneling is generally no more a straight line in the field space connecting the 
involved vacua. Finding the optimal path is a computationally intensive job and 
even a single evaluation of $\sthreet$ along the same may be rather slow. 
To know $\topt$ beforehand, $\sthreet$ is to 
be estimated for a few temperatures within a relevant range and the quicker 
option would be to compute these along straight paths for a given set of input 
parameters. Minimization of the fitted $\sthreet$ would solve for $T=\topt$.
This also implies minimization of $P(T_i=T, \, T_f=0)$ appearing on the 
left hand side of equation \ref{eq:upper-bound-survival}. The 
right hand side of equation \ref{eq:upper-bound-survival} evaluated at $T=\topt$ 
would then give the upper bound of the survival property of the DSB vacuum.
Thereon, one can look for the optimal path and resort to the same, if warranted. 

In \veva ~a false (metastable) DSB vacuum is flagged short-lived (hence not 
viable) if it could tunnel to the panic vacuum in three giga-years (the default
setting). This 
corresponds to a probability of 1\% or less for the DSB vacuum to survive 
through the age of the Universe ($\approx 13.8$ giga-years). The threshold 
decay-time (amounting to above figures) is supplied to \veva ~as an (input) 
fraction of the age of the Universe. 
Operationally, estimation of the upper bound on the survival probability is 
taken up in the following stages with increasing degree of complication. 
These result in a gradual lowering of this upper bound. Each subsequent stage is 
invoked only if, for the current one, the DSB vacuum 
still has the upper bound of the survival probability 
greater than the threshold set, i.e., can still be stable:
  {\bf (i)} zero temperature potential with tunneling in the straight path (uses $S_4$),
 {\bf (ii)} temperature-corrected potential (uses $\sthreet$) but still adhering to the 
      straight path,
 {\bf (iii)} reverting to zero temperature configuration (uses $S_4$) but now adopting 
      the optimal tunneling path and finally
 {\bf (iv)} switching on again the finite temperature effect (uses $\sthreet$) and 
      sticking to the optimal path.
This approach ensures a robust rejection of a parameter point in the quickest possible way. 
%
\section{Vacuum structure of the $Z_3$-symmetric NMSSM}
\label{sec:vacuum-structure}
%
The vacuum structure of 
the $Z_3$-symmetric\footnote{It is well known that once a discrete symmetry like the $Z_3$ breaks spontaneously in the early Universe, dangerous cosmological domain walls \cite{Vilenkin:1984ib} would be generated. However, to retain the essential features of $Z_3$-symmetric NMSSM intact, these terms should be small in magnitude. Even then, these can give rise to quadratically divergent tadpole terms in the singlet superfield \cite{Ellwanger:1983mg, Bagger:1993ji, Jain:1994tk, Bagger:1995ay, Abel:1995wk}. It has been shown in references \cite{Abel:1996cr,Kolda:1998rm,Panagiotakopoulos:1998yw} that these can be avoided by imposing some discrete $R$-symmetry on the non-renormalisable terms that break the $Z_3$-symmetry. Thus, a small $Z_3$-symmetry breaking term that is necessary to avoid the domain wall problem would not have any impact on the phenomenology including that involves vacuum structure.} 
NMSSM have been studied in the past in the constrained
NMSSM (CNMSSM) \cite{Ellwanger:1996gw, Ellwanger:1999bv}. In recent times 
similar studies were taken up in the phenomenological (weak-scale) NMSSM (pNMSSM) 
where no explicit  assumption on physics at a high scale is made 
\cite{Kanehata:2011ei, Kobayashi:2012xv}, albeit within a limited scope. 
It may be reiterated that these limitations arise from the presence of
increased (three) number of Higgs fields (when compared to two, 
in the case of the MSSM) which quickly makes the task of finding the exhaustive 
set of vacua not only daunting but also virtually impossible via purely 
analytic means. 

Situations like this were already encountered in the studies of CCB vacua in 
various SUSY scenarios including the MSSM and the CMSSM
\cite{AlvarezGaume:1983gj,Gunion:1987qv,Komatsu:1988mt,Casas:1995pd,Bordner:1995fh,
Strumia:1996pr,Baer:1996jn,Abel:1998cc,Abel:1998wr,Ferreira:2000hg,LeMouel:2001ym,
LeMouel:2001sf,Brhlik:2001ni,Cerdeno:2003yt,Ferreira:2004yg,Brandenberger:1984cz,
Kusenko:1995jv,Riotto:1995am,Falk:1996zt,Kusenko:1996jn, Kusenko:1996xt,
Kusenko:1996vp,Cohen:2013kna,Datta:2000xy,Datta:2001dc,Datta:2001qa,
Gabrielli:2001py,Datta:2004hr,Kobayashi:2010zx,Hisano:2010re}
where additional scalars (squarks and sleptons) carrying charge and color enter
analyses. As indicated in the Introduction, in the absence of a suitable 
numerical approach, these studies were restricted to a few specific (flat) 
directions in the field 
space. Packages like {\tt SuSpect} 
\cite{Djouadi:2002ze} for the MSSM and {\tt NMSSMTools} 
\cite{Ellwanger:2004xm, Ellwanger:2005dv, Ellwanger:2006rn} for the NMSSM scenarios
followed suit\footnote{Only in the recent past \cite{Camargo-Molina:2014pwa} an 
exhaustive study of such CCB vacua had been undertaken for the MSSM using the 
package \veva ~\cite{Camargo-Molina:2013qva}.}. 

Incidentally, however, these cannot be guaranteed to be the only (dangerous) 
directions in the field space along which false vacua might appear. In the next 
subsection we briefly review the outcomes in some such directions for the NMSSM 
which were studied earlier
\cite{Ellwanger:1996gw, Ellwanger:1999bv, Kanehata:2011ei, Kobayashi:2012xv}.
Understanding them would be useful for our present study. One particular direction 
which may closely correspond to the actual situation (under certain circumstances) 
is picked up for an in-depth analytical study. We then carry out a semi-analytical 
({\tt Mathematica}-based \cite{mathematica}) study of situations by opening up to arbitrary directions 
in the field space. This will be followed by a general (numerical) study of 
the viable vacuum configurations of the $Z_3$-symmetric NMSSM using \veva. There, 
we first consider those field directions along which only the neutral scalars are 
non-vanishing. The inclusion of the charged and the colored scalars (sleptons and 
squarks) and the study of the CCB vacua thereof are deferred to section 
\ref{sec:ccb-vacua}. 
%
\subsection{False vacua appearing in specific field directions}
\label{subsec:specific-directions}
%
Choosing one or the other direction(s) in the field space in search for false 
vacua is primarily guided by the form of the Higgs potential of equation 
\ref{eq:higgs-pot-tree-org}. As can be gleaned from these equations, it is not 
difficult to find some specific directions where minima deeper than the 
DSB vacuum are likely to appear. A general observation \cite{Kobayashi:2012xv}
is that such minima tend to appear as the positive quartic terms in the Higgs 
potential balance against the negative quadratic and trilinear terms. Deeper
minima show up when the former lose out to the latter. 

Thus, the simplest possible approach might be to ensure that the direction 
$|\higgsd| = |\higgsu| = |S|$ does not yield minima deeper than the DSB minimum for which 
$\higgsd$ and $\higgsu$ must be non-vanishing (and hence also the same for 
$|S|$ since, as discussed earlier, either one or all three of $\higgsd$, 
$\higgsu$ and $S$ only could be non-vanishing at the minima). In fact, 
{\tt NMSSMTools} explores this direction along with three other possible ones, 
viz., $\higgsd=S=0$, $\higgsu=S=0$ and $\higgsd=\higgsu=0$. 
In reference \cite{Kanehata:2011ei} the authors, in addition, analyzes the 
direction $|\higgsd|=|\higgsu| \neq 0$ along with $S \neq 0$. Note that with 
$|\higgsd|=|\higgsu|$, the $D$-term $V_D^H$ in equation \ref{eq:higgs-pot-tree-org} 
or \ref{eq:higgs-pot-tree-sub} vanishes (the so-called $D$-flat direction).
It was further observed that the deepest minimum might arise in the direction in which 
trilinear couplings are negative and the $F_S$-term vanishes (the so-called 
$F_S$-flat direction). Additionally, reference \cite{Kanehata:2011ei} also 
discusses situations with only one of $\higgsu$, $\higgsd$ and `$S$' acquiring 
\vev. On the other hand, reference \cite{Kobayashi:2012xv} picks up an altogether 
new set of directions based on combinations of (vanishing or not) $V_D$ and/or 
$F_S$. 

To the best of our knowledge, a thorough study of the appearance of 
panic vacua in arbitrary field directions and hence the fate of the the DSB 
vacuum in a scenario like the pNMSSM (by taking into account the radiative and 
thermal effects to the scalar potential and estimating the tunneling time) is 
still lacking. The present work addresses these issues. However, before 
moving on to such a study,
it would be instructive to analyze in some detail a particular situations with
$\higgsd=\higgsu=0$ and $S \neq 0$ that has been briefly discussed in the 
aforementioned literature. It would be also instructive to study in some detail
how the problem gets complicated in the presence of non-vanishing $\higgsd$ and
$\higgsu$ (even if we stick to some fixed directions in the field space) and then 
with the inclusion of radiative correction to the potential.
%
\subsubsection{$S \neq 0$, vanishing $\higgsd$ and $\higgsu$}
\label{subsubsec:single-field-potential}
%
In the limit $S \neq 0$ while $\higgsd$ and $\higgsu$ are both vanishing, 
the neutral scalar (Higgs) potential, from equation 
\ref{eq:higgs-pot-tree-org} (leaving out the radiative corrections), can be 
written as
\beq
V^{^{\mathrm{neutral}}}_\mathrm{Higgs}|_{_{\scriptsize\begin{aligned} \higgsd &= \higgsu=0 \\ 
                   S &\neq 0 \end{aligned}}} = m_S^2 S^2 + \kappa^2 S^4 
+ {2 \over 3} \kappa \akappa S^3 \, .
\label{eq:higgs-pot-large-s}
\eeq
This potential is quartic in `$S$'. It thus possesses three extrema. The form of 
the potential guarantees a minimum at $S=0$ for $m_S^2 >0$. However, this 
could not be the DSB 
minimum since it would not lead to an acceptable value for $\mueff$. Hence there 
must be a deeper (global) minimum which is the DSB vacuum. This is represented 
by one of the other two extrema solutions given by
\beq
s_\pm =  \frac{-\akappa}{4\kappa} \left( 1 \pm \sqrt{1 - \frac{8 m_S^2}{\akappa^2}} \right) ,
\label{eq:sol-for-global-mimimum}
\eeq
provided $\akappa^2 \geq 9 m_S^2$ \cite{Ellwanger:1996gw, Ellwanger:1999bv, 
Kanehata:2011ei, Kobayashi:2012xv}\footnote{Note that only one or the other
solution, and not both, has been indicated to represent the global minimum in 
these works.}. The other one of these two, then, represents the only maximum of 
the potential. Further, given the form of equation \ref{eq:sol-for-global-mimimum}, 
a metastable minimum can appear for $8 m_S^2 < \akappa^2 \leq 9 m_S^2$. 
Using the tadpole equation for the above potential, $m_S^2$ can be fixed in 
terms of $\vs$ ~(which is going to be the desired \vev ~of `$S$') as
\beq
m_S^2 = -\kappa \akappa \vs - 2 \kappa^2 \vs^2
\label{eq:ms-square}
\eeq
and thus can be eliminated from equation \ref{eq:higgs-pot-large-s}. 
The value of the potential at the desired vacuum is given by
\beq
V^{^{\mathrm{neutral}}}_\mathrm{Higgs}|_{_{\scriptsize\begin{aligned} \higgsd &= \higgsu=0 \\
                   S &\neq 0 \end{aligned}}} = -\kappa^2 \vs^4
   - \frac{1}{3} \kappa \akappa \vs^3 \, ,
\label{eq:higgs-pot-global-minimum} 
\eeq
where $\vs$ is the value of the field `$S$' at that point.
Note, however, the subtle issue that the potential profile as a
function of field `$S$' is still determined by equation \ref{eq:higgs-pot-large-s}
with 
$m_S^2$ computed from equation \ref{eq:ms-square} and with
$\vs=\mueff/\lambda$. Furthermore, substituting the expression of 
$m_S^2$ from equation \ref{eq:ms-square} into the condition for 
appearance of a global minimum (mentioned earlier) reduces the latter to
\beq
(\akappa + 3 \kappa \vs) (\akappa + 6 \kappa \vs) \geq 0 \, .
\label{eq:condition-global-minimum}
\eeq
Note that a maximum might appear as well at $S=0$ but for $\mssq <0$. Under such 
a circumstance, the inequalities discussed above would not hold. Thus, an 
exhaustive analytical understanding of such a setup would be rather revealing.
Fortunately, this is relatively simple for the case in hand. We refer to the 
$\kappa \vs - \akappa$ plane; the choice being obvious from the form of equation 
\ref{eq:condition-global-minimum}. Its actual relevance (along with the 
ranges studied), however, would be apparent in the subsequent sections.
%
\begin{table}[hbt]
\begin{center}
{\footnotesize
  \begin{tabular}{cc|c|c|}
                  \cline{3-4}
                    & & & \\
                    &  & {\large Solution $s_+$ at the DSB} & {\large Solution $s_-$ at the DSB} \\
                    & & & \\
\hline
\hline
\multicolumn{1}{|c||}{} & \multicolumn{1}{c|}{}  
                        & $\akappa \geq 0$ \& $\kappa \vs >0$
                        & $\akappa \leq 0$  \& $\kappa \vs <0$ \\
\multicolumn{1}{|c||}{} & \multicolumn{1}{c|}{Stability}                       
                        & $||$ 
                        & $||$ \\
\multicolumn{1}{|c||}{} & \multicolumn{1}{c|}{only} 
                        & ($0 \leq \akappa \leq -3 \kappa \vs \; || \; \akappa \geq -6 \kappa \vs$) 
                        & ($-3 \kappa \vs \leq  \akappa \leq 0 \; || \; \akappa \leq -6 \kappa \vs$) \\
\multicolumn{1}{|c||}{} & \multicolumn{1}{c|}{} 
                        & \& $\kappa \vs <0$
                        & \& $\kappa \vs >0$ \\
                       \cline{2-4}
\multicolumn{1}{|c||}{\large Global}  &  \multicolumn{1}{c|}{+ Non-tachyonic}
                        & $\kappa \vs < 0$ 
                        & $\kappa \vs > 0$ \\
\multicolumn{1}{|c||}{\large (stable)}         &  \multicolumn{1}{c|}{$CP$-odd}
                        & \& 
                        & \& \\
\multicolumn{1}{|c||}{\large minimum} & \multicolumn{1}{c|}{scalar singlet}
                        & $(0 < \akappa \leq -3 \kappa \vs \; || \; \akappa \geq -6 \kappa \vs)$
                        & $(\akappa \leq -6 \kappa \vs \; || \; -3 \kappa \vs \leq  \akappa < 0)$ \\
                          \cline{2-4}
\multicolumn{1}{|c||}{} & \multicolumn{1}{c|}{+ Non-tachyonic}
                        & $\kappa \vs < 0$ 
                        & $\kappa \vs > 0$ \\
\multicolumn{1}{|c||}{} & \multicolumn{1}{c|}{$CP$-even}
                        & \&
                        & \& \\
\multicolumn{1}{|c||}{} & \multicolumn{1}{c|}{scalar singlet}
                        & $0 < \akappa \leq -3 \kappa \vs$
                        & $-3 \kappa \vs \leq \akappa  < 0$ \\
\hline
\hline
\multicolumn{1}{|c||}{} &  \multicolumn{1}{|c|}{}  
                        & $\akappa < 0$ \& $\kappa \vs >0$  
                        & $\akappa > 0$ \& $\kappa \vs <0$ \\
\multicolumn{1}{|c||}{} & \multicolumn{1}{|c|}{Metastability}                       
                        & $||$ 
                        & $||$ \\
\multicolumn{1}{|c||}{} & \multicolumn{1}{|c|}{only} 
                        & ($-3 \kappa \vs < \akappa < -6 \kappa \vs \;\, \& \;\, \kappa \vs \leq 0$) 
                        & ($-6 \kappa \vs < \akappa < -3 \kappa \vs \;\, \& \;\, \kappa \vs >0$)\\ 
                       \cline{2-4}
\multicolumn{1}{|c||}{\large Local}  &  \multicolumn{1}{c|}{+ Non-tachyonic}
                        & ($\kappa \vs > 0 \;\, \& \;\,  \akappa < 0$) 
                        & ($\kappa \vs < 0 \;\, \& \;\, \akappa > 0$) \\
\multicolumn{1}{|c||}{\large (metastable)}          &  \multicolumn{1}{c|}{$CP$-odd}
                        & $||$ 
                        & $||$ \\
\multicolumn{1}{|c||}{\large minimum} & \multicolumn{1}{c|}{scalar singlet}
                        & ($\kappa \vs < 0 \;\, \& \;\, -3 \kappa \vs < \akappa < -6 \kappa \vs$)
                        & ($\kappa \vs > 0 \;\, \& \;\, -6 \kappa \vs < \akappa < -3 \kappa \vs$) \\
                          \cline{2-4}
\multicolumn{1}{|c||}{} & \multicolumn{1}{c|}{+ Non-tachyonic}
                        & ($\kappa \vs < 0 \;\, \& \;\, -3 \kappa \vs < \akappa < -4 \kappa \vs$) 
                        & ($\kappa \vs > 0 \;\, \& \;\,  -4 \kappa \vs < \akappa < -3 \kappa \vs$) \\
\multicolumn{1}{|c||}{} & \multicolumn{1}{c|}{$CP$-even}
                        & $||$
                        & $||$ \\
\multicolumn{1}{|c||}{} & \multicolumn{1}{c|}{scalar singlet}
                        & ($\kappa \vs > 0 \;\, \& \;\,  -4 \kappa \vs < \akappa < 0$)
                        & ($\kappa \vs < 0 \;\, \& \;\, 0 < \akappa < -4 \kappa \vs$) \\
\hline
  \end{tabular}
}
\end{center}
\caption{Conditions that are to be satisfied for a global (stable) or a local 
(metastable) minimum to appear with the two solutions indicated in equation 
\ref{eq:sol-for-global-mimimum}.}
\label{tab:analytical-ranges}
\end{table}
%
%

\begin{figure}[t]
\centering
\includegraphics[height=0.2\textheight, width=0.32\columnwidth , clip]{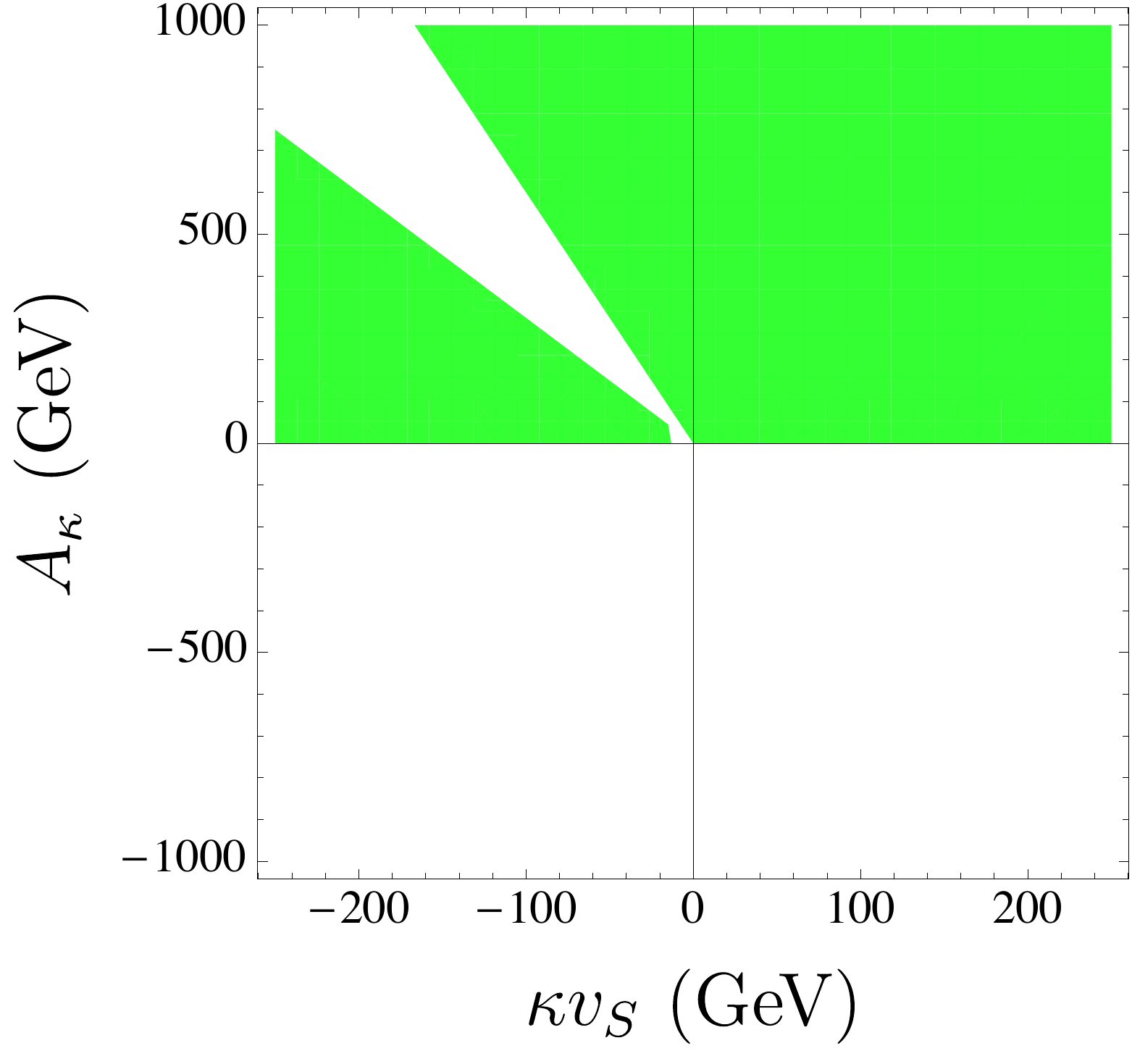}
\includegraphics[height=0.2\textheight, width=0.32\columnwidth , clip]{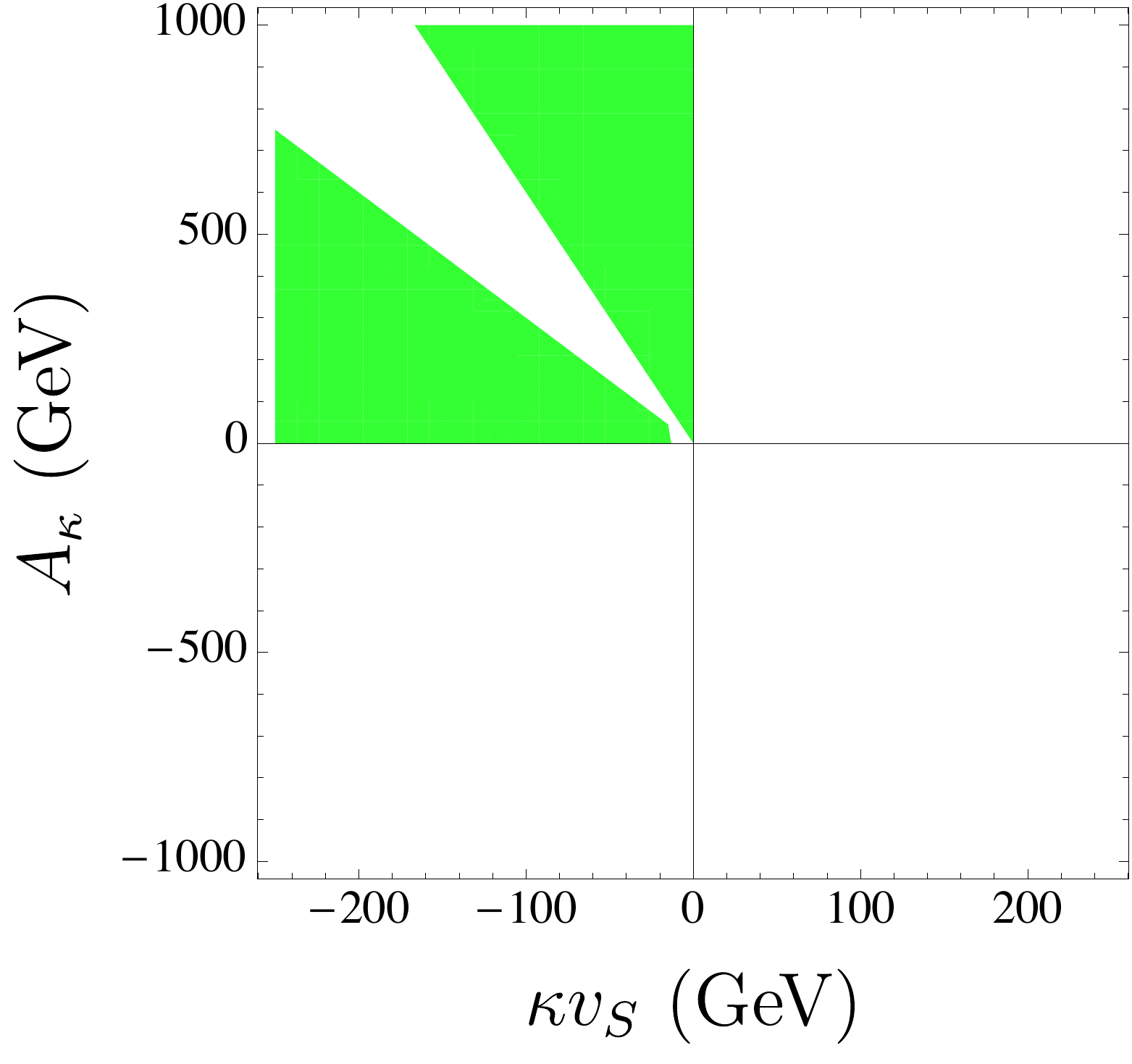}
\includegraphics[height=0.2\textheight, width=0.32\columnwidth , clip]{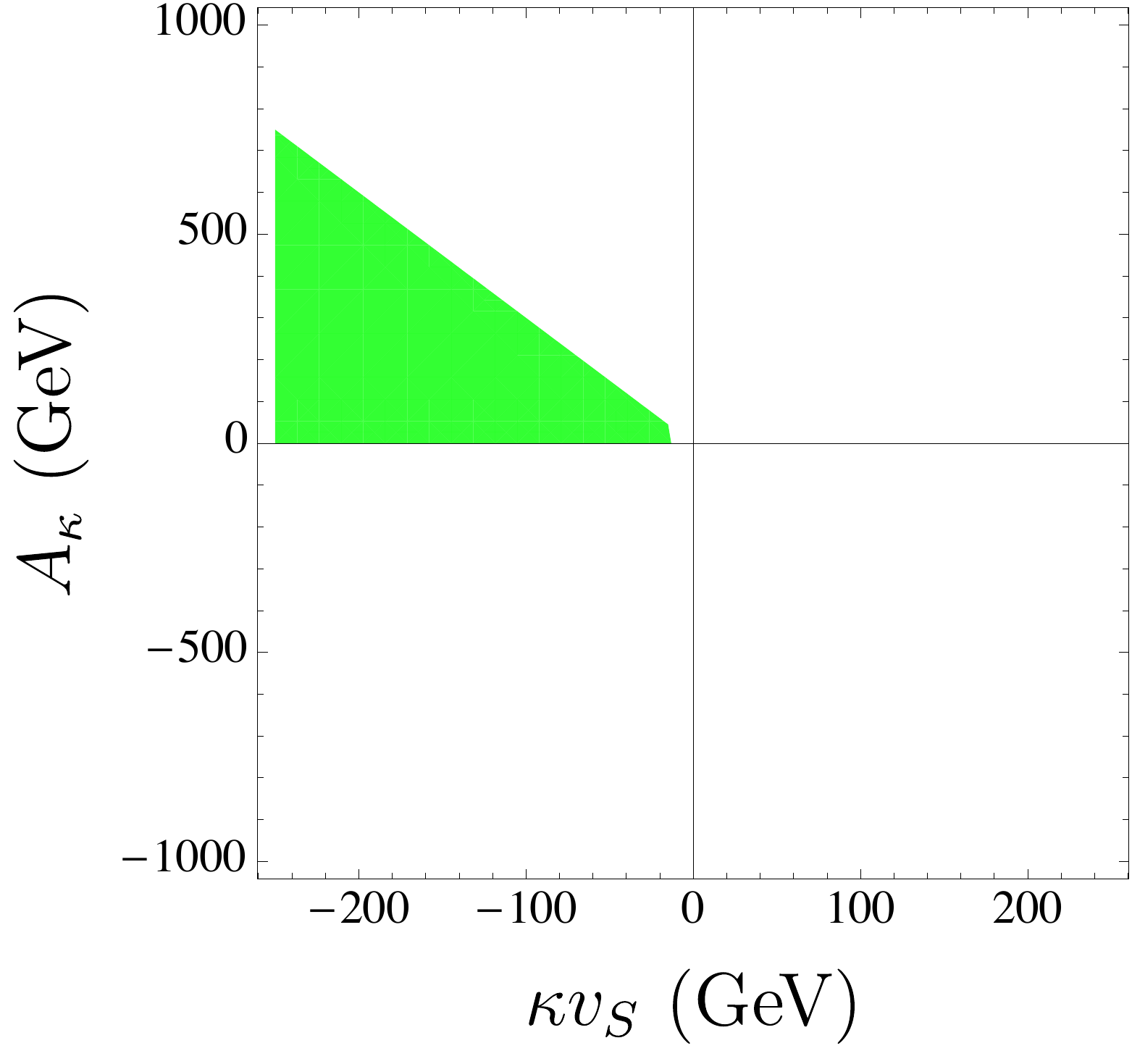}
\vskip 0.1in
\includegraphics[height=0.2\textheight, width=0.32\columnwidth , clip]{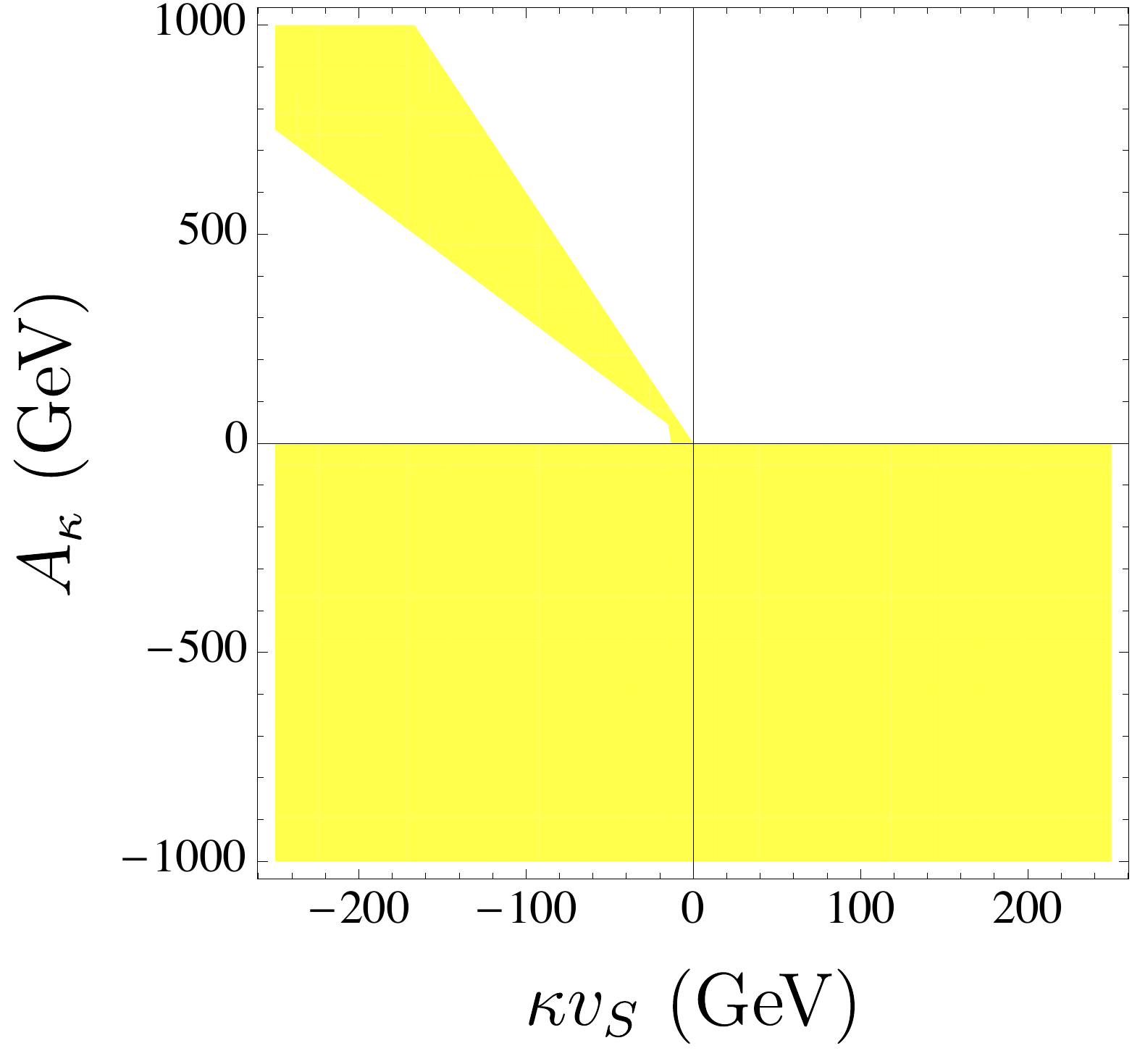}
\includegraphics[height=0.2\textheight, width=0.32\columnwidth , clip]{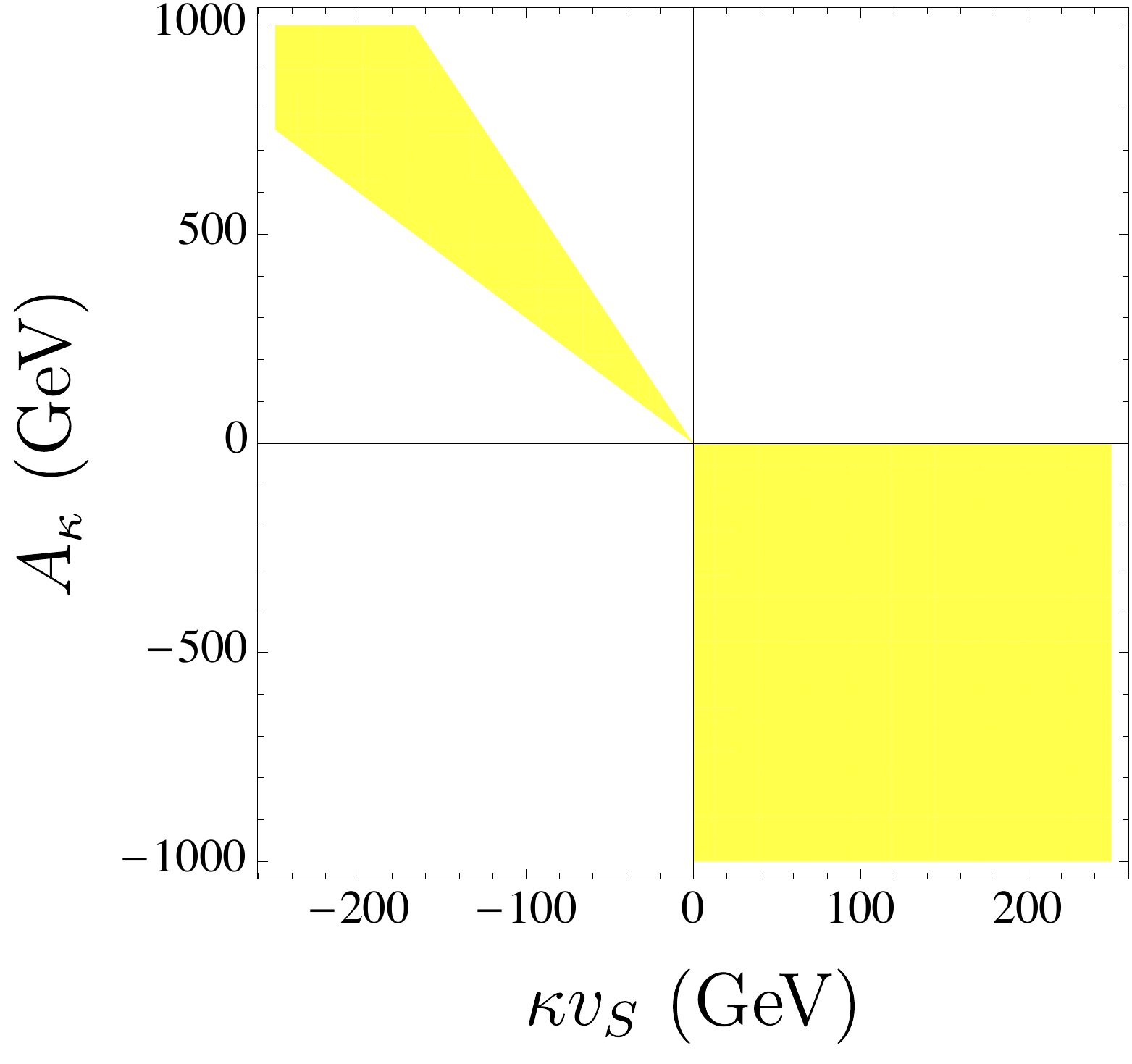}
\includegraphics[height=0.2\textheight, width=0.32\columnwidth , clip]{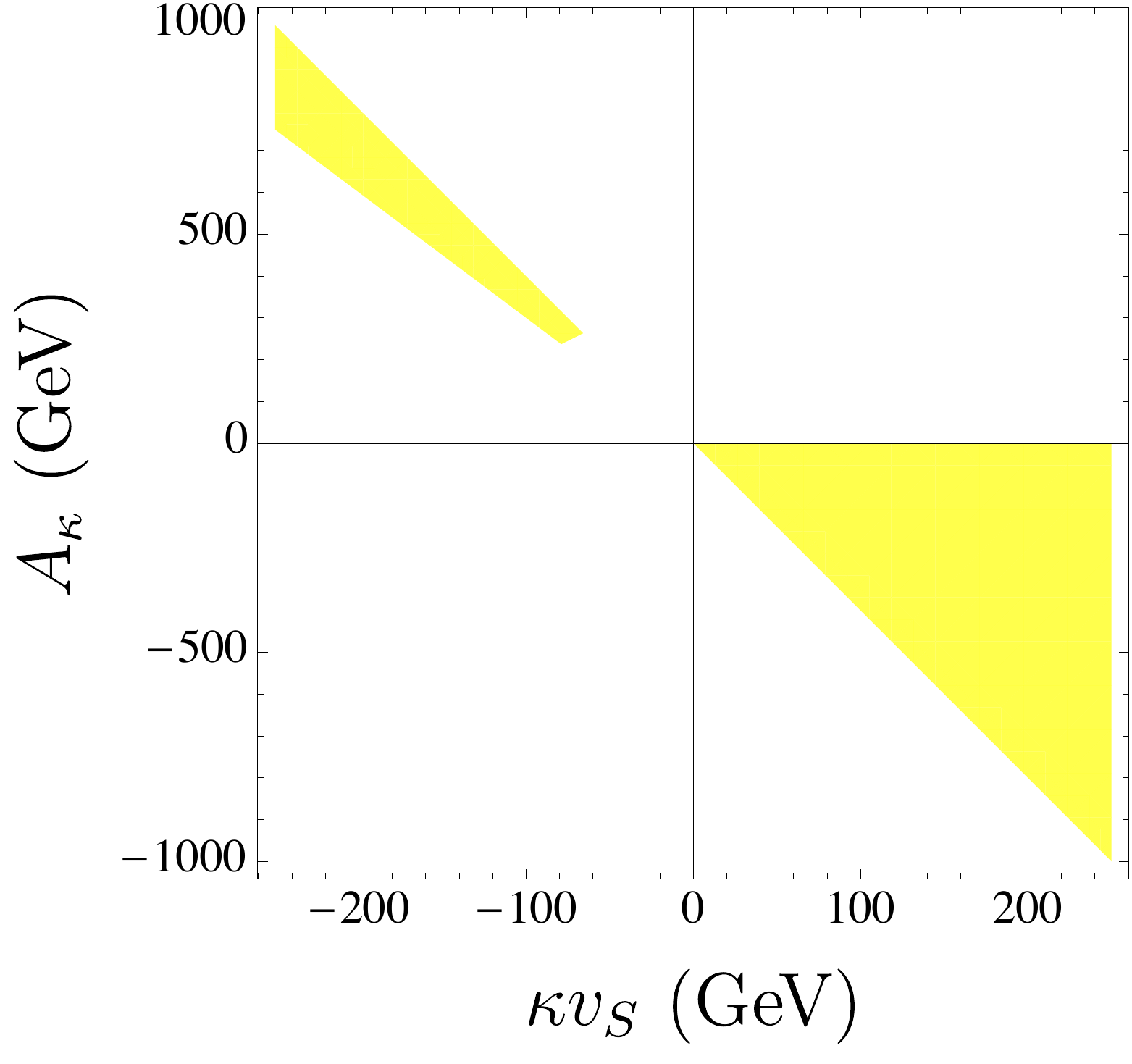}

\vskip 15pt
\caption{Allowed regions in the $\kappa \vs-\akappa$ plane (left column) with 
the solution $s_+$ compatible with stable (top panel, in green) and metastable 
(bottom panel, in yellow) DSB minimum. Plots in the second and the third columns 
demonstrate the effects of requiring successively non-tachyonic singlet scalars
(see text for details).}
\label{fig:allowed-ranges-solution-1}
\end{figure}

In table \ref{tab:analytical-ranges} we present the sets of conditions that are 
to be satisfied for a global (stable) or a local (metastable) minimum to appear 
with the two solutions indicated in equation \ref{eq:sol-for-global-mimimum}.
These results are from a {\tt Mathematica}-based \cite{mathematica} analysis.
The allowed ranges are obtained by first demanding appearance of a 
stable/metastable minimum for the given potential and then requiring, in 
addition, the singlet $CP$-odd and $CP$-even scalars to be non-tachyonic,
at the tree level. Squared masses for these two states, in the limit
$\higgsd=\higgsu=0$ and $S \neq 0$, are given by (from equations \ref{eq:msq-cp-odd} 
and \ref{eq:msq-cp-even}, respectively)
\beq
\mathcal{M}_{P,22}^2|_{_{\scriptsize{\begin{aligned} 
\higgsd &= \higgsu =0 \\
S &\neq 0
\end{aligned}}}} =  - 3 \kappa \akappa \vs, 
\qquad \qquad
\mathcal{M}_{S,33}^2|_{_{\scriptsize{\begin{aligned}
\higgsd &= \higgsu =0 \\
S &\neq 0
\end{aligned}}}} =  \kappa \vs (\akappa + 4 \kappa \vs) .
\label{eq:msq-odd-even}
\eeq
For the first solution ($s_+$) these result in figure 
\ref{fig:allowed-ranges-solution-1}. The top (bottom) panel corresponds to the 
case of stable (metastable) minimum. From left to right, the plots illustrate
the effects of requiring successively non-tachyonic singlet scalars, as 
mentioned above. 
Clearly, with $s_+$, stable minima occur only for $\akappa \geq 0$. 
As can be clear from equation \ref{eq:msq-odd-even}, demanding a non-tachyonic 
$CP$-odd singlet scalar (i.e., $\mathcal{M}_{P,22}^2 \geq 0$) further restricts 
this region to $\kappa \vs <0$ in the second column. Requiring on top of that the 
$CP$-even singlet scalar to be non-tachyonic (i.e., $\mathcal{M}_{S,33}^2 \geq 0$) 
retains only the lower green wedge in the rightmost plot of the first row.

It may be noted that for the plots in the first column, the shaded regions in 
green (first row) and yellow (second row) are complementary in nature. This is 
trivially expected since 
if one of the minima is the global one, the other one has to be local in nature. 
The complementarity is carried over to the second column for which the singlet 
$CP$-odd scalar is required to be non-tachyonic. However, this is no more 
the case when one demands, in the third column, the same for the singlet $CP$-even 
scalar. There, a narrow (yellow) slice indicating metastability for $\kappa \vs <0$ 
and $\akappa >0$ survives along with the yellow wedge 
that directly complements a similar green region but with opposite signs on
$\kappa \vs$ and $\akappa$, for the first solution. 

\begin{figure}[t]
\centering
\vskip 15pt
\includegraphics[height=0.20\textheight, width=0.32\columnwidth , clip]{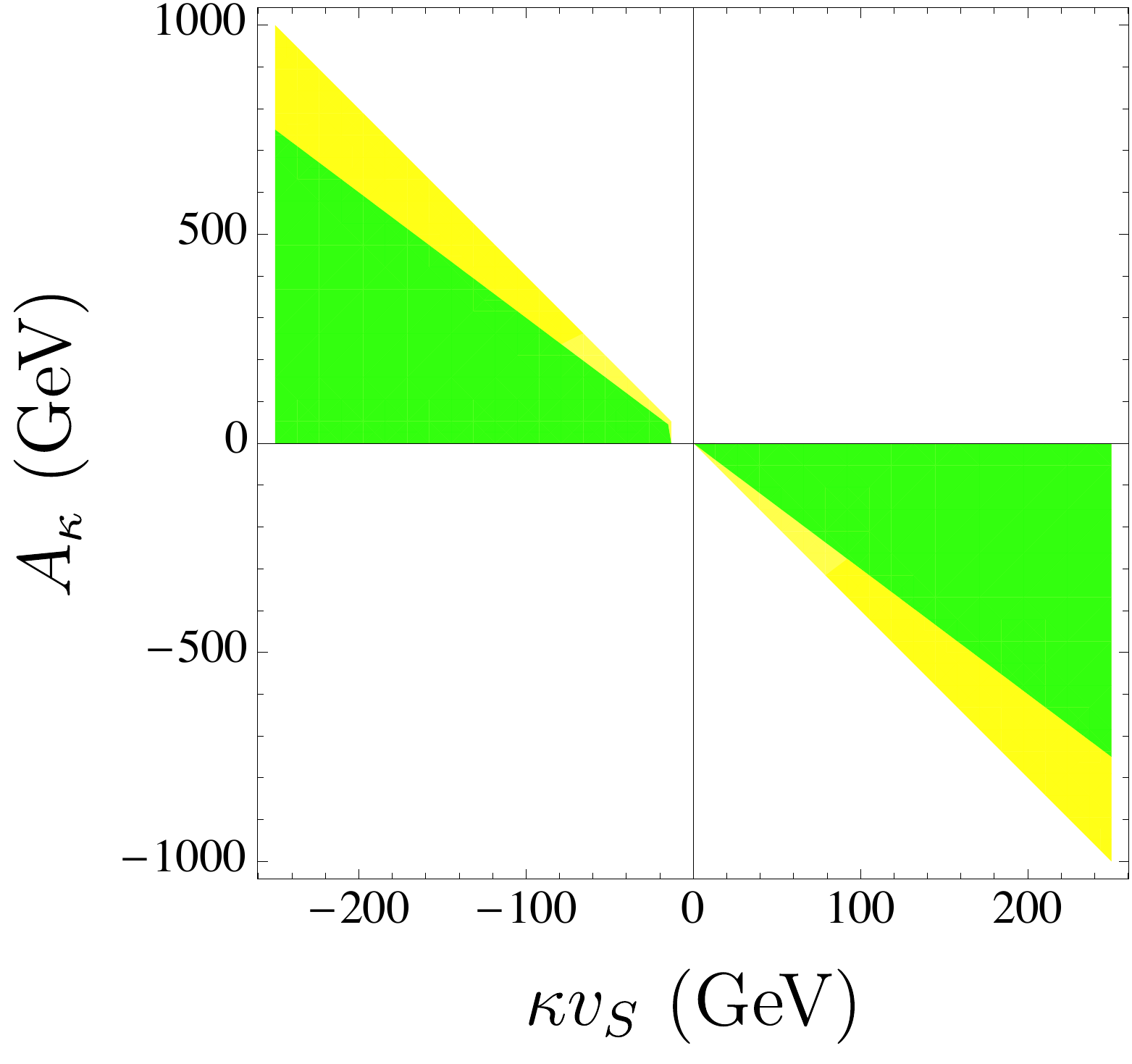}
\caption{Consolidated allowed regions in the $\kappa \vs - \akappa$ plane
including both solutions $s_+$ and $s_-$ and on imposing the requirement 
of non-tachyonic scalars.} 
\label{fig:allowed-ranges-consolidated}
\end{figure}

Furthermore, it can be found from table \ref{tab:analytical-ranges} that 
perfectly viable (corresponding) regions with flipped signs on the coordinates 
show up with the other non-vanishing solution $s_-$. This can be 
understood if we take note, in conjunction, of equation 
\ref{eq:higgs-pot-large-s}, in particular, its last term. Such flips
may have important phenomenological implications. Note that existing works
\cite{Ellwanger:1996gw, Ellwanger:1999bv, Djouadi:2008uj} discuss one or the 
other of these solutions. We instead
include both solutions in our present analysis. 
Figure \ref{fig:allowed-ranges-consolidated} shows the consolidated situation 
thereof. Global minima with non-tachyonic states 
are possible over the green 
regions. Regions of metastability are in yellow. The latter also spread 
below whole of the green parts. As a check, it may be noted that the gradients
of the edges of various wedge-shaped regions appearing in all these plots
perfectly derive from the respective conditions collected in table
\ref{tab:analytical-ranges}. It may be noted here that not all of the conditions
indicated in table \ref{tab:analytical-ranges} would hold in generic
situations where the doublet Higgs fields also acquire non-vanishing values.
%
%
%
%
%
\subsubsection{Non-vanishing $\higgsd$ and $\higgsu$ and radiatively corrected 
               potential}
\label{subsubsec:nonzero-hd-hu}
%
This is perhaps the point where we might rightly like to pause and understand 
the role of two unavoidable `complications' that are going to be present beyond 
the simplistic analysis presented above. These are the presence of non-vanishing 
$\higgsd$ and $\higgsu$ and the inclusion of radiative correction (at the
1-loop level only, though) to the potential. As pointed out earlier, inclusion 
of additional directions in field space makes the analysis rather tedious and
extracting useful information thereof becomes challenging. 

At the tree level itself, the general scenario with three non-vanishing neutral 
Higgs fields differs from the case with only one of them acquiring non-zero value. 
This is since the former may receive non-negligible contribution from terms  
trilinear and quartic in fields (as can be seen in the expression for the
potential in equation \ref{eq:higgs-pot-tree-org}) even in the vicinity of the 
DSB vacuum given that `$S$' could turn out to be large 
($\mathcal{O}(10 \, \mathrm{TeV})$) there.

Away from the DSB vacuum, with increasing magnitude of $\higgsd,\higgsu$, the 
effect may alter the potential profile giving rise to deeper minima. 
This would then have a non-trivial bearing on the stability of the DSB vacuum. 
Note that some of these terms further contain `$\lambda$' multiplied to them. 
Hence in scenarios with small $\lambda$, the effect can only be moderate. 
For larger $\lambda$, the impact could be drastic. 
The inclusion of radiative correction further complicates the issue. As a result, 
the setup develops an intricate dependence on the NMSSM parameters. 

Nonetheless, interestingly enough, all these complications can still be handled 
semi-analytically, to a very good extent. Such an approach would help understand 
the behaviour of the Higgs potential in arbitrary directions in the field space, 
away from the DSB. We undertake such an analysis via {\tt Mathematica}\footnote{In 
finding the extrema of the potential, {\tt Mathematica} resorts to numerical optimization.}
which is similar to the one described in the previous 
subsection, this time the starting point being the full tree-level Higgs potential 
of equation \ref{eq:higgs-pot-tree-org}.

For a given set of input parameters, at each stage, we use the appropriate set 
of tadpole equations to find the values for $\mhiggsdsq$, $\mhiggsusq$
and $\mssq$ 
by requiring a DSB minimum at $\langle S \rangle =\vs={\mueff / \lambda}$.
All parameters appearing in the potential are thus `locked' and one can now 
find the potential profile as a function of the three neutral scalar fields.
This reveals if there exists a minimum deeper than the DSB one (the panic
situation)\footnote{Crucially, it is 
ultimately only the experimental data that would tell us if the DSB minimum
is indeed a 
global (stable) minimum \cite{Barroso:2013awa} of the potential. 
Thus, to start with, the DSB minimum cannot be set to be 
just so.}.
Radiative correction to the tree-level potential at the 1-loop level (see equation 
\ref{eq:radcorr}) is estimated using the field-dependent mass-squared matrices 
that are extracted from the tree-level potential\footnote{The analysis 
can be made more rigorous by considering renormalization-group (RG)-improved 
effective potential (see, for example, references
\cite{Gamberini:1989jw, Quiros:1999jp, Einhorn:2007rv, Martin:2014bca, 
Andreassen:2014eha}).}. 
Thus, corrected $\mhiggsdsq$, $\mhiggsusq$ and $\mssq$, 
obtained further as 
solutions to the tadpoles of the 1-loop effective potential, bear implicit 
field-dependence.

It is important to realize at this point that a general study of false vacua 
involving all three neutral scalar fields of the
NMSSM gets to be rather involved and tedious. Along with the fields, a multitude
of input parameters also enter the picture and hence add to the complication.
It is thus necessary to have an optimal approach in exploring the scenario. This
would help understand the underlying physics better and extract systematically 
information that are phenomenologically relevant and/or interesting.
To this end, we divide the NMSSM parameter space into two broad categories:
\begin{itemize}
\item  a scenario with small $\lambda (\lesssim 0.1)$, thus requiring large radiative corrections to have
       the mass of the SM-like Higgs boson in the right ballpark. In this sense,
       the situation is akin to the MSSM and hence the scenario is dubbed here
       as `MSSM-like'\footnote{Note that, as discussed in section 
       \ref{subsec:higgs-neutralino-sectors}, the mass of the SM-like Higgs boson,
       under certain circumstances, might receive a negative correction from 
       singlet-doublet mixing. Under such a circumstance, even for large values of 
       $\lambda$, reasonably
       large radiative contribution to the mass of the SM-like Higgs boson may be 
       required similar to what happens in the MSSM. Hence large $\lambda$ values 
       are eventually included in its allowed range under this category\label{foot:sd-mixing}.}.
       A large radiative contribution is ensured primarily by setting the soft 
       parameters pertaining to the third generation at $\mathcal{O}$(TeV);
\item  a scenario with somewhat large $\lambda \, (\lesssim 0.7)$. As is well known, such large
       values of $\lambda$ could push the mass of the SM-like Higgs boson
       substantially up at the tree level itself thus diminishing the need of radiative
       contributions to the same that are so indispensable in the MSSM. Naturally, the
       resulting scenario is dubbed `NMSSM-like' in the present work.
\end{itemize}
A further classification is done based on a phenomenologically interesting
possibility that the LSP (lightest supersymmetric particle)-neutralino in the 
NMSSM could essentially be a mixture of only the singlino and the higgsinos. 
The corresponding sub-classes are indicated in this work as scenarios which 
are `singlino-like' and `higgsino-like' requiring at least 75\% admixture of 
singlino and higgsinos in the LSP, in the respective cases. For a general but
concise understanding of the involved neutralino and the Higgs states, in
particular reference to their singlet sectors, we refer the reader to section 
\ref{subsec:higgs-neutralino-sectors}.
%
%
\begin{table}[t]
\begin{center}
{\scriptsize
\begin{tabular}{|c|c|c|c|}
\hline
Parameters & Quantity & {\tt Mathematica v10} & {\tt SPheno v3.3.8} \\
\hline
 & & & \\
 & \multicolumn{1}{r|}{$V^{^{\mathrm{neutral}}}_{\mathrm{Higgs|1-loop}}$ @ DSB (GeV$^4$)}  
  & $-6.18*10^{12}$ & $-6.08*10^{12}$ \\
Set A      &                                  &         &       \\
           & \multicolumn{1}{l|}{$\left\{\mhiggsusq/ \, \mhiggsdsq/ \, \mssq \right\}_\mathrm{1-loop}$ (GeV$^2$)}  
  & $1.79*10^5/ 8.96*10^6/ 4.99*10^3$ & $1.78*10^5/ 8.96*10^6/ 4.99*10^3$ \\
   \hline
 & & & \\
 & \multicolumn{1}{r|}{$V^{^{\mathrm{neutral}}}_{\mathrm{Higgs|1-loop}}$ @ DSB (GeV$^4$)}  
  & $-6.76*10^{12}$ & $-6.53*10^{12}$ \\
Set B      &                                  &         &       \\
           & \multicolumn{1}{l|}{$\left\{\mhiggsusq/ \, \mhiggsdsq/ \, \mssq \right\}_\mathrm{1-loop}$ (GeV$^2$)}  
  & $2.45*10^5/ 1.29*10^7/ 1.86*10^4$ & $2.42*10^5/ 1.29*10^7/ 1.86*10^4$ \\
   \hline
 & & & \\
 & \multicolumn{1}{r|}{$V^{^{\mathrm{neutral}}}_{\mathrm{Higgs|1-loop}}$ @ DSB (GeV$^4$)}  
  & $1.88*10^{12}$ & $1.85*10^{12}$ \\
Set C      &                                  &         &       \\
           & \multicolumn{1}{l|}{$\left\{\mhiggsusq/ \, \mhiggsdsq/ \, \mssq \right\}_\mathrm{1-loop}$ (GeV$^2$)}  
  & $8.96*10^4/ 4.46*10^5/ 2.22*10^4$ & $9.68*10^4/ 4.51*10^5/ 2.24*10^4$ \\
   \hline
 & & & \\
 & \multicolumn{1}{r|}{$V^{^{\mathrm{neutral}}}_{\mathrm{Higgs|1-loop}}$ @ DSB (GeV$^4$)}  
  & $1.88*10^{12}$ & $1.85*10^{12}$ \\
Set D      &                                  &         &       \\
           & \multicolumn{1}{l|}{$\left\{\mhiggsusq/ \, \mhiggsdsq/ \, \mssq \right\}_\mathrm{1-loop}$ (GeV$^2$)}  
  & $-2.82*10^4/ 2.50*10^5/ 8.16*10^4$ & $-2.35*10^4/ 2.56*10^5/ 8.17*10^4$ \\
   \hline
  \end{tabular}
\label{tab:numerical-comparison}
}
\end{center}
\caption{Numerical comparison of the values of the 1-loop radiatively-corrected
neutral scalar potential and the resulting
squared soft masses as obtained from a {\tt Mathematica}-based analysis and
\veva ~for the following sets of inputs parameters: \newline
\hspace*{1cm}
Set A: 
\hspace*{0.2cm}
$\lambda=0.15, \, \kappa=0.01, \, \alambda=3000 \, \mathrm{GeV}, \, 
\akappa=0, \, \mueff=100 \, \mathrm{GeV}, \, \tan\beta=30$; \newline
\hspace*{1cm}
Set B:
\hspace*{0.2cm}
$\lambda=0.06, \, \kappa=0.05, \, \alambda=5000 \, \mathrm{GeV}, \, 
\akappa=-500 \, \mathrm{GeV}, \, \mueff=250 \, \mathrm{GeV}, \, \tan\beta=10$;
\newline
\hspace*{1cm}
Set C:
\hspace*{0.2cm}
$\lambda=0.675, \, \kappa=0.1, \, \alambda=800 \, \mathrm{GeV}, \, 
\akappa=-450 \, \mathrm{GeV}, \, \mueff=400 \, \mathrm{GeV}, \, \tan\beta=1.65$; \newline
\hspace*{1cm}
Set D:
\hspace*{0.2cm}
$\lambda=0.75, \, \kappa=0.5, \, \alambda=200 \, \mathrm{GeV}, \, 
\akappa=-850 \, \mathrm{GeV}, \, \mueff=400 \, \mathrm{GeV}, \, \tan\beta=2$. \newline
The soft mass-squared parameters and the soft trilinear parameters for the third generation
squark sector are fixed at the following values: for sets `A' and `B', 
$\msQthree=\msUthree=\msDthree=3$ TeV
and $A_t=3$ TeV; for sets `C' and `D',  $\msQthree=\msUthree=750$ GeV, $\msDthree=3$ TeV and
$A_t=0$, where $\msQthree$, $\msDthree$ and $\msUthree$ 
stand for the soft masses for the doublet,
the up-type singlet and the down-type singlet squarks and $A_t$ is the trilinear
coefficient for the stop sector. All other trilinear parameters are set to zero while
the other scalar masses are taken to be heavy enough. Soft masses in the gaugino sector
are all set to 2 TeV. The cutoff/renormalization scale $Q$
 (equation \ref{eq:radcorr-mhiggs})
is chosen to be $Q=\sqrt{\msQthree \msUthree}$.
}
\label{tab:spheno-math}
\end{table}

In table \ref{tab:spheno-math} we compare some relevant quantities obtained from 
our {\tt Mathematica}-based analysis with 
the ones from {\tt SPheno} (which \veva ~uses). A few representative sets of 
parameters are chosen (indicated in the 
table-caption) which we would put to context later and use them further. The 
level of overall agreement is rather compelling (within 5\%; except for the value
of $\mhiggsusq$ at the 1-loop level in Case C where the deviation is about 8\%). 
This would serve as a robust basis when we try to make sense of results obtained
from \veva ~(modulo the thermal effects) in terms of our semi-analytical 
approach to the problem.
%
%

\begin{figure}[t]
\centering
\includegraphics[height=0.16\textheight, width=0.32\textwidth]{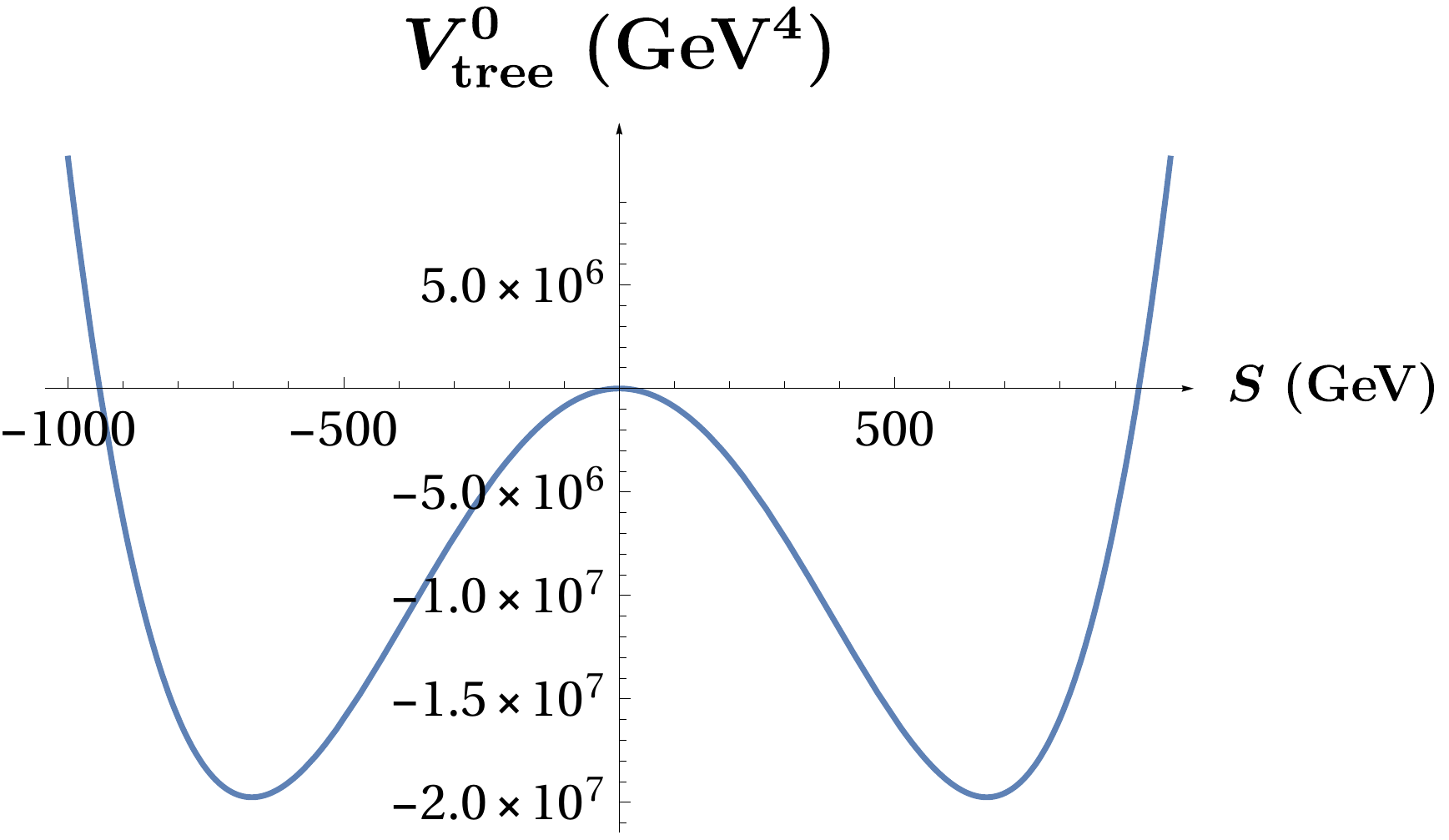}
\includegraphics[height=0.16\textheight, width=0.32\textwidth]{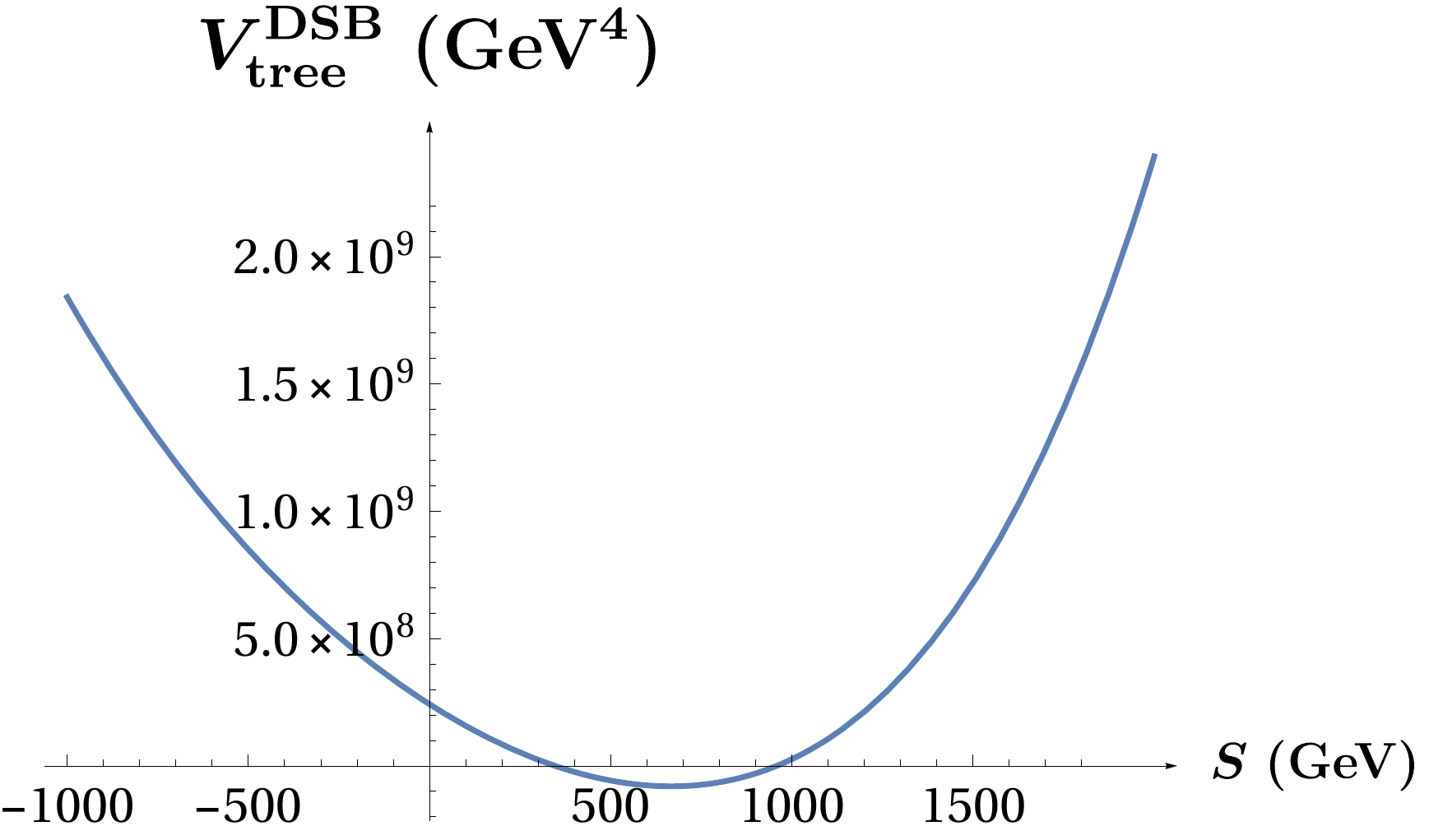}
\includegraphics[height=0.16\textheight, width=0.32\textwidth]{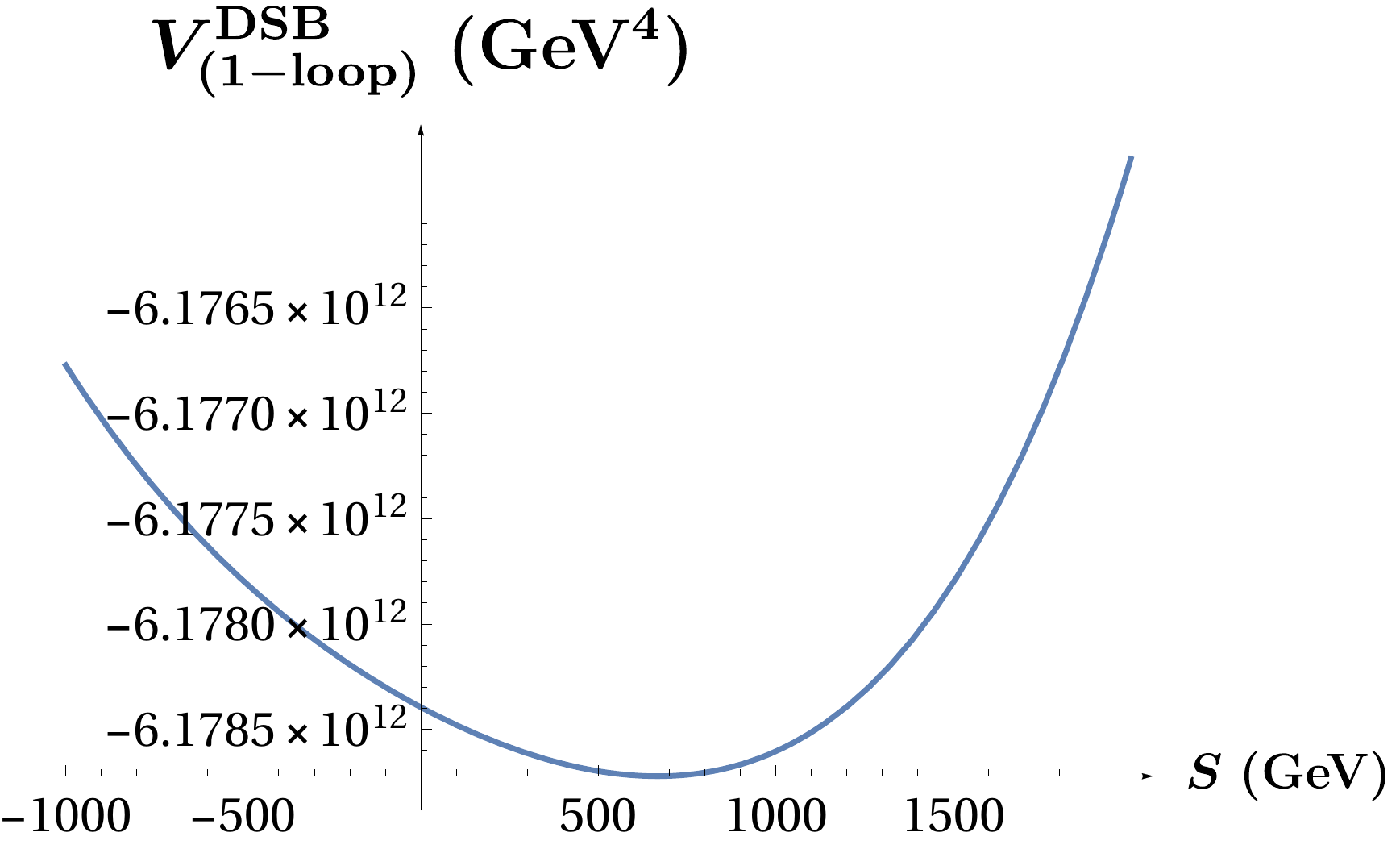}
\vskip 0.4in
\includegraphics[height=0.16\textheight, width=0.32\textwidth]{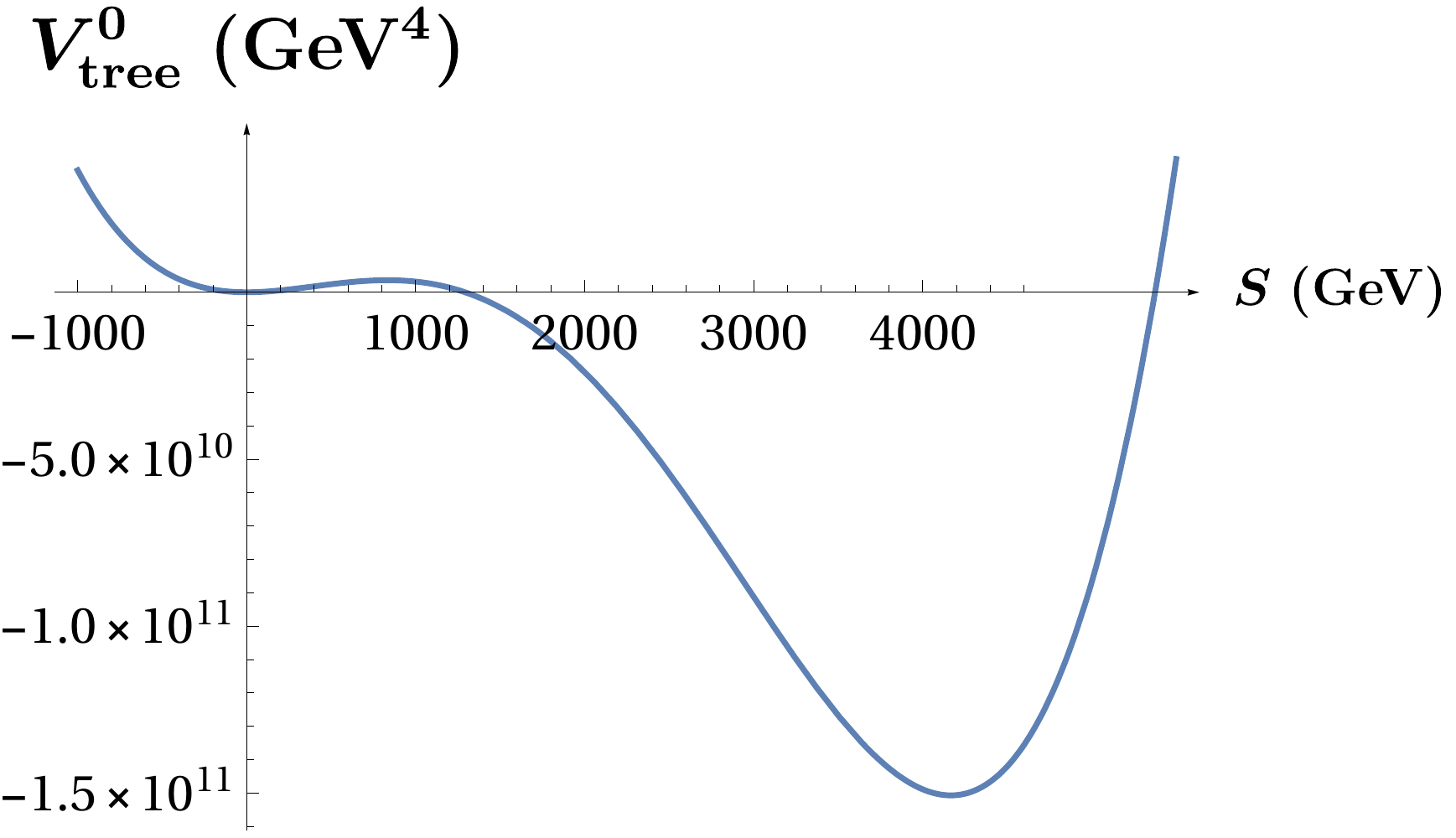}
\includegraphics[height=0.16\textheight, width=0.32\textwidth]{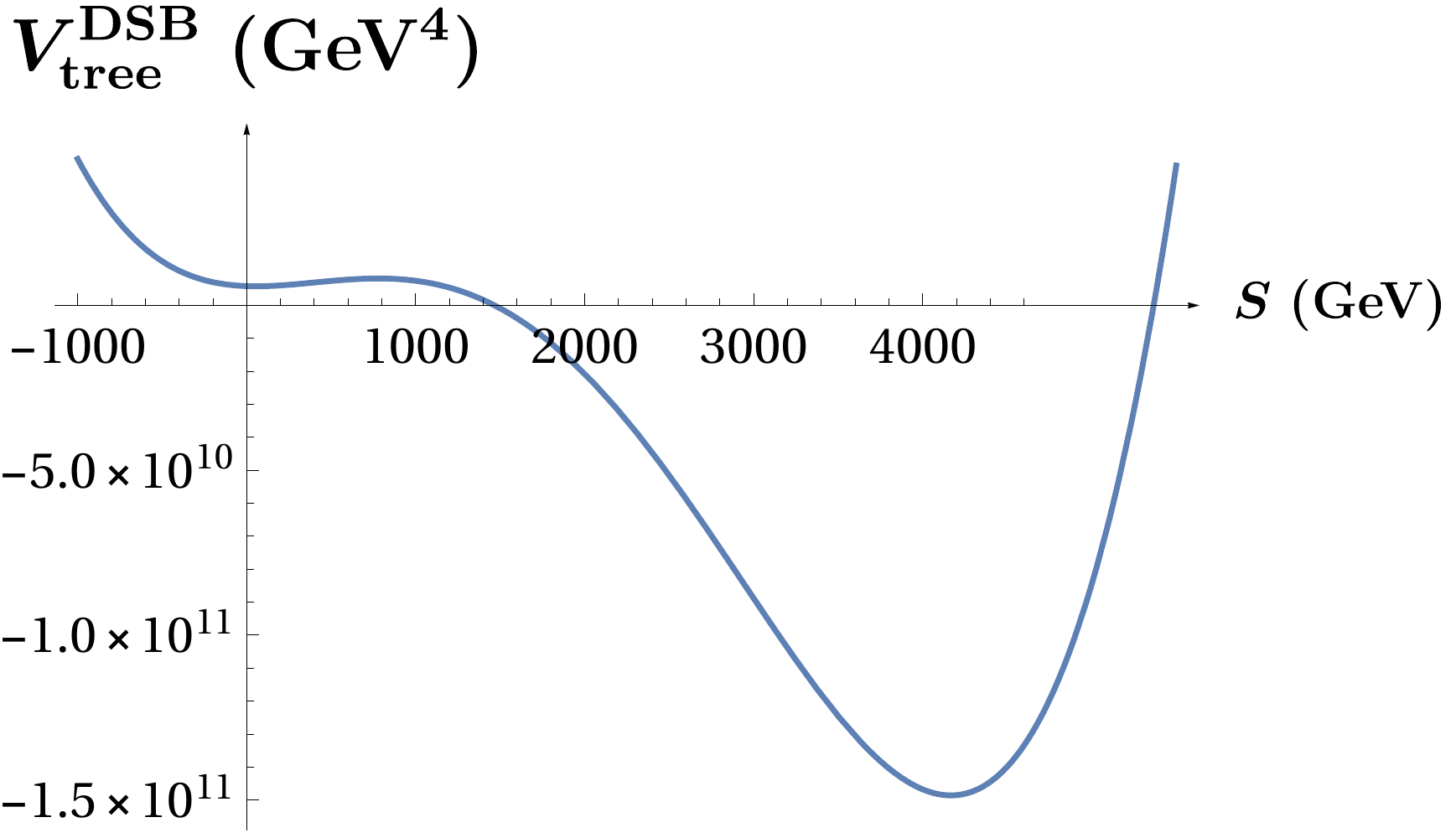}
\includegraphics[height=0.16\textheight, width=0.32\textwidth]{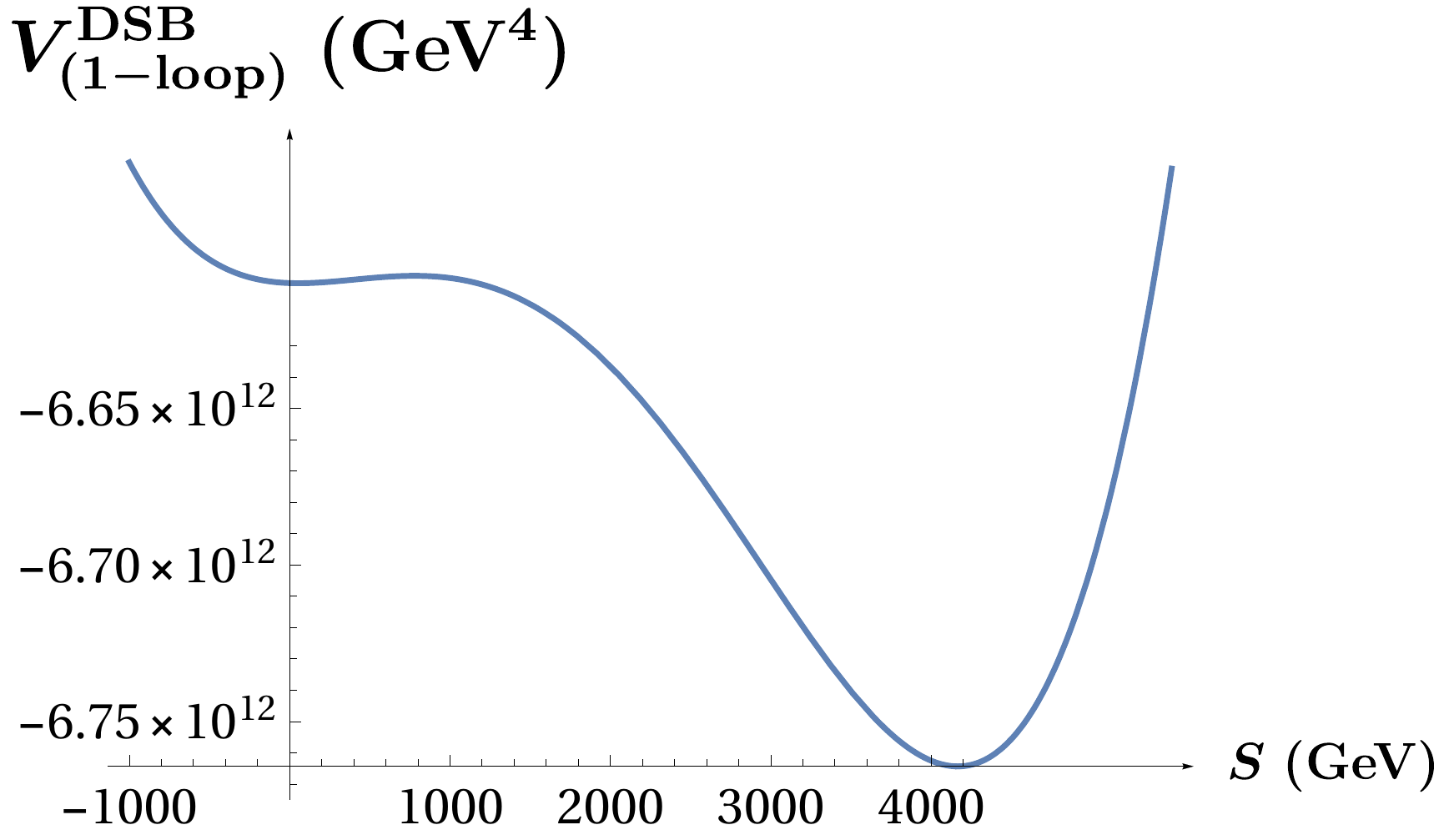}
\caption{Profiles of the neutral scalar potential as a function of the field `$S$' 
in the MSSM-like scenario. The top (bottom) panel stands for the case with a 
singlino- (higgsino)-like LSP. Left panel shows the tree-level profiles when 
`$S$' is the only scalar field present. Plots in the middle panel illustrate the 
situation with non-vanishing $\higgsd$ and $\higgsu$ while their values are fixed 
at the respective \vevs ~they assume at the DSB minimum. Further, profiles in the 
rightmost panel result from the consideration of 1-loop radiatively-corrected 
potential. $V^0$ is evaluated with $H_u^0=H_d^0=0$ whereas $V^\mathrm{DSB}$ is computed by fixing the Higgs fields at their DSB values. }
\label{fig:pot-profiles-mssm}
\end{figure}

\begin{figure}[t] 
\centering		
\includegraphics[height=0.16\textheight, width=0.32\textwidth]{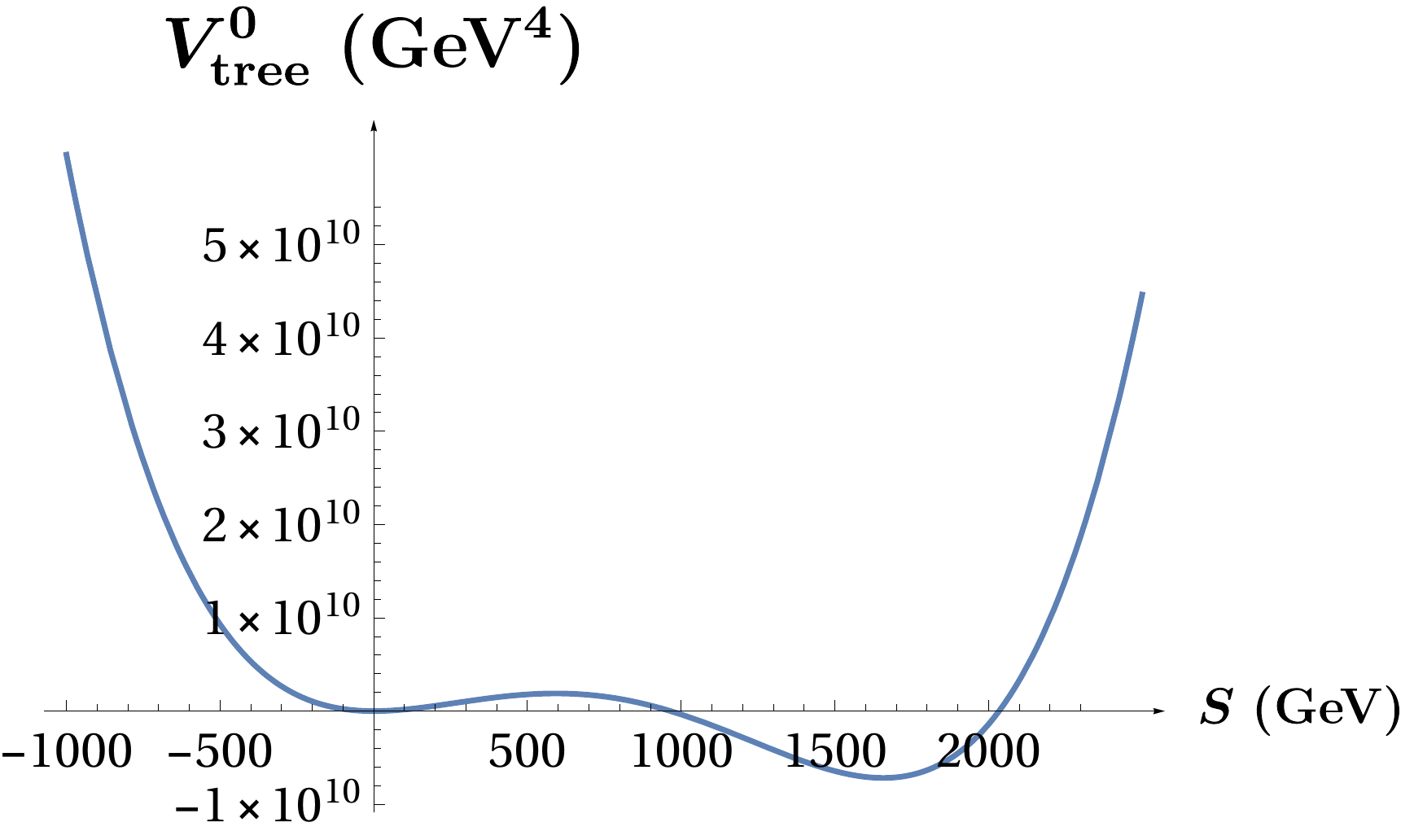}
\includegraphics[height=0.16\textheight, width=0.32\textwidth]{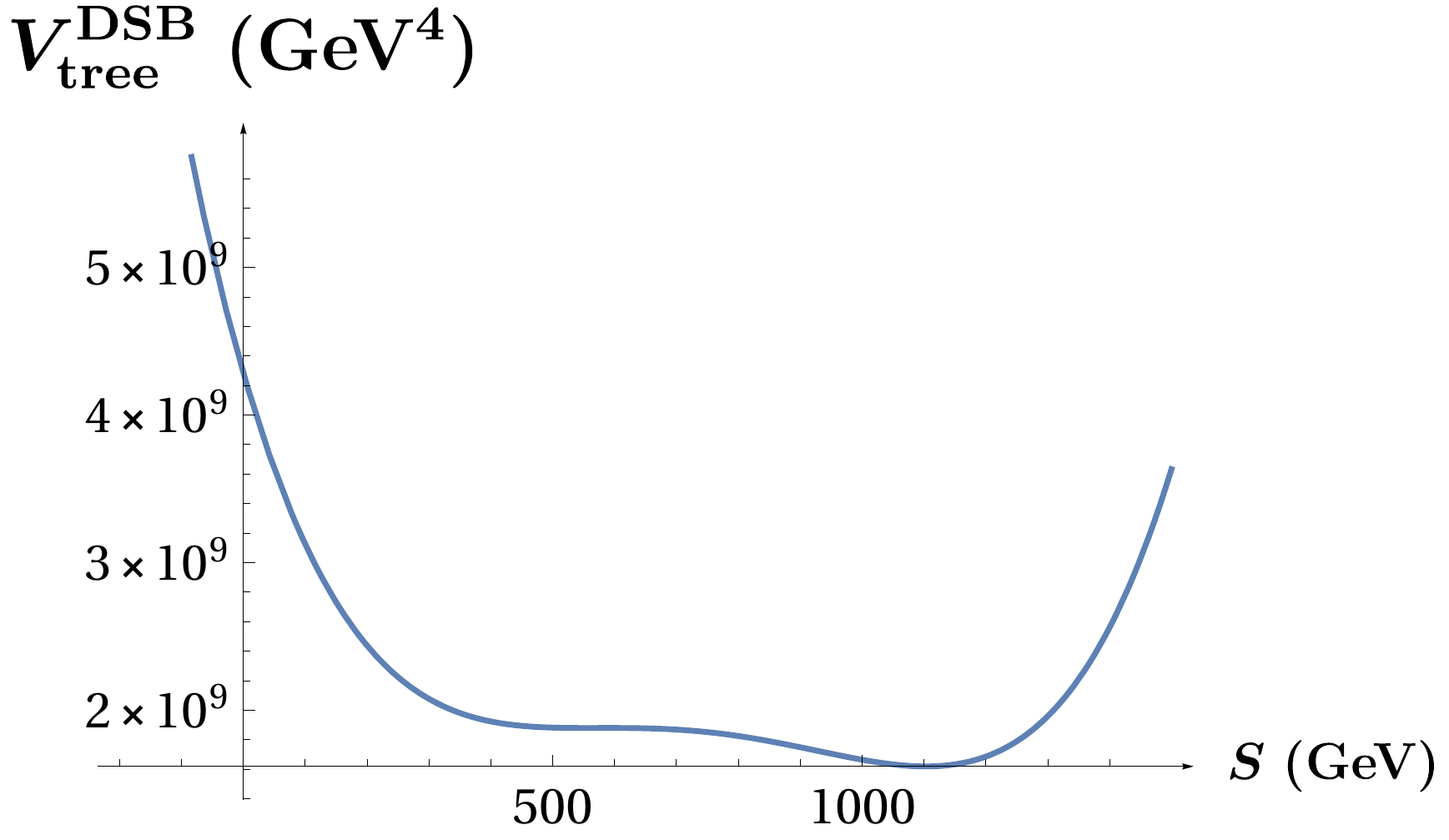}
\includegraphics[height=0.16\textheight, width=0.32\textwidth]{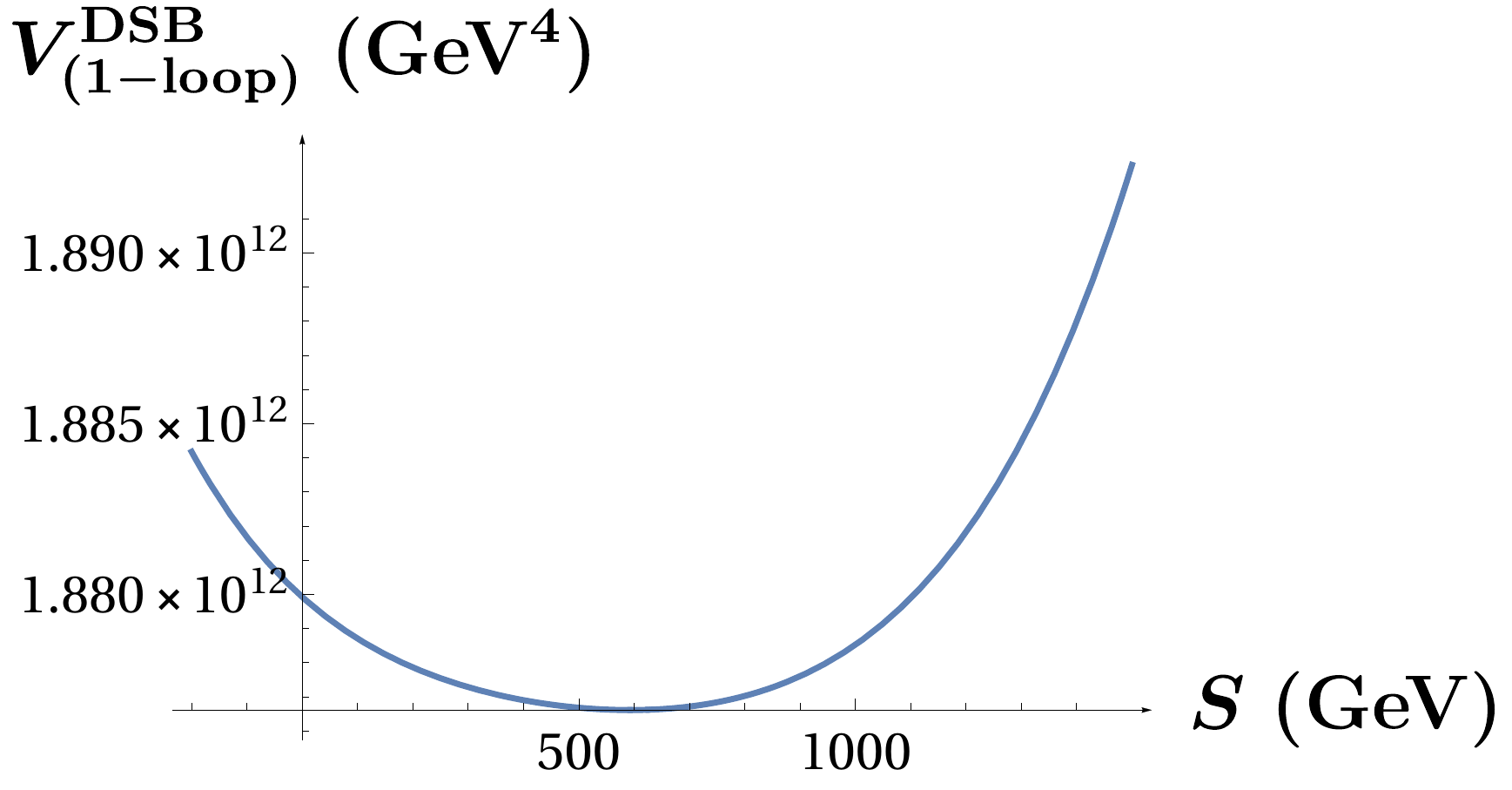}
\vskip 0.4in
\includegraphics[height=0.16\textheight, width=0.32\textwidth]{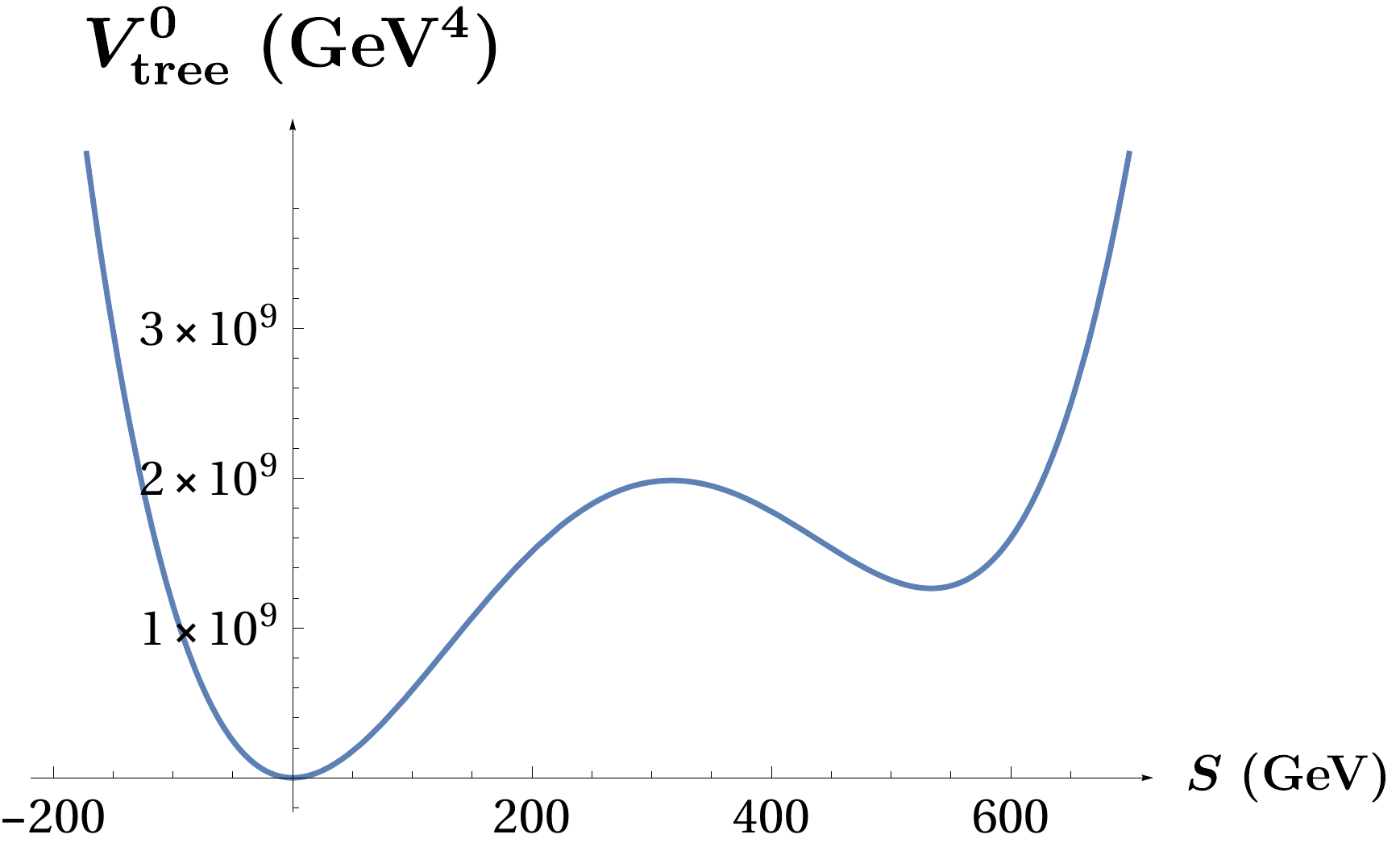}
\includegraphics[height=0.16\textheight, width=0.32\textwidth]{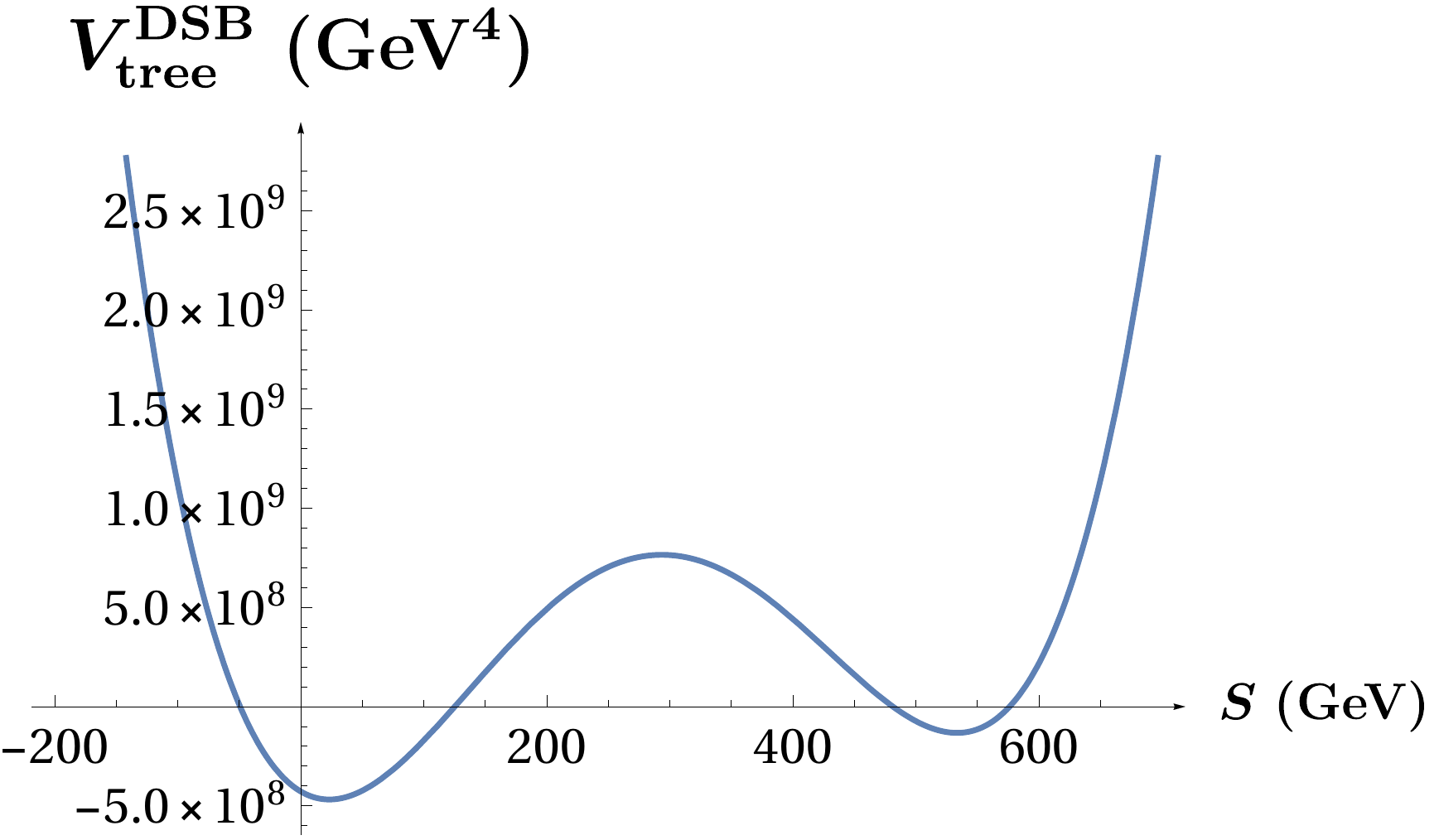}
\includegraphics[height=0.16\textheight, width=0.32\textwidth]{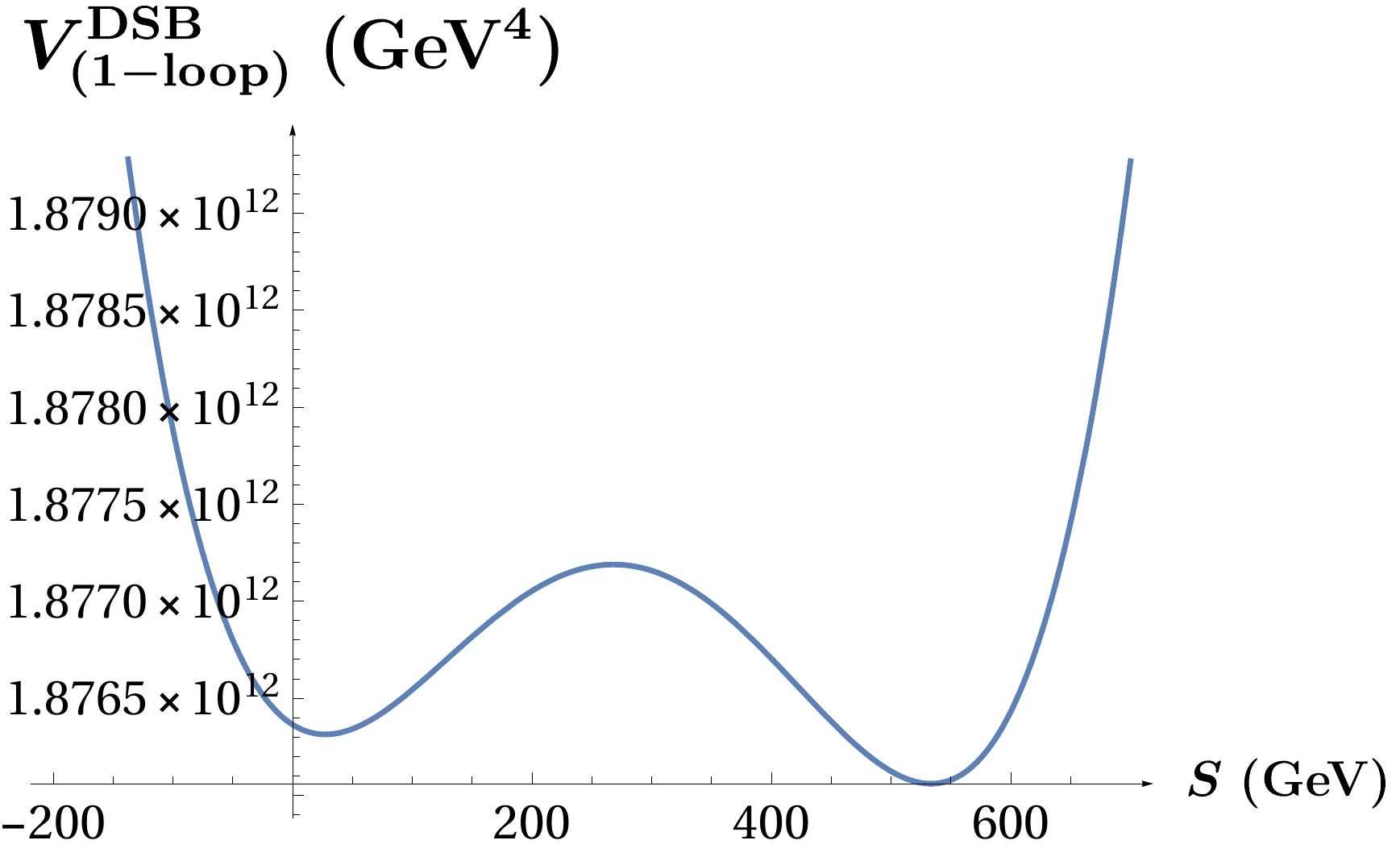}
\caption{Same as in figure \ref{fig:pot-profiles-mssm} but for the 
NMSSM-like case.}
\label{fig:pot-profiles-nmssm}
\end{figure}

In figures \ref{fig:pot-profiles-mssm} and \ref{fig:pot-profiles-nmssm} we
illustrate the aforesaid effects via a set of potential profiles for scenarios
presented in table \ref{tab:spheno-math}. For each figure the top (bottom) row 
corresponds to the case with a singlino-like (higgsino-like) LSP 
In each row the leftmost plot is with
$\higgsd=\higgsu=0$ and only `$S$' is non-vanishing 
(equation \ref{eq:higgs-pot-large-s}). The middle one is with non-vanishing 
$\higgsd$ and $\higgsu$ (equation \ref{eq:higgs-pot-tree-org}) fixed at their 
values at the DSB vacuum
Note that this choice confines us to a direction normal to the $\higgsd$-$\higgsu$ 
plane, passing through it at a point with polar coordinates ($v=\sqrt{\vd^2+\vu^2}, 
\, \beta=\tan^{-1}{\vu \over \vd}$) and extended in the $\pm S$ direction. Clearly, 
it should then be noted that these profiles do not reveal the actual `terrain' of 
the potential in higher field dimensions and deeper minima could appear away from 
the chosen direction. Finally, plots in the rightmost column demonstrate the effect 
of adding radiative correction to the potential.
%
%
%
%

The leftmost plot in the first row of figure \ref{fig:pot-profiles-mssm} has a 
symmetric profile under $S \rightarrow -S$ with a degenerate pair of minima. 
This is understood in terms of vanishing $\akappa$ that we set for this row 
which makes the only term with odd (cubic) power on `$S$' (for vanishing 
$\higgsd$ and $\higgsu$) in equations \ref{eq:higgs-pot-tree-org} also vanishing. 
As can be seen from the middle plot, switching on finite values of $\higgsd$ and
$\higgsu$ lifts this degeneracy. Thus,  the only minimum appears for a positive
value of `$S$' ($\sim 670$ GeV). This is indeed where the DSB vacuum should appear
given our choices of $\mueff$ (=100 GeV) and $\lambda$ (=0.15) for the case in 
hand. The rightmost plot in this panel shows that the inclusion of radiative 
correction roughly preserves the shape of the profile. However, as expected, the 
amount of (negative) correction is appreciable. Plots in the second row of figure
\ref{fig:pot-profiles-mssm} reveal that there may be a situation when the profile 
may not get distorted at all, although, radiative correction does alter the scale 
of the potential. As discussed earlier, introducing non-vanishing $\higgsd$ and
$\higgsu$ has hardly any effect on the potential in this case since $\lambda$ is 
set to a small value (=0.06). 
%
%
%
%

The plot in the left column of figure \ref{fig:pot-profiles-nmssm} refers to a
situation where the would be DSB vacuum starts off as a local maximum of the 
tree-level potential (with $\higgsd=\higgsu=0$) at $S={\mueff \over \lambda} 
\approx 590$ GeV. Finite $\higgsd$ and $\higgsu$ turn this, to be precise, into 
a point of inflection (middle plot) before radiative effects make this the DSB 
vacuum which is also the global minimum. On the other hand, the leftmost plot
in the second row of figure \ref{fig:pot-profiles-nmssm} describes a situation
where the DSB minimum is a local one. Finite $\higgsd$ and $\higgsu$ affect
the profile to a moderate extent by decreasing the relative depths between the
DSB and the global minimum (middle plot). Further, radiative correction turns
the DSB minimum into the global minimum of this (single-field) potential.    
%
%
\begin{figure}[htb!]
\centering		

\includegraphics[height=0.4\textheight, width=0.44\textwidth]{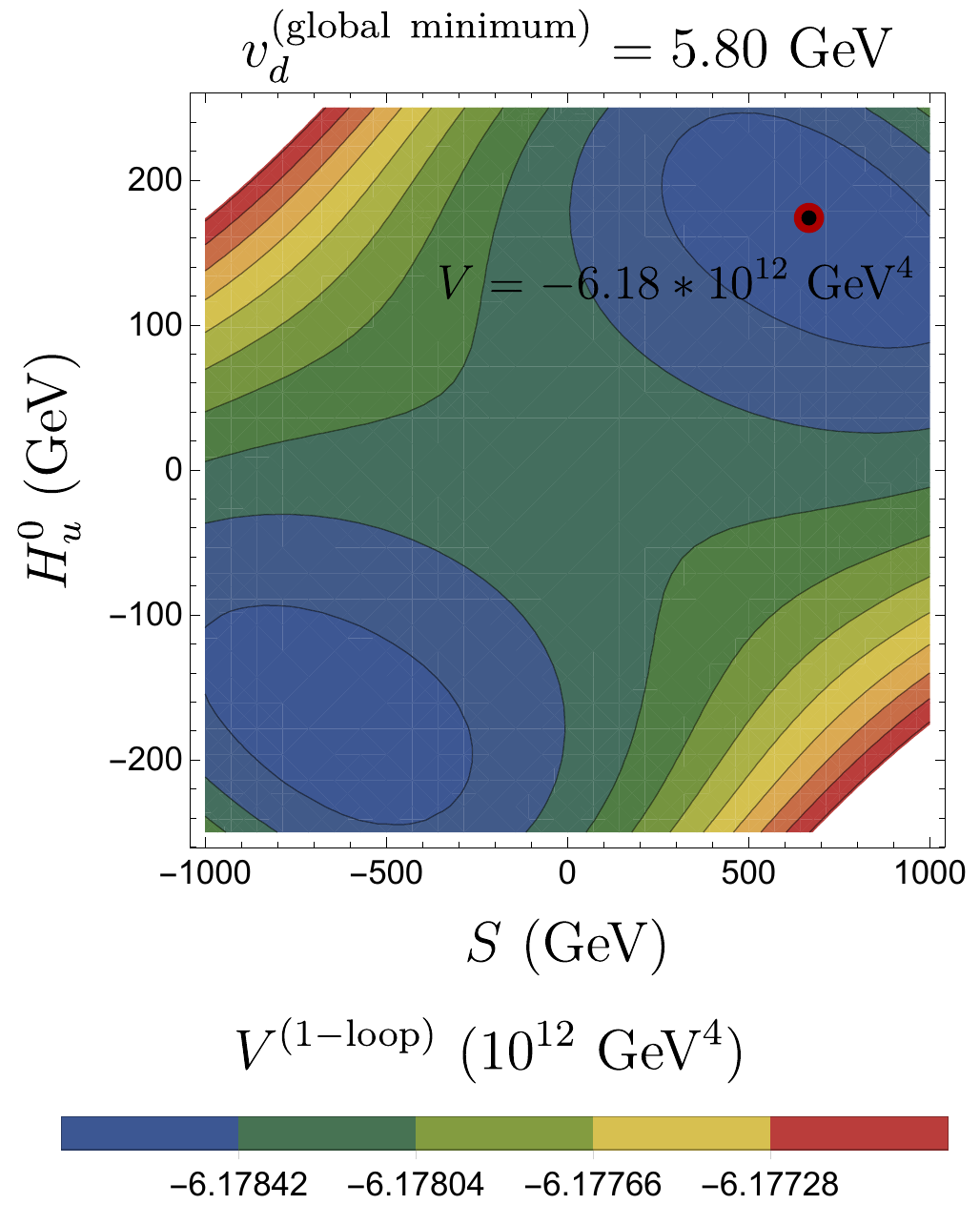}
\hskip 15pt
\includegraphics[height=0.4\textheight, width=0.44\textwidth]{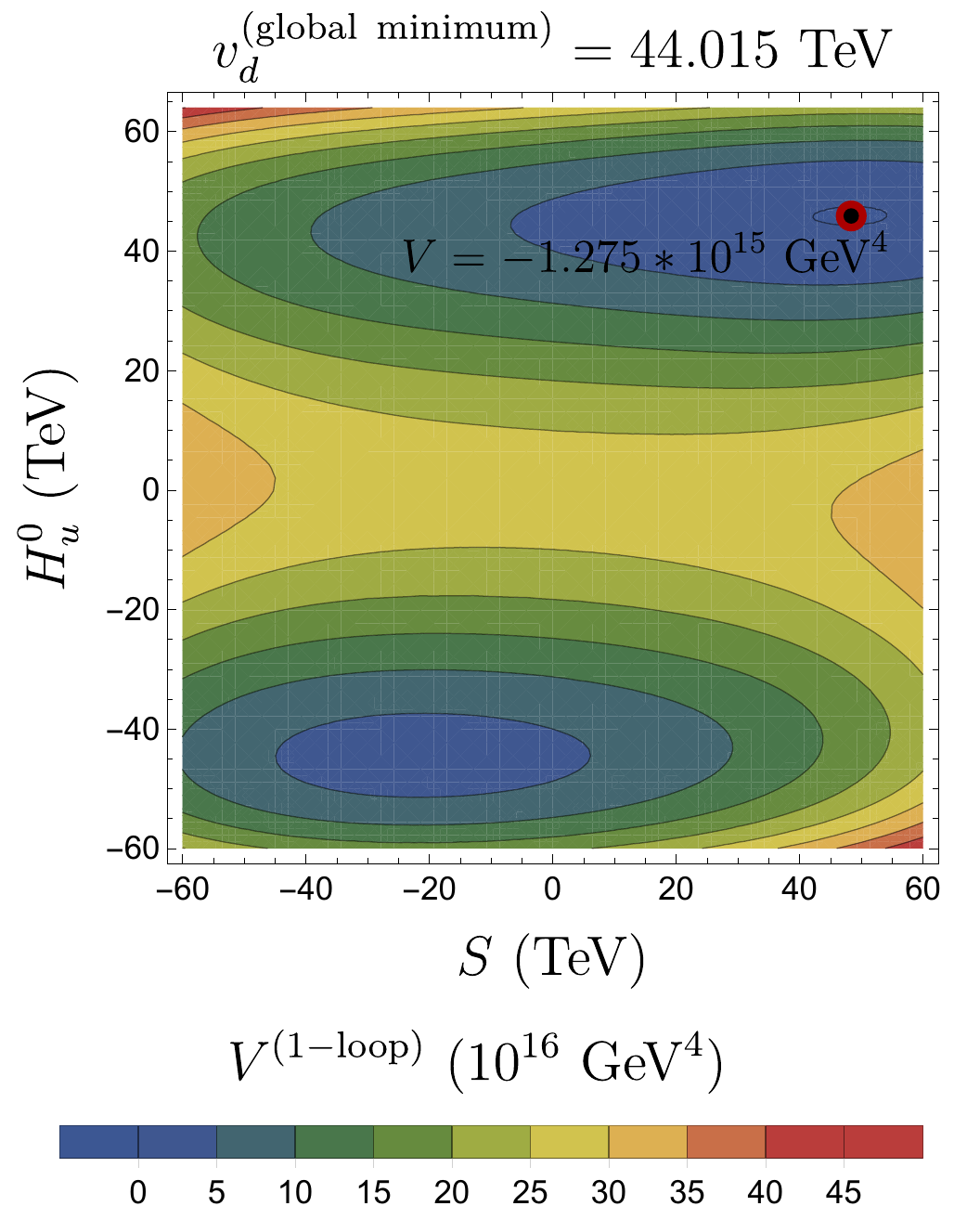}
\vskip 0.1in
\includegraphics[height=0.38\textheight, width=0.44\textwidth]{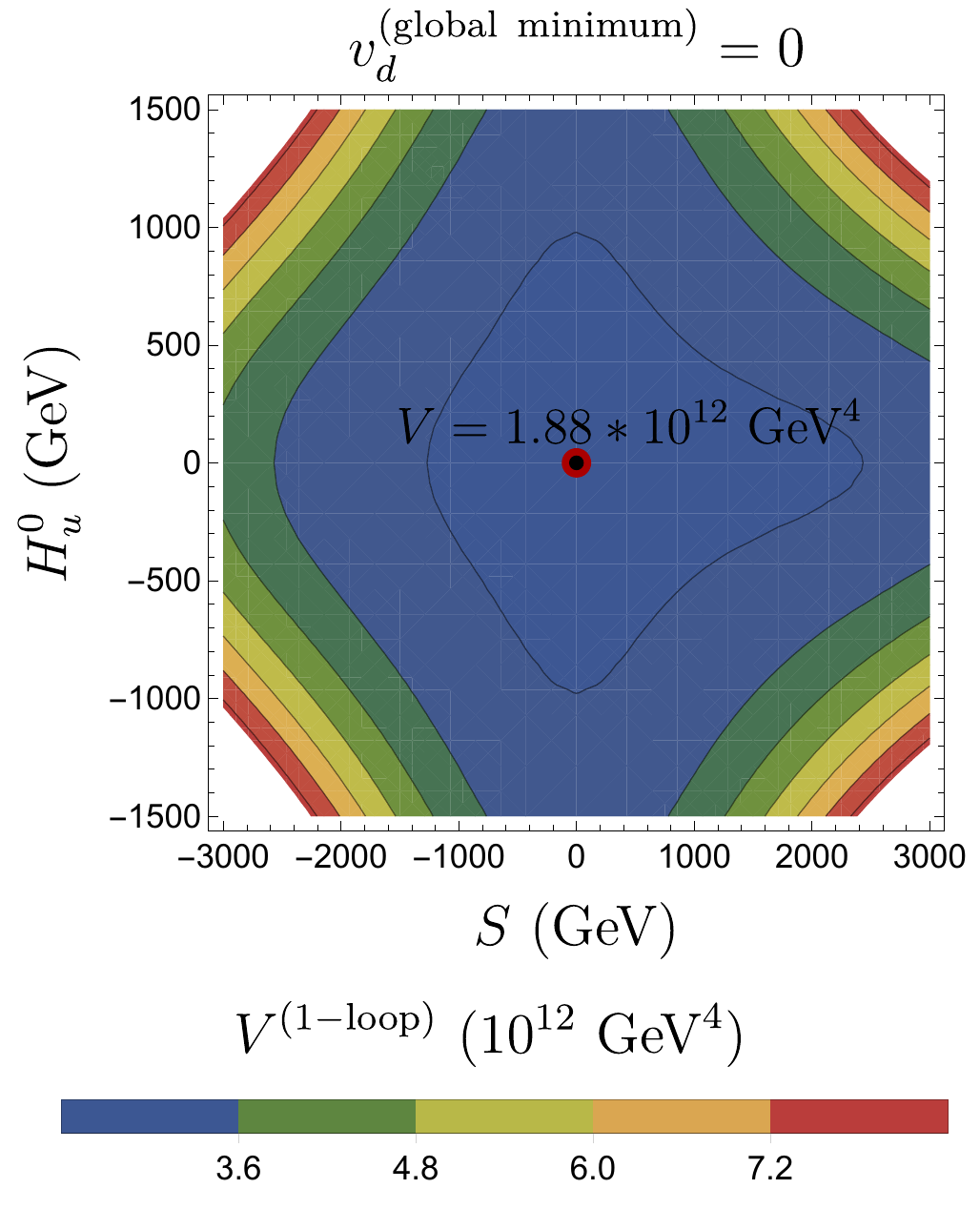}
\hskip 15pt
\includegraphics[height=0.4\textheight, width=0.44\textwidth]{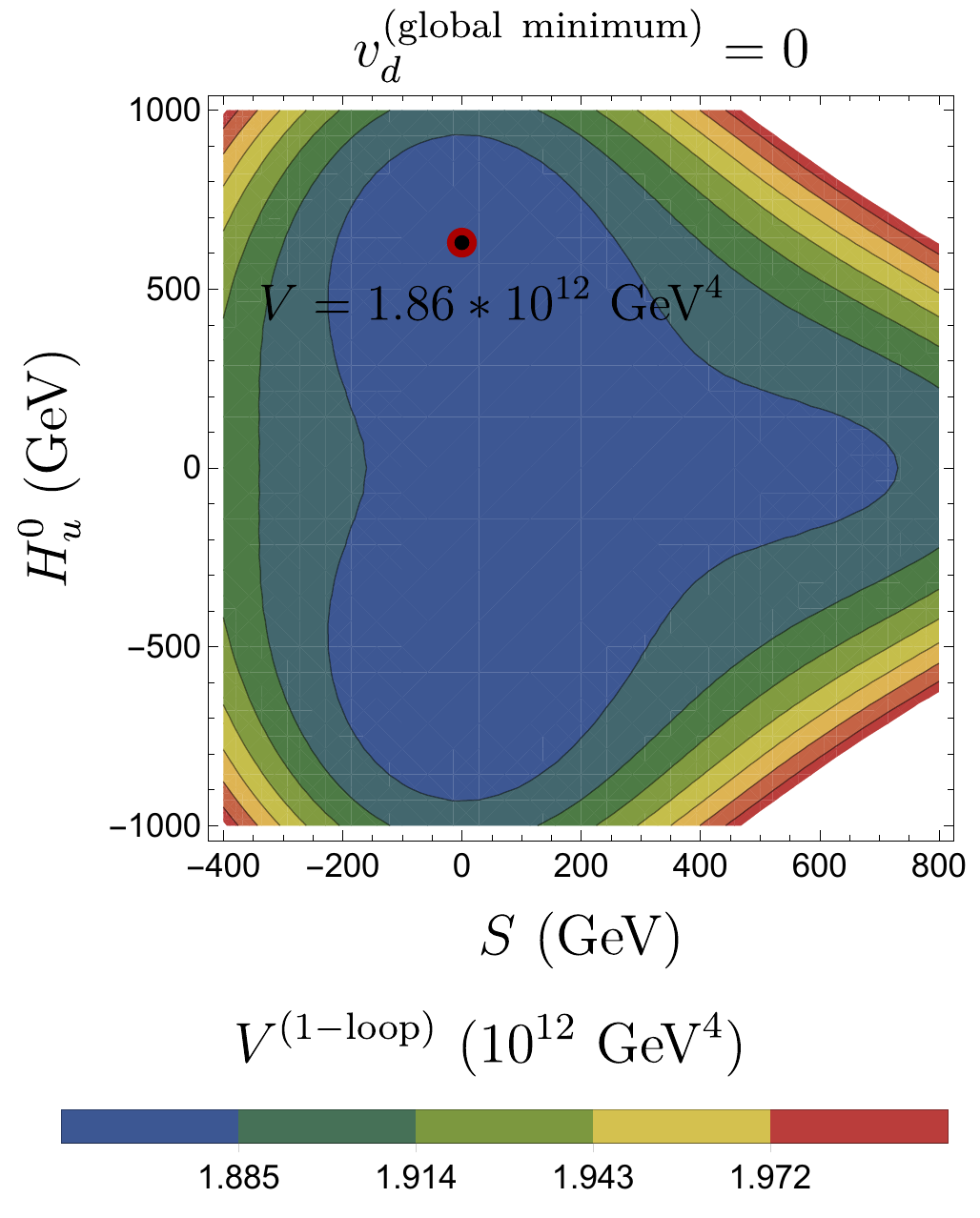}
\caption{Contour profiles of the 1-loop radiatively-corrected potential
(containing all three neutral scalar fields) $S-\higgsu$ plane along the fixed
$\higgsd$ directions that yield the deepest (global) minima in respective cases.
Top (bottom) panel presents the MSSM-like (NMSSM-like) scenario. In each case,
the plot on left (right) illustrates the situation with a singlino- (higgsino)-like LSP.
Pallets below the plots shows the color codes indicating the magnitude of the effective
potential. The bullet in each plot points to the actual location of the global minimum
in the field space.}
\label{fig:contours} 
\end{figure}
%
%

An important purpose of the present study is to have the idea about where exactly
the extrema appear in the field-space. With this in mind, in figure 
\ref{fig:contours} we present the `iso-potential' (radiatively corrected) 
contours for the four sets of parameters used in figures 
\ref{fig:pot-profiles-mssm} and \ref{fig:pot-profiles-nmssm}. However, this time 
we move away from the fixed directions adopted there and choose the ones along 
which the global (deepest) minimum of the potential occur for each case. As can be
seen, all these are either along or very close by to some flat directions.
We indicate the location of the global minimum on the $S-\higgsu$ plane 
in each case and mention the magnitudes of the potential.
These values can be compared straightaway to the ones associated with the 
corresponding DSB minima shown in the rightmost column of 
\ref{fig:pot-profiles-mssm} and \ref{fig:pot-profiles-nmssm}. Furthermore, plots 
in figure \ref{fig:contours} also convey that regions in the field space with
deeper potential are always surrounded from all sides by regions with higher
values of the potential. This implies that in all these cases the potential is
bounded from below as it should be in a realistic situation. 
However, presence of dangerous directions along which the 
potential is unbounded from below (UFB) are still possible 
\cite{Ellwanger:1999bv, Abel:1998ie}.
This is in spite of having positive contributions to the potential from $\vs$ which
generally come in rescue.
%
\subsection{False vacua along arbitrary field directions: a semi-analytical study}
\label{subsec:arbitrary-directions-semi}
%
Given the discussion in section \ref{subsubsec:nonzero-hd-hu}, it would now be
easier to appreciate how involved a study of false vacua arising along arbitrary 
directions in the NMSSM field space could get. Such a study not only involves
a larger volume of a multi-dimensional parameter space but also opens up to
variations of all relevant NMSSM fields. Thus, a reasonable scan over the
parameter space gets rather
time-consuming even in a state-of-the-art (multi-processing) computing 
environment. Hence, in this section and in the rest of this work, 
we would adhere to the broad scenarios introduced in section \ref{subsubsec:nonzero-hd-hu}.
In order to deal with the emerging intricacies, we also generalize the scope
of our {\tt Mathematica}-based code in an appropriate manner.

\begin{figure}[t]
\centering		
\includegraphics[height=0.22\textheight, width=0.45\textwidth]{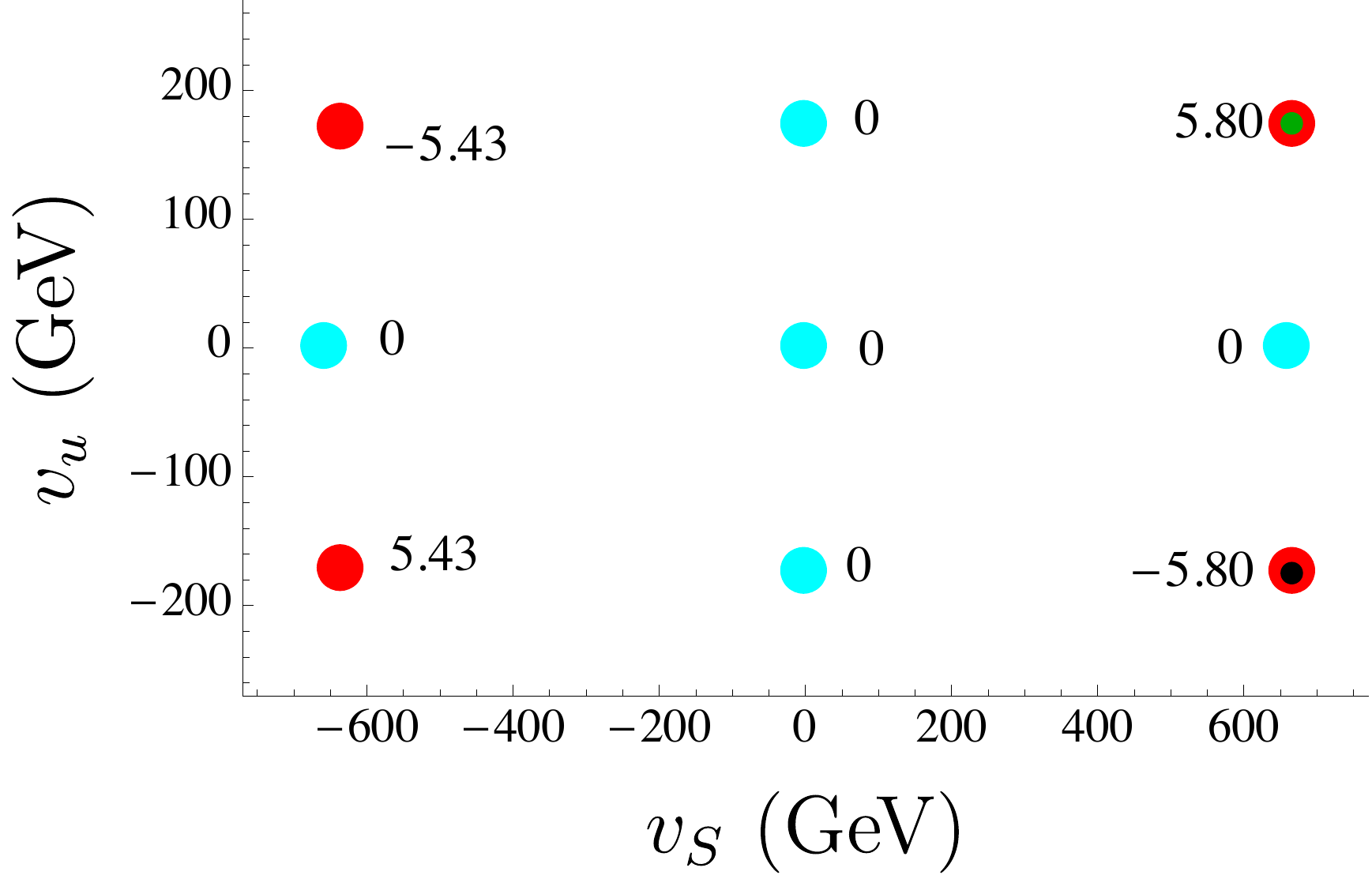}
\hskip 25pt
\includegraphics[height=0.22\textheight, width=0.45\textwidth]{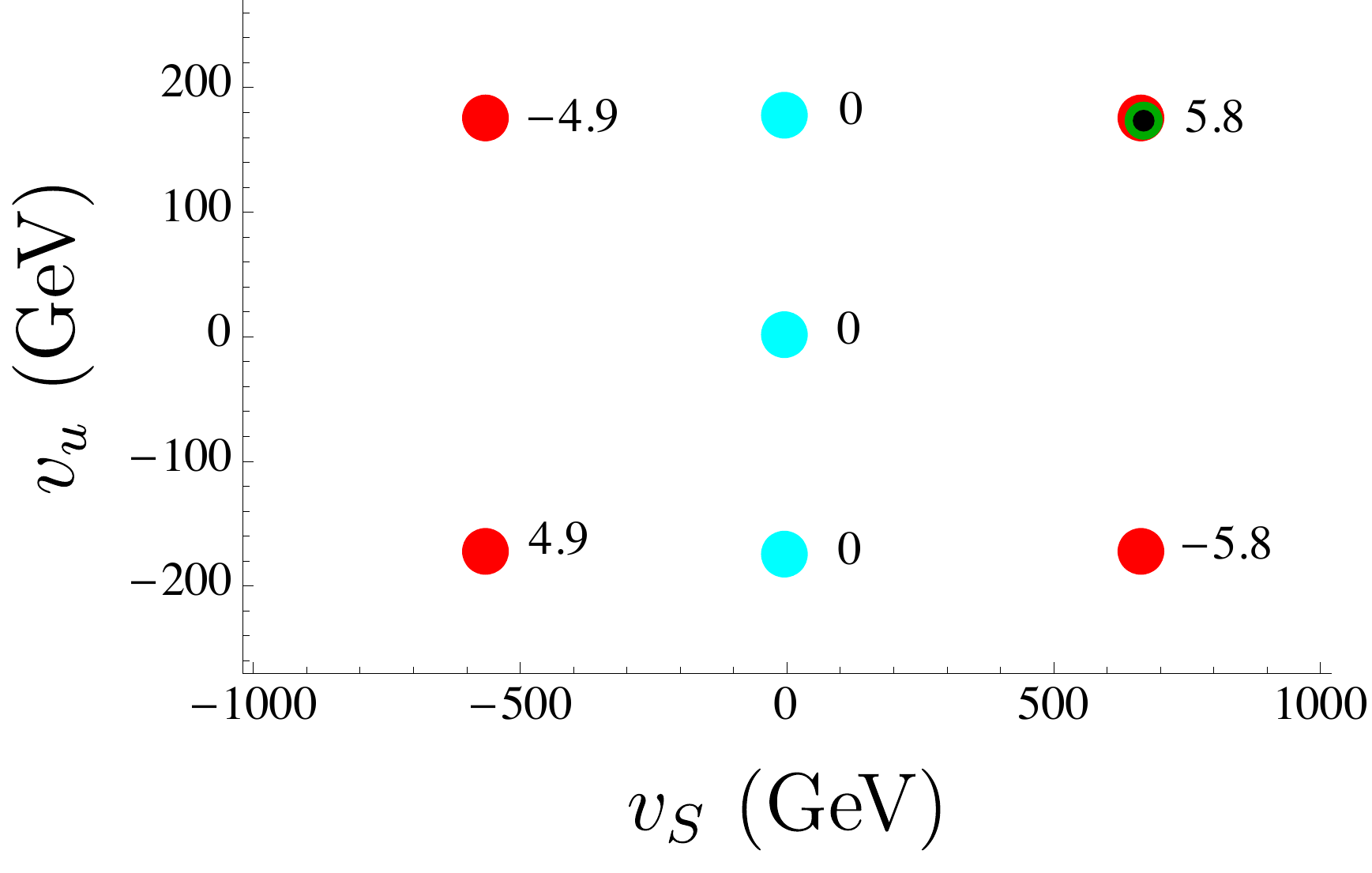}
\vskip 25pt
\includegraphics[height=0.22\textheight, width=0.45\textwidth]{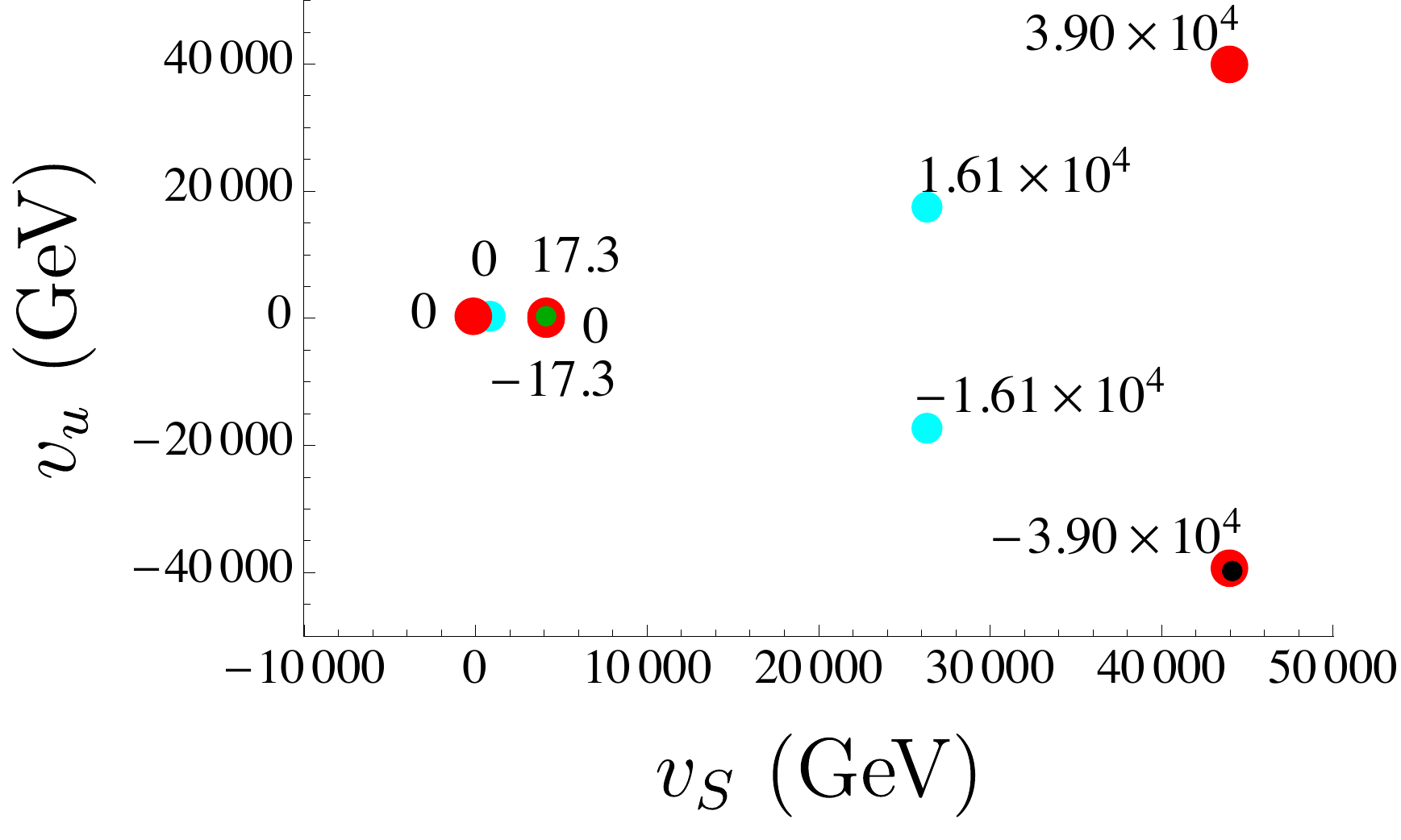}
\hskip 25pt
\includegraphics[height=0.22\textheight, width=0.45\textwidth]{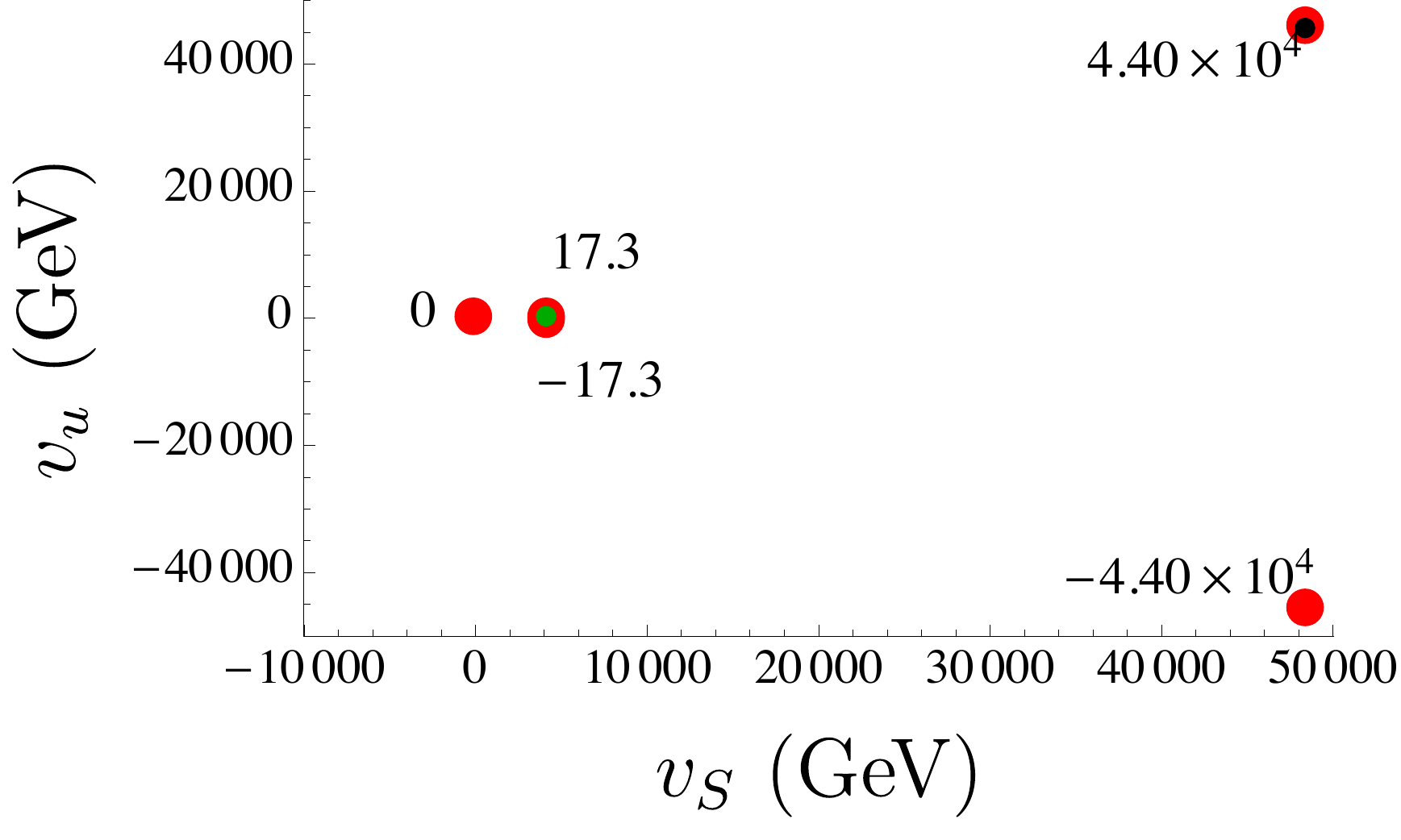}
\caption{Locations of all the extrema (in the $S$-$\higgsu$-$\higgsd$ space) found 
in the MSSM-like scenario with a singlino-like (top panel) and higgsino-like 
(bottom panel) LSP and with the tree-level (left panel) and the 1-loop corrected 
(right panel) potential. Red bullets represent (local) minima, black-centered
red bullets indicate global minimum, green-centered red bullets stand for DSB
minimum and bullets in cyan point to saddle points of the potential.
Numbers alongside the bullets indicate the corresponding magnitudes of $\higgsd$
(in GeV).}
\label{fig:mssm-bullet}
\end{figure}

\begin{figure}[htb!]
\centering	
\includegraphics[height=0.22\textheight, width=0.45\textwidth]{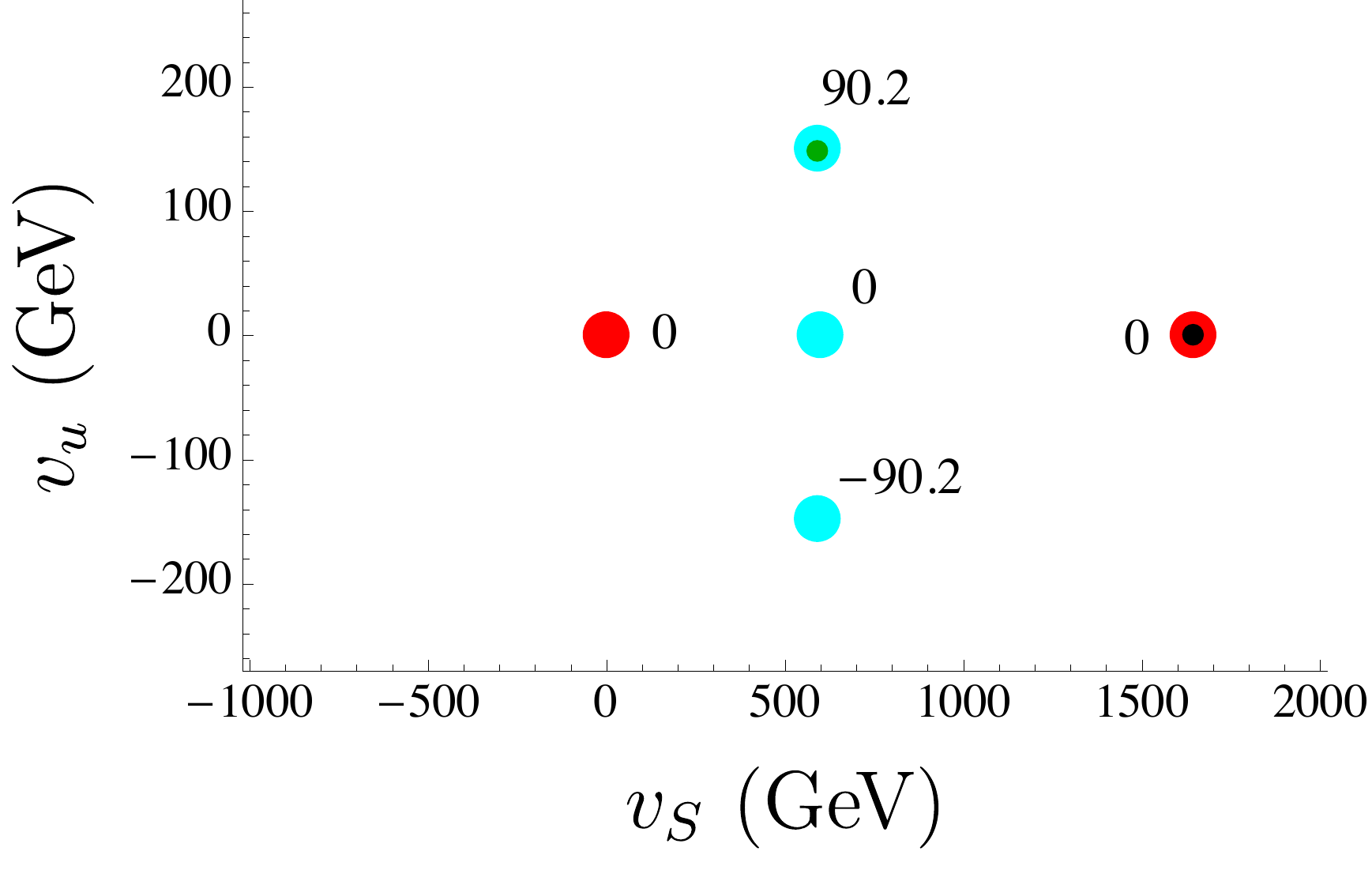}
\hskip 25pt
\includegraphics[height=0.22\textheight, width=0.45\textwidth]{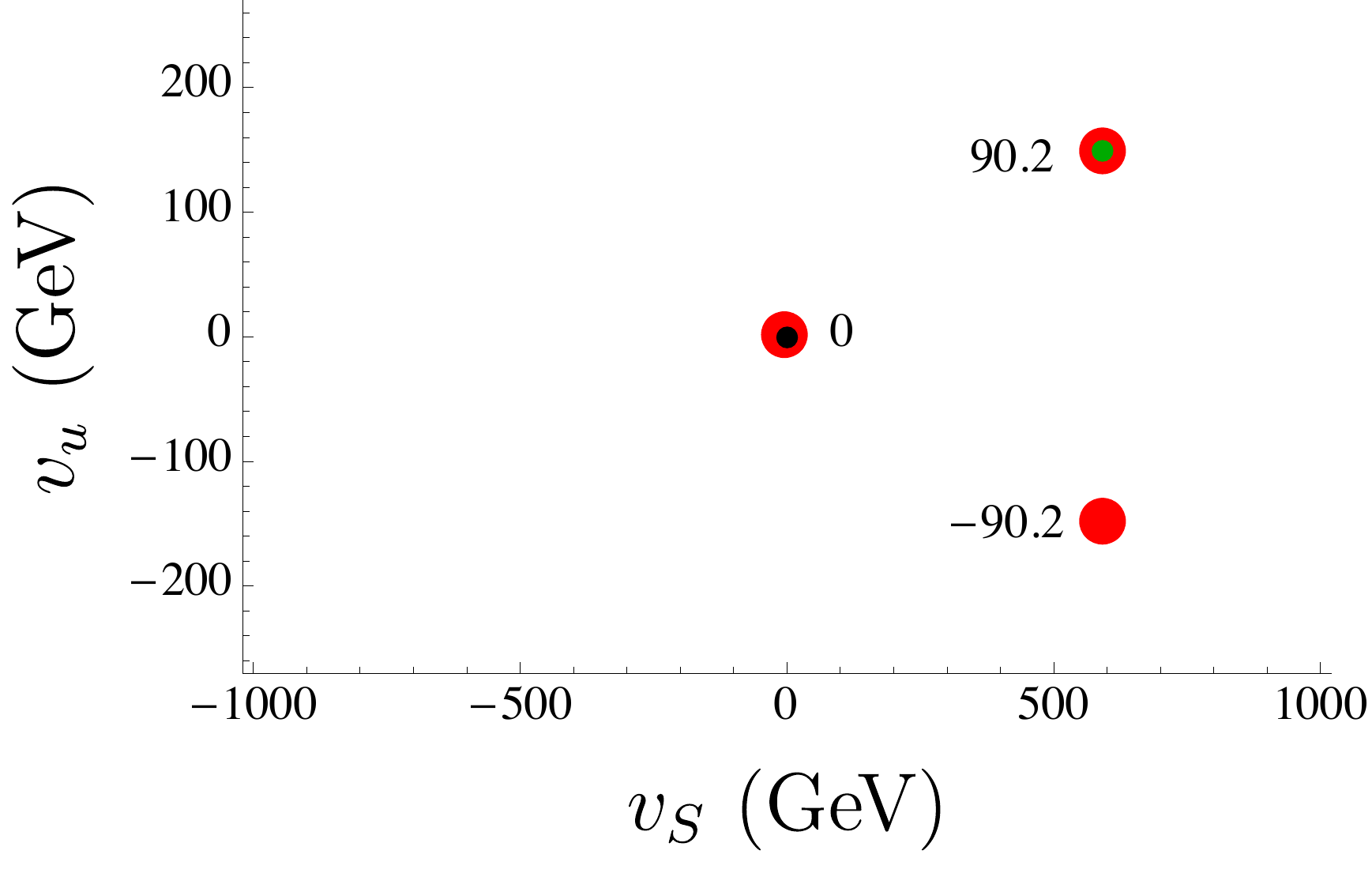}
\vskip 25pt
\includegraphics[height=0.22\textheight, width=0.45\textwidth]{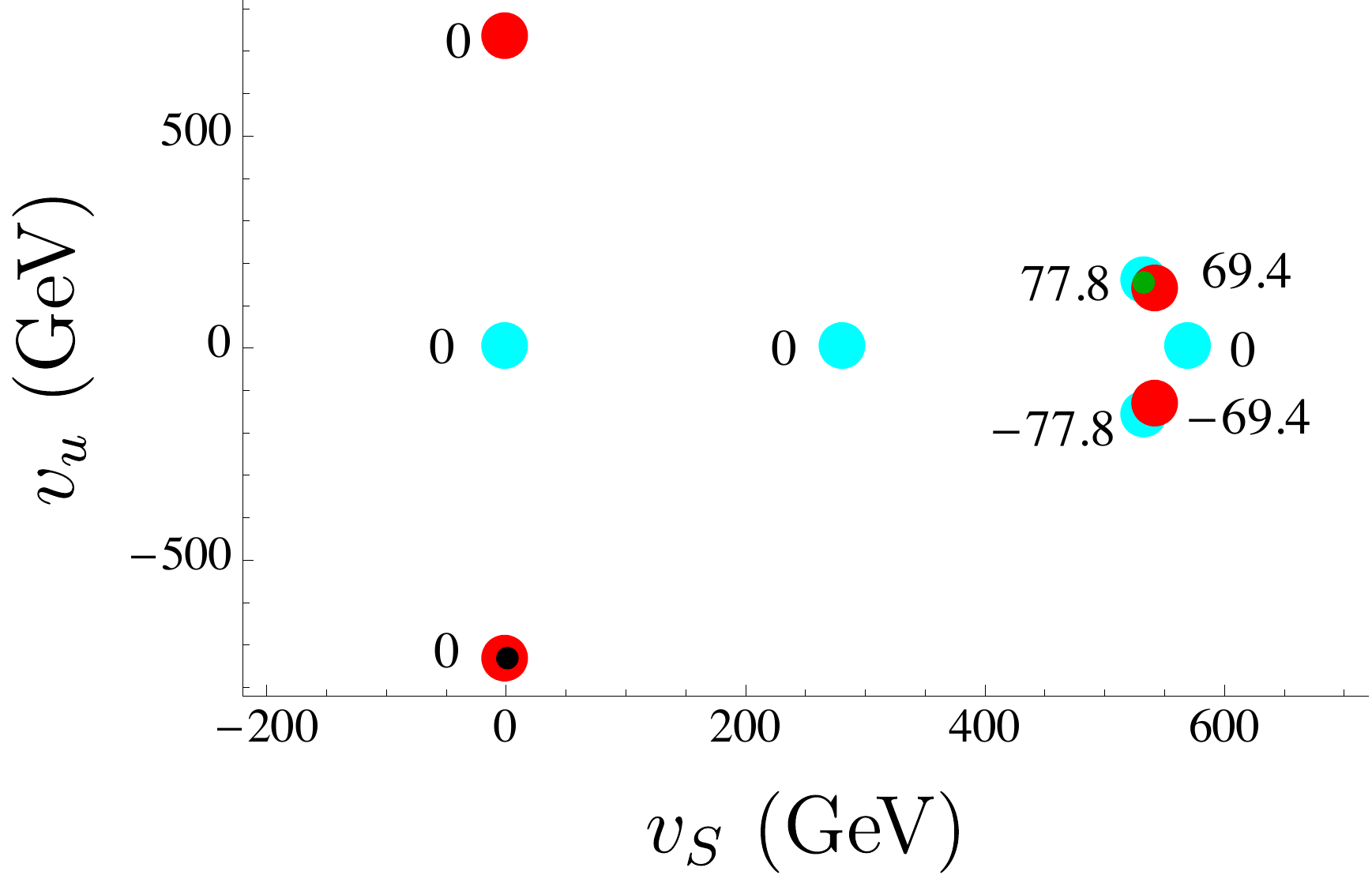}
\hskip 25pt
\includegraphics[height=0.22\textheight, width=0.45\textwidth]{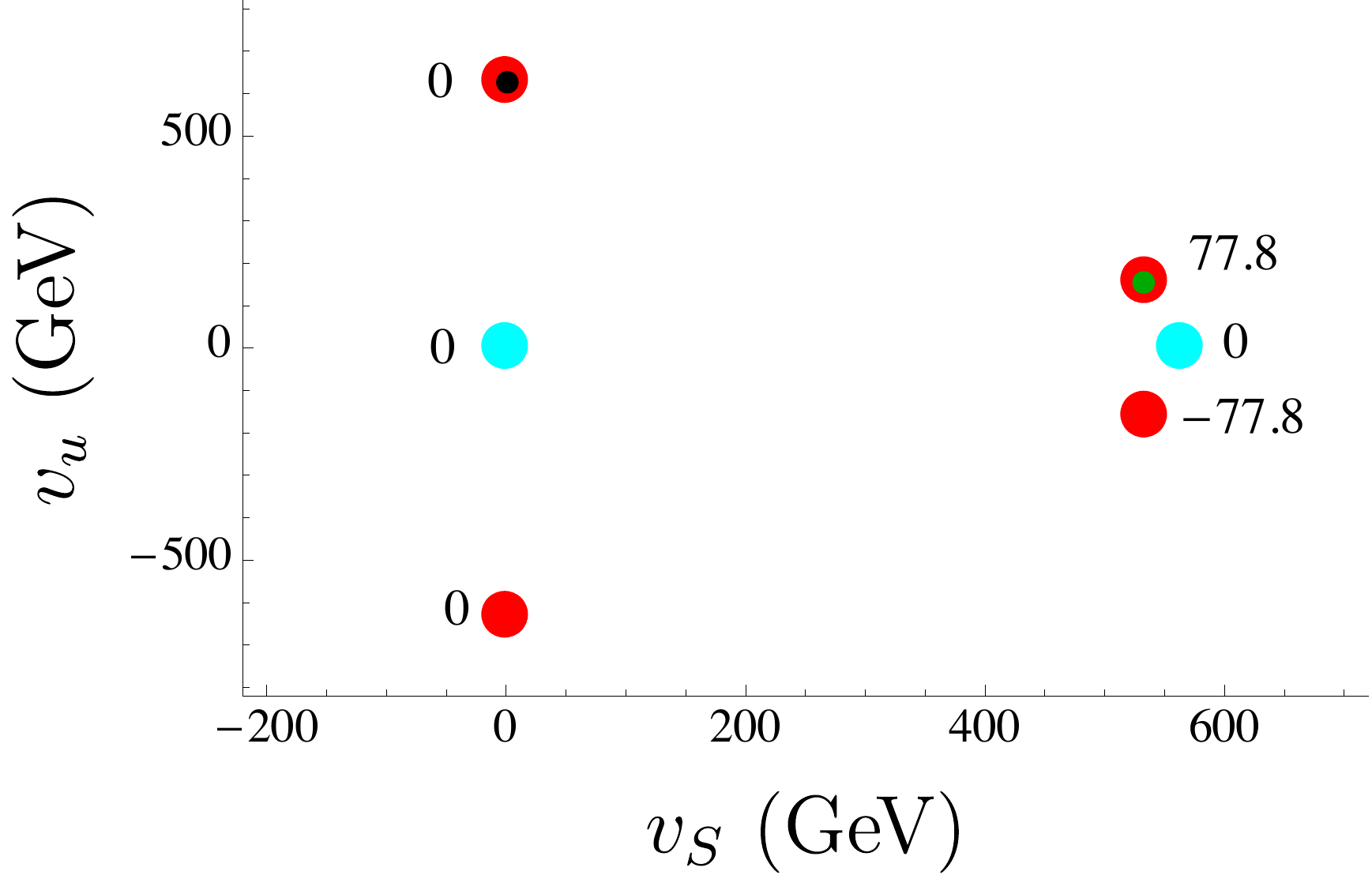}
\caption{Same as in figure \ref{fig:mssm-bullet} but for the NMSSM-like scenario.}
\label{fig:nmssm-bullet} 
\end{figure}

The complete picture pertaining to the locations of the extrema of the 
potential, in particular, those of the DSB minimum and the global minimum, is 
illustrated through figures \ref{fig:mssm-bullet} and \ref{fig:nmssm-bullet} for 
the MSSM- and the NMSSM-like cases, respectively. The top (bottom) rows 
correspond to the cases with a singlino- (higgsino)-like LSP. The left 
(right) columns stand for the situations with the tree-level (1-loop effective)
potential with non-vanishing $\higgsd$, $\higgsu$ and `$S$'. All three fields are 
allowed to vary simultaneously. Coordinates of the bullets point to the locations
 of these extrema in the $S-\higgsu$ plane. Values assumed by $\higgsd$ are 
indicated (in GeV) alongside the bullets. Following color convention is adopted 
to indicate the nature of the extrema: red for minima, black within red for the 
global minimum, green within red for the DSB minima and cyan for the saddle 
points. Thus, a bullet having red, green and black all appearing in it represents 
a DSB minimum which, at the same time, is the global minimum of the potential. 
The following general remarks/observations would be in place.
%
%
\begin{itemize}
\item Search for extrema of the 1-loop effective potential is performed in
{\tt Mathematica} with the tree-level extrema as the guess values. 
\item An exact symmetry under $\higgsd, \higgsu \to -\higgsd, -\higgsu$ is
apparent at the tree level which is mostly holding with the 1-loop effective
potential.
\item There is no general correspondence among various extrema found for the
tree-level and the corrected potential, except for the DSB extrema for which 
the nature may get altered, though. For example, the DSB extremum in the 
NMSSM-like scenario with a singlino-like LSP (top, left plot of figure 
\ref{fig:nmssm-bullet}) appears to be a saddle point with tree-level potential. 
This becomes a likely global minimum when the corrected potential is used.
\end{itemize}
%

It may also be noted that figure \ref{fig:mssm-bullet} 
(figure \ref{fig:nmssm-bullet}) connects to figure \ref{fig:pot-profiles-mssm}
(figure  \ref{fig:pot-profiles-nmssm}) via the locations of the DSB minima.
On the other hand, the right plots of the former set of figures connect to 
the plots in the top (bottom) row of figure \ref{fig:contours} via the 
coordinates of the global minima which are not necessarily the DSB minima. 
These connections are explicitly indicated on individual plots of figures 
\ref{fig:mssm-bullet} and \ref{fig:nmssm-bullet}. In particular, it 
is noted that in the MSSM-like scenario with a higgsino-like LSP (bottom, right
plot of figure \ref{fig:mssm-bullet} and top, right plot in figure
\ref{fig:contours}) the global minimum occurs for large values of `$S$' with
$\higgsd \approx \higgsu$, i.e., approximately along the $D$-flat direction.
It would be interesting to see if our subsequent analysis using \veva
~corroborates this fact along with other findings from our studies so far.

Thus, the left plots in
figure \ref{fig:mssm-bullet} (figure \ref{fig:nmssm-bullet}) are to be compared
with the middle plots of figure \ref{fig:pot-profiles-mssm} 
(figure \ref{fig:pot-profiles-nmssm}). Similarly, the right plots in figure
\ref{fig:mssm-bullet} (figure \ref{fig:nmssm-bullet}) are to be compared with
the rightmost plots of figure \ref{fig:pot-profiles-mssm} 
(figure \ref{fig:pot-profiles-nmssm}). The latter set of four plots (having the
radiative effects included) are also to be compared with the corresponding ones
from a similar set in figure \ref{fig:contours}.
%
%
%
%
%
%

The MSSM-like scenario is broadly realized by choosing small values of $\lambda$
($\lesssim 0.1$). However, in view of the possibility discussed in footnote 
\ref{foot:sd-mixing}, we allow for a little wider range of $\lambda$ spanning 
over an order of magnitude that also includes somewhat large values of $\lambda$. 
Thus, scan is done over the following ranges of the relevant set of parameters:
%
%
\begin{gather}
10^{-4} < \lambda <0.75, \quad 
10^{-3}< |\kappa|  < 0.75, \quad
|\alambda| < 7.5 \, \mathrm{TeV}, \quad 
|\akappa| < 7.5 \, \mathrm{TeV}, \nonumber  \\ 
\quad |\mueff| < 500 \, \mathrm{GeV}, \quad 1< \tan\beta < 50  \quad .
\label{eq:mssm-like-ranges}
\end{gather}
%
%
Note that for a smaller $\lambda$ ($\sim 0.01$), to obtain
$\mueff$ in the phenomenologically right ballpark ($|\mueff| \sim \cal{O}$(100 GeV)), 
$\vs={\mueff \over \lambda}$ could turn out to be rather large 
($\gtrsim \cal{O}$(10 TeV)). Since for the DSB minimum the \vevs ~$\vd$ and
$\vu$ cannot be too large ($\lesssim \cal{O}$(100 GeV)), the potential in the
vicinity of the same can be approximated as a function of only one scalar field, 
`$S$'. Thus, the previous analysis in section \ref{subsubsec:single-field-potential} 
based on equation \ref{eq:higgs-pot-large-s} could be largely applicable for the 
MSSM-like scenario. 

On the other hand, we confine the NMSSM-like scenario in a region with relatively
larger values of $\lambda$, as it should be by its definition. Thus, we set the 
following ranges for scanning the parameter space:
\begin{gather}
0.5< \lambda <0.75, \quad 
10^{-3} < |\kappa|  < 0.75, \quad
|A_{\lambda}| < 2.5 \, \mathrm{TeV}, \quad
|A_{\kappa}| < 2.5 \, \mathrm{TeV} \nonumber \\
\quad |\mueff| < 500 \, \mathrm{GeV}, \quad 1< \tan\beta < 5 \quad .
\label{eq:nmssm-like-ranges}
\end{gather}
%
%
For the set of various fixed input parameters used in this part of the analysis,
the reader is referred to the caption of table \ref{tab:spheno-math}.
%
Note that ranges of the NMSSM parameters that are varied in these two
broad scenarios are rather different. These are primarily so to ensure acceptable
(non-tachyonic) values for various Higgs masses.
%
%
\begin{figure}[t]
\centering
\includegraphics[height=0.20\textheight, width=0.45\columnwidth , clip]{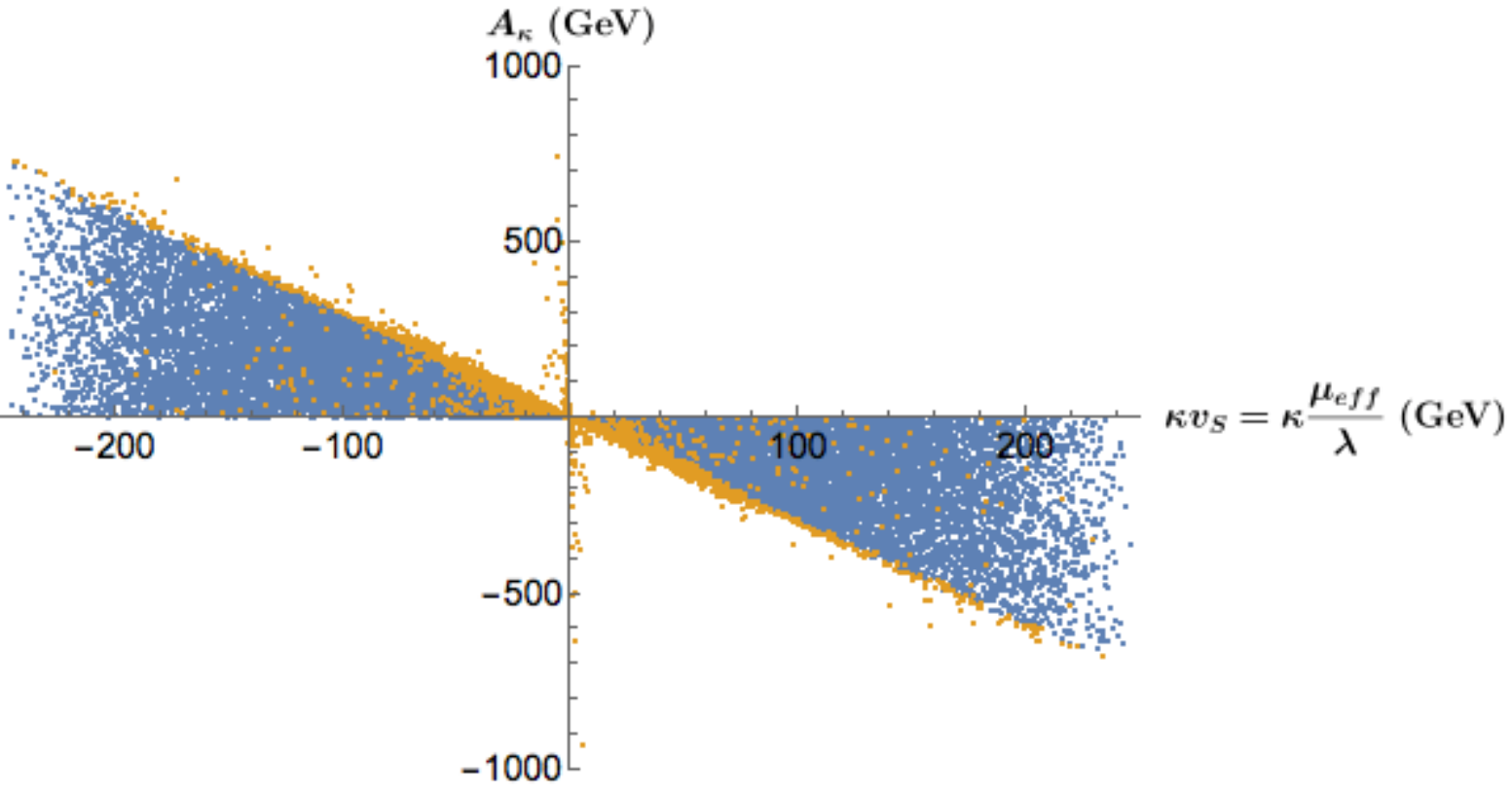}
\hskip 0.3 in
\includegraphics[height=0.20\textheight, width=0.45\columnwidth , clip]{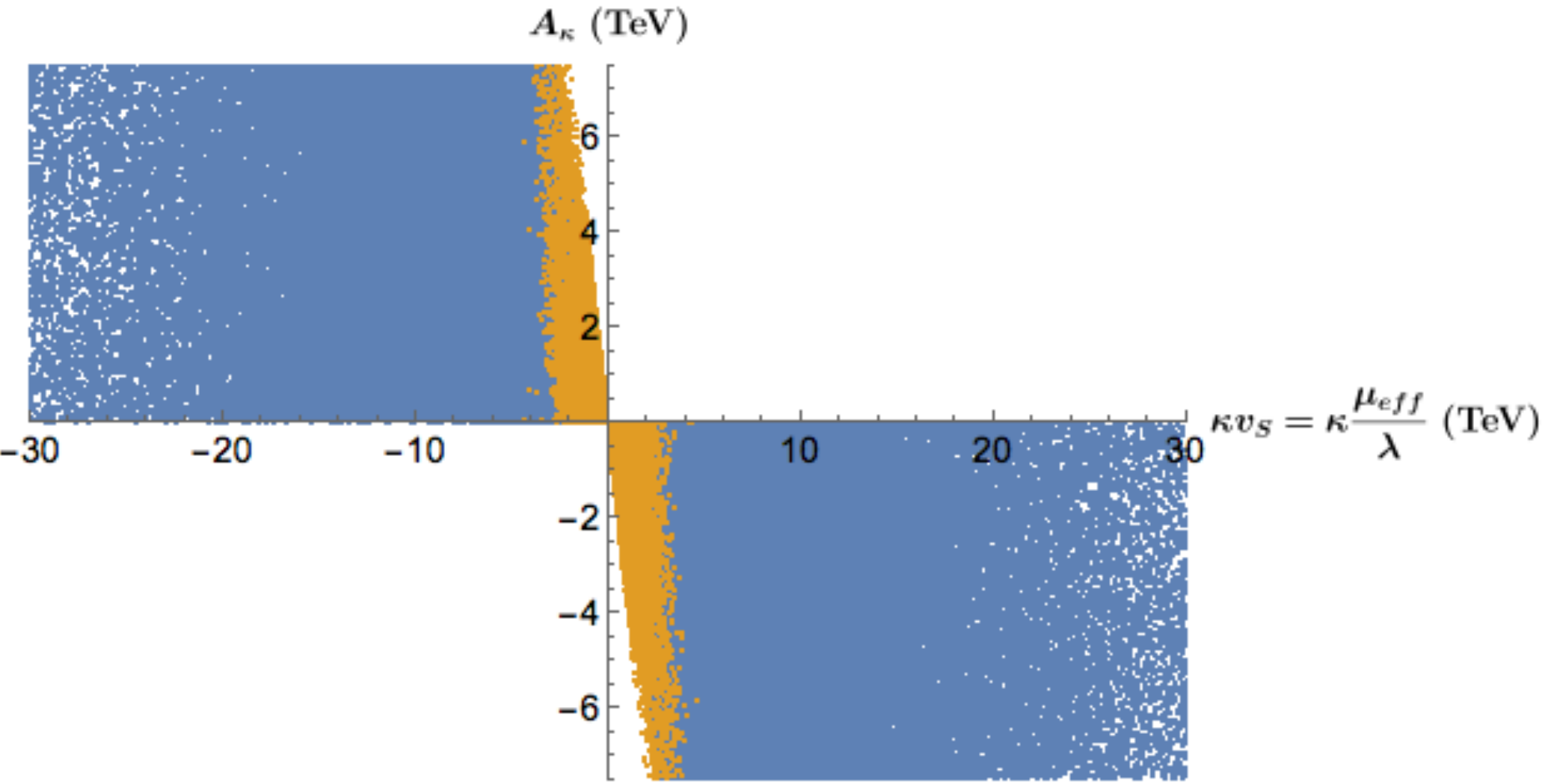}
\vskip 0.4 in
\includegraphics[height=0.20\textheight, width=0.45\columnwidth , clip]{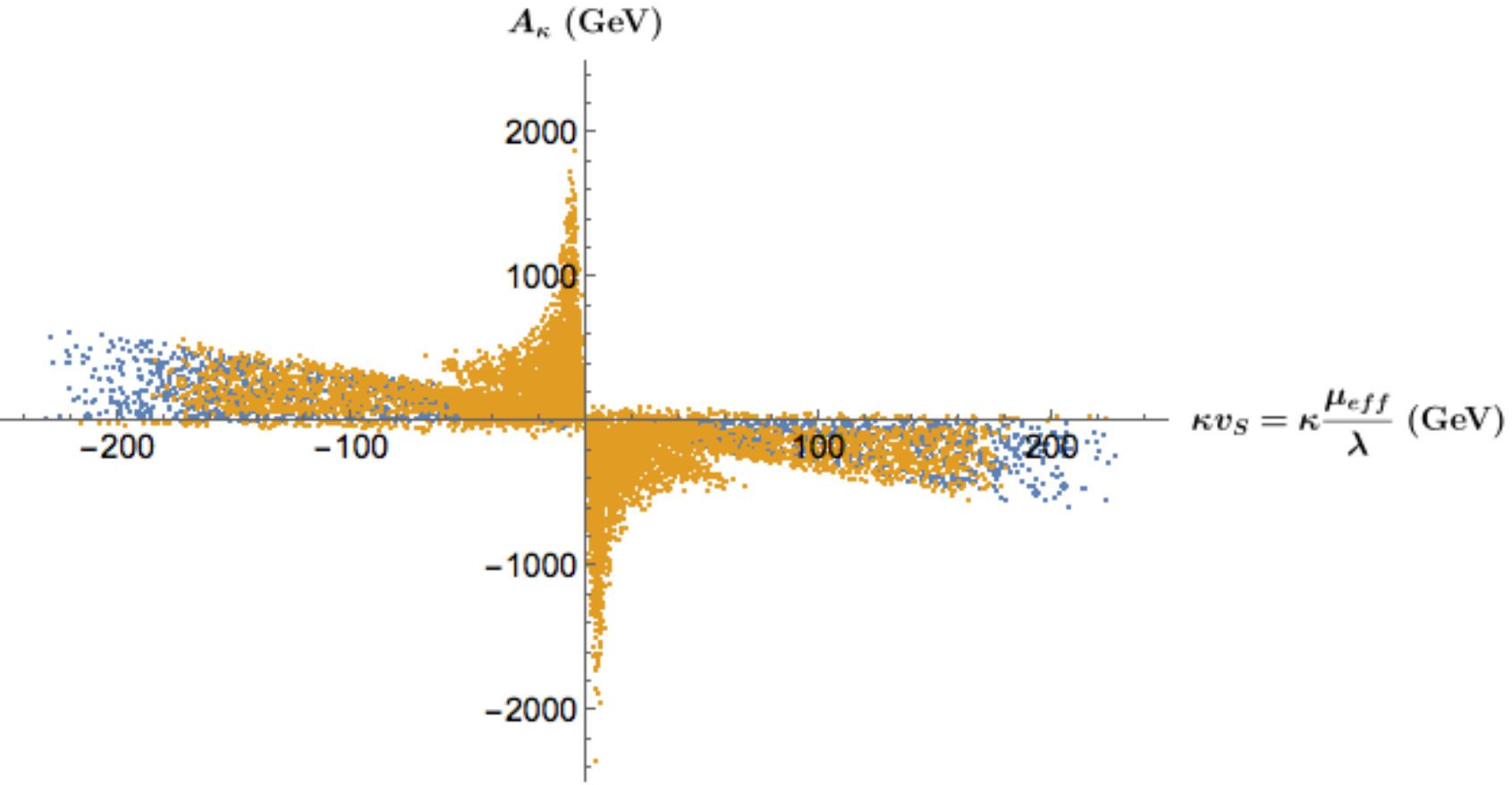}
\hskip 0.3 in
\includegraphics[height=0.20\textheight, width=0.45\columnwidth , clip]{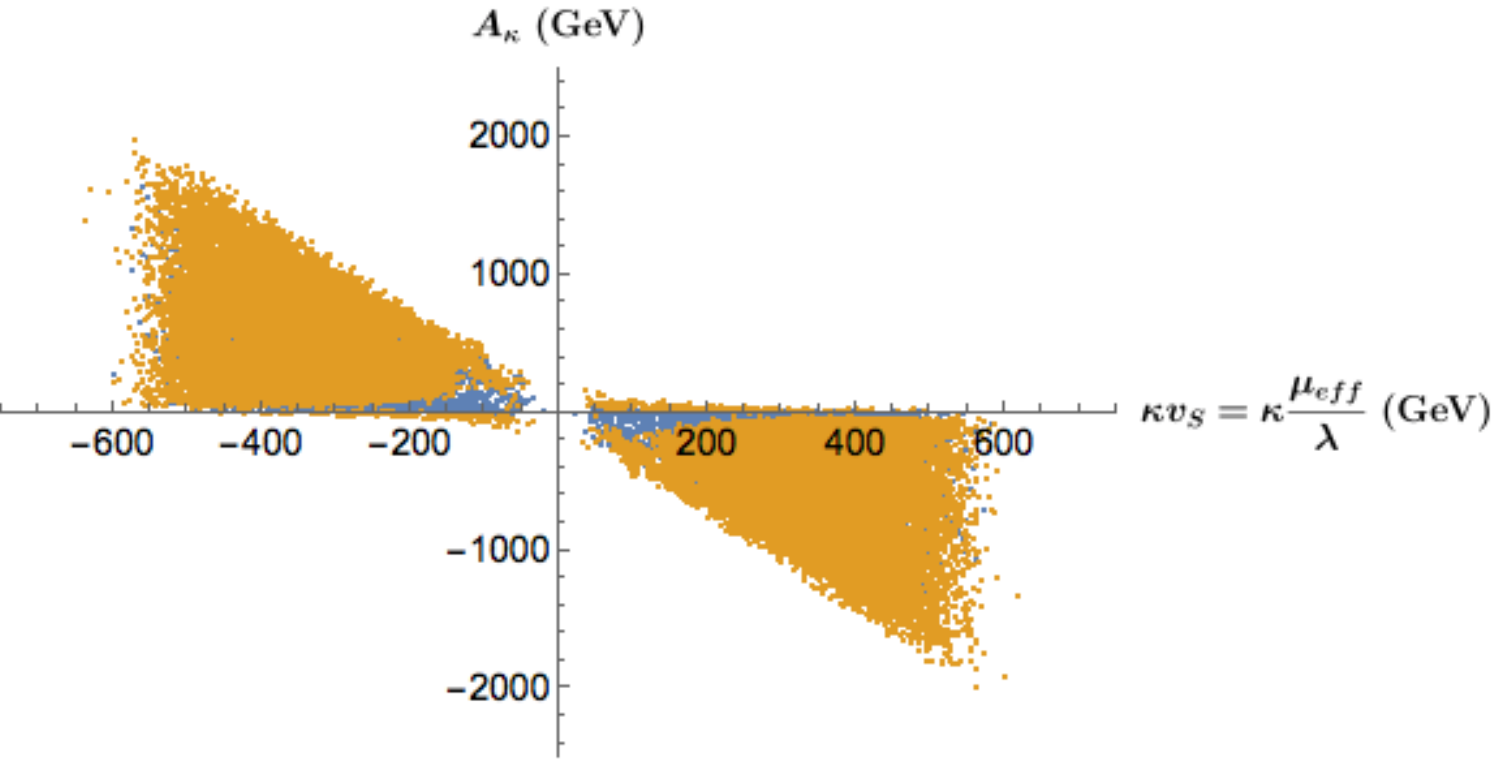}
\caption{Scatter plot from the {\tt Mathematica}-based analysis showing allowed 
regions in the $\kappa \vs - \akappa$ plane compatible with stability (in blue)
and metastability (in golden yellow) of the DSB vacuum and also with 
non-tachyonic neutral scalar states. See text for the ranges of various input 
parameters used in the scan. The upper (lower) panel stands for the
MSSM-like (NMSSM-like) scenario. The left (right) column presents the case
with a singlino-like (higgsino-like) LSP.}
\label{fig:ks-akappa-math}
\end{figure}
%
%

We now perform random scans of the NMSSM-parameter space with the 
{\tt Mathematica}-based analysis routine we develop. Scans are done for the two 
scenarios discussed above and over the respective sets of ranges for various 
NMSSM inputs as indicated in equations \ref{eq:mssm-like-ranges} and 
\ref{eq:nmssm-like-ranges}. While scanning, we confine ourselves to the 
tree-level potential since inclusion of radiative correction makes the scan 
prohibitively slow.

In figure \ref{fig:ks-akappa-math} we present the results of the scan in the
$\kappa \vs - \akappa$ plane in the form of scatter plots. The upper (lower)
panel illustrates the situation in the MSSM-like (NMSSM-like) case while the
left (right) column stands for the situation where the LSP is singlino-like
(higgsino-like). Over the regions in blue the DSB minimum emerges as the global
minimum and hence stable. The golden-yellow region stands
for a situation where the DSB minimum is the local minimum of the potential and
hence metastable. Among these plots, the upper, left one representing the
MSSM-like case and having a singlino-like LSP, can be taken a particular note of.
This approximately corresponds to the setup with the single field (`$S$') potential
elaborated in section \ref{subsubsec:single-field-potential} and is illustrated
in figure \ref{fig:allowed-ranges-consolidated}. It is rather convincing to find
how decently the regions obtained through a purely analytical treatment
(see figure \ref{fig:allowed-ranges-consolidated}) are now reproduced in this
scan which is semi-analytic in nature. In general, figure \ref{fig:ks-akappa-math}
would serve as a useful reference for a subsequent analysis taken up in section
\ref{subsec:arbitrary-directions-veva} using \veva.
%
\subsection{False vacua along arbitrary field directions: a \veva-based study}
\label{subsec:arbitrary-directions-veva}
%
In this subsection we discuss issues pertaining to the stability of the DSB 
vacuum in the $Z_3$-symmetric NMSSM by subjecting its parameter space to a
thorough analysis via \veva ~\cite{Camargo-Molina:2013qva}, a state-of-the-art
package dedicated for the purpose. Its working principle has been 
briefly outlined in section \ref{subsec:veva}.

An analysis of stability of the NMSSM vacuum using \veva ~would take us beyond 
the {\tt Mathematica}-based analysis we presented in section 
\ref{subsec:arbitrary-directions-semi}
in two important aspects. As pointed out in section \ref{subsec:veva},
\veva ~includes both quantum and thermal corrections to the 
potential (which is proven to be crucial 
for the purpose) and computes the tunneling time of the DSB vacuum to a deeper 
(panic) minimum.
Furthermore, such a study can be seamlessly 
interfaced to other external packages for an in-flight analysis of the viability 
of a given point in the 
parameter space against other experimental constraints. These make the 
analysis all the more realistic. 
In the following, we outline the broad categories into which \veva ~classifies
the fate of the DSB vacuum. These are later exploited in our presentation.
%
\begin{itemize}
\item {\it Stable:} when the DSB vacuum is the global minimum of the potential;
\item {\it Metastable but long-lived:} when there is a minimum deeper (global 
      minimum) than the DSB vacuum (local minimum) but the decay time of the 
      DSB vacuum to this is large enough in reference to 
      the age of the Universe (not only at zero temperature but also when
      thermal effects are included) so that it could safely be considered 
      viable;
\item {\it Metastable but short-lived at zero temperature:} when the decay time 
      of the DSB vacuum to such a deeper minimum is short enough under 
      a rapid quantum tunneling at zero temperature thus making the former 
      unstable and thus, inviable; 
\item {\it Metastable but short-lived only at finite temperatures:} when the 
      instability of the DSB vacuum is triggered only at finite temperatures. 
\end{itemize}
%
In table \ref{tab:color-number-codes} we collect the corresponding number codes
used by \veva ~to flag each of these situations along with the color codes we
use for these in some of the subsequent scatter plots.
%
%
\begin{table}[hbt]
\begin{center}
{\footnotesize
  \begin{tabular}{|c|c|c|l|}
   \hline
Stability/Longevity of &             & Viability of    & Color-code \\
the DSB vacuum         & \veva ~code & the DSB vacuum  &            \\
   \hline
   \hline
Stable                            &    1       & Viable     &  \cgreen{$\bullet$ Green}  \\ 
   \hline
Metastable but long-lived         &    0       & Viable     &  \cblue{$\bullet$ Blue}   \\ 
   \hline
Metastable but short-lived        &   -1       & Not Viable &  $\bullet$ Black  \\ 
            (tunneling at zero temperature) &  &            &         \\ 
   \hline
Metastable but short-lived        &   -2       & Not Viable &  \cred{$\bullet$ Red}    \\ 
           (tunneling at finite temperature) & &            &         \\ 
   \hline
  \end{tabular}
}
\end{center}
\caption{\veva ~numeric codes flagging specific status and viability of the DSB 
vacuum and the corresponding color codes used subsequently in this work.}
\label{tab:color-number-codes}
\end{table}
%
%

Scans are done over the same set of parameters, over the same ranges 
(see expressions \ref{eq:mssm-like-ranges} and \ref{eq:nmssm-like-ranges} 
and the caption of table \ref{tab:spheno-math}) and referring 
to the same set of scenarios as have been adopted for the {\tt Mathematica}-based scan 
presented in section \ref{subsec:arbitrary-directions-semi}.
SLHA \cite{Allanach:2008qq} files containing the spectrum and other important 
information obtained from {\tt SARAH v4.5.8}-generated \cite{Staub:2013tta, Staub:2015kfa}
{\tt SPheno v3.3.8} \cite{Porod:2003um, Porod:2011nf} are fed into \veva.

We also keep track of the following issues which are of phenomenological 
importance. One of the $CP$-even Higgs states is required to have a mass within 
the range $122$-$128$ GeV and should behave like the SM-like Higgs boson. 
This is ensured by subjecting the analysis to treatments by {\tt HiggsBounds v4.3.1}
\cite{Bechtle:2013wla} and {\tt HiggsSignals v1.4.0} \cite{Bechtle:2013xfa} in parallel. 
Furthermore, we subject the analysis to important current bounds from the
flavor sector in the form of $B \to X_s \gamma$, allowing for 
the range $2.77 \times 10^{-4} \leq \mathrm{BR}[B \to X_s \gamma] \leq 4.09 \times 10^{-4}$ 
at $3\sigma$ level \cite{Amhis:2012bh} via {\tt SPheno}, from the dark matter sector (in 
the form of a $5\sigma$ upper bound on its relic density which is taken to be 
$\Omega_\mathrm{CDM} h^2 <0.127$ \cite{Ade:2015xua}) via {\tt micrOMEGAs v4.2.5} 
\cite{Belanger:2014vza} and to the lower bound of the chargino mass of 103 GeV 
\cite{lepsusy} which is close to the kinematic threshold of the LEP experiments.
However, unless specifically  mentioned, the figures we present
subsequently are obtained without imposing these constraints (except for
respecting the bound on the chargino mass and requiring a Higgs boson with a
mass in the range mentioned above). This is to ensure
that the basic features and constraints obtained from the bare analysis of the
possible vacua first get clarified.
%
\subsubsection{Scanning of the MSSM-like scenario with \veva}
\label{subsubsec:mssm-like-veva}
%
To ensure an optimally large radiative correction to the mass of the SM-like
Higgs boson that defines the MSSM-like scenario in this work, we set the soft 
masses of all the sfermions at a somewhat large value of 3 TeV. The soft 
trilinear coupling $A_t$ in the top squark sector is also fixed at 3 TeV
while the same for the bottom ($A_b$) and the tau ($A_\tau$) sectors are set
to zero.

\begin{figure}[t]
\centering
\includegraphics[height=0.19\textheight, width=0.369\columnwidth , clip]{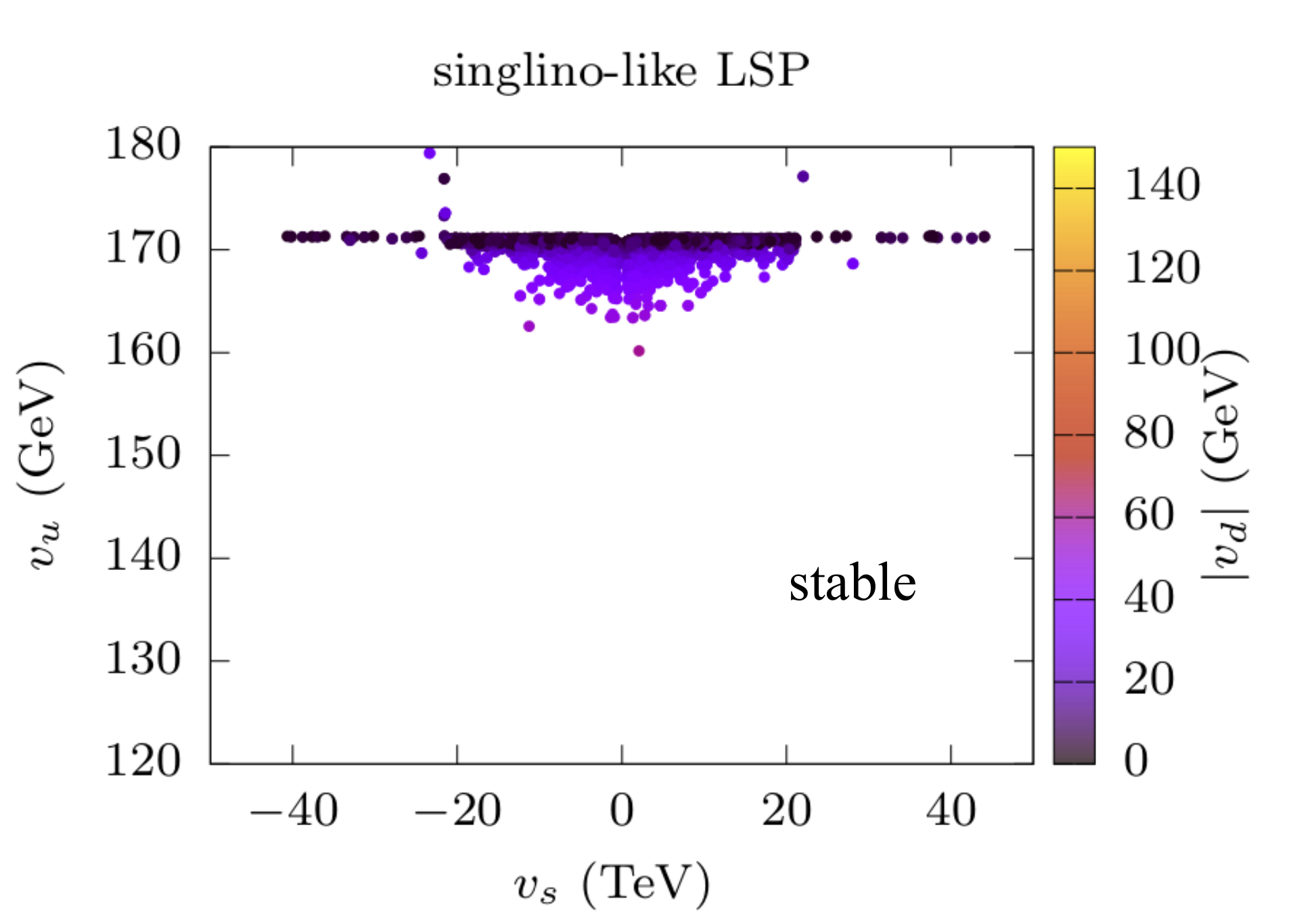}
\includegraphics[height=0.19\textheight, width=0.369\columnwidth , clip]{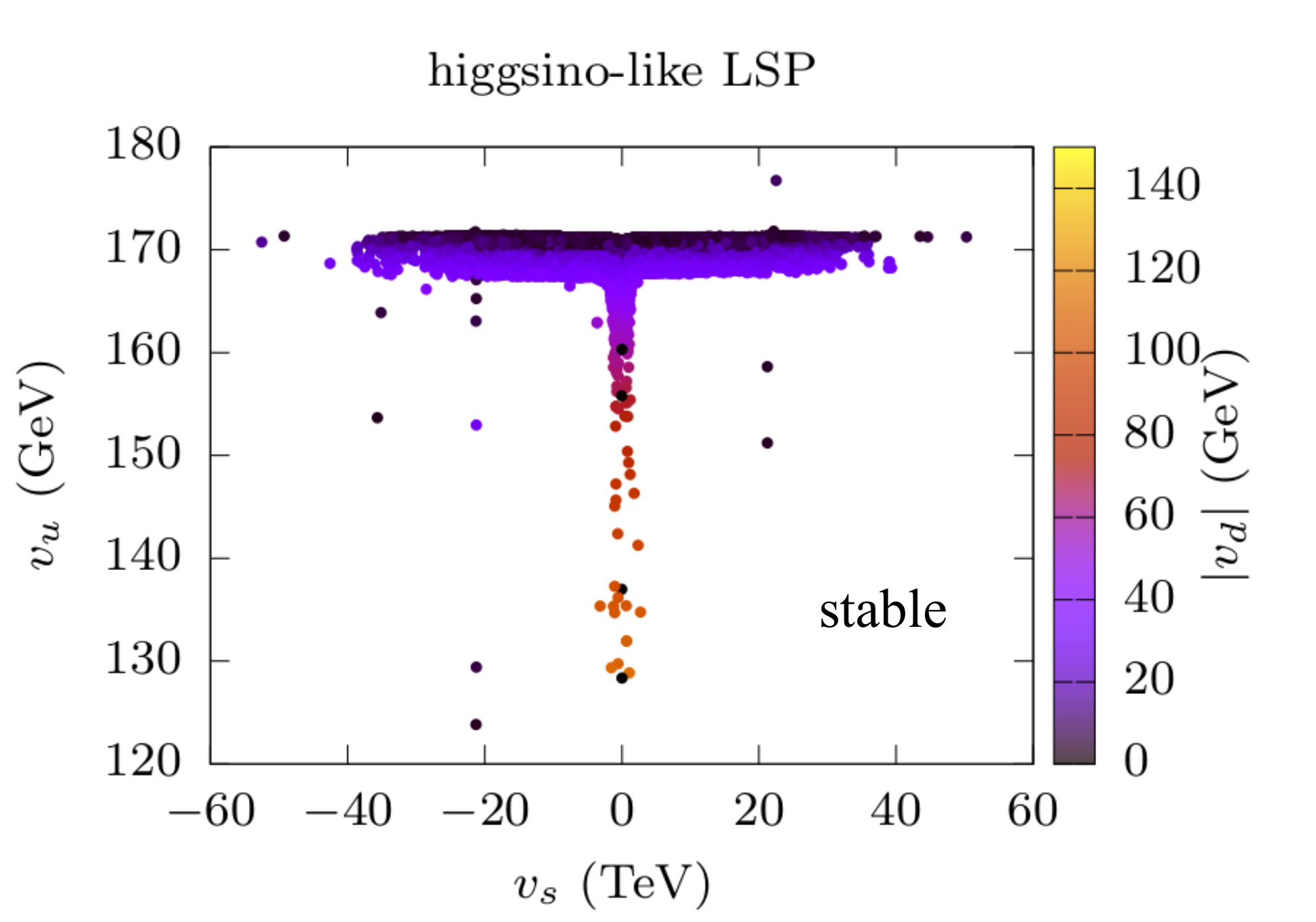}
\vskip 0.1 in
\includegraphics[height=0.16\textheight, width=0.369\columnwidth , clip]{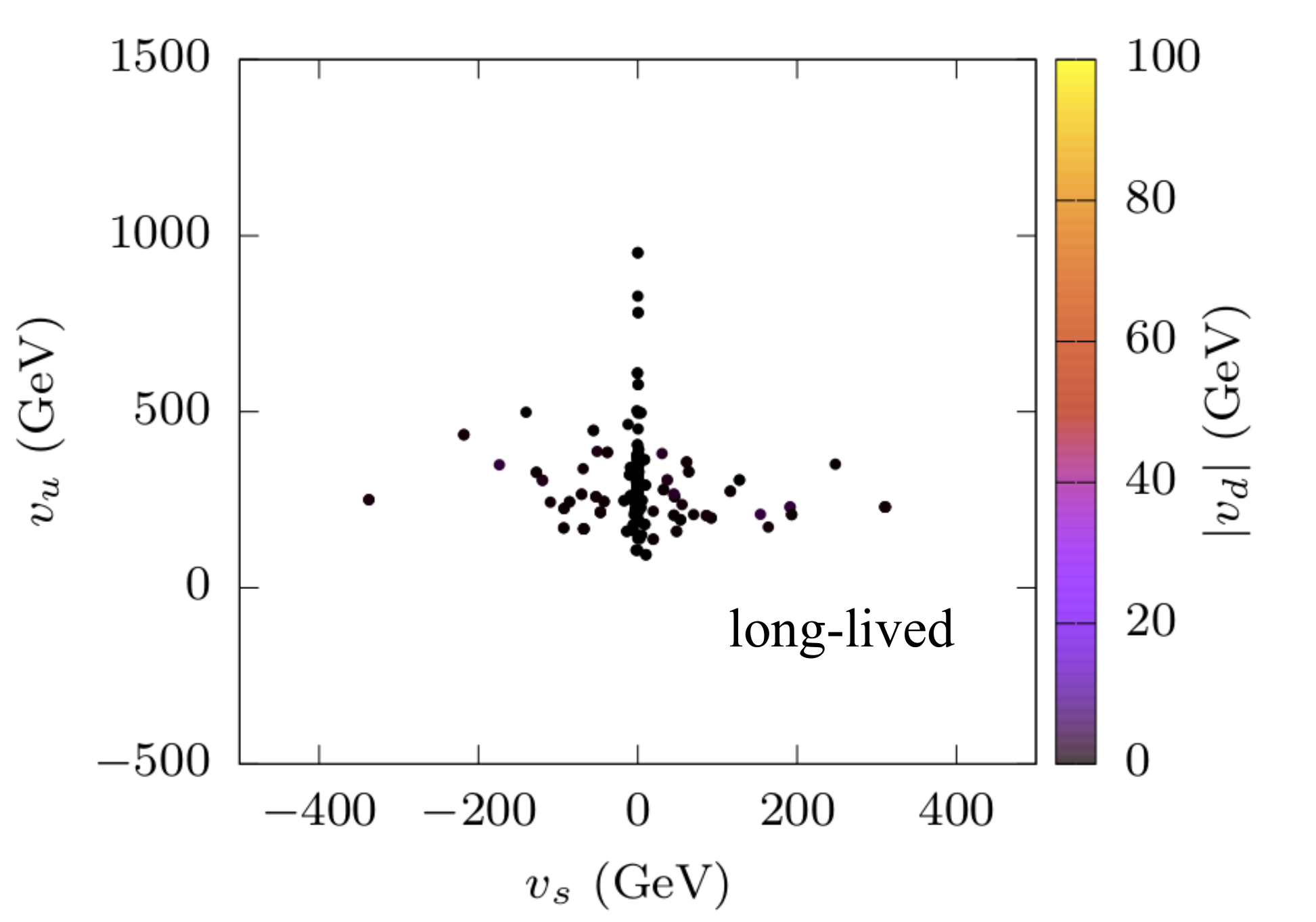}
\includegraphics[height=0.16\textheight, width=0.369\columnwidth , clip]{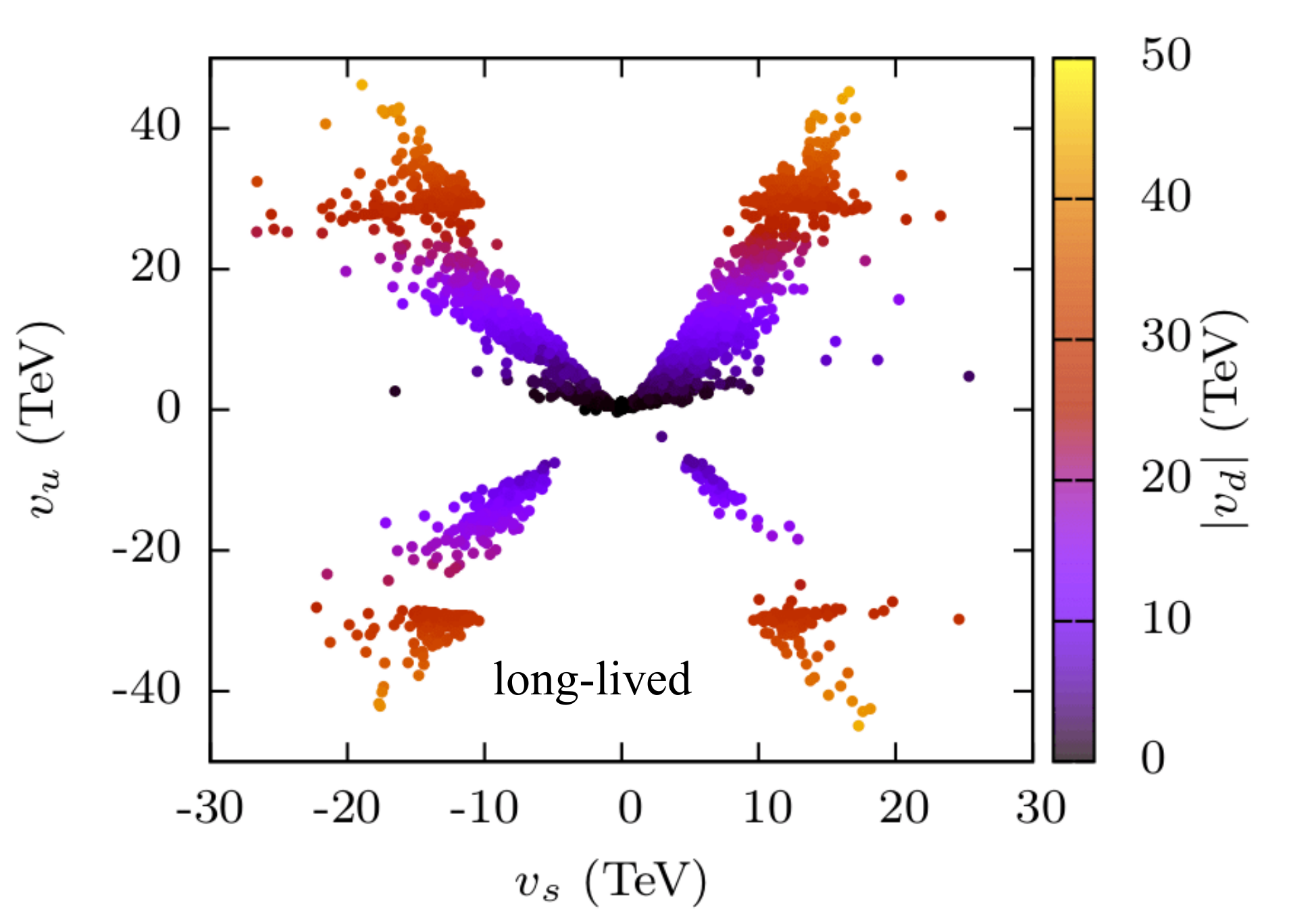}
\vskip 0.1 in
\includegraphics[height=0.16\textheight, width=0.369\columnwidth , clip]{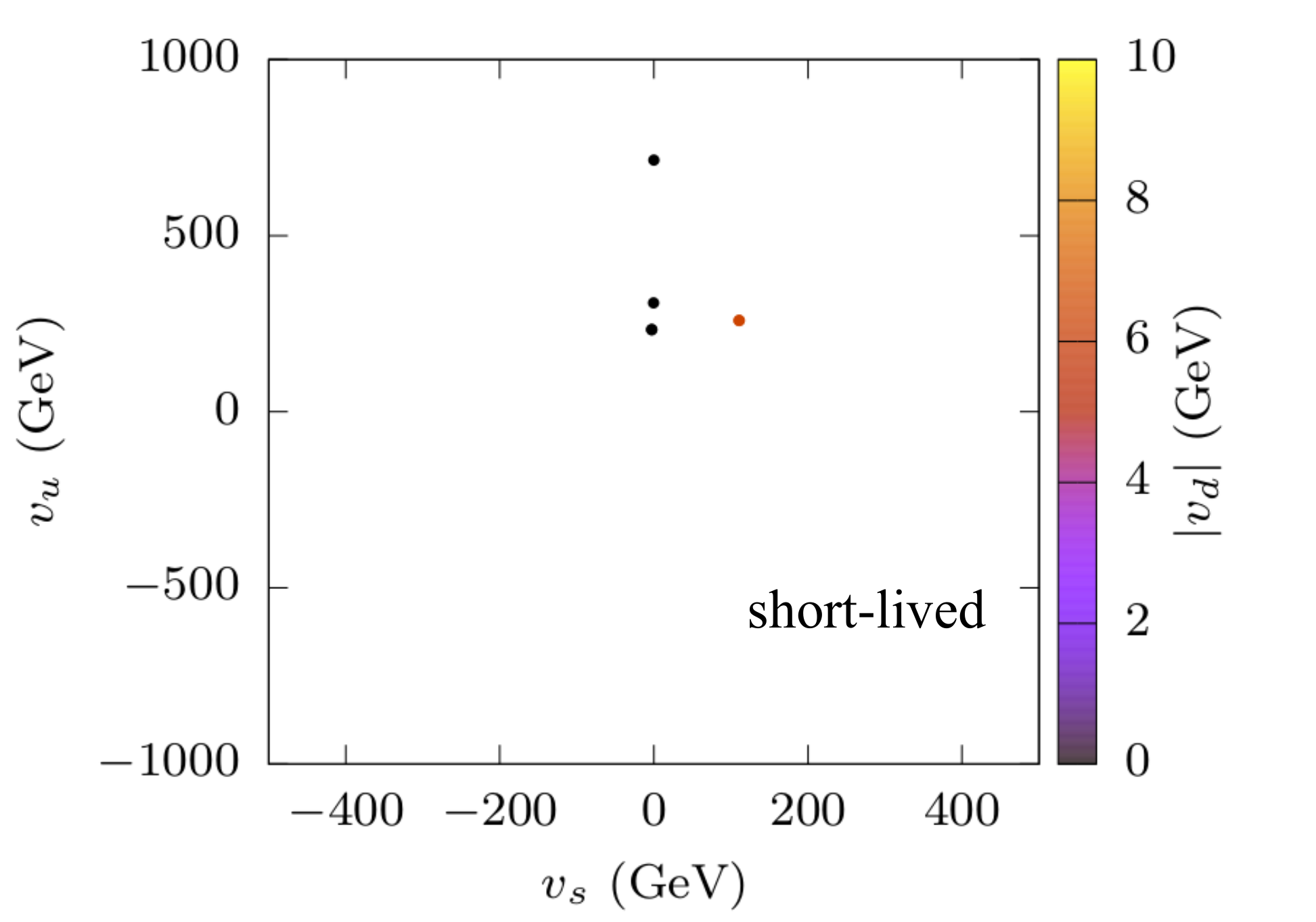}
\includegraphics[height=0.16\textheight, width=0.369\columnwidth , clip]{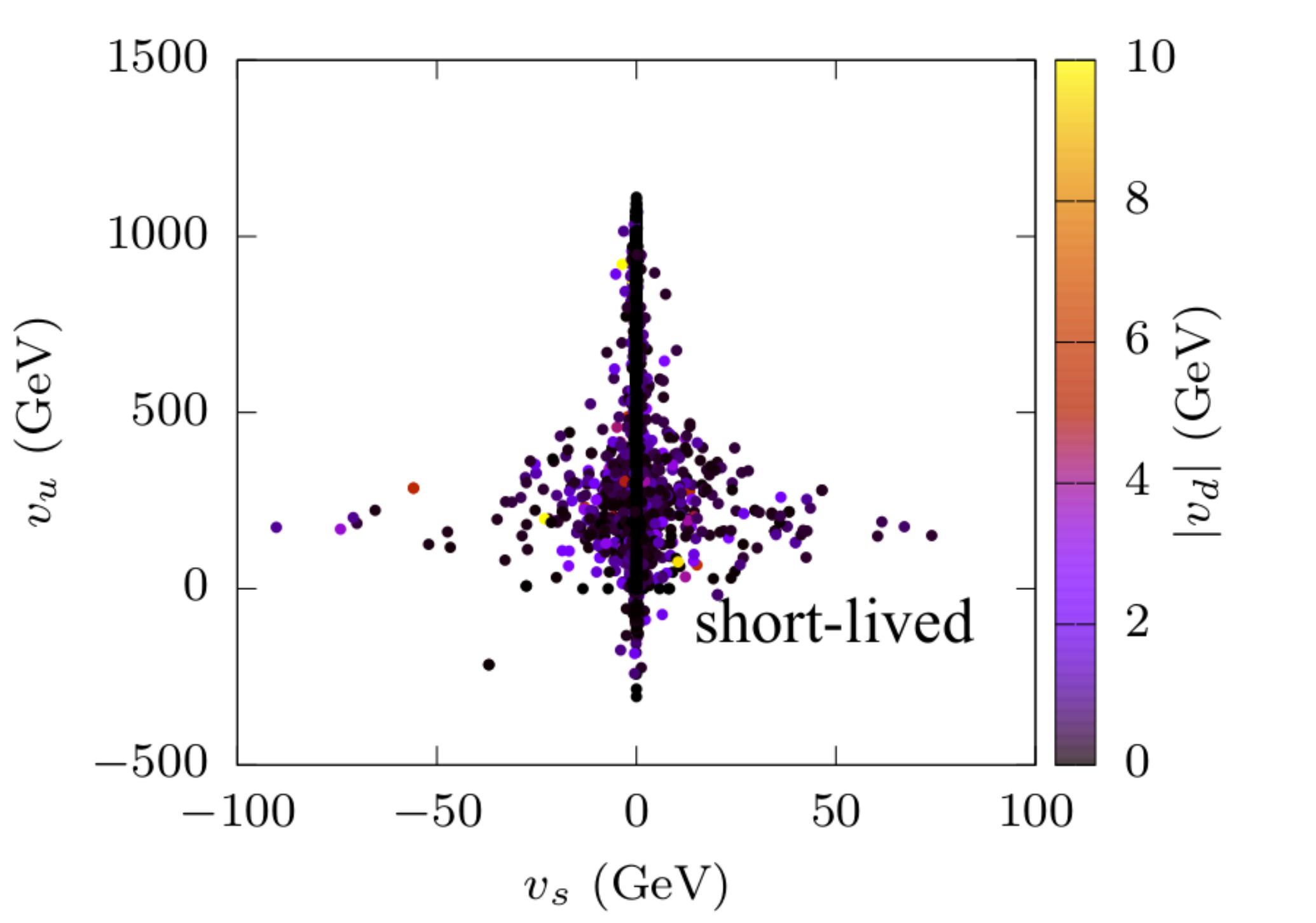}
\vskip 0.1 in
\includegraphics[height=0.16\textheight, width=0.369\columnwidth , clip]{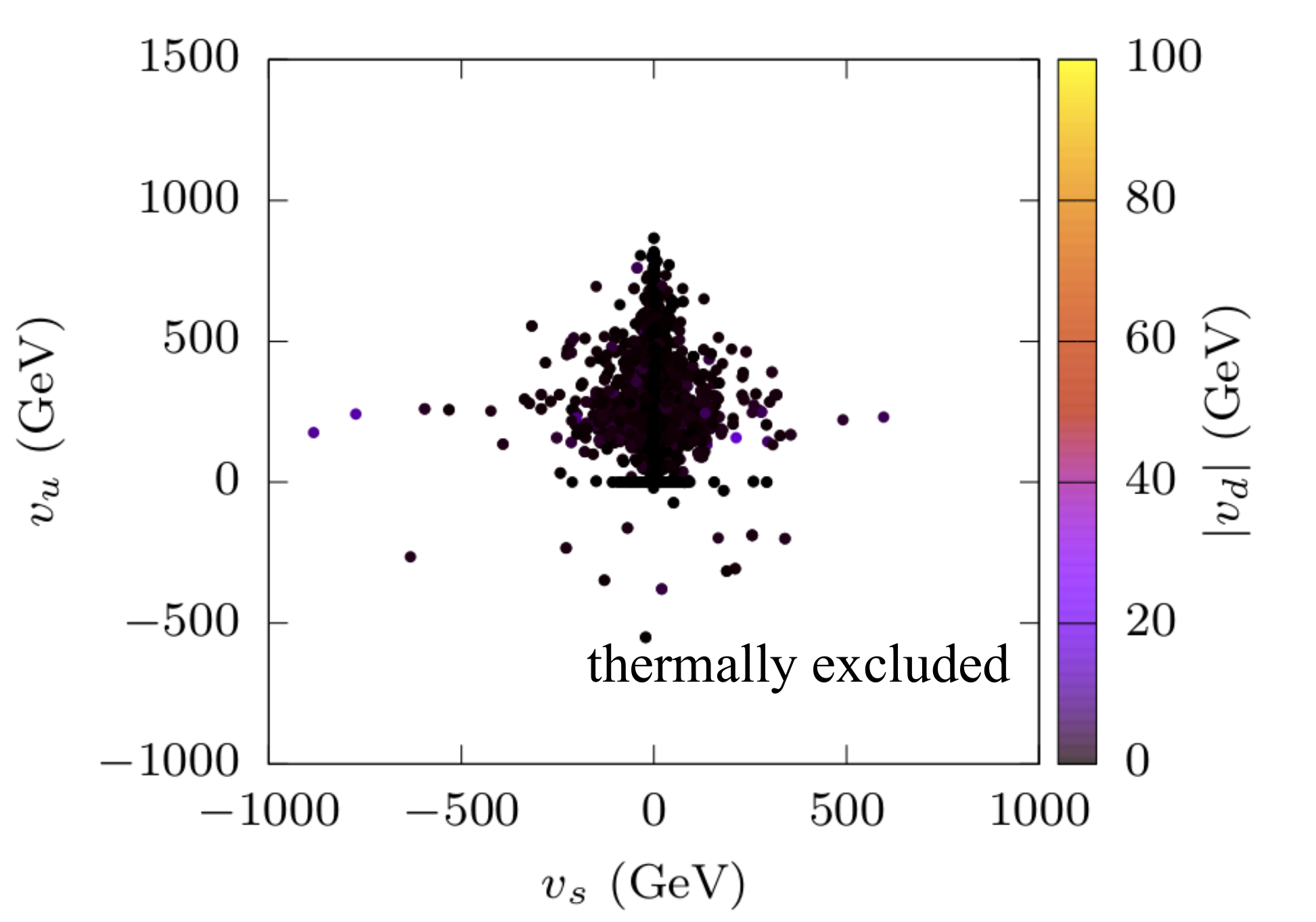}
\includegraphics[height=0.16\textheight, width=0.369\columnwidth , clip]{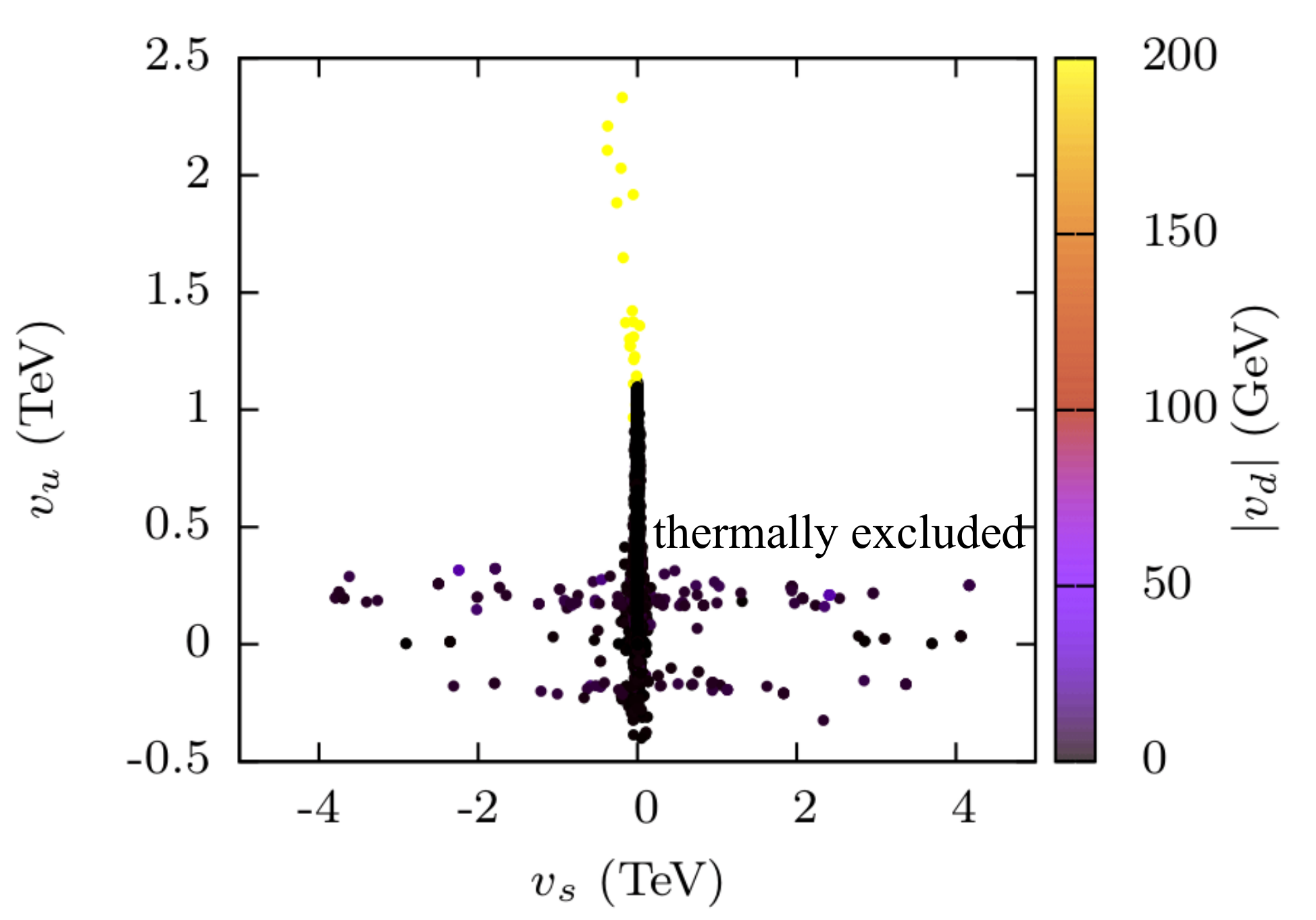}
\caption{Regions in the $\vs$-$\vu$-$\vd$ space in the MSSM-like scenario obtained
from \veva ~scan that
lead to (i) a viable DSB vacuum which is the global minimum of the potential 
(top row), (ii) a global minimum that is deeper than the (metastable) DSB vacuum 
but the latter being still long-lived enough and hence viable (second row), 
(iii) the same but the DSB vacuum being short-lived under zero-temperature 
quantum tunneling and hence not viable (third row) and (iv) the same but when 
the DSB vacuum is short-lived and thus, not viable, only if thermal effects to 
the potential are included (bottom row). The left (right) panel stand for the 
cases with a singlino- (higgsino)-like LSP. Colors represent values taken
by $|\vd|$ as shown in the accompanying pallets. 
The ranges of various parameters used in the scan are as given in expression 
\ref{eq:mssm-like-ranges}.
}
\label{fig:mssm-like-vs-vu}
\end{figure}

In figure \ref{fig:mssm-like-vs-vu} we illustrate the nature of stability of the 
DSB vacuum in terms of the \vevs ~acquired by the fields `$S$' ($\vs$) and 
$\higgsu$ ($\vu$) at the DSB vacuum and other deeper minima. 
The absolute value of the \vev ~of the 
third neutral scalar field, $\higgsd$ ($|\vd|$) is indicated via adjacent color palettes. 
Columns stand for scenarios with a singlino-like (left) and a higgsino-like (right) LSP. 
Rows present possible fates of the DSB vacuum which 
are summarized in section \ref{subsec:arbitrary-directions-veva}.
From top 
to bottom, in that order, the plots show the field values consistent with 
(i) the DSB vacua which are stable (top row), (ii) the deeper minima with which 
the DSB minima become metastable but long-lived (second row), (iii) the same 
which make the DSB vacua metastable and short-lived under zero-temperature 
quantum tunneling (third row) and (iv) the same that render the DSB vacua 
metastable and short-lived only with thermal effects factored in (last row).
Thus, parameter points in the first two rows yield viable DSB vacua while, in 
the latter two, they do not. The following set of observations are in order.
%
%
%
\begin{itemize}
\item It can be clearly recognized that the sets of three 
\vev-values that emerge for each scattered point in the plots in the first 
row of figure \ref{fig:mssm-like-vs-vu} are
in compliance with a DSB vacuum
and are consistent with the
values of the supplied NMSSM input parameters, in particular, those for $\mueff$ 
and $\lambda$.
\item For the rest three rows the \vevs ~refer to minima deeper than the 
DSB vacuum thus making the latter metastable. These reveal that deeper 
minima might appear for rather high values (reaching up to tens of TeV) for all 
three fields. These are in addition to deeper minima that occur along some flat 
directions in the field space which are already known to be the `dangerous' for 
the stability of the DSB vacuum. In particular, the plots indicate that the 
direction $\higgsd=S=0$ is such a flat direction. 
\item Furthermore, a closer look at the right plot in the second row (long-lived 
DSB vacuum in the higgsino-dominated LSP case) reveals that deeper minima causing 
the DSB vacua to be metastable but long-lived occur along the diagonals.
These correspond to directions which are both (approximately; at the 1-loop level)
$D$-flat ($\higgsd \approx \higgsu$) and $F_S$-flat 
($S^2 \approx \frac{\lambda}{\kappa} \higgsd \higgsu$; see section 
\ref{subsec:specific-directions}) in the field space.
\item It is important to note from the plot on the left in the third row 
(singlino-like LSP case) that tunneling at zero temperature does note pose a
threat to the stability of the DSB vacuum. 
Plots from the fourth row clearly reveal that the thermal effect indeed plays 
an important role in determining the fate of metastable DSB vacua. 
Had it not been for the thermal contributions, these regions
would have given rise to a deeper minimum which could still ensure long-lived DSB
vacua. Some of these appear distinctively for somewhat large (compared to the
values it may assume at the DSB, i.e., $\sim 125-175$ GeV) positive values of 
$\higgsu$ in the MSSM-like case and along the flat direction $S=\higgsd=0$ with
magnitudes of $\higgsu$ comparable to its DSB value.
\end{itemize}
%
%
%
\begin{figure}[t]
\centering
\includegraphics[height=0.24\textheight, width=0.49\columnwidth , clip]{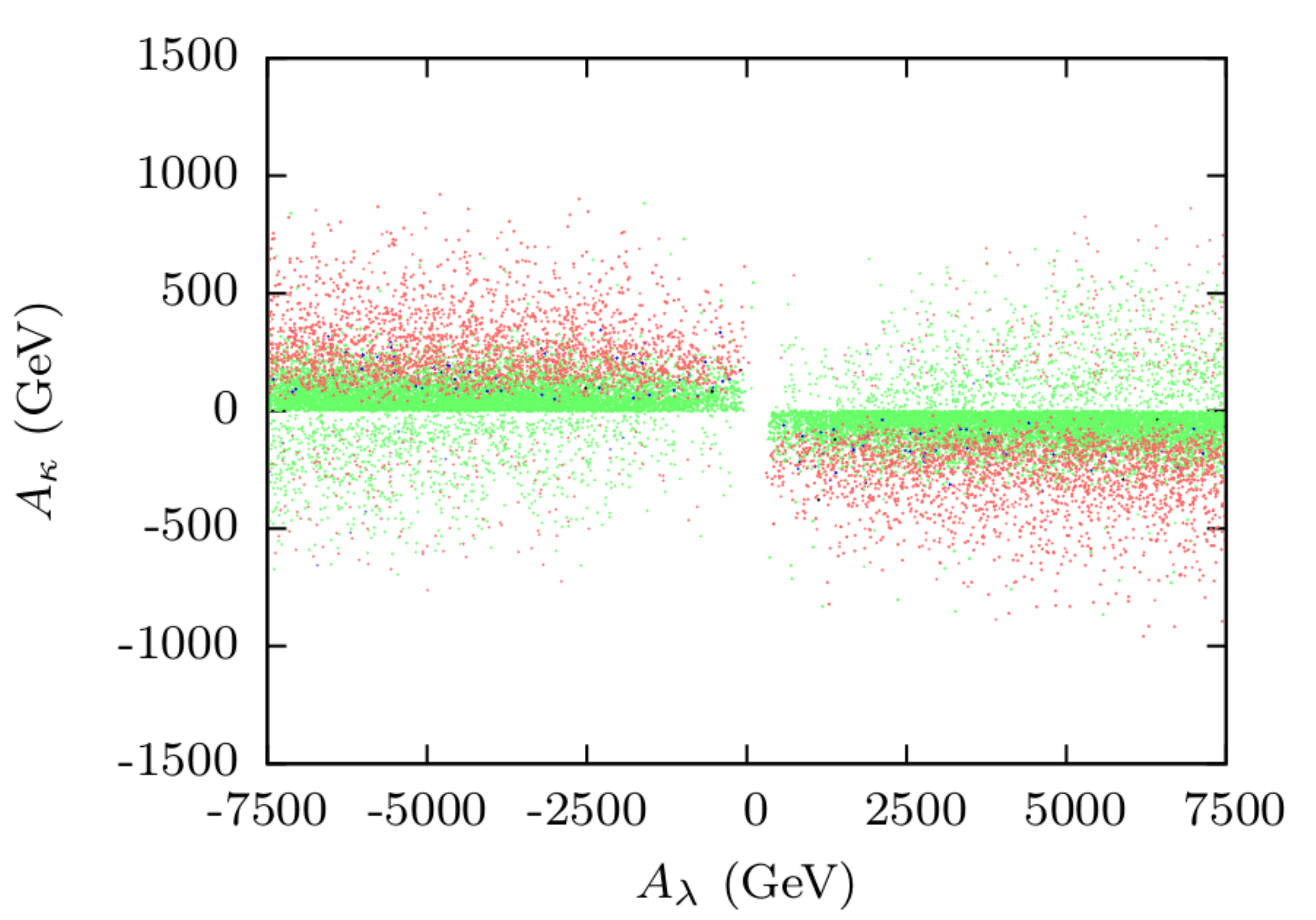}
\includegraphics[height=0.24\textheight, width=0.49\columnwidth , clip]{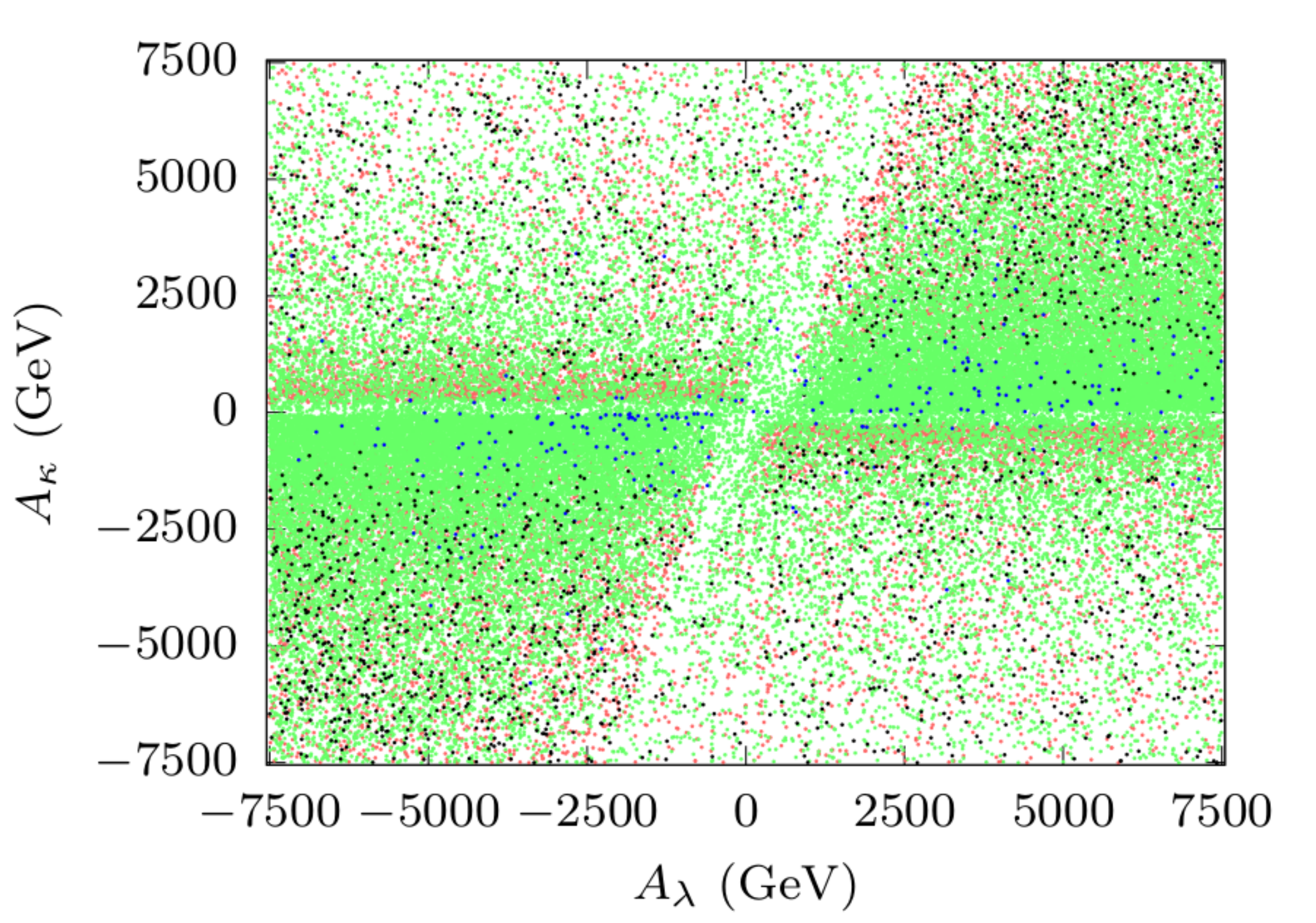}
\caption{Regions in the $\alambda-\akappa$ plane showing the fate of the 
DSB vacuum for the MSSM-like scenario with a singlino-like LSP (left) and with a 
higgsino-like LSP (right). Refer to table \ref{tab:color-number-codes} for the 
color-code used.
}
\label{fig:mssm-like-alambda-akappa-veva}
\end{figure}
As may be expected from earlier studies in the MSSM, the trilinear terms in the NMSSM sector 
as well, are likely to play some important role in determining the stability
of the DSB vacuum. In figure \ref{fig:mssm-like-alambda-akappa-veva} we illustrate
the stability pattern of the DSB vacua in the $\alambda-\akappa$ plane in the 
MSSM-like scenario and for the cases where the LSP is singlino-like (left plot) 
and higgsino-like (right plot).
An analysis of a similar effect but applied in the context of dark matter phenomenology
has been carried out earlier in reference \cite{Cerdeno:2004xw}.
The features of these plots, which we will discuss shortly, would be easier to
understand once we consider the following issues.

The singlet $CP$-even and $CP$-odd scalar masses in the MSSM-like scenario,
for large $\vs$, can be written down as (from equations \ref{eq:msq-cp-even} and 
\ref{eq:msq-cp-odd})
\begin{subequations}
\bea
\label{eq:ms33-large-vs}
\mathcal{M}_{S,33}^2 &\approx & \kappa \vs (\akappa + 4 \kappa \vs) \\
%
\label{eq:mp11-large-vs}
\mathcal{M}_{P,11}^2 &=& {{2 \mueff \, (\alambda + \kappa \vs)} \over {\sin 2\beta}} \\
%
\label{eq:mp22-large-vs}
\mathcal{M}_{P,22}^2 &\approx&  - 3 \kappa \akappa \vs  \quad .
\eea 
\label{eq:mscalar-large-vs}
\end{subequations}
This is reminiscent of the discussion we had in the context of the study of
single-field ($S$) potential in section \ref{subsubsec:single-field-potential}.
Requiring $\mathcal{M}_{P,22}^2 >0$ yields $\kappa \vs \akappa <0$. When this is
combined with the demand of $\mathcal{M}_{S,33}^2 >0$, one finds
$|\akappa| < 4 |\kappa \vs|$, i.e., the magnitude of $\akappa$ is bounded from
above. This is clearly seen in the case with a singlino-like LSP (left plot)
for which $|\kappa \vs|$ cannot exceed $\frac{|\mueff|}{2}$. Given that we scan
over the range $|\mueff|<500$ GeV, $|\akappa|$ cannot exceed 1 TeV. On the other
hand, in the case with a higgsino-like LSP, $|\kappa \vs|$ can be legitimately 
large compared to $\frac{|\mueff|}{2}$ and thus $\akappa \sim \mathcal{O}(\mathrm{TeV})$
are also allowed. 
As may be apparent (and we will see soon) from the set of expressions in 
equation \ref{eq:mscalar-large-vs}, demanding non-tachyonic scalar 
states might eventually lead to only a few discrete possibilities of relative 
signs and magnitudes among various NMSSM parameters.
%
%
\begin{figure}[tbh]
\centering
\includegraphics[height=0.24\textheight, width=0.49\columnwidth , clip]{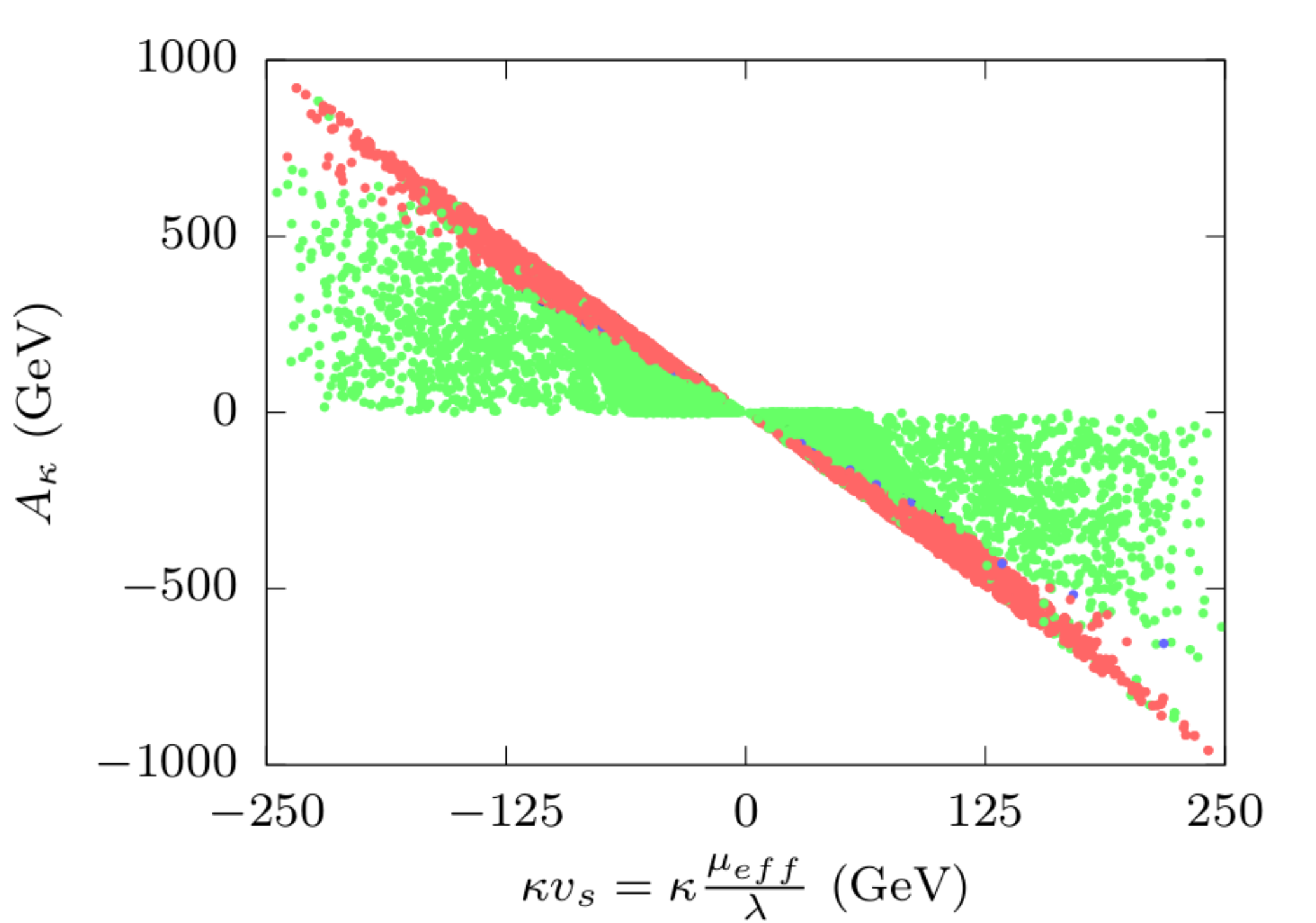}
\includegraphics[height=0.24\textheight, width=0.49\columnwidth , clip]{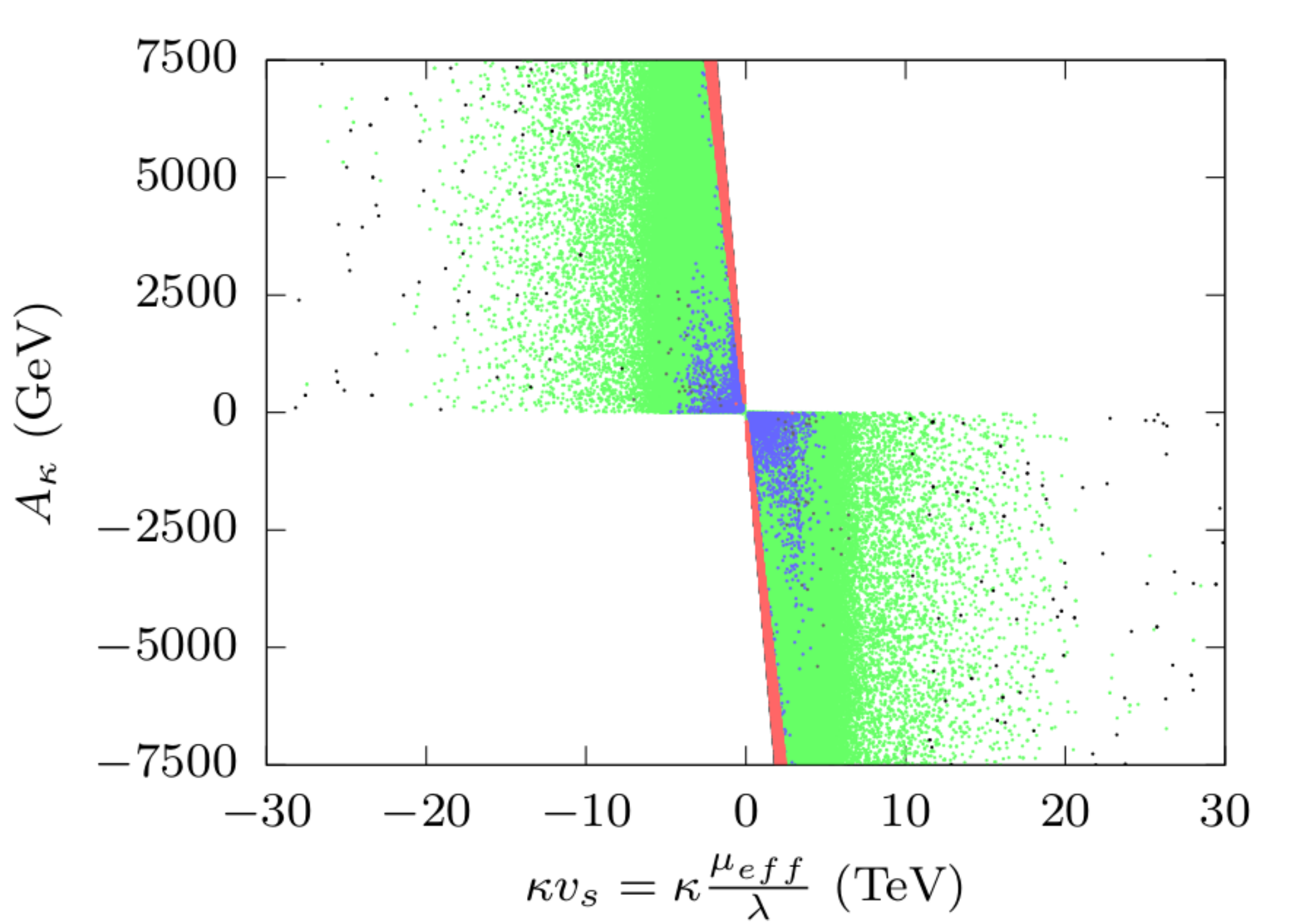}
\caption{Regions in the $\kappa \vs-\akappa$
indicating the fate of the DSB vacuum in the MSSM-like scenarios with a 
singlino-like (left) and a higgsino-like (right) LSP. Color-code employed is 
as summarized in table \ref{tab:color-number-codes}.}
\label{fig:mssm-like-ks-akappa-veva}
\end{figure}
%
%

In the left plot (the case with a singlino-like LSP) of figure 
\ref{fig:mssm-like-alambda-akappa-veva} the first (second) and the
third (fourth) quadrants with $\alambda \akappa >0 \; (<0)$ correspond to $\kappa <0 \; (>0)$.  
This is because of the following reason. Note that for the singlino-like case the product
$\kappa \vs$ is small and thus can be neglected. Demanding the $CP$-odd doublet scalar
boson to be non-tachyonic, it follows from equation \ref{eq:mp11-large-vs} that $\mueff$ 
(i.e., $\vs$) and 
$\alambda$ are of the same sign. Given that (see above) $\akappa$ and $\kappa \vs$ are of
opposite signs, a positive `$\kappa$' requires $\akappa$ and $\alambda$ to be of opposite
sign. The reverse is true for `$\kappa$' being negative. A similar analysis mostly holds in
the case with a higgsino-like LSP (right plot) with the exception that one 
may not be 
able to neglect the magnitude of $\kappa \vs$ here. In this case its cancellation
against $\alambda$ on the right-hand side of equation \ref{eq:mp11-large-vs} is a 
possibility which could lead to a tachyonic
doublet $CP$-odd scalar. This possibility shows up in a tilted edge 
(in contrast to a vertical one in
the left plot) that separates the region with $\mueff >0$ on the right side of the plot 
from  that with $\mueff <0$ on the left. In fact, we are able to predict correctly the 
magnitude of the tilt by using the set of above equations.
Apart from that, we observe that DSB vacua with diverse stability 
properties could appear over the entire $\alambda$-$\akappa$ plane as shown in
this plot.

Here we recall (see section \ref{subsec:arbitrary-directions-semi}) that the free parameter 
$\akappa$ plays a subtle role in the fate of the DSB vacuum in the MSSM-like 
scenario for which $\vs >> \vd, \vu$. In figure \ref{fig:mssm-like-ks-akappa-veva} 
we illustrate this through a \veva-based analysis. Color-code used is already 
introduced in table \ref{tab:color-number-codes}. We compare these plots with 
the plots in the top row of figure \ref{fig:ks-akappa-math} (obtained from a 
{\tt Mathematica}-based analysis) which also represent the MSSM-like scenario. For 
both sets plots on the left (right) correspond to the cases with singlino 
(higgsino-like) LSP. An impressive level of agreement does not escape notice.
%
%
\begin{figure}[t]
\centering
\includegraphics[height=0.20\textheight, width=0.44\columnwidth , clip]{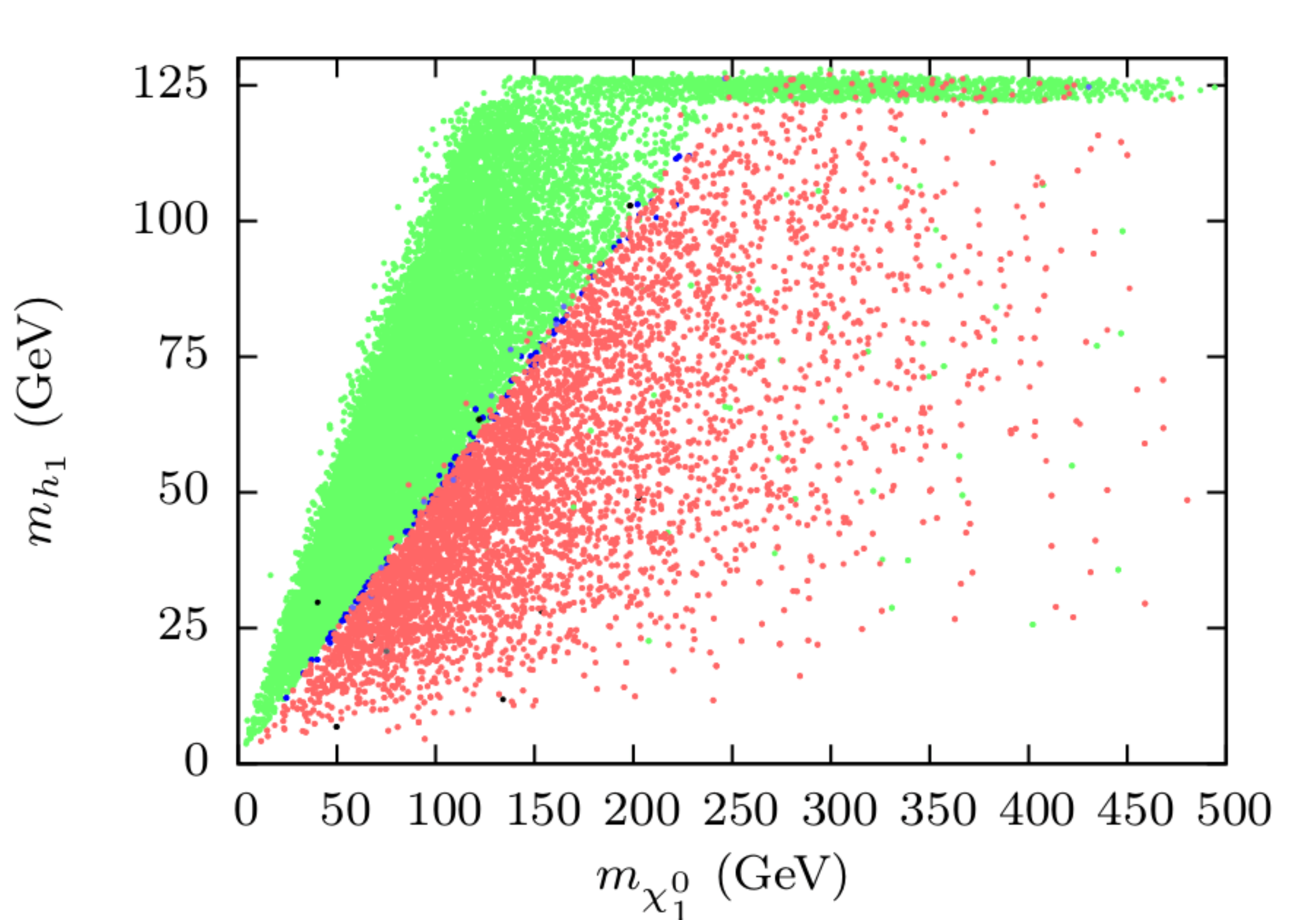}
\includegraphics[height=0.20\textheight, width=0.42\columnwidth , clip]{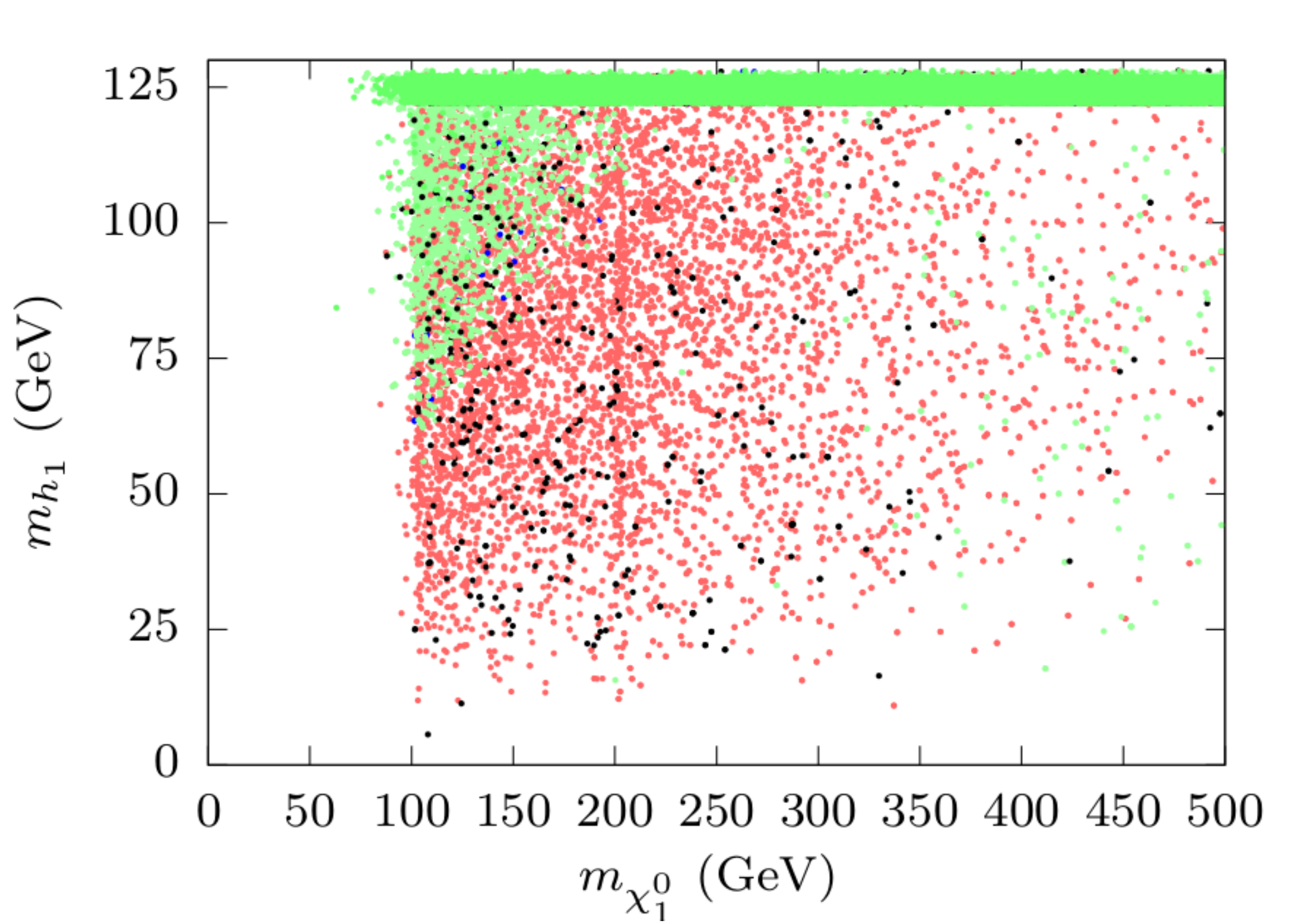}
\includegraphics[height=0.20\textheight, width=0.44\columnwidth , clip]{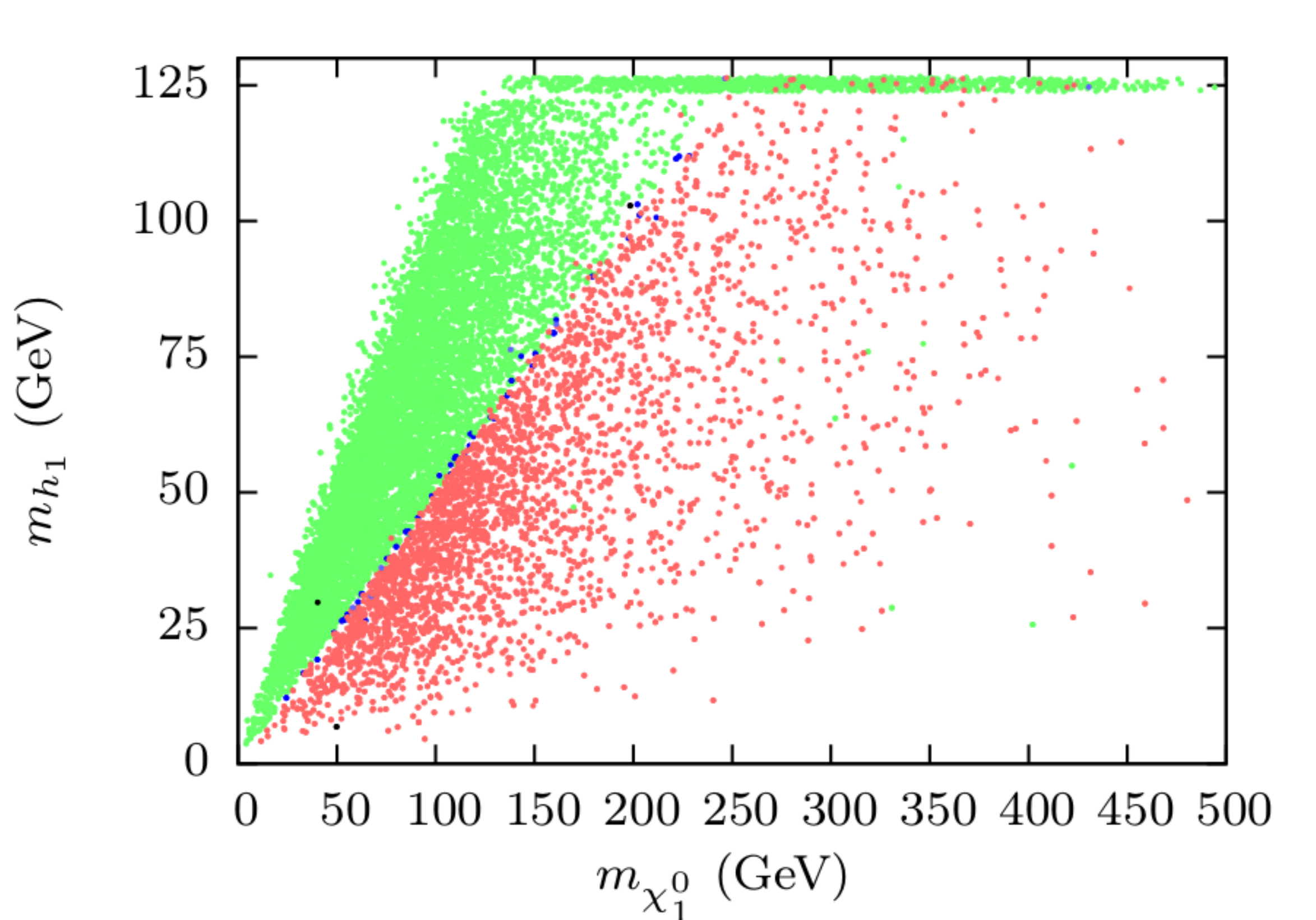}
\includegraphics[height=0.20\textheight, width=0.42\columnwidth , clip]{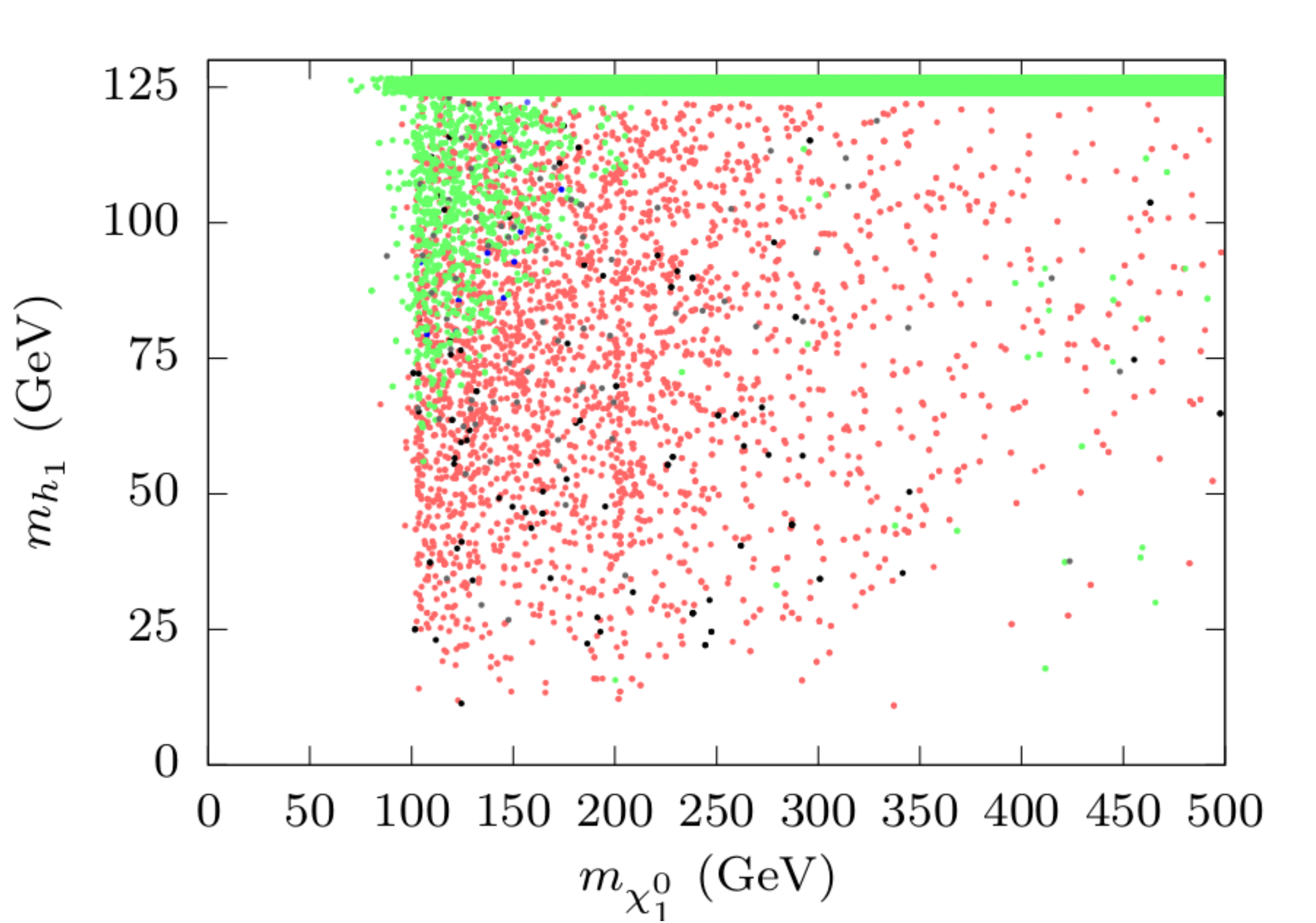}
\includegraphics[height=0.20\textheight, width=0.44\columnwidth , clip]{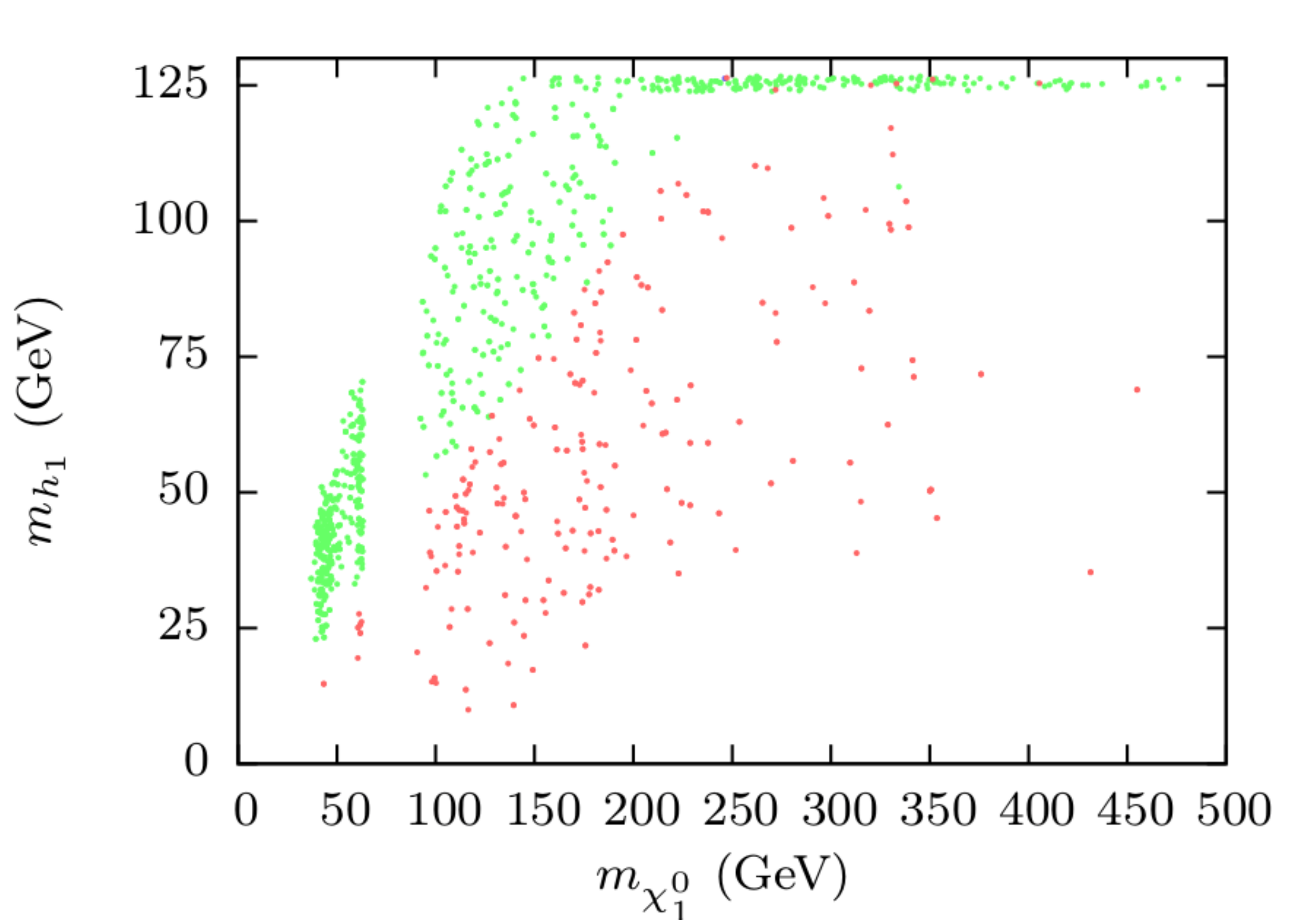}
\includegraphics[height=0.20\textheight, width=0.42\columnwidth , clip]{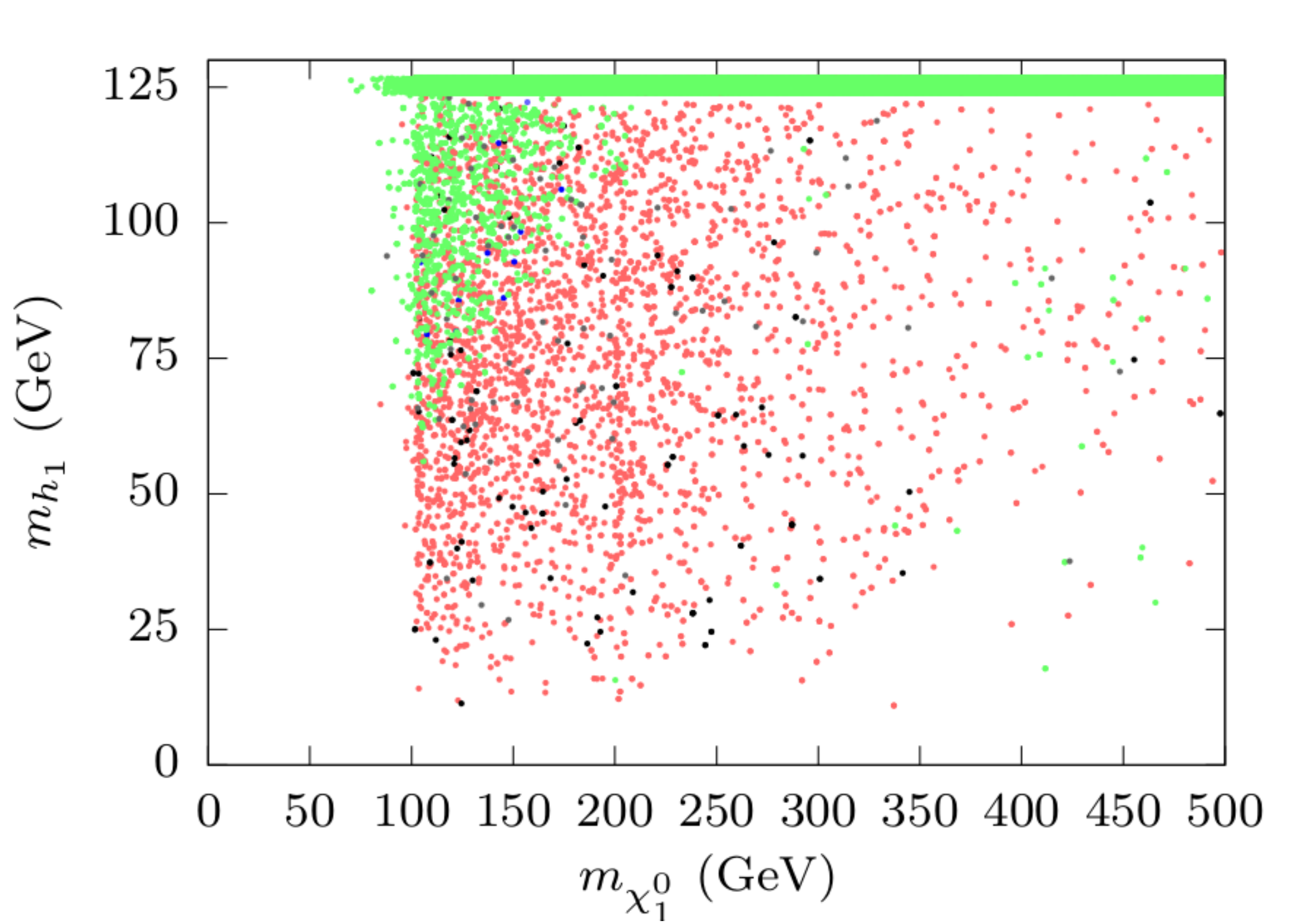}
\caption{Regions obtained from \veva ~scan in the $\mntrlone-\mhone$ 
space indicating the fate of the DSB vacuum in the MSSM-like scenario with a
singlino-like LSP (left panel) and a higgsino-like LSP (right panel). Plots in 
the top row incorporate all basic constraints along with those coming from
{\tt HiggsBounds} and {\tt HiggsSignals}. Plots in the middle (bottom) row include, in 
addition, bounds from the flavor (dark matter) sector. Color-code employed is
as summarized in table \ref{tab:color-number-codes}.}
\label{fig:mssm-like-flav-dm}
\end{figure}
%

We now move on to discuss how the nature of the stability of the DSB vacuum is 
related to phenomenology. Issues that are of immediate interest in the NMSSM 
context pertain to the Higgs sector, especially the lighter Higgs bosons of the 
singlet kind and the lighter neutralinos which can have significant singlino 
admixture. Hence in figure \ref{fig:mssm-like-flav-dm} we illustrate how the 
stability of the DSB vacuum manifests itself in the $\mntrlone-\mhone$ plane.
As before, the left (right) column presents the singlino-like (higgsino-like) 
case. The first row uses the same set of data as for figures 
\ref{fig:mssm-like-vs-vu} and \ref{fig:mssm-like-ks-akappa-veva}.
Plots in the second row are obtained after subjecting these data sets to the
scrutiny by {\tt HiggsBounds} and {\tt HiggsSignals} to ensure an overall 
conformity to the experimental findings on the SM-like Higgs boson. In addition, 
plots in this row are also in compliance with the constraint from the flavor 
sector as mentioned in section \ref{subsec:arbitrary-directions-veva}. Note, 
however, that in the scenario under discussion these do not have much bearing on 
the allowed region of parameter space.
Plots in the third row depicts the situation further on the imposition of the upper 
bound on the DM relic density again as indicated in section
\ref{subsec:arbitrary-directions-veva}.

It is clear that only the constraint pertaining to DM relic density and that 
also solely for the case with a singlino-like LSP have a noticeable bearing on 
the allowed region in the $\mntrlone-\mhone$ plane.
This is since a singlino-like LSP has in general a rather feeble interaction 
with other particles and thus its
annihilation rates are suppressed giving rise to an unacceptable level of
relic density. This is more so when $\lambda$ is small,
which is generally the case with the MSSM-like scenario, thus suppressing its 
couplings with higgsino-like states in the spectrum. In contrast, a
higgsino-like LSP can have a substantial couplings to other relevant states
which may aid its rapid annihilation. In addition, for a higgsino-like LSP
there is always a nearby light neutralino and chargino states which are 
also higgsino-like. These facilitate efficient coannihilation of the LSP.

Note that such an efficient coannihilation is also possible in the case with
a singlino-like LSP once its mass comes closer to the higgsino-like states.
In the left plot from the last row of figure \ref{fig:mssm-like-flav-dm}
this is clearly the case when the LSP mass reaches 100 GeV.
This is close to the minimum mass considered for the near-degenerate
higgsino-like states with $|\mueff| \geq 100$ GeV prompted by the lower
bound (103 GeV) on the mass of the lighter chargino \cite{lepsusy} as indicated
in section \ref{subsec:arbitrary-directions-veva}.
Furthermore, on the left of this plot there are
also two green (vertical) strands separated by a gap. The left (right) 
strand occurs at an LSP-mass of $~45$ ($\sim 62-63$) GeV which clearly points 
to $s$-channel annihilation of a pair of LSPs via a resonant $Z$-boson
(SM-like Higgs boson). 

The message to take home from figure \ref{fig:mssm-like-flav-dm} is that 
the region of stability (in green) of the DSB vacuum is segregated from the region 
where it is metastable only for the cases with 
a singlino-like LSP. 
Crucially enough, the dominance of the color `red' in this metastable regime
indicates that it falls out of favor primarily on the inclusion of 
thermal effects. 
For cases with a higgsino-like LSP, the instability of the DSB vacuum 
is more pervasive. The region with a stable DSB vacuum for this case is mostly concentrated
at lower masses for the LSP and for a singlet-like $CP$-even Higgs boson
which is not much lighter than the SM-like Higgs boson. 
It may also be noted that for cases with a singlino-like LSP one mostly finds the 
region with $h_1$ as the SM-like
Higgs boson to yield a stable DSB vacuum (the green horizontal bands about
$\mhone \sim 125$ GeV). However, for scenarios with a higgsino-like
LSP this is not guaranteed. This is so since, in these cases, we notice that red horizontal bands
(implying thermal instability) are hiding underneath the green ones about 
$\mhone \sim 125$ GeV.

In general, we observe that the thermal effects become more significant with 
increasing $|\akappa|$ 
which turn the global (DSB) minimum to a local one by altering their depths.
This may be quite expected given the discussion in the beginning of section
\ref{subsec:specific-directions}.
Had it not been for the thermal corrections, the red (unstable) region would have 
become viable being home to metastable but long-lived DSB vacua. This underscores 
the importance of a \veva-based analysis. A careful look at the plots in the left
column for the
first two rows also reveals thin streaks of regions (in blue) where the DSB vacua 
are indeed `long-lived' and separate the green and the red regions.
The role of thermal correction in this particular respect could be 
generically summarized as follows. We find that this introduces contributions cubic in the fields 
via a term $-{\pi \over 6} ({m^2 \over T^2})^{3/2}$ originating in the
expansion of the (bosonic) quantity $J_{+}({m^2 \over T^2})$ of equation \ref{eq:jpm}.
This might dilute the role of the tree-level term ${2 \over 3} \kappa \akappa S^3$
and acts towards decreasing the barrier-height between the false (DSB) vacuum and
the deeper (panic) vacuum thus facilitating the tunneling process.

It may be noted that since $\lambda$ and `$\kappa$' are relatively 
small in the MSSM-like scenario, the singlet $CP$-even scalar can be naturally 
light. Thus, the SM-like Higgs boson (with a mass $\sim 125$ GeV) turns out to 
be predominantly a heavier $CP$-even Higgs boson over a significant part of the 
MSSM-like parameter space.
Furthermore, the gradient of the left (outer) edge of 
the green patch for the plots in the left column can be understood by studying equation 
\ref{eq:singlet-singlino-mass} that relates the masses of the $CP$-even singlet
scalar and the singlino. In contrast, there appears a vertical edge at around 100 GeV for
the LSP-mass in the plots in the right column. This is simply because
these plots represent the case of a higgsino-like LSP which is roughly degenerate with the
lighter chargino thus attracting a similar lower mass-bound ($\sim 103$ GeV) as for the
latter.

An accompanying light $CP$-odd scalar would be phenomenologically (in particular,
in the collider context) rather interesting. We find that in the MSSM-like scenario that we are discussing, 
a relatively light singlino-like LSP could appear along with light singlet-like
$CP$-even and $CP$-odd scalars. All of them could have masses ($\mntrlone$, $\mhone$ and
$\maone$, respectively) in the order of a few tens of a GeV and can be consistent with 
all relevant experimental constraints and with the
requirement of a stable DSB vacuum. This is possible since $\kappa \vs$ is small for
such a scenario and hence a small value of $|\akappa|$ suffices to yield a light
$CP$-even scalar (see equation \ref{eq:mscalar-large-vs}). This in turn keeps
the mass of the singlet-like $CP$-odd scalar light. 
Interestingly, it is rather straightforward to find that the same set of equations 
predicts an anti-correlation between the compatible masses of these two states for 
cases with a higgsino-like LSP where $\kappa \vs$ is required to be large. In this
case it is noted that while $\maone \approx 50$ GeV can be compatible with
$\mhone \gtrsim 100$ GeV, $\mhone \lesssim 50$ GeV could only be accompanied by
$\maone \gtrsim 200$ GeV. These roughly summarize what could be the implications
for the collider experiments of such a scenario with a stable vacuum.
%
\subsubsection{Scanning of the NMSSM-like scenario with \veva}
\label{subsubsec:nmssm-like}
\begin{figure}[t]
\centering
\includegraphics[height=0.16\textheight, width=0.369\columnwidth , clip]{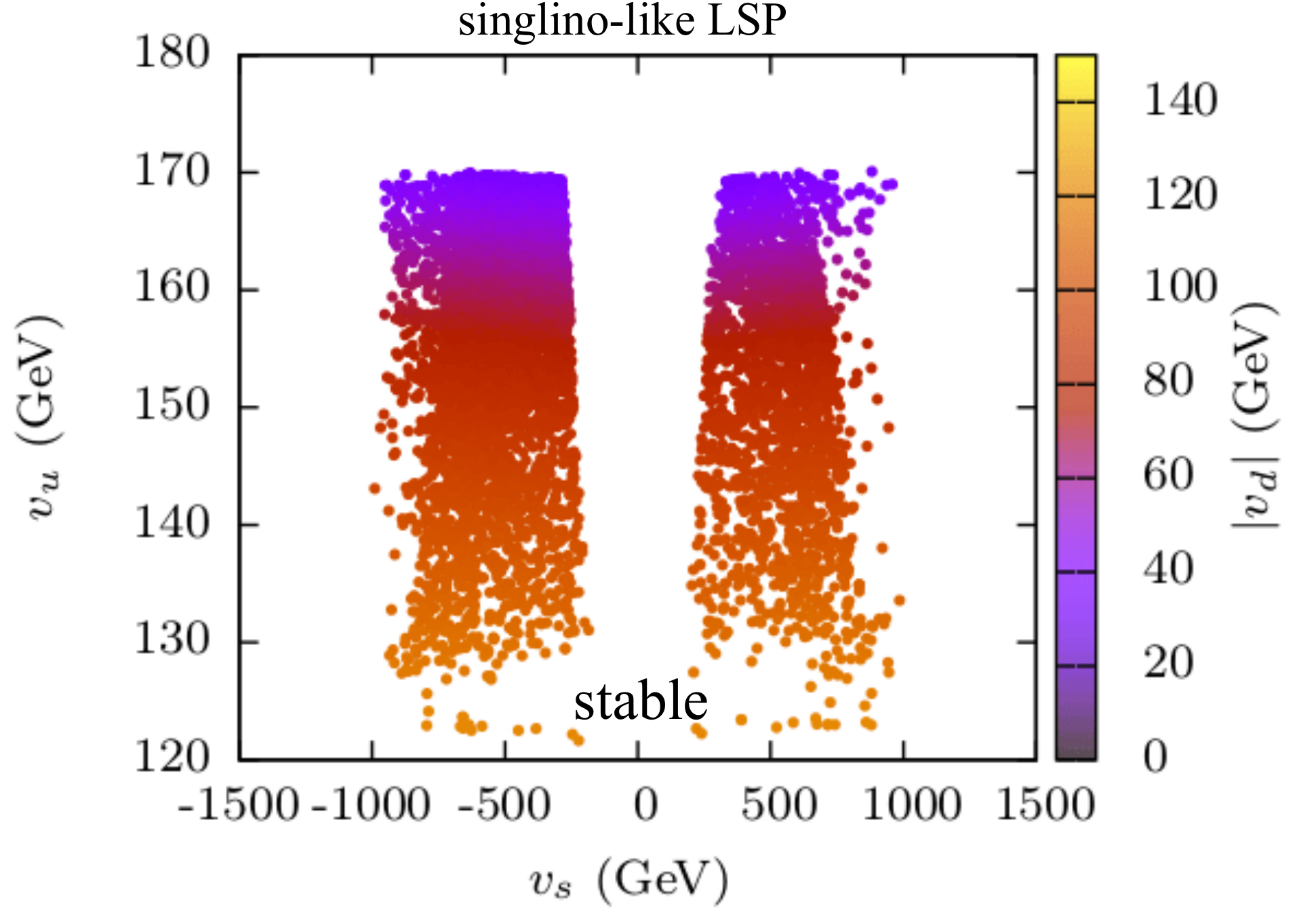}
\includegraphics[height=0.16\textheight, width=0.369\columnwidth , clip]{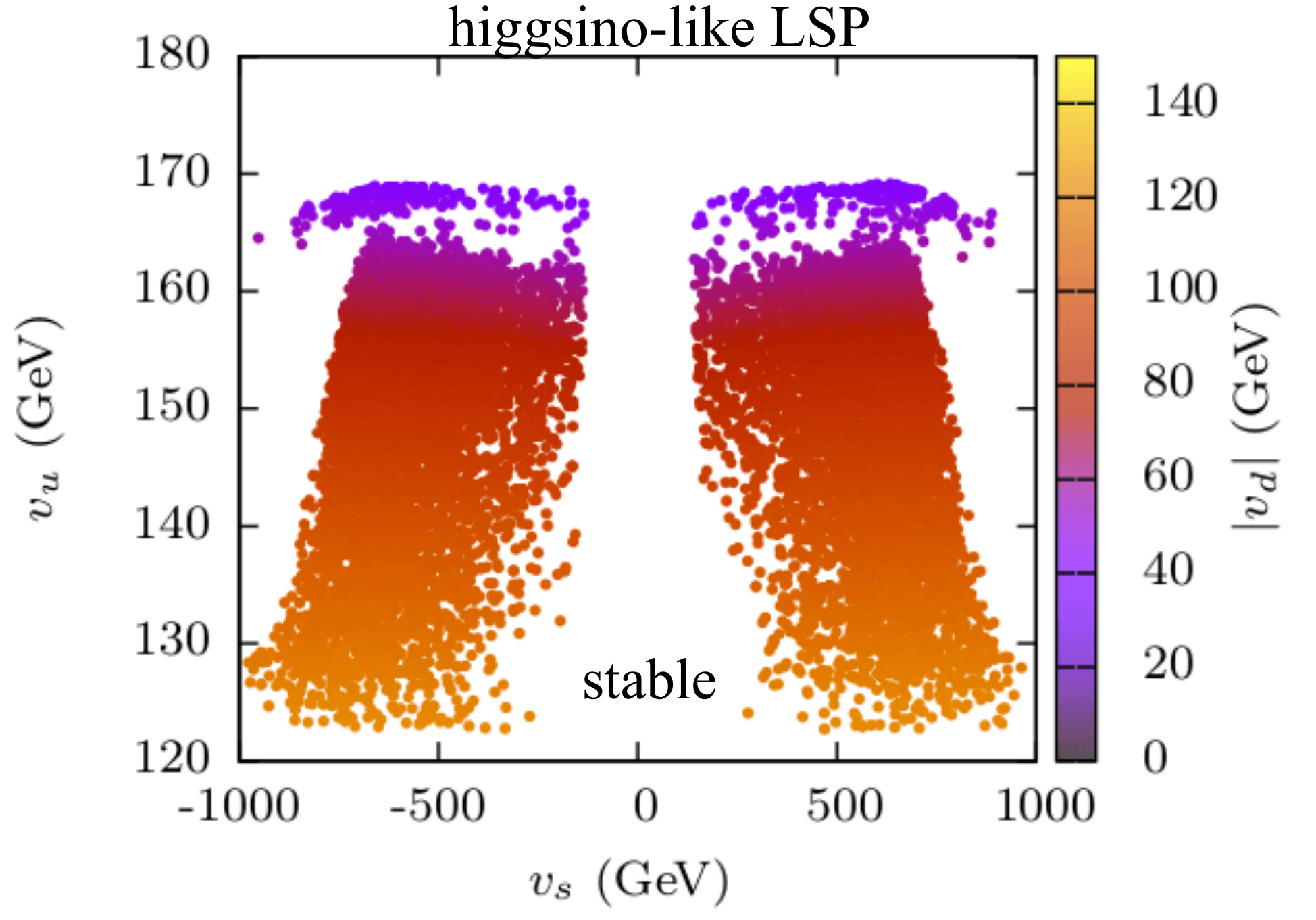}
\vskip 0.1 in
\includegraphics[height=0.16\textheight, width=0.369\columnwidth , clip]{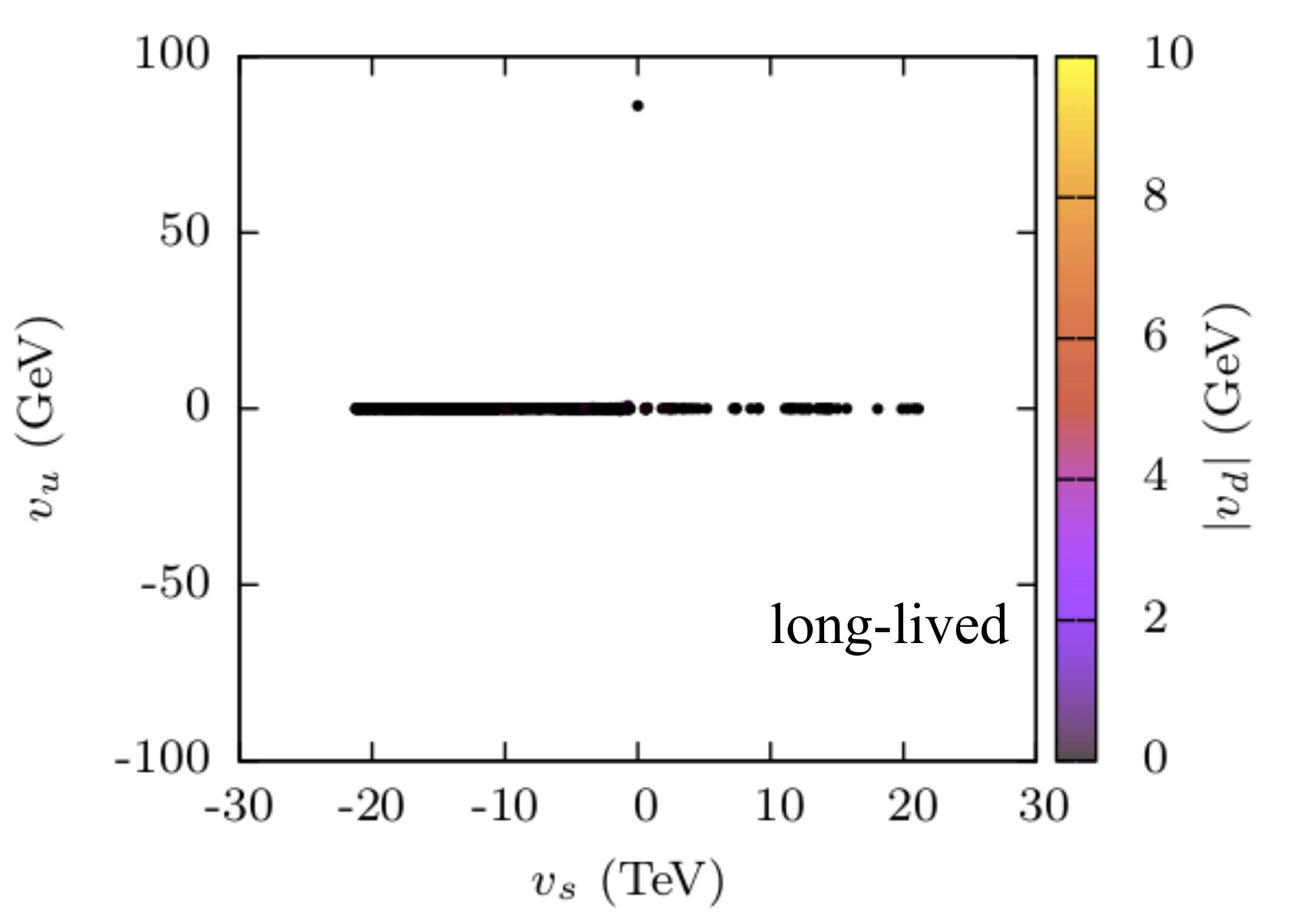}
\includegraphics[height=0.16\textheight, width=0.369\columnwidth , clip]{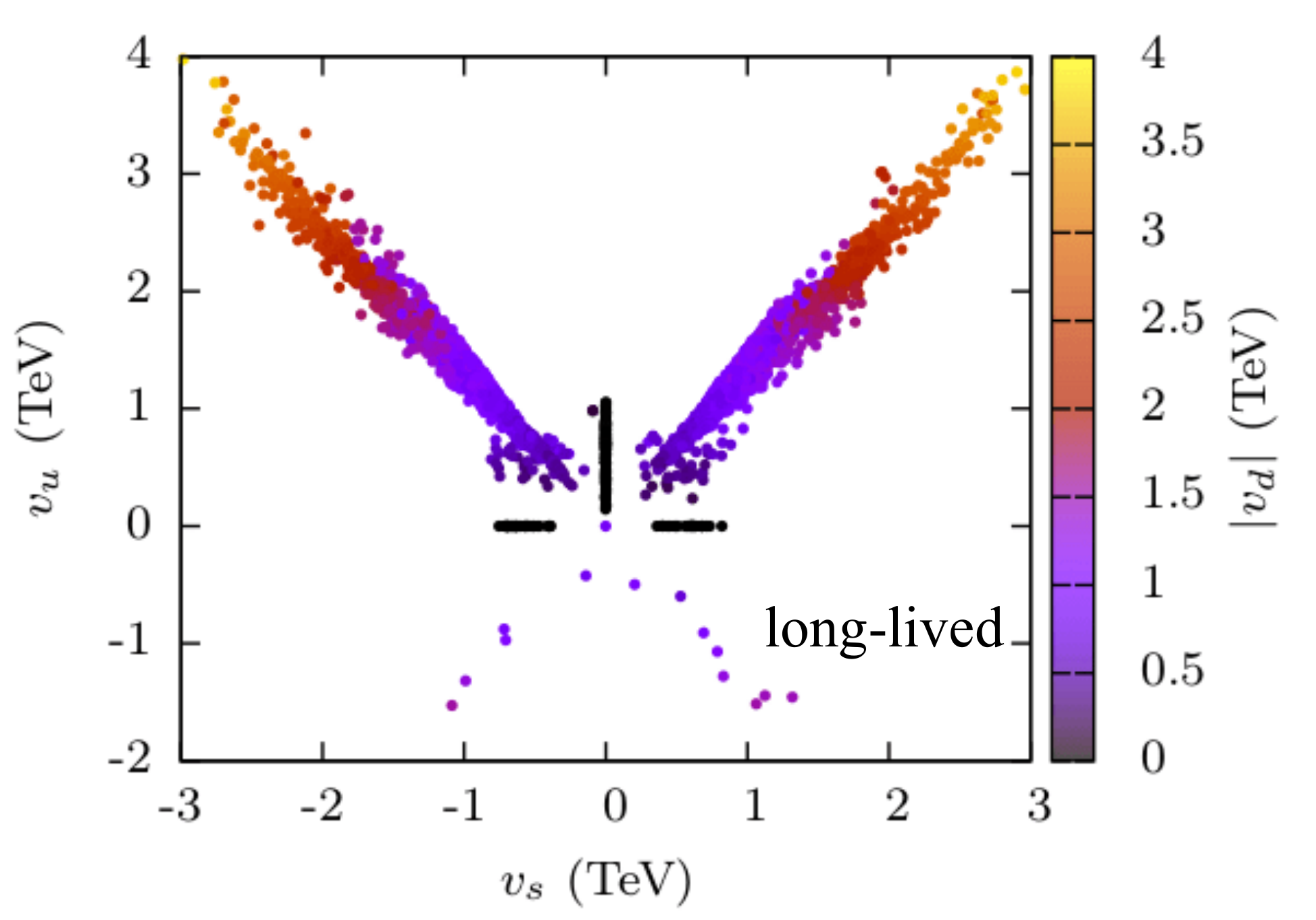}
\vskip 0.1 in
\includegraphics[height=0.16\textheight, width=0.369\columnwidth , clip]{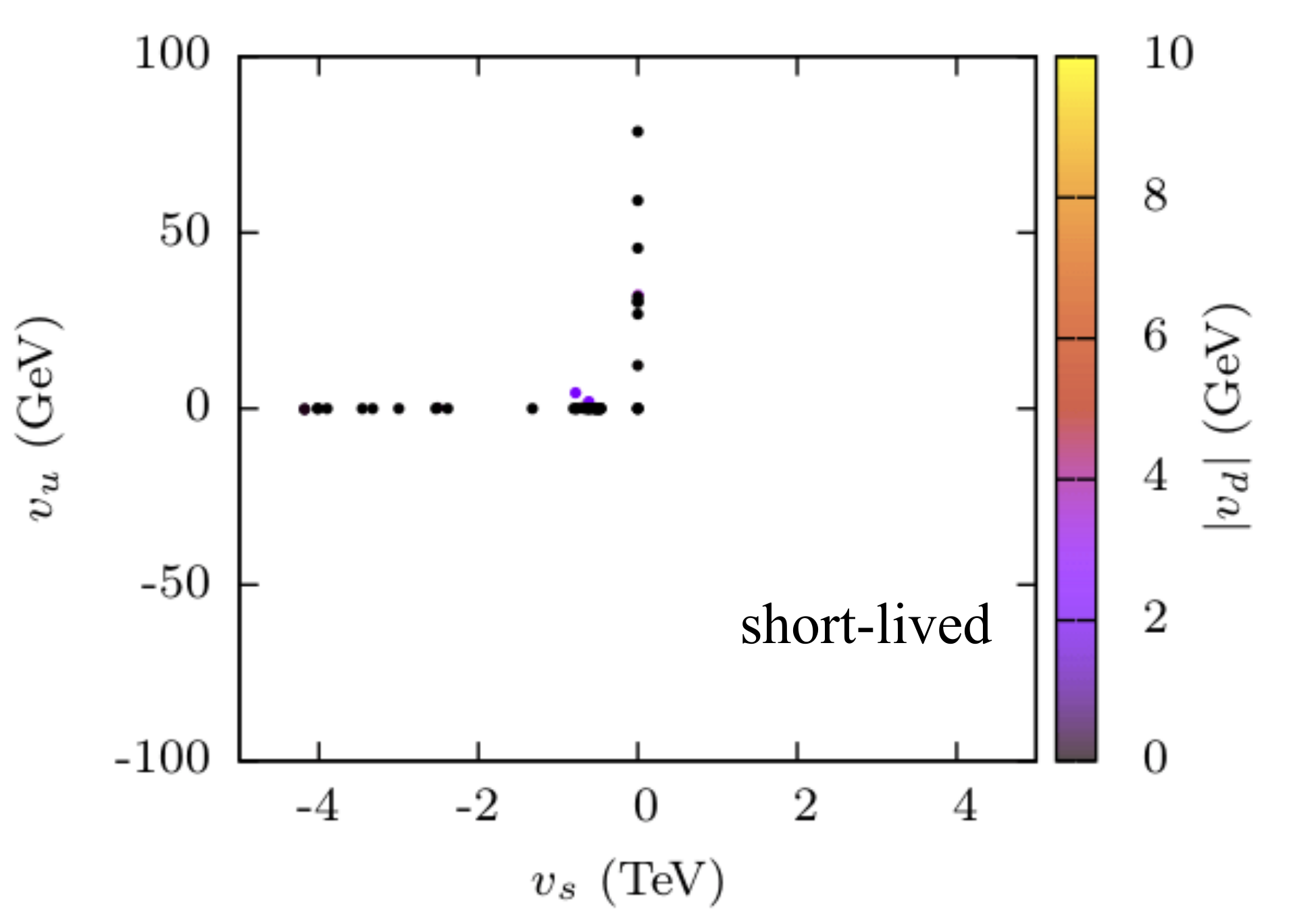}
\includegraphics[height=0.16\textheight, width=0.369\columnwidth , clip]{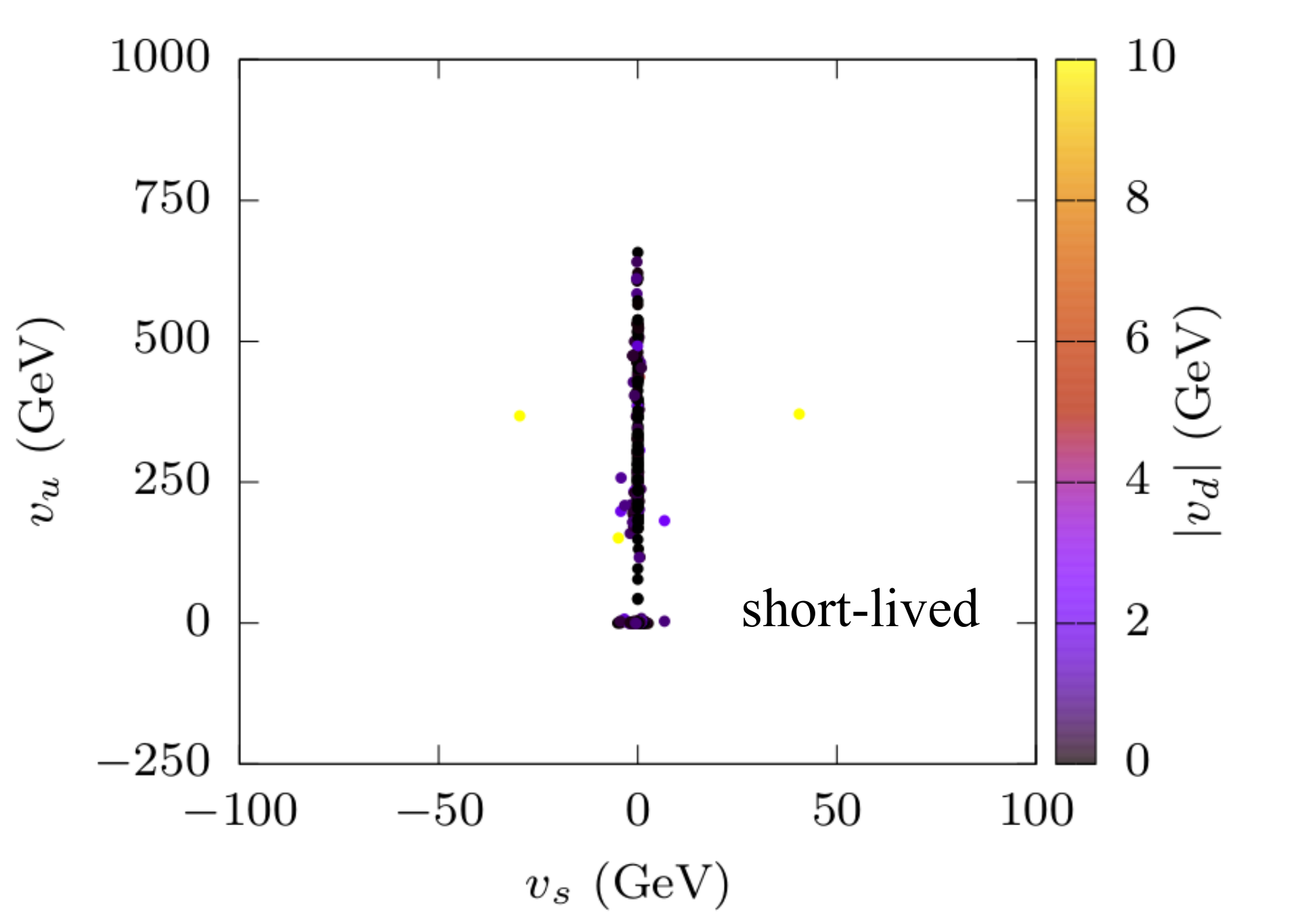}
\vskip 0.1 in
\includegraphics[height=0.16\textheight, width=0.369\columnwidth , clip]{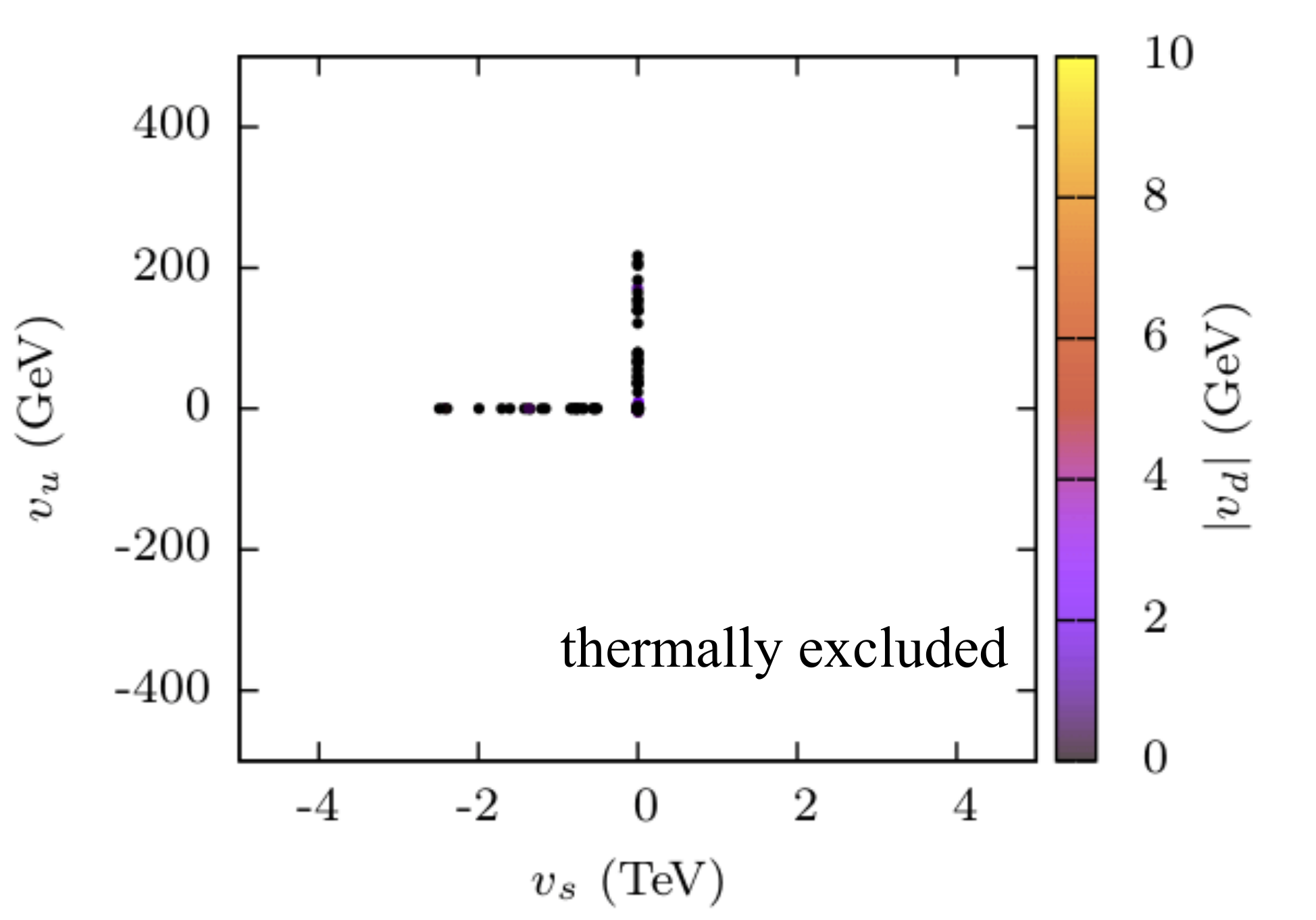}
\includegraphics[height=0.16\textheight, width=0.369\columnwidth , clip]{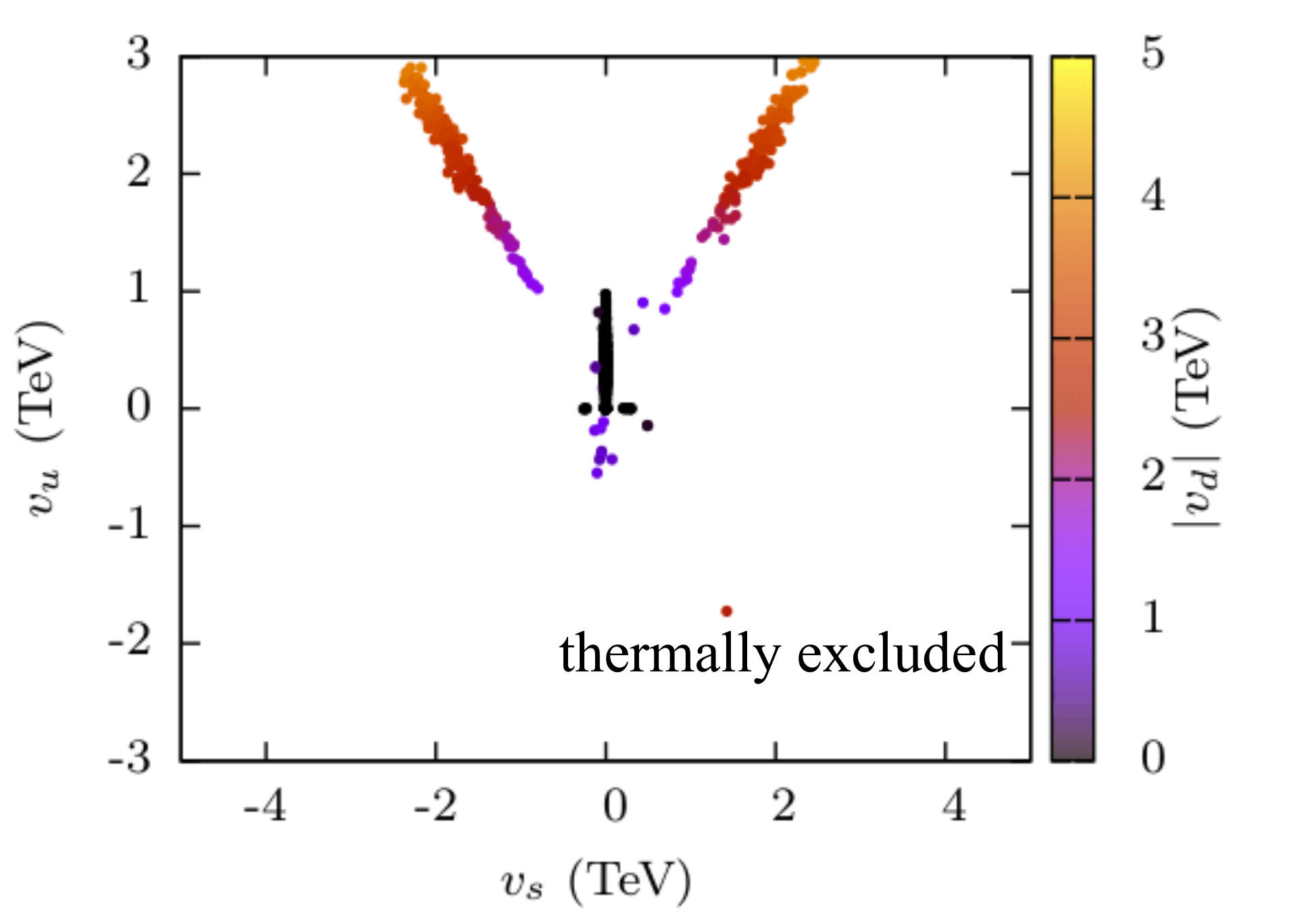}
\caption{The same as in figure \ref{fig:mssm-like-vs-vu} but for the NMSSM-like
scenario with ranges of various parameters used in the scan as given by
expression \ref{eq:nmssm-like-ranges}.
}
\label{fig:nmssm-like-vs-vu}
\end{figure}

%
We take up a similar study using \veva ~on the nature of stability of the DSB 
vacuum in the NMSSM-like scenario and its phenomenological implications. In 
figure \ref{fig:nmssm-like-vs-vu} we present the characteristic ranges of 
field-values. Plots are arranged in the same way as in figure 
\ref{fig:mssm-like-vs-vu}. 
In this regard, some salient features of the NMSSM-like scenario, when compared 
to the MSSM-like case (figure \ref{fig:mssm-like-vs-vu}) are as follows.
%
%
%
%
%
\begin{itemize}
\item Stable DSB vacua (first row) now appear more frequently over a little bigger ranges of 
$\higgsd$ and $\higgsu$ thanks to possible low values of $\tan\beta$. 
The relevant range for $\vs$ at the stable DSB vacua 
expectedly shrinks to around a TeV to comply with an input 
$\mueff \sim {\mathcal O}(100)$ GeV and large values of $\lambda$ that is 
characteristic of the NMSSM-like scenario. Clear vertical gaps about $S=0$ 
delineates the range in `$S$' which is incompatible with 
the input values of
$\lambda$ and $\mueff$. Note that such gaps also appear with the corresponding
situations in the MSSM-like case (top row of figure \ref{fig:mssm-like-vs-vu}). 
However, because of the range of `$S$' that has to be covered there,
the gaps shrink and fail to become visible. 
\item The region over which long-lived metastable DSB vacua (second row) appear 
is somewhat different from the MSSM-like case, in both qualitative and quantitative 
terms. Note that, for the singlino-like case (left), the flat direction 
$\higgsu=\higgsd=0$ now appears as the (only) relevant one and deeper minima
occurring along this may not destabilize the DSB vacuum. 
In the higgsino-like case (right), such DSB vacua occur with deeper minima 
arising in the $D$- and $F_S$-flat directions but now predominantly 
for positive values of $\higgsu$. 
Moreover, flat directions along $\higgsu=\higgsd=0$ and $S=\higgsd=0$ 
may also lead to a long-lived DSB vacuum. 
\item Metastable DSB vacua that are found to be short-lived already under tunneling at
zero temperature (third row) are yet again found to be scanty. These tend to 
appear along the flat directions $\higgsd=\higgsu=0$ and $S=\higgsd=0$. 
This is mostly true for the DSB vacua which are found to be short-lived only when
thermal effects are incorporated (fourth row) are also encountered along these directions.
For cases with a
higgsino-like LSP, sparsely populated regions with such deeper minima occur
about the $D$- and $F_S$-flat directions. Again, as pointed in the context of
figure \ref{fig:mssm-like-vs-vu}, in the absence of thermal correction to
the potential these parameter points would have 
long-lived vacua and hence perfectly viable.
\end{itemize}
\begin{figure}[t]
\centering
\includegraphics[height=0.24\textheight, width=0.49\columnwidth , clip]{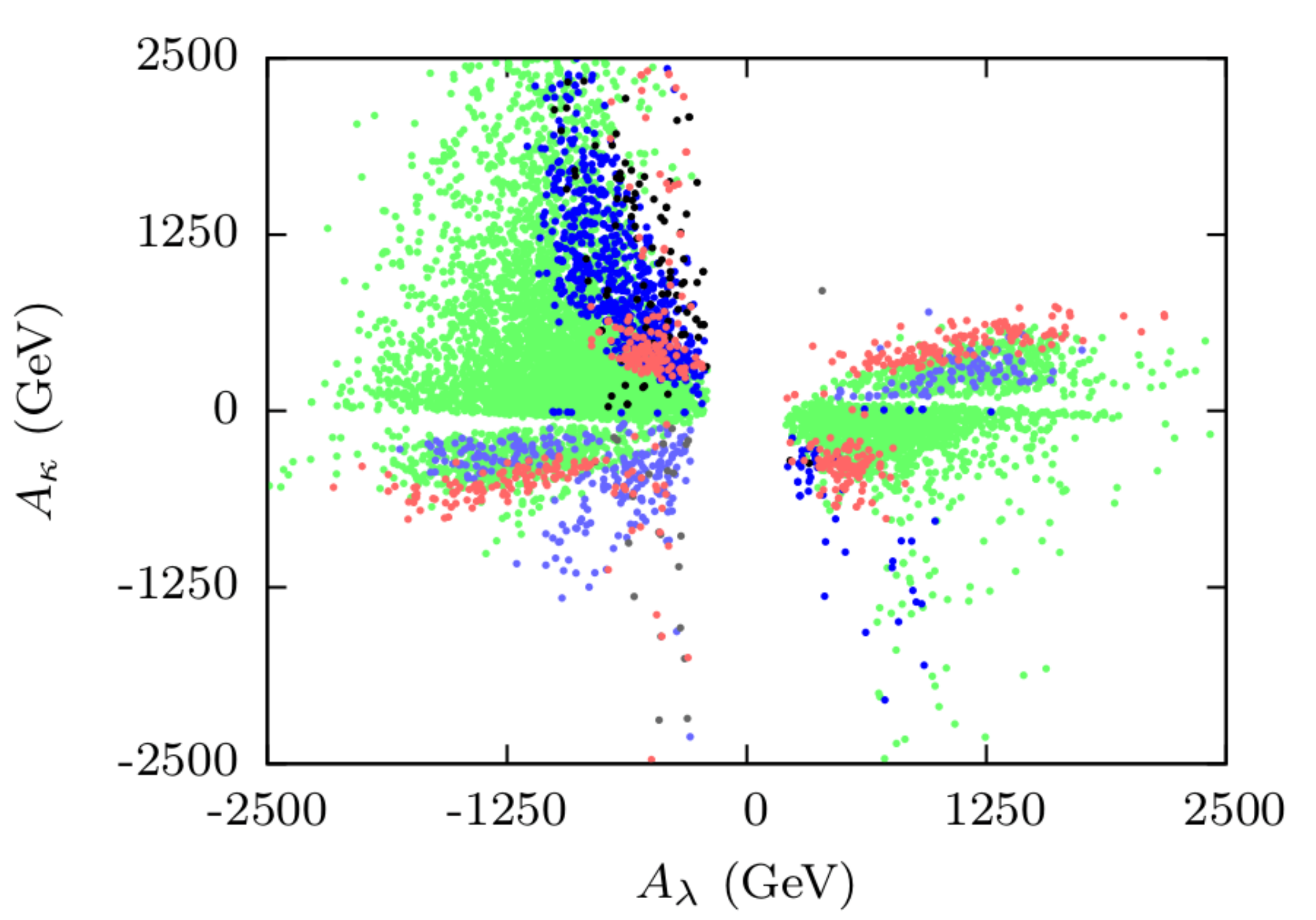}
\includegraphics[height=0.24\textheight, width=0.49\columnwidth , clip]{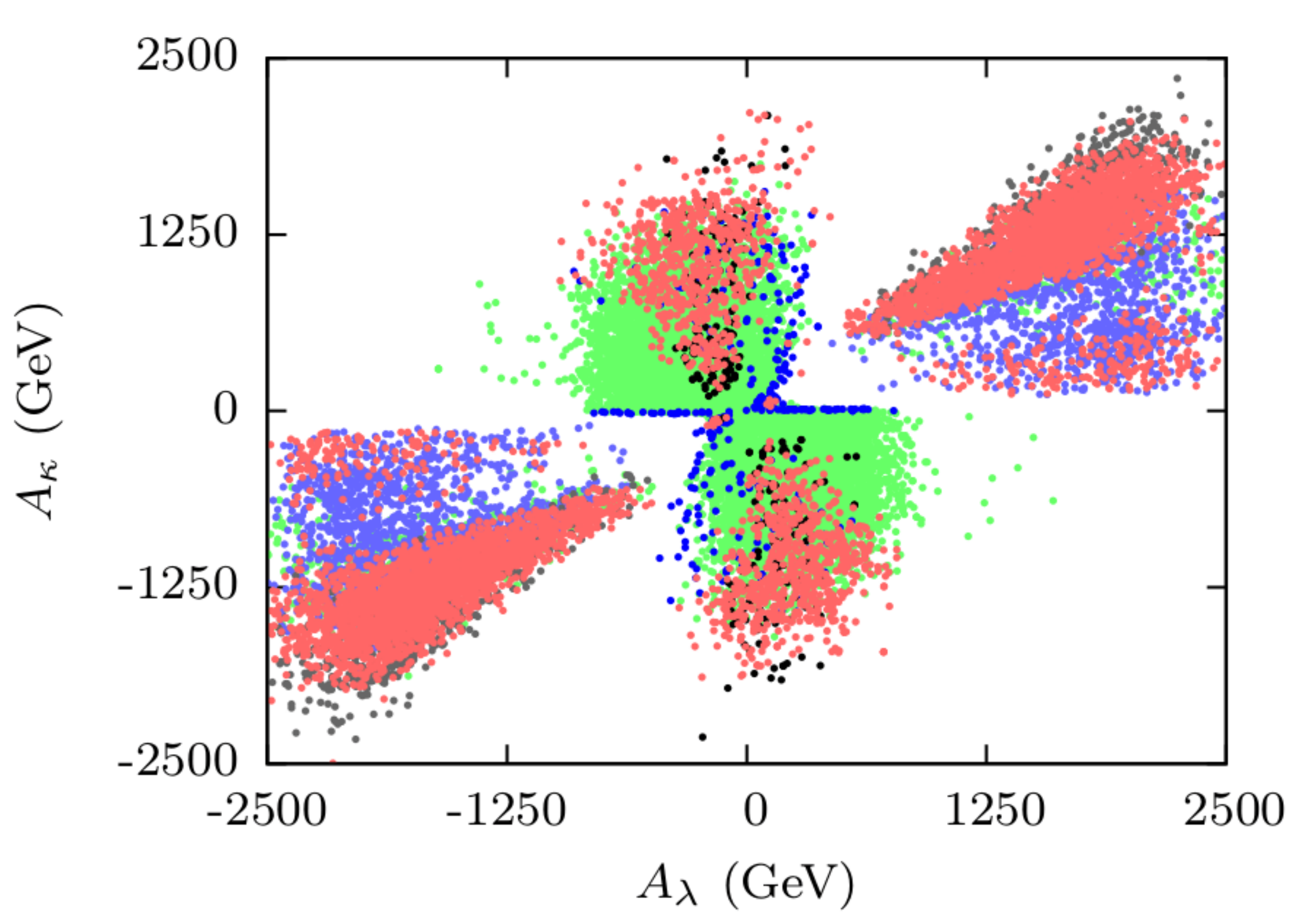}
\caption{Same as in figure \ref{fig:mssm-like-alambda-akappa-veva} but for the NMSSM-like scenario.}
\label{fig:nmssm-like-alambda-akappa-veva}
\end{figure}
%
%

In line with figure \ref{fig:mssm-like-alambda-akappa-veva}, we present in 
figure \ref{fig:nmssm-like-alambda-akappa-veva} the
the status of the DSB vacuum in the $\alambda-\akappa$ plane for the 
NMSSM-like regime. Similar 
considerations as discussed in the context of figure 
\ref{fig:mssm-like-alambda-akappa-veva} apply except for we now have to 
retain the terms containing the product $\vd \vu$ in the relevant expressions
of equations \ref{eq:msq-cp-even} and \ref{eq:msq-cp-odd} and cannot just 
use their approximated versions presented in equation\ref{eq:mscalar-large-vs}. 
As may be recalled, this is 
since in the NMSSM-like scenario $\lambda$ is taken to be relatively large
thus rendering $\vs$ to be on the smaller side for a given value of $\mueff$.
A little algebra with equation \ref{eq:msq-cp-odd} then reveals that the 
magnitude of $\akappa$ gets quickly bounded from above 
when one demands the
singlet $CP$-odd scalar to be non-tachyonic. Subsequently, an inequality
connecting $\akappa$ and $\alambda$ can be derived by demanding the $CP$-even
singlet scalar to be non-tachyonic as well using equation \ref{eq:msq-cp-even}. 
This reveals that $\alambda$ also eventually receives an
upper bound. The phenomenon is reflected in both plots of figure 
\ref{fig:nmssm-like-alambda-akappa-veva} that correspond to the cases with
singlino-like (left) and higgsino-like (right) LSP, respectively. 
However, such a restriction is more evident in the first (second) and the third 
(fourth) quadrants with $\kappa <0$ ($\kappa > 0$) in the cases with a singlino-
(higgsino)-like LSP.
Note that a characteristic dependence on $\alambda$ in the NMSSM-like scenario is in contrast 
to the MSSM-like case for which the dependence on $\alambda$ fades out because 
of large values of $\vs$.
For the case with a singlino-like LSP it is observed that a stable (in green) 
DSB vacuum appears more frequently for a larger magnitude of $\akappa$ when $\akappa <0$. 
The case with a higgsino-like LSP is rather symmetric in $\alambda$ and $\akappa$.
In both cases  regions having a DSB vacuum with varied stability-status are broadly
demarcated.
%
%
\begin{figure}[t]
\centering
\includegraphics[height=0.24\textheight, width=0.49\columnwidth , clip]{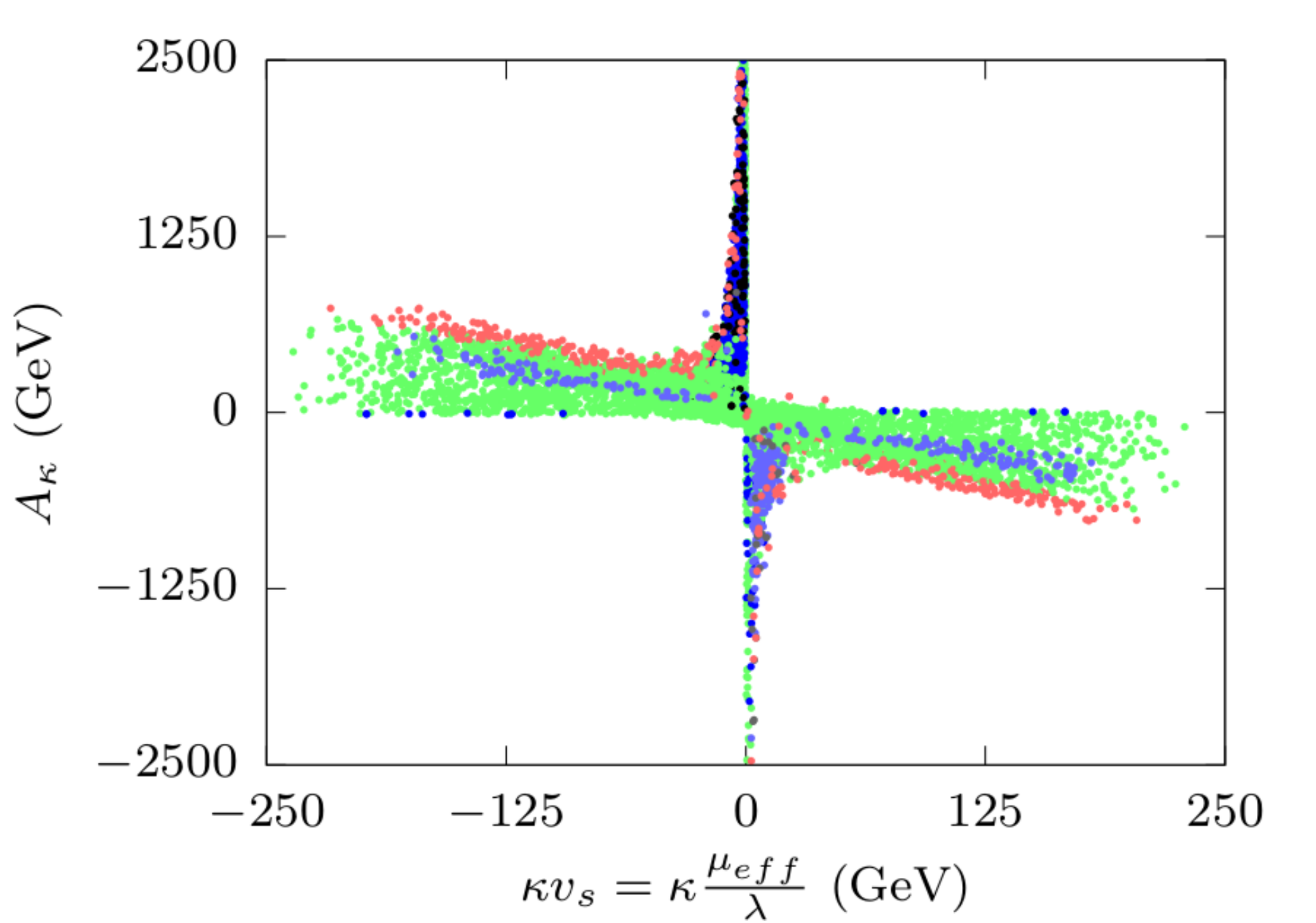}
\includegraphics[height=0.24\textheight, width=0.49\columnwidth , clip]{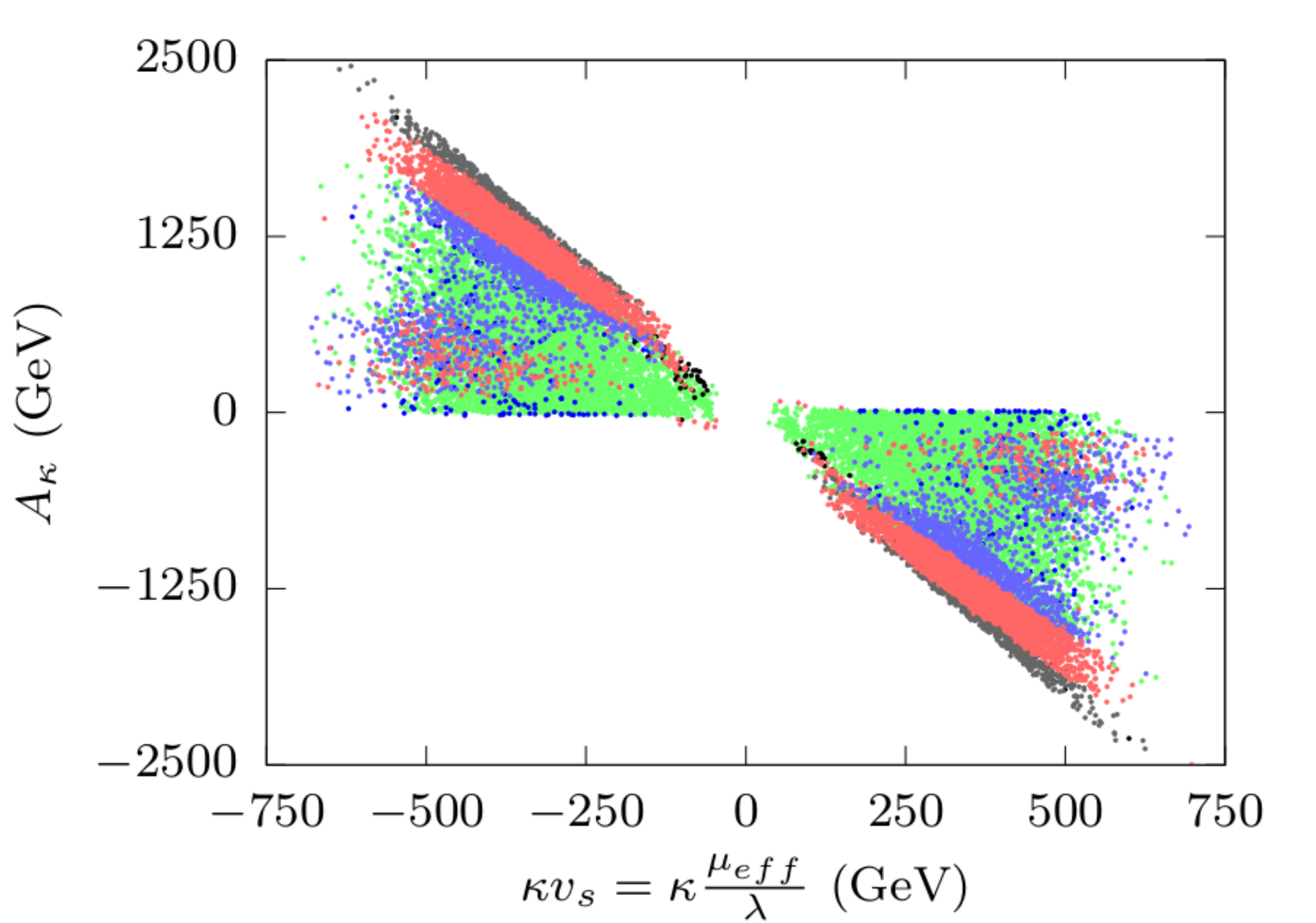}
\caption{Same as in figure \ref{fig:mssm-like-ks-akappa-veva} but for the 
NMSSM-like scenario.}
\label{fig:nmssm-like-ks-akappa-veva}
\end{figure}
%
%

Figure \ref{fig:nmssm-like-ks-akappa-veva} illustrates the results of a
{\tt \veva} scan projected on the $\kappa \vs-\akappa$ plane for the NMSSM-like
case and is the counterpart of figure \ref{fig:mssm-like-ks-akappa-veva} 
presented earlier for the MSSM-like case. Along with regions with stable (in 
green) and thermally unstable (in red) DSB vacua that appear in figure
\ref{fig:mssm-like-ks-akappa-veva}, here we also find clean
regions where long-lived DSB vacua appear. Furthermore, regions in black
(unstable DSB vacuum under tunneling at zero temperature) also appear clearly beyond 
the red regions. As in the MSSM-like case, here also
the agreement with the results from a {\tt Mathematica}-based scan is rather
impressive. The \veva ~scan additionally reveals metastable regions that
ultimately qualify for long-lived DSB vacua (the region in blue) 
and makes clear distinction between the regions in which the DSB vacuum is
long-lived and thermally unstable. 
It can safely be concluded that over a bulk region of the NMSSM-like parameter 
space, the DSB vacuum turns out to be stable (the global minimum).

We find that, in the present case, the role of $\akappa$ in (thermally) 
destabilizing the DSB vacuum is also very prominent. It may be further observed 
that for cases with a singlino-like LSP, the role of $\akappa$ is somewhat 
diluted and a region with a metastable DSB vacuum eventually survive tunneling
to a deeper minimum and becomes long-lived. For cases with a higgsino-like
LSP, $\akappa$ has a more active role to play and it renders a good part of
the region of metastability to that with unstable DSB vacuum.

It is noteworthy that such a level of agreement could be achieved in practice 
even when we go beyond the single-field approximation for the potential (that 
is valid only in the MSSM-like case). This is since both our 
{\tt Mathematica}-based analysis and the \veva ~formulation do consider the 
effects of non-vanishing $\higgsd$ and $\higgsu$, though at a varied level of 
sophistication. An implication of such an agreement is that this 
now provides a genuine scope (and its credibility) of  possible, in-depth 
probes to such multi-scalar scenarios for their underlying analytical nuances 
in the study of vacuum stability. This is something which is 
beyond the scope of \veva ~in its present incarnation.
%
\begin{figure}[t]
\centering
\includegraphics[height=0.20\textheight, width=0.44\columnwidth , clip]{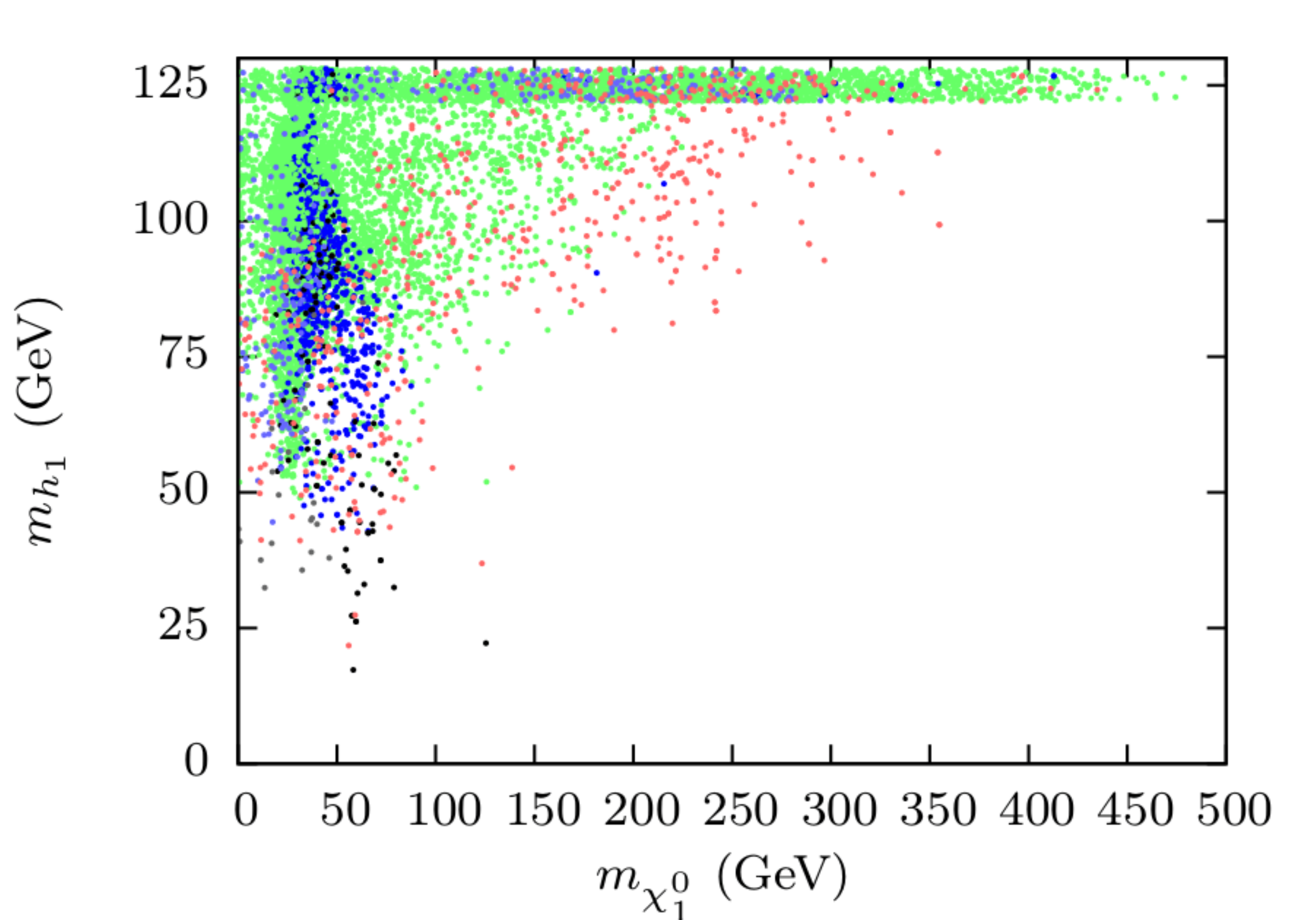}
\includegraphics[height=0.20\textheight, width=0.42\columnwidth , clip]{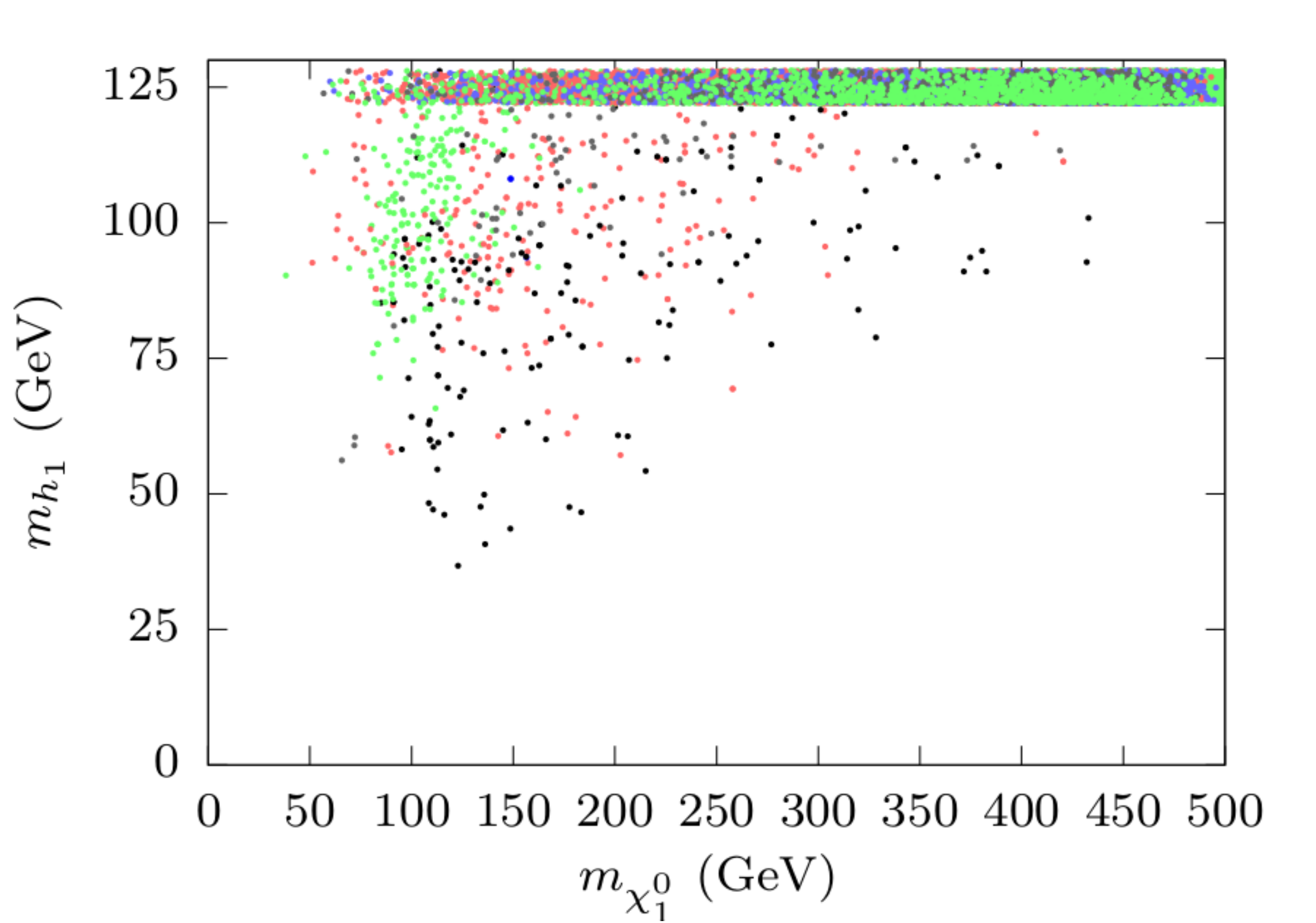}
\includegraphics[height=0.20\textheight, width=0.44\columnwidth , clip]{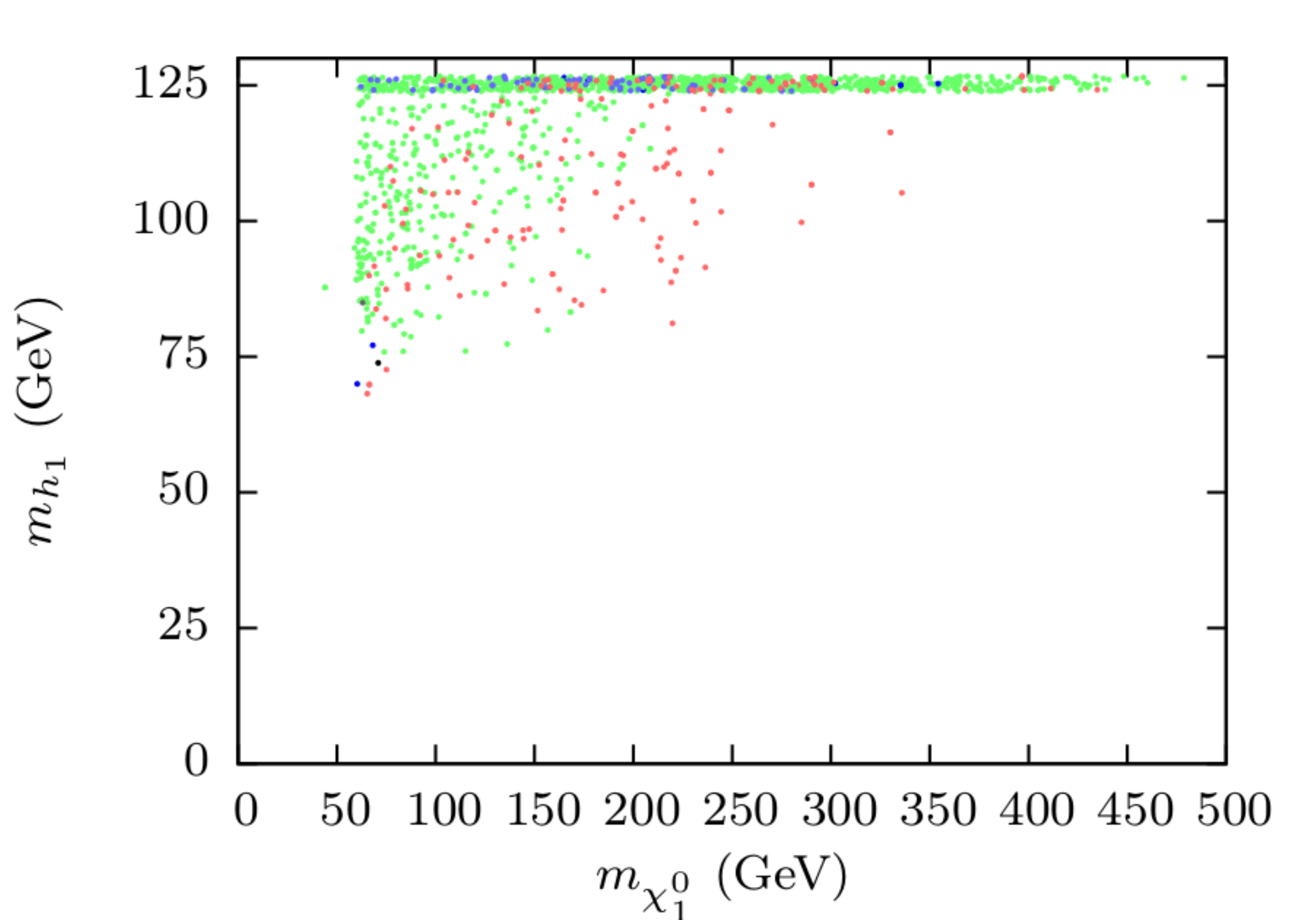}
\includegraphics[height=0.20\textheight, width=0.42\columnwidth , clip]{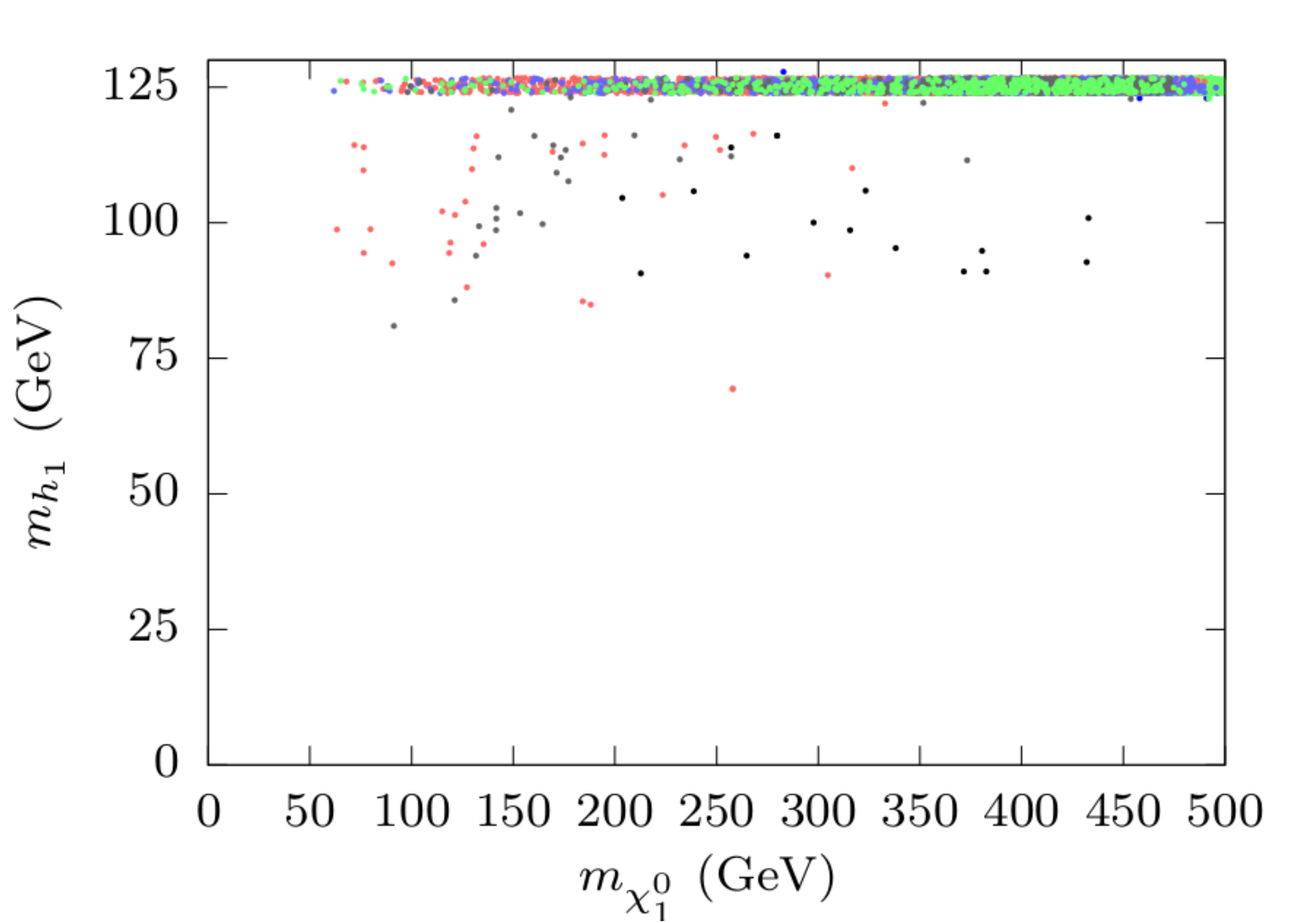}
\includegraphics[height=0.20\textheight, width=0.44\columnwidth , clip]{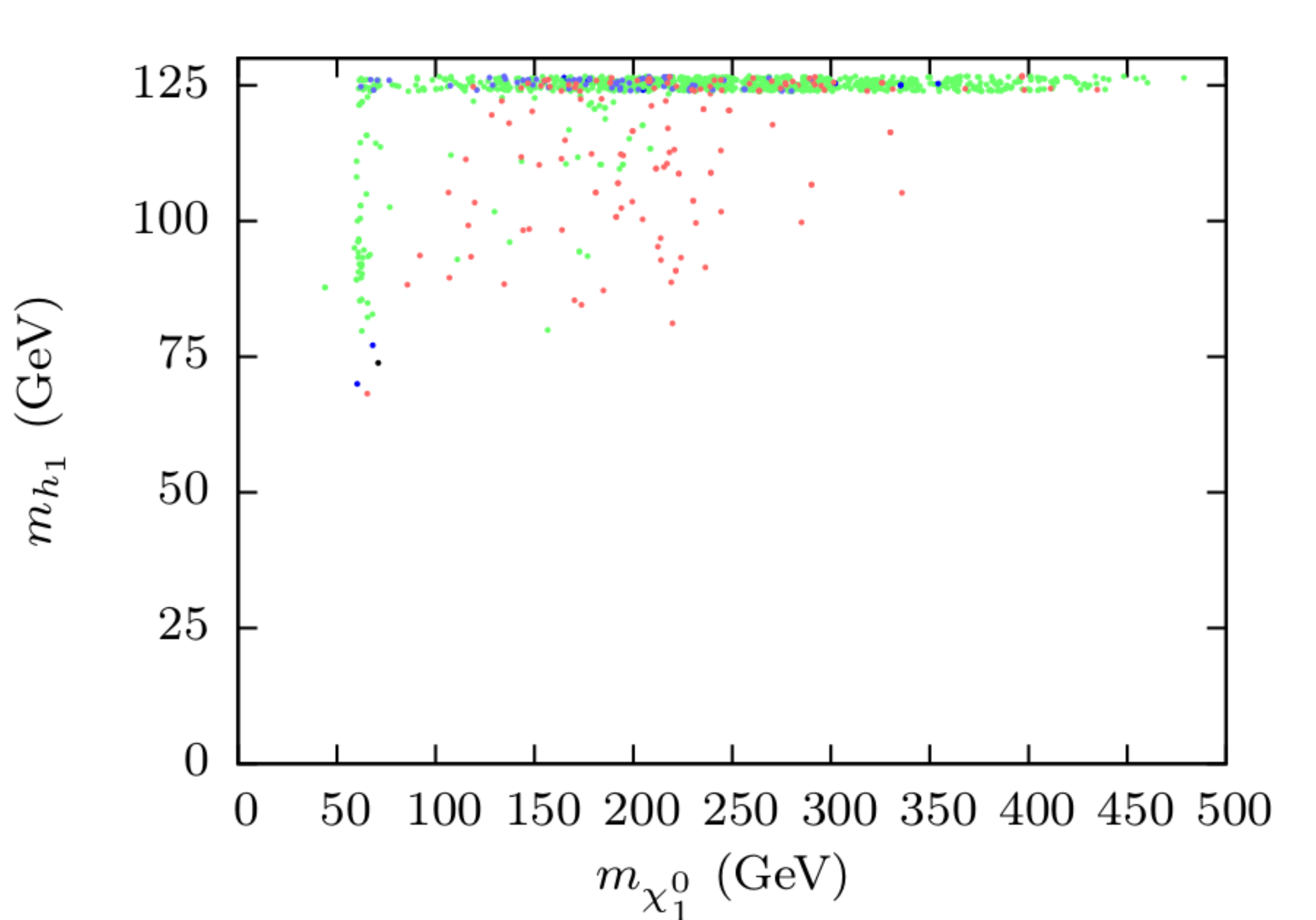}
\includegraphics[height=0.20\textheight, width=0.42\columnwidth , clip]{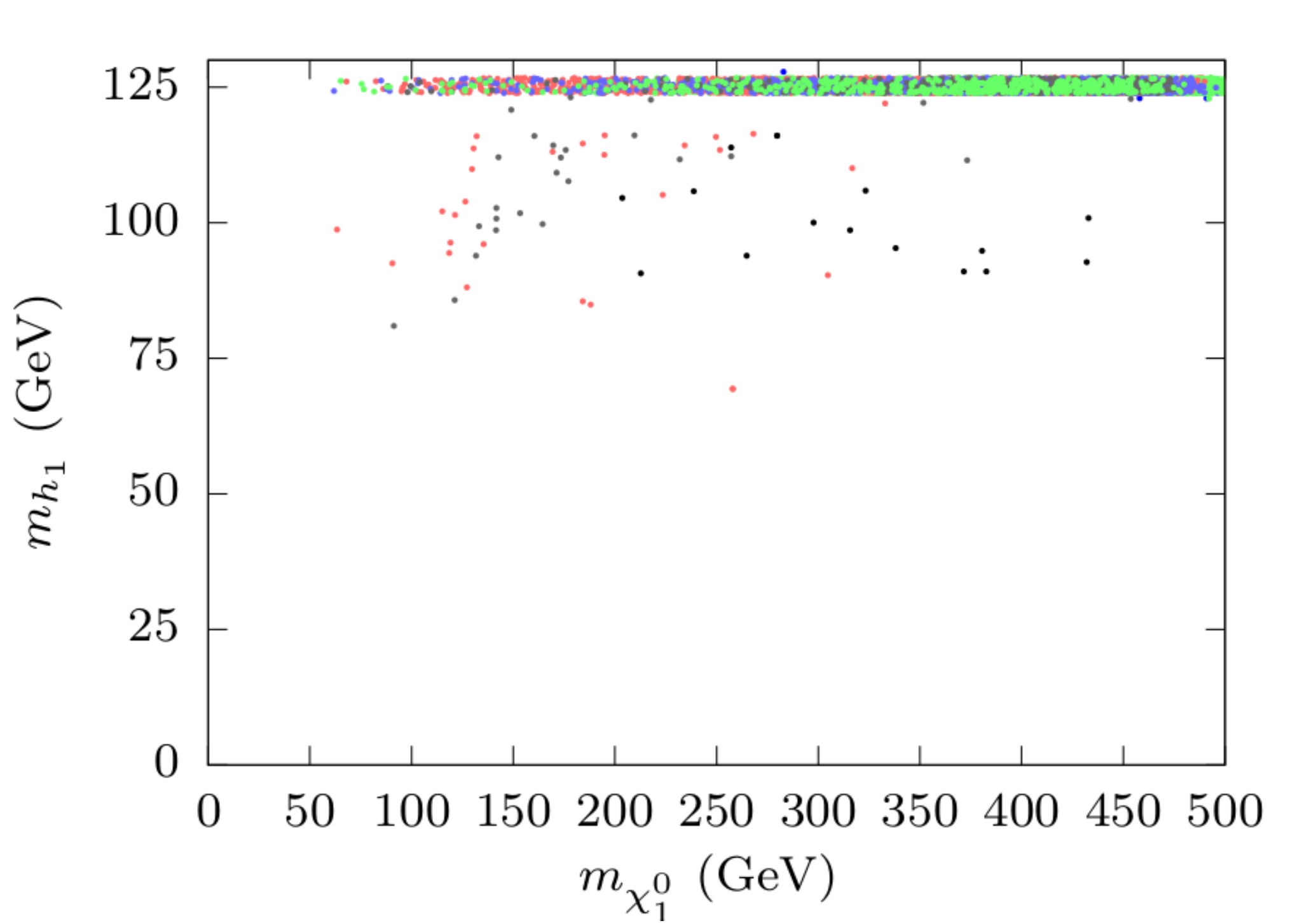}
\caption{Same as in figure \ref{fig:mssm-like-flav-dm} but for  the NMSSM-like
scenario.}
\label{fig:nmssm-like-flav-dm}
\end{figure}

In figure \ref{fig:nmssm-like-flav-dm} we present the same set of scan-data
projected yet again on the $\mntrlone-\mhone$ plane for the cases with 
singlino-like (left) and higgsino-like (right) LSP. 
As for figure \ref{fig:mssm-like-flav-dm}, here also constraints from
{\tt HiggsBounds} and {\tt HiggsSignals} are included in the plots in the second and the third rows.
Furthermore, plots in the second row also receive bounds from the flavor sector. 
Plots in the third row, in addition, include constraints from the dark matter 
sector in the form of an upper bound on the relic density as discussed earlier 
in the context of figure \ref{fig:mssm-like-flav-dm}.
In this case, it is noted that constraints coming from the use of 
{\tt HiggsBounds} and {\tt HiggsSignals} 
have  more drastic effects in the case with a singlino-like LSP. These exclude much 
of the regions with a long-lived DSB vacuum (in blue) that appear in the plots in the 
first row. The reason is that for the NMSSM-like scenario $\lambda$ is relatively large
and this induces significant mixing between the singlino and the higgsino states
in the LSP. This, in turn, results in larger decay rates of the SM-like Higgs 
boson and the $Z$-boson to a pair of LSP leading to an unacceptable level of
invisible widths for them. The sharp edge appearing along an LSP mass of $\sim 60$ GeV
is just the kinematic manifestation of the fact that the invisible decay of the 
SM-like Higgs boson with mass around 125 GeV is totally forbidden as it immediately
yields too large an invisible width. Note that this is in clear contrast with the
MSSM-like scenario (figure \ref{fig:mssm-like-flav-dm}) where such a light LSP is a possibility. 
On the other hand, for the case with higgsino-like LSP, the flavor constraints 
remove the region with a stable (in green) DSB vacuum appearing for lower values 
of $m_{h_1}$ in the first row. Use of {\tt HiggsBounds} and {\tt HiggsSignals}
further render the region with a heavier LSP only sparsely populated.

With relatively large $\lambda$ in the NMSSM-like scenario, another associated effect 
is the enhanced coupling among the SM-like Higgs and a pair of lighter singlet-like
Higgs state that might lead to an unacceptable level of decay rate for the SM-like
Higgs to the latter states. This results in a (kinematic) edge along 
$m_{h_1} \sim 60$ GeV.

Constraints from the dark matter sector (bottom row) work in a way similar to
the MSSM-like scenario in figure \ref{fig:nmssm-like-flav-dm}. Clearly, a thin
vertical strand at around $\mntrlone \sim 62-63$ GeV survives. A palpable
depletion in allowed points at moderate value of $\mntrlone$ is a possible
artifact of conspiring parameters that lead to reduced couplings of the LSP
to other higgsino-like states thus enhancing the relic density to unacceptable
levels. 

As far as the the occurrence of a light singlet $CP$-odd scalar is concerned,
the considerations are very similar to that in the case of the MSSM-like scenario
as dicussed at the end of section \ref{subsubsec:mssm-like-veva}
(except for more parameters now entering the picture). The anticorrelation
between the masses of the $CP$-odd and $CP$-even singlet scalars in cases
with a higgsino-like LSP stil exists. However, in the present case,
it is difficult to find an accompanying light (lighter than the SM-like Higgs
boson) singlet $CP$-even scalar when its $CP$-odd cousin is light.
%
\section{Charge and color breaking vacua in $Z_3$-symmetric NMSSM}
\label{sec:ccb-vacua}
%
In this section we discuss the issue of possible appearance of CCB vacua in the
$Z_3$-symmetric NMSSM. It may be quickly appreciated that the phenomenon acquires
some added degree of intricacy in the presence of the all crucial singlet neutral
scalar, when compared to the same in the MSSM. Traditional CCB bounds arise from
various combinations of so-called $D$- and $F$-flat directions that now include
the colored scalars \cite{Ellwanger:1999bv, Djouadi:2008uj}. 
In the present study we use \veva ~and take a minimal but representative approach by only allowing for 
the left- and 
the right-handed top squarks developing \vevs. Also, in \veva ~ we consider only one color degree of freedom developing \vev ~at a time.

Generic dangerous directions in the field space, in theories with an extra 
(singlet) scalar and in the CCB context, have been listed in reference 
\cite{Ellwanger:1999bv}. Traditional CCB bounds arise out of negative 
contribution from some trilinear coupling $A_{ijk} \phi_i \phi_j \phi_k$ where 
$\phi$-s are the various relevant scalar fields. $D$-flat directions appear 
along $\phi_i=\phi_j=\phi_k$. Directions flat in both `$D$' and `$F$' are 
generally the ones along which 
one comes across a UFB scalar 
potential\footnote{It is known that a $D$-flat direction involving the
scalar electrons fields and $\higgsd$ could emerge as a potentially very dangerous CCB 
direction \cite{Ellwanger:1999bv} depending upon the value of the corresponding trilinear coupling
$A_e$. However, in the present study since we fix $A_e=0$ and the scalar electron masses
to a value as high as 3 TeV, we are safe against developing of such a dangerous direction.
On the same ground, with typically larger values of $A_t$ and smaller masses of the top 
squarks that we use, it is important to have a safeguard against dangerous directions
involving the top squark fields.}. 
There the improved CCB bounds are also derived in the framework of constrained
NMSSM (CNMSSM) along directions other than 
the above two. These involve non-trivial combinations of uncorrelated 
field values and exploit the increased
number of free parameters in  the
scenario.   

The $D$-flat direction involving the top squarks is given by 
$|\stleft|=|\stright|=|\higgsu|$. In the MSSM, CCB minima develop when
$A_t^2 > 3(\mstleft^2 + \mstright^2 + \mhiggsusqhat)$ 
and the depth of the minima is of 
$\mathcal{O}\left({A_t^4 \over y_t^2} \right)$ \cite{Ellwanger:1999bv} where 
$\mhiggsusqhat=\mhiggsusq + \mu^2$, 
$A_t$ is the trilinear coupling in the top squark sector and $y_t$ is
the top Yukawa coupling. In the NMSSM, presence of non-vanishing singlet scalar 
(`$S$') turns the minima further deeper to the tune of  
$\mathcal{O}\left({A_k^4 \over \kappa^2} \right)$ \cite{Ellwanger:1999bv}.
Thus, `$\kappa$' and $\akappa$ take central roles in a study of CCB vacua
in the NMSSM. At the same time, it is observed that $\lambda$ and $\alambda$ 
affect the proceedings somewhat indirectly. This will become clear as we proceed.

We find that the optimal approach to present the CCB phenomenology is to stick 
to the broad scenario we adopt in section \ref{sec:vacuum-structure}. However,
given our present purpose, we choose the parameters in a way such that CCB 
effects are pronounced, i.e., deeper minima appear along the top squark field
direction. This, in general, requires a somewhat larger values of $A_t$ and 
$\tan\beta$. In table \ref{tab:ccb-inputs} we list the set of parameter for various representative 
scenarios adopted in this study of the  CCB vacua.
%
\begin{table}[]
\centering
\begin{tabular}{|c|c|c|c|}
\hline
\multicolumn{2}{|c|}{}                                                                                                     & Singlino-like LSP                          & Higgsino-like LSP                          \\ \hline
\multirow{2}{*}{MSSM-like}  & \multirow{2}{*}{\begin{tabular}[c]{@{}c@{}}$\lambda$ = 0.05\\ $\tan\beta$ = 20\end{tabular}} & $\kappa$ = 0.02                            & $\kappa$ = 0.10                             \\ 
                            &                                                                                              & -2500 GeV \textless $A_\kappa$ \textless 0 & -500  GeV \textless $A_\kappa$ \textless 0 \\ \hline
\multirow{2}{*}{NMSSM-like} & \multirow{2}{*}{\begin{tabular}[c]{@{}c@{}}$\lambda$ = 0.55\\ $\tan\beta$ = 10\end{tabular}} & $\kappa$ = 0.20                             & $\kappa$ = 0.55                            \\ 
                            &                                                                                              & -1500 GeV \textless $A_\kappa$ \textless 0 & -500 GeV \textless $A_\kappa$ \textless 0  \\ \hline
\end{tabular}
\caption{The fixed input parameters and the ranges of others used in the scan for studying
the CCB minima for the MSSM- and NMSSM-like scenarios with a singlino-like and higgsino-like
LSP. Throughout, $A_t$ is varied over the range $|A_t|<7.5$ TeV and we take $\mueff=300$ GeV.}
\label{tab:ccb-inputs}
\end{table}
%
%

In the light of the discussion above, we explore how the stability of the DSB 
vacua gets affected by changing values of the soft trilinear parameters
$A_t$, $\akappa$ and $\alambda$ and the soft masses of the top squarks. For
clarity, we keep other parameters fixed at values representative of the
scenarios presented. 
%
\begin{figure}[t]
\centering
\includegraphics[height=0.24\textheight, width=0.49\columnwidth , clip]{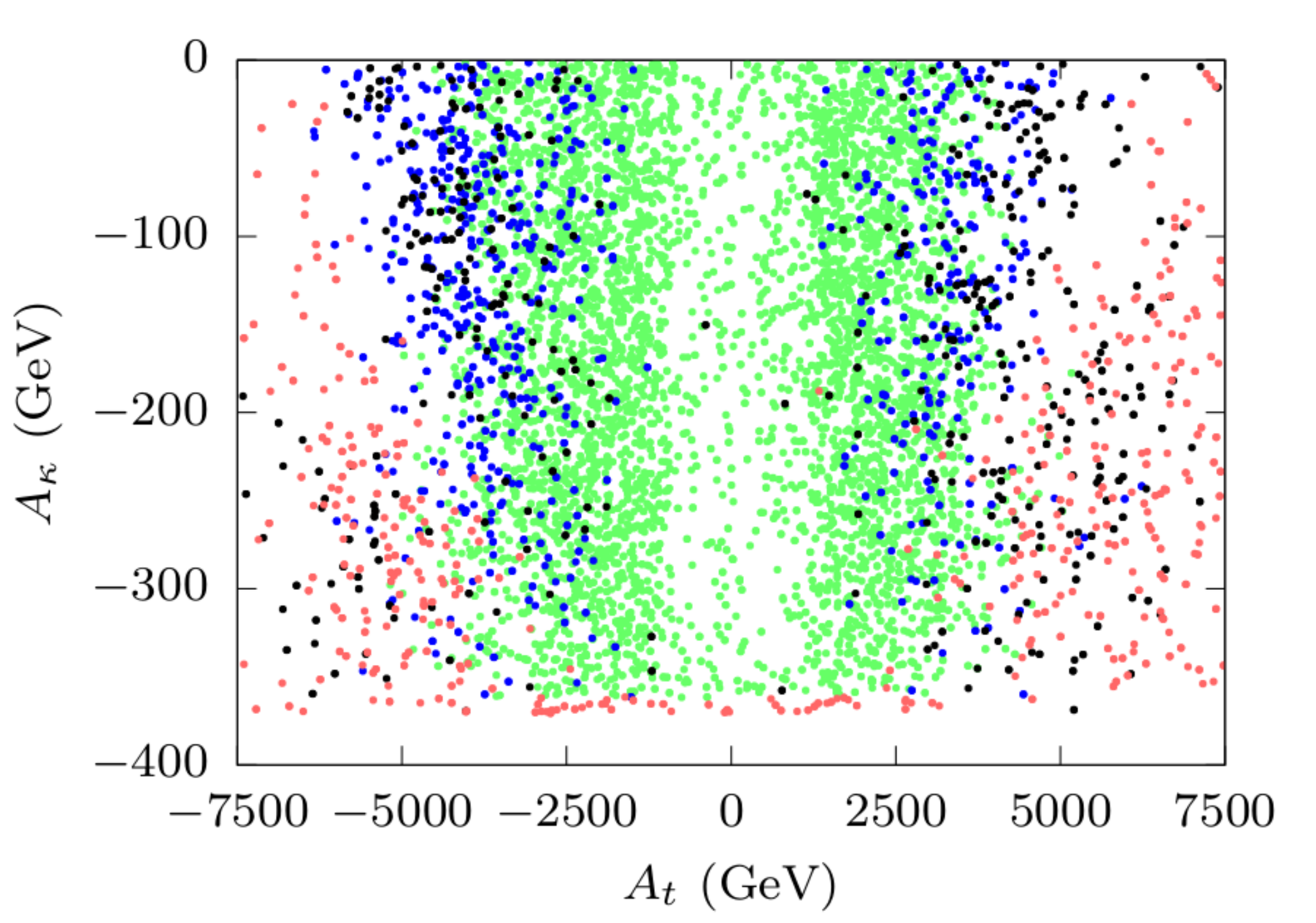}
\hskip -5pt
\includegraphics[height=0.24\textheight, width=0.49\columnwidth , clip]{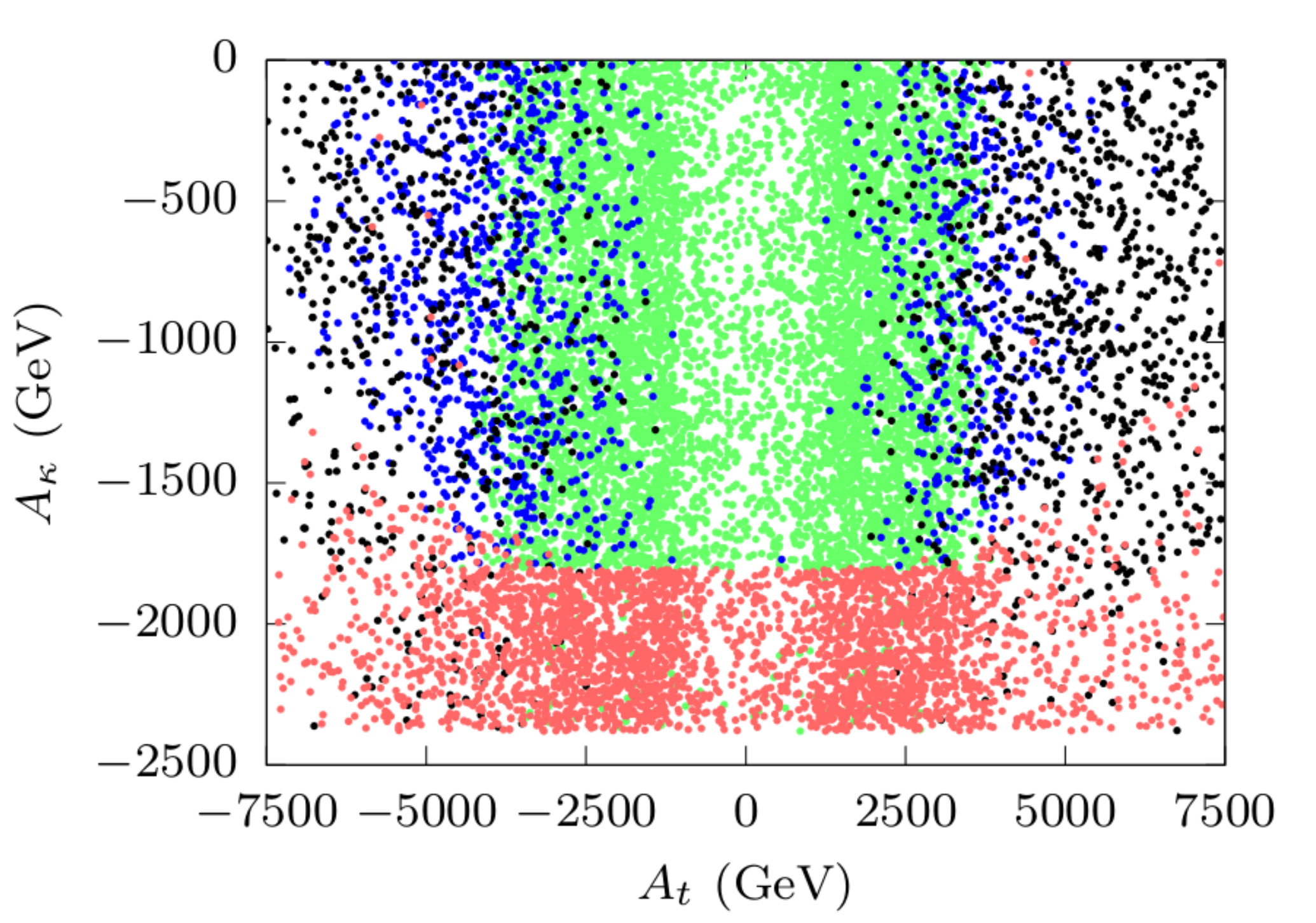}
\includegraphics[height=0.24\textheight, width=0.49\columnwidth , clip]{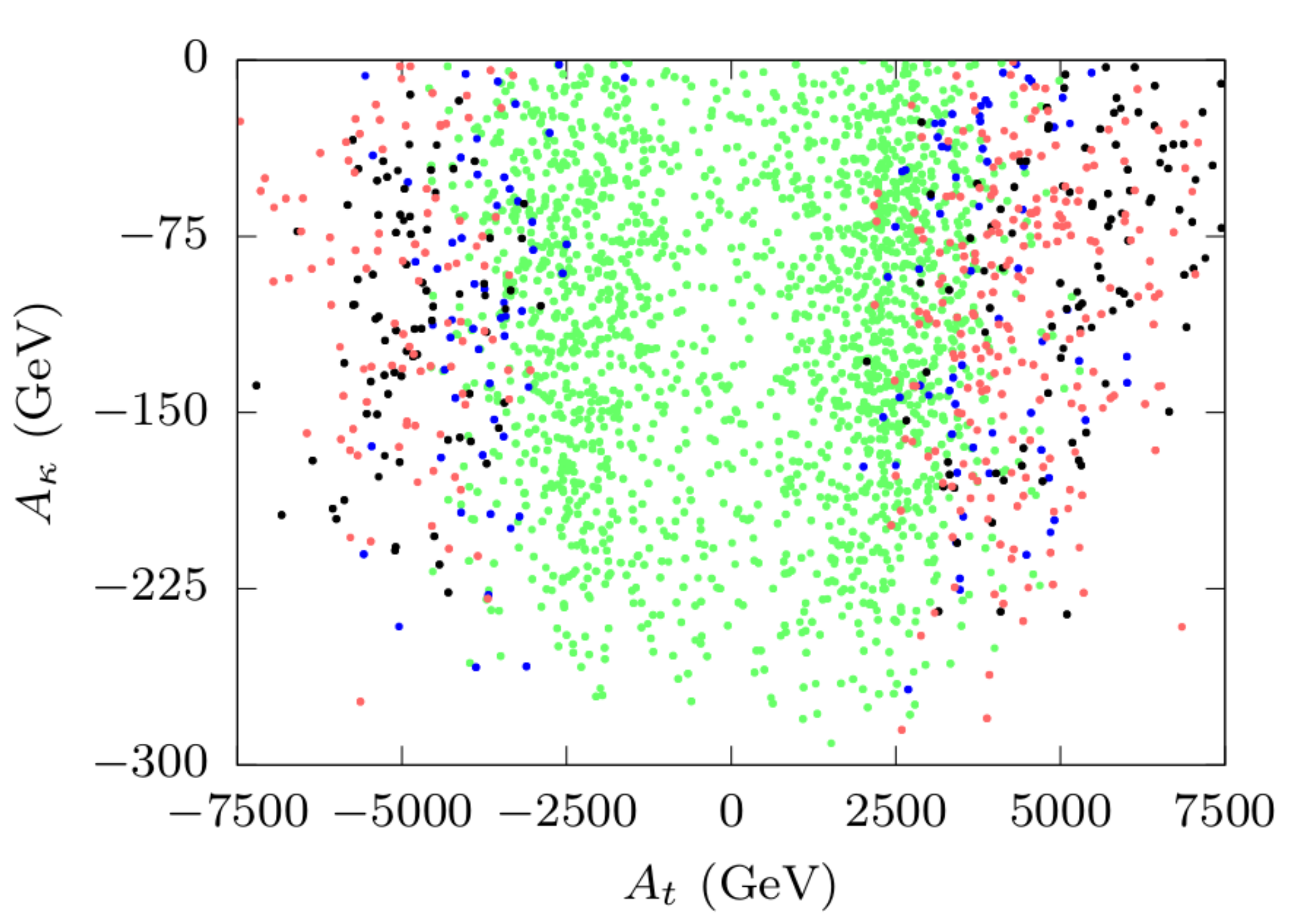}
\hskip -5pt
\includegraphics[height=0.24\textheight, width=0.49\columnwidth , clip]{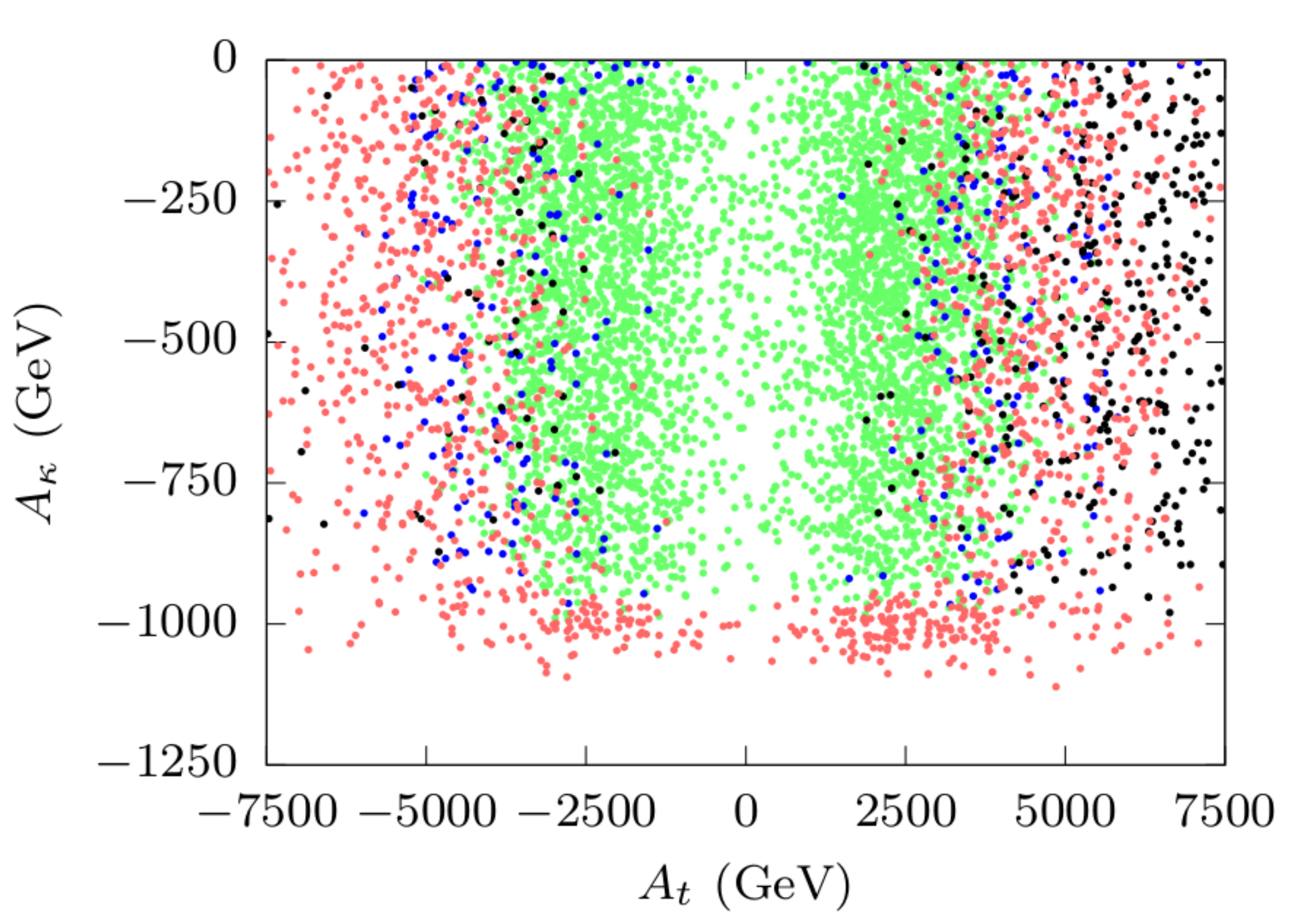}
\caption{Regions obtained from \veva ~scan in the $A_t-\akappa$ plane
indicating the fate of the DSB vacuum in the MSSM-like (top panel) and the 
NMSSM-like (bottom panel) scenarios with a singlino-like (left) and a 
higgsino-like LSP (right). Color-code used is as summarized in table 
\ref{tab:color-number-codes}. The fixed and varying input parameters are as
shown in table \ref{tab:ccb-inputs}.}
\label{fig:ccb-At-akappa}
\end{figure}
%

In figure \ref{fig:ccb-At-akappa} we explore how stable the DSB vacua are in the
$A_t-\akappa$ plane for the MSSM-like (top panel) and the NMSSM-like (bottom
panel) scenarios and with a singlino-like (left) and higgsino-like (right) LSP.
It shows that, for not too large an $\akappa$, the fate of 
the DSB vacua is broadly determined by $|A_t|$, as is the case in the MSSM. As 
expected, a smaller $|A_t|$ would still leave the DSB minima to be the global 
ones and thus stable (green points). However, with increasing $|A_t|$, the scalar 
potential develops deeper minima in the direction of top squark field(s). Thus, 
the DSB vacua become metastable. For an intermediate range of $|A_t|$, the 
tunneling time of the DSB vacuum to a deeper minimum is still long enough thus 
making the latter long-lived (blue points). Even larger $|A_t|$ values, however, 
render the DSB vacua unstable (black and red points).

The role of $|\akappa|$ becomes apparent as soon as it becomes
larger than a critical value that depends on the case in hand. Such effects are
marked by abrupt onset of regions in red at the bottom of (three of) these plots.
This can be understood in terms of deeper minima developing in the direction of
the singlet scalar field `$S$'\footnote{It may be noted that the nature of such
deeper minima and other pertinent model parameters are such that once metastability 
develops there is always a fast enough thermal tunneling of the DSB vacua to the
deeper minima. Thus, situations with long-lived DSB vacua just may not arise.}.
Given that the depth of such deeper minima goes as 
$\mathcal{O}\left({\akappa^4 \over \kappa^2} \right)$, these minima could get 
quickly deeper than the ones developed in the top squark direction(s) for smaller
values of $\kappa$. Note that, in a given broad scenario (say, the MSSM-like one;
the first row) the larger the value of `$\kappa$', the bigger an $|\akappa|$ is 
required for the onset of such a deeper minimum in the `$S$' direction. It may also
be expected that once there is an onset of such a minimum, increasing $|\akappa|$
would only render the same even deeper before it reaches the critical value that 
turns the singlet $CP$-even scalar tachyonic (see equations \ref{eq:msq-cp-even}, 
\ref{eq:msq-odd-even}). The expression for the mass of the singlet $CP$-odd scalar 
in the latter equation and the requirement of this state to be non-tachyonic 
explain the sign on $\akappa$ for our choice of $\kappa,\lambda > 0$ and 
$\mueff (=\lambda / \vs)>0$.
%
%
\begin{figure}[t]
\centering
\includegraphics[height=0.24\textheight, width=0.49\columnwidth , clip]{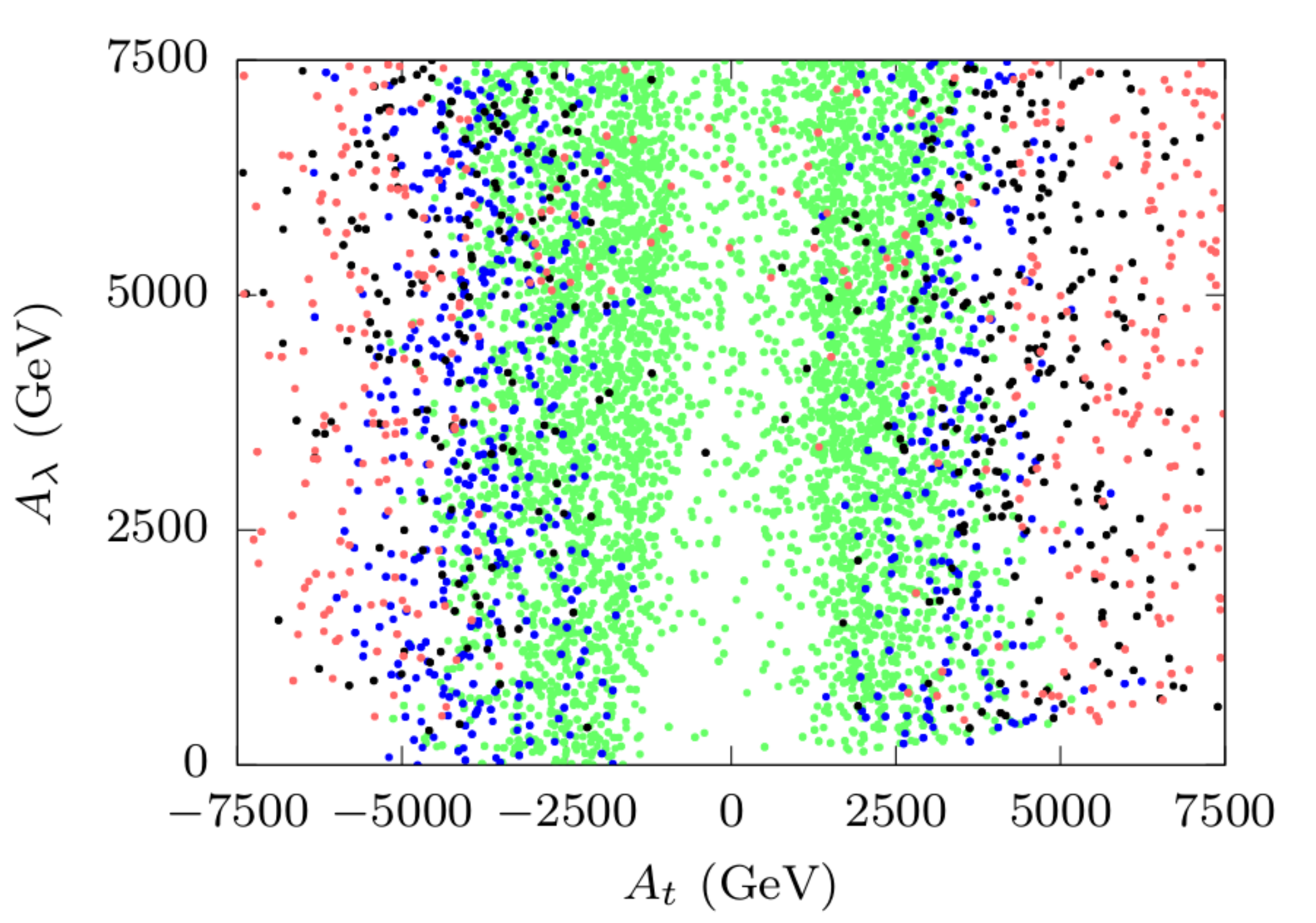}
\hskip -5pt
\includegraphics[height=0.24\textheight, width=0.49\columnwidth , clip]{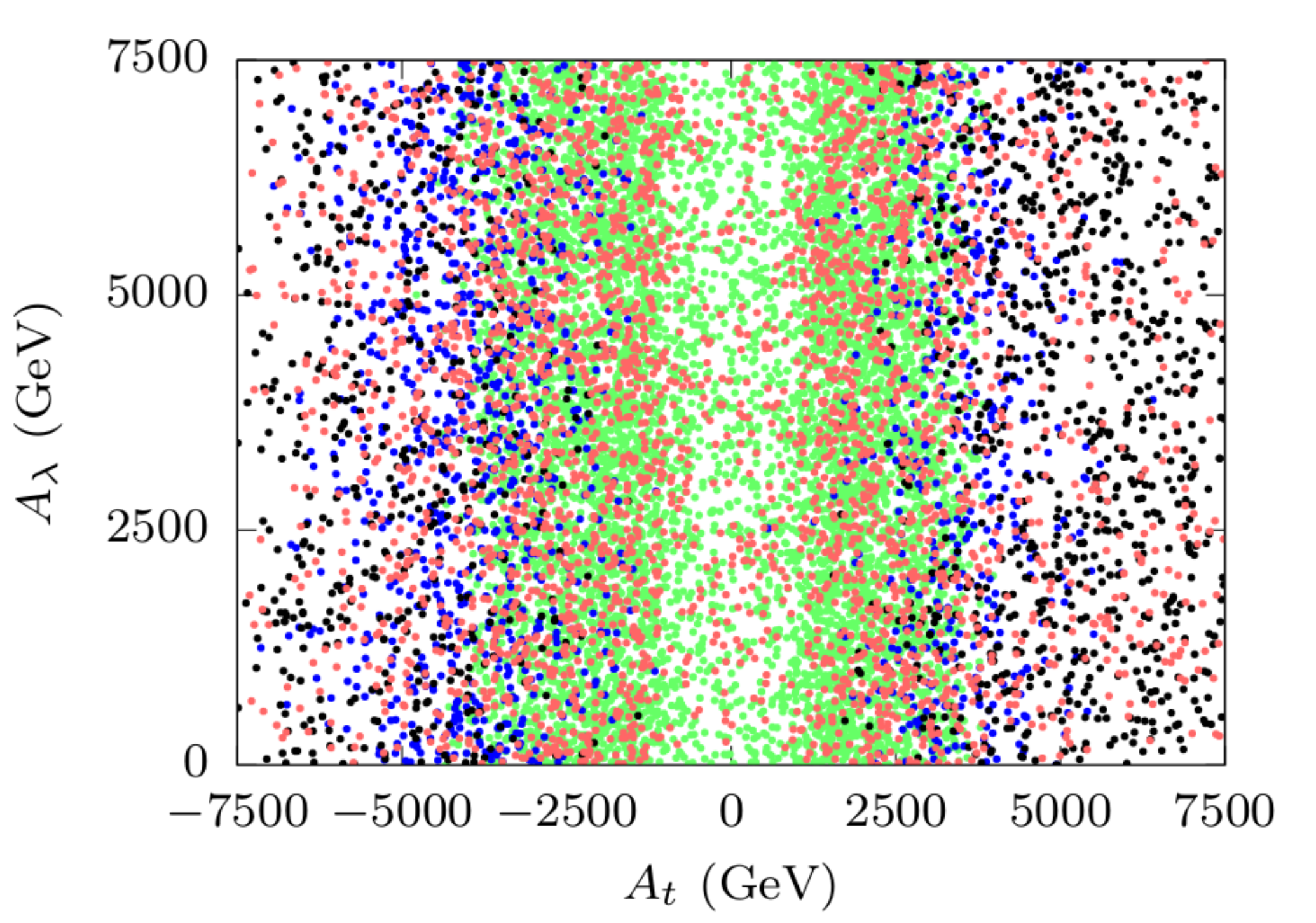}
\includegraphics[height=0.24\textheight, width=0.49\columnwidth , clip]{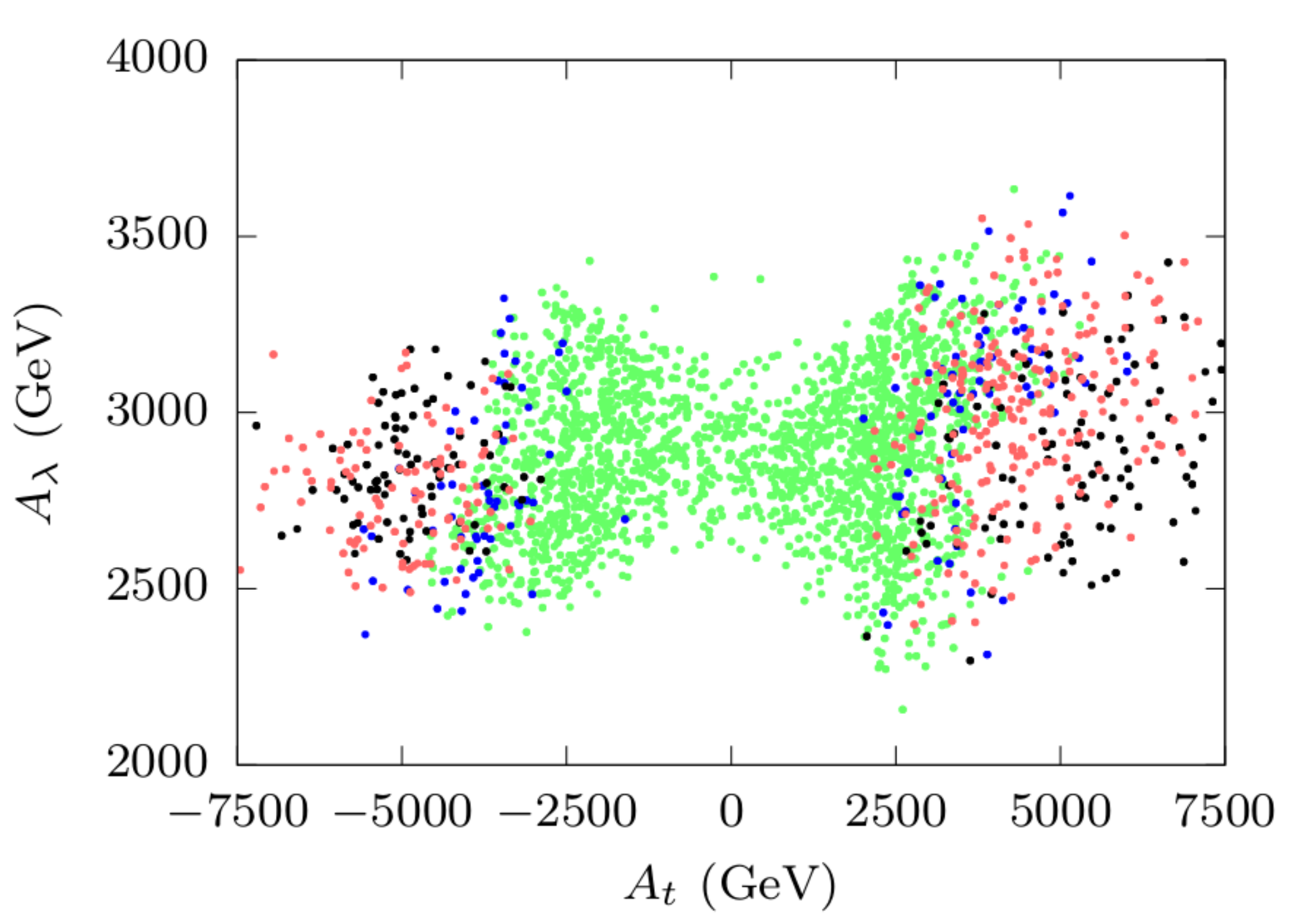}
\hskip -5pt
\includegraphics[height=0.24\textheight, width=0.49\columnwidth , clip]{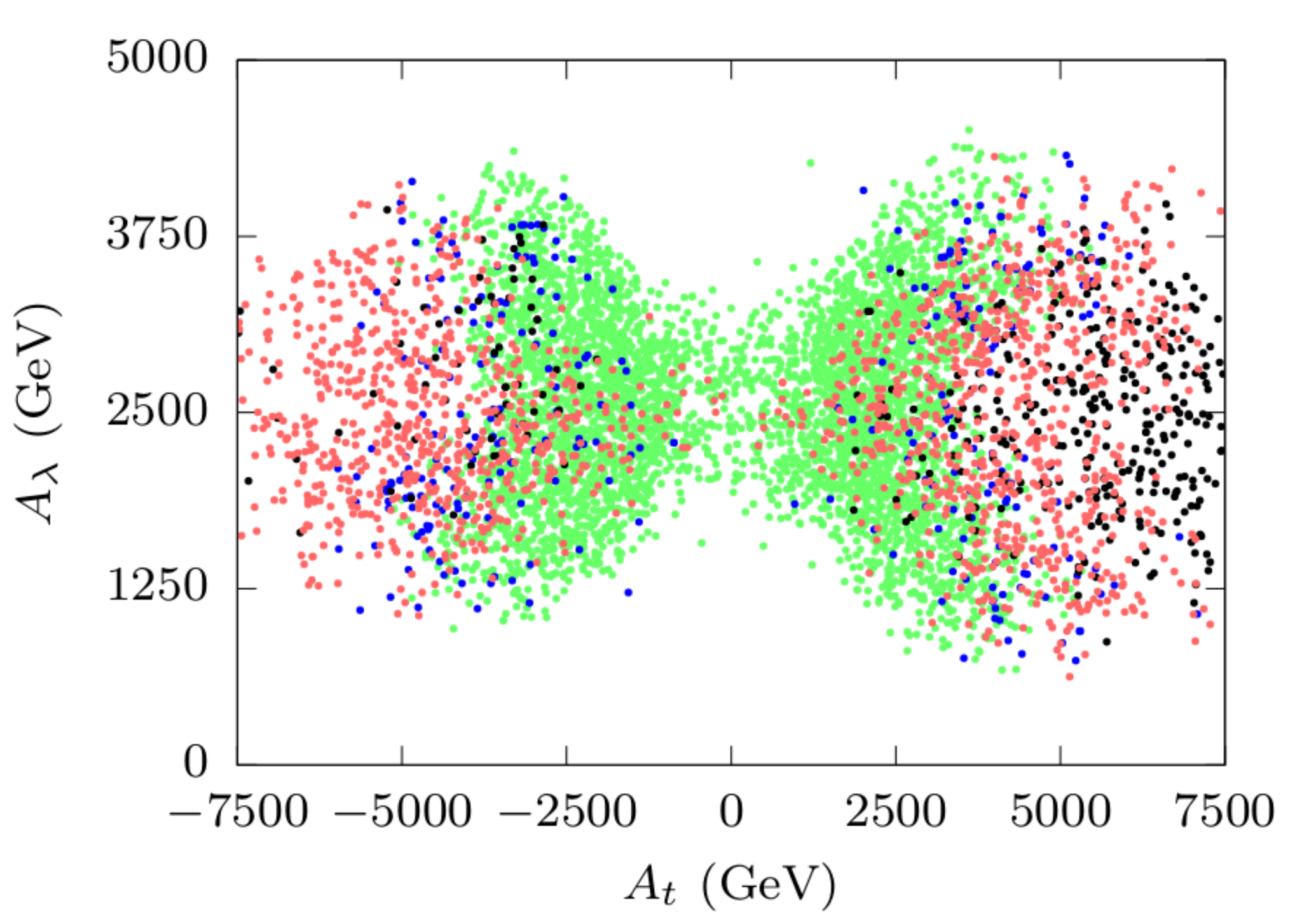}
\caption{Same as in figure \ref{fig:ccb-At-akappa} but in the $A_t-\alambda$ plane.}
\label{fig:ccb-At-alambda}
\end{figure}
%
%

The vertical depletion of points with smaller values of $A_t$ can, in general,
be understood in terms of not big enough a correction that it provides to the
mass of the SM-like Higgs boson, in the MSSM- and NMSSM-like cases alike. For
the MSSM-like scenario this is already expected while for the NMSSM-like case
this is an artifact of choosing not large enough a value of $\lambda$. 

In figure \ref{fig:ccb-At-alambda} an analogous study is presented in the
$A_t-\alambda$ plane. In the MSSM-like scenario with a singlino-like LSP 
(top panel, left) the issue of stability of the DSB vacuum is almost solely
determined by $A_t$, i.e., on whether (and what kind of) CCB minima develop.
For the MSSM-like scenario with a higgsino-like LSP (top panel, left) the
various stability regions are mixed up. Given the scan-range of $\akappa$ for 
this case (which includes values giving rise to the thermally unstable (red) 
region in a similar scenario in figure \ref{fig:ccb-At-akappa}), it may not 
be difficult to recognize that the red points appearing at small values of
$A_t$ are not due to deeper CCB minima but because of deeper minima appearing
in the `$S$' direction. Also, it may be noted, that stable (green) and long-lived
(blue) DSB vacua exist for comparatively smaller $A_t$ values. For the
NMSSM-like scenario (bottom panel) the stability of the DSB vacua hardly 
depends on $\alambda$ and is solely governed by the nature of the CCB minima
which in turn depends on $|A_t|$. However, some rather sparsely populated
red (thermally unstable) regions may be seen in the bottom, right plot
which is again due values of $\akappa$ that lead to deeper minima in the
`$S$' direction.
We find that with larger values of $\lambda$ in the NMSSM-like scenario
it is difficult to obtain the mass of the SM-like Higgs boson in the correct
range unless $\alambda$ is large.

\begin{figure}[t]
\centering
\includegraphics[height=0.24\textheight, width=0.49\columnwidth , clip]{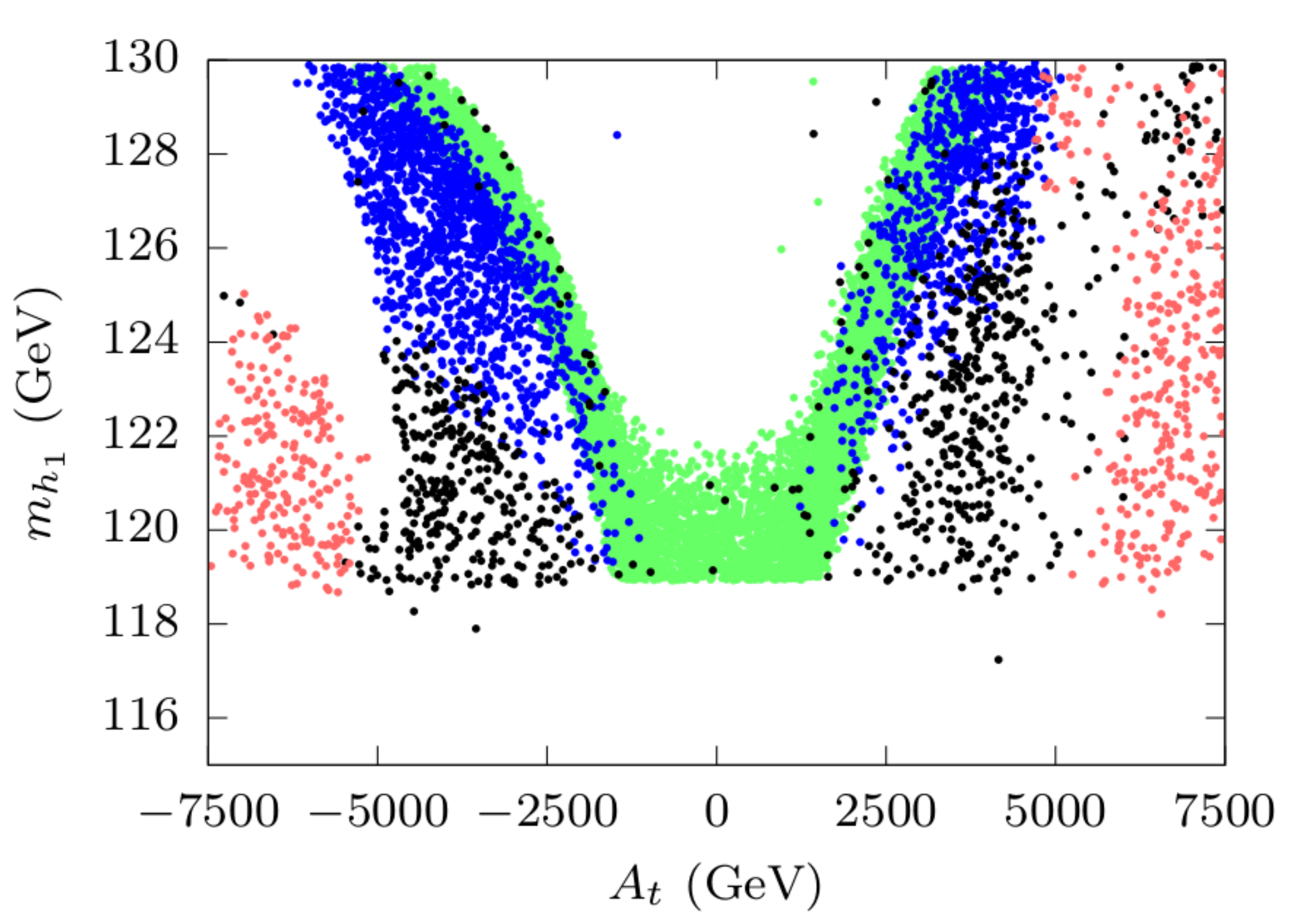}
\hskip -5pt
\includegraphics[height=0.24\textheight, width=0.49\columnwidth , clip]{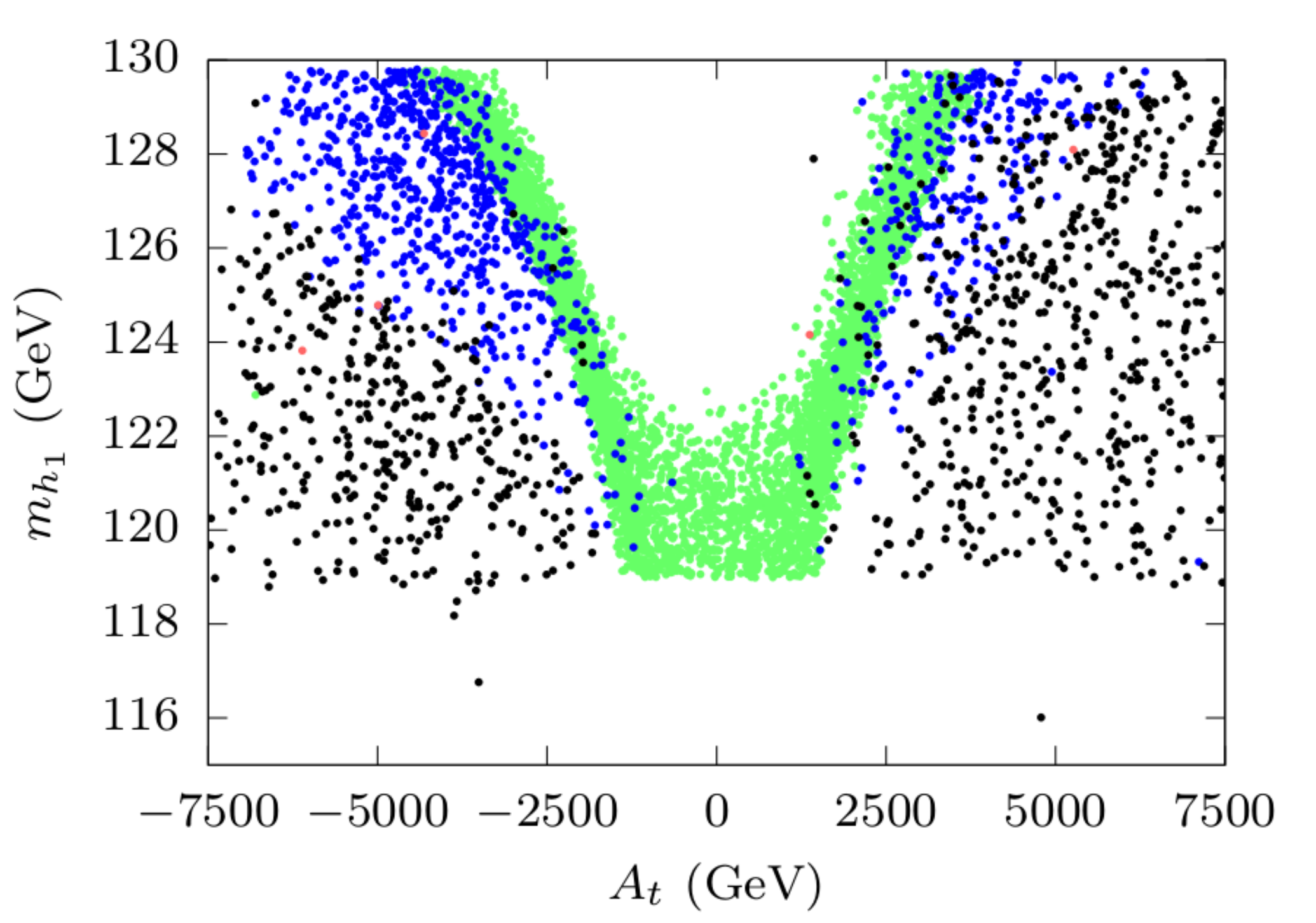}
\includegraphics[height=0.24\textheight, width=0.49\columnwidth , clip]{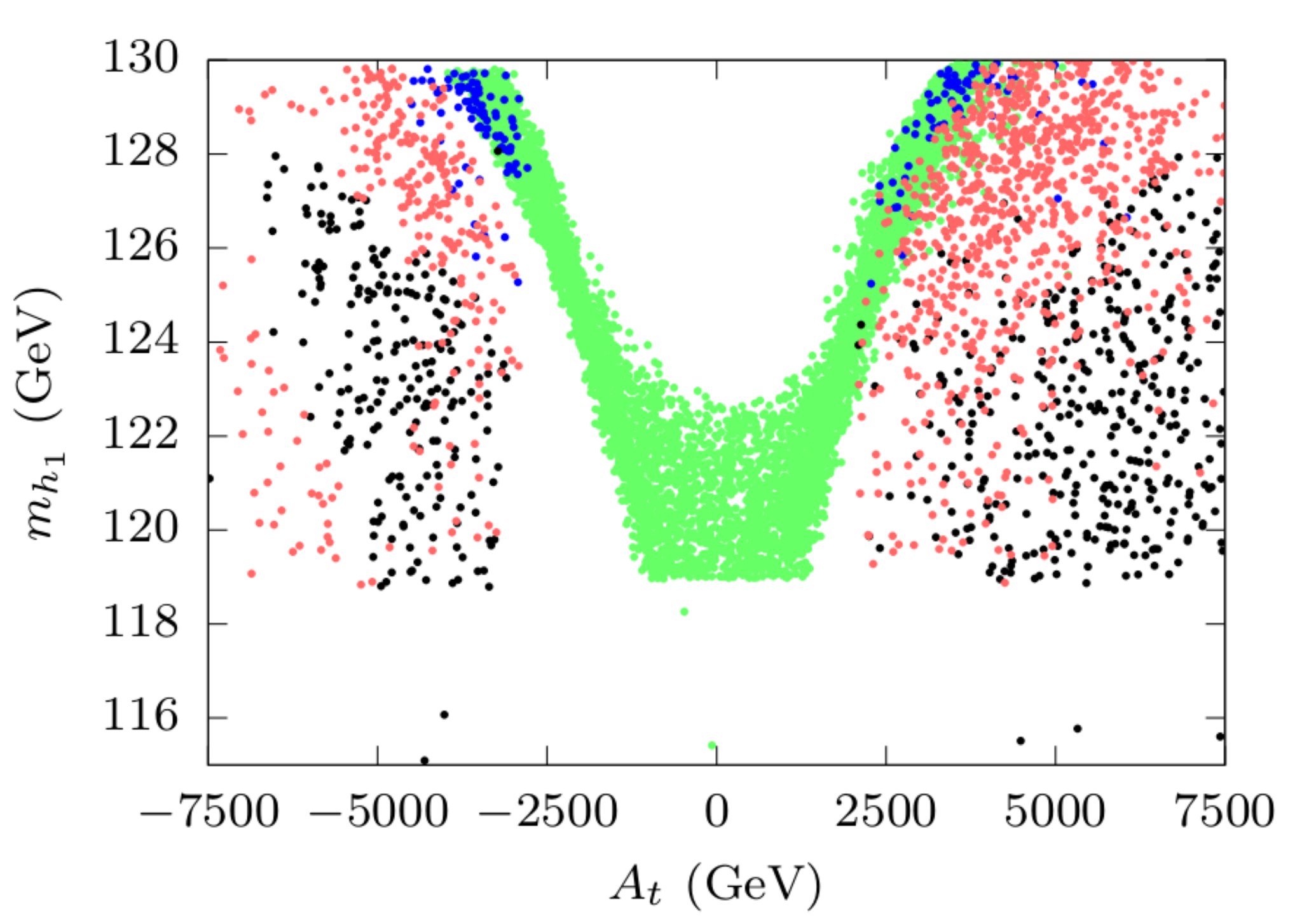}
\hskip -5pt
\includegraphics[height=0.24\textheight, width=0.49\columnwidth , clip]{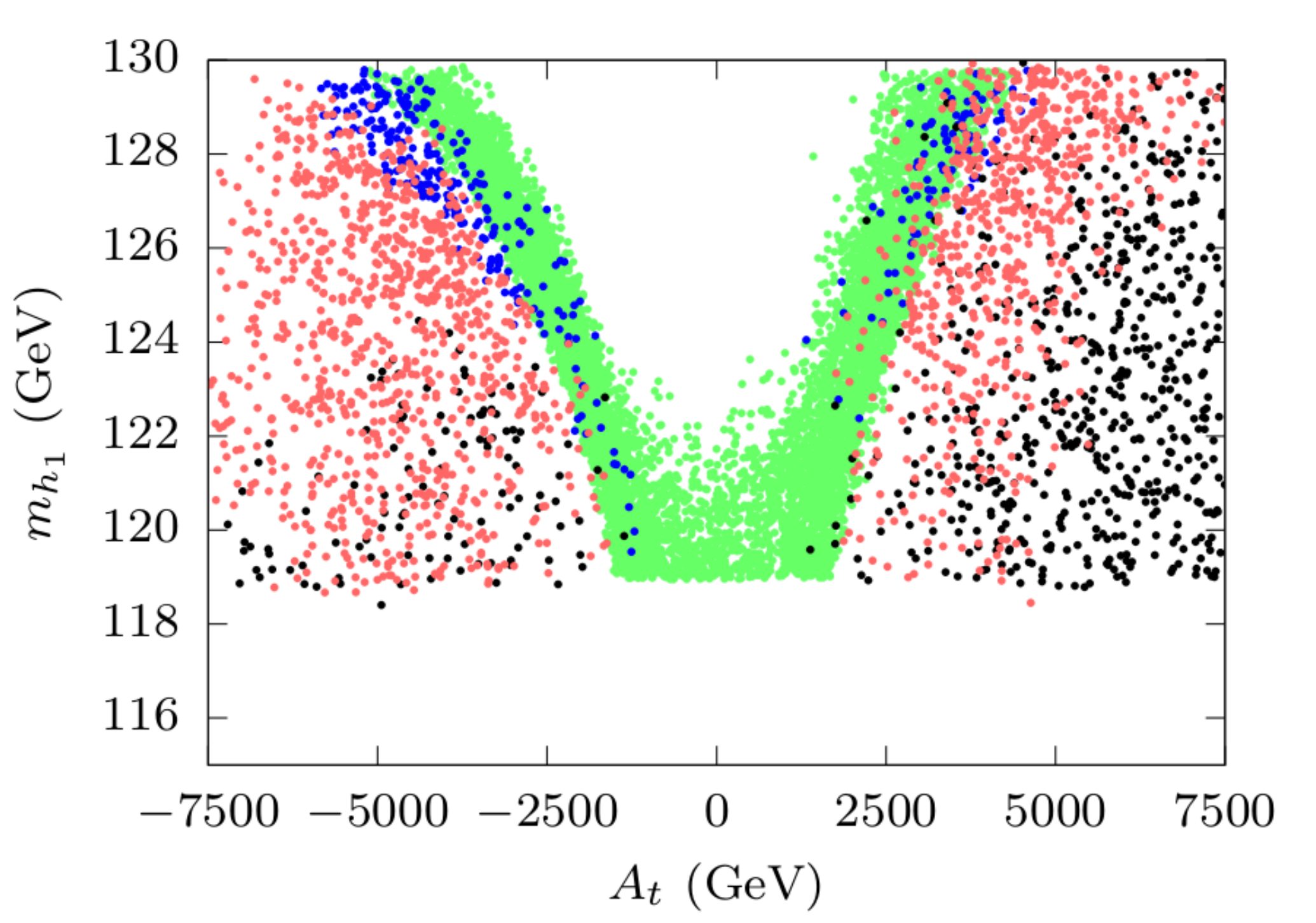}
\caption{Same as in figure \ref{fig:ccb-At-akappa} but in the $A_t-\mhone$ plane
for a fixed value of $\akappa= -100$ GeV. The values of $\alambda$ are chosen as follows:
$\alambda=2500 \, \mathrm{GeV} \; 
(500 \, \mathrm{GeV})$ for the case with a singlino- (higgsino)-like LSP in the MSSM-like
scenario and $\alambda=2500 \, \mathrm{GeV}  \;
(2000 \, \mathrm{GeV})$ for the case with a singlino- (higgsino)-like LSP in the NMSSM-like
scenario.}
\label{fig:ccb-fixed-akappa-alambda}
\end{figure}

Figure \ref{fig:ccb-fixed-akappa-alambda} presents the results of \veva ~scan in 
the familiar $A_t-\mhone$ plane with varying top squark masses and $A_t$ while 
$\akappa$ and $\alambda$ are kept fixed at values shown in its caption. 
In all the cases (mostly) non-overlapping 
regions with different kinds of stability property of the DSB vacuum are observed. 
The choice of $\akappa=-100$ GeV, as can be gleaned from figure 
\ref{fig:ccb-At-akappa}, rules out the possibility of development of deeper minima
in the `$S$' direction. Thus, the issue of stability of the DSB vacuum exclusively
refers to the appearance of deeper CCB minima as $|A_t|$ grows and hence should be
consistent with the traditional CCB constraints (in the MSSM) 
\cite{Chattopadhyay:2014gfa}. Indeed, if we take a close look at these set of 
plots this is found to be the case.
%
%
\section{Conclusions}
%
In this work we study in detail the occurrence and nature of possible vacua of 
the $Z_3$-symmetric phenomenological NMSSM. The study presented here goes beyond 
what is available in the existing literature in quite a few aspects. 

First, we break away from the fixed (flat) directions in the field space which 
hitherto were adhered to. We develop a {\tt Mathematica}-based framework that 
efficiently deals with the emerging complications when the neutral scalar fields 
are allowed to vary simultaneously. Furthermore, the study incorporates radiative 
correction at the 1-loop level to the tree-level neutral scalar potential.
Given that it is technically impossible to present a sensible visual map of the 
terrain of the scalar potential 
in more than two field dimensions (which is the case here), we illustrate some
key features of such a scenario by projecting the potential on suitable slices
of low dimensionality. The in-built flexibility of the framework facilitates 
targeted probe (sometimes purely analytical in nature) and thus, is capable of 
shedding important light on the intricacies such a system offers.
This semi-analytical framework is used further to derive numerical constraints on 
the free parameters of the scenario. 

Second, such an analysis is thoroughly
backed up by a detailed scan of the NMSSM parameter space using a dedicated
tool like \veva. This further includes the thermal correction to the 1-loop
effective potential and estimates the stability of the DSB vacuum in the presence
of a deeper minimum. This is, to the best of our knowledge, has not previously
been considered in the studies of viable vacua of the NMSSM scenarios.
The two modes of analyses are thus
complimentary in nature; the former helps comprehend the results from the
latter with a broad brush while the latter comes up with the state-of-the-art
refinements that could factor in all known and applicable, theoretical and  
experimental constraints from various pertinent sectors of particle physics and 
cosmology. Nonetheless, there is no denying the fact that the general problem at 
hand is ultimately one of a hard-core numerical nature. The semi-analytical approach 
has a limited scope and in practice, feasible when dealing with only a small number 
of scalar fields.

We carry out our analysis by dividing the NMSSM parameter space into two broad 
regions based primarily on the magnitude of $\lambda$; a small value of $\lambda$ leads
to an `MSSM-like' scenario while a larger value for it results in an `NMSSM-like'
one. Furthermore, for a given value of $\mueff$, a relatively small `$\kappa$' 
yields a scenario having a `singlino-like' LSP while for a larger `$\kappa$' one 
ends up with an LSP which is `higgsino-like'. As for the magnitude of $\mueff$ 
is concerned, we confine ourselves to somewhat smaller values for the same and thus
settle for $|\mueff| <500$ GeV. Such a choice is generally claimed to be friendly 
to (enhanced) `naturalness' via mitigation of the so-called `little hierarchy 
problem'. 

It appears in general that the `NMSSM-like' scenario with a higgsino-like LSP could 
have absolutely stable DSB vacuum over a large region of parameter space.

Thermal effects are found to play a crucial role in determining 
the fate of the DSB
vacuum. The issue appears to be more important in the MSSM-like case in which had 
it not been for the thermal contributions, some DSB vacuum configurations would
have survived as viable ones. 
In the MSSM-like scenario, minima deeper than the DSB vacuum that are still safe 
(thus turning the latter a `metastable' one) arise mostly along the $D$- and 
$F_S$-flat directions in the field space. In the NMSSM-like scenario these appear
along other flat directions as well.

With only three neutral scalars allowed to have \vevs ~we find that in the
MSSM-like scenario with a higgsino-like LSP, singlet scalars lighter than the
SM-Higgs boson are mostly disfavored on the thermal ground. For a situation 
with a singlino-like LSP in such a scenario, similar restrictions come in for 
larger values of the LSP mass. On the other hand, in the NMSSM-like scenario 
with a higgsino-like LSP is the singlet Higgs is prone to be heavier than the 
SM-like Higgs boson. Attempting to have a lighter singlet-like scalar makes 
this region thermally inviable. In contrast, since the NMSSM-like scenario 
with a singlino-like LSP has a comparatively low value of `$\kappa$', finding 
a somewhat lighter (than the SM-like Higgs boson) singlet scalar is easier by 
tweaking $\akappa$.  However, this comes with a price. The region with light 
singlet scalars tend to get metastable but long-lived (short-lived) for lighter 
(heavier) LSP. We also find that regions with $\mhone < {m_{H_{\mathrm{SM}}} \over 2}$ 
are now experimentally disfavored because of an unacceptably large 
decay rate for $H_{\mathrm{SM}} \to h_1 h_1$ when $\lambda$ is large. 

As far as the CCB sector is concerned, we observe that it is sensitive to $A_t$ 
(as in the MSSM) `$\kappa$' and $\akappa$ and their interplay. The latter two
determine the depth of the minima in the `$S$' direction which may easily surpass
the depths of the CCB minima and hence play the all important role in determining
the fate of the DSB vacuum. Traditional CCB bounds hold only when $\akappa$ is 
on the smaller side. Beyond a critical value which depends upon other free 
parameters, the dominance of $\akappa$ sets in. Under its spell, it is mostly the 
thermal effect that makes the DSB vacuum short-lived. 

We find that light singlet scalars are rather common over a significant region 
of the parameter space of phenomenological interest that is also compatible with 
a viable DSB vacuum. They can also be accompanied by a light LSP having 
significant singlino admixture. Hence their search at the LHC would continue to 
be rather interesting. This usually includes looking for production of a pair of 
such light scalars (or their production in cascades of heavier particles) 
followed by their decays to final states with multiple $b$-jets and/or photons 
and/or leptons.
%
\acknowledgments
%
JB is partially supported by funding available from the Department of 
Atomic Energy, Government of India for the Regional Centre for 
Accelerator-based Particle Physics, Harish-Chandra Research Institute.
He acknowledges helpful discussions with K. Das, F. Staub and T. Stefaniak.
UC acknowledges the hospitality of the Theoretical Physics Division
of CERN, Switzerland during the course of this work.
The authors acknowledge the use of the cluster computing setup available 
at RECAPP and at the High-Performance Computing facility of HRI.
They also thank Amit Khulve and Ravindra Yadav for technical support.
%
%

%

\begin{thebibliography}{999}
%
\bibitem{Aad:2012tfa}
  G.~Aad {\it et al.} [ATLAS Collaboration],
  ``Observation of a new particle in the search for the Standard Model Higgs boson with the ATLAS detector at the LHC,''
  Phys.\ Lett.\ B {\bf 716} (2012) 1
  [arXiv:1207.7214 [hep-ex]].



\bibitem{Chatrchyan:2012xdj}
  S.~Chatrchyan {\it et al.} [CMS Collaboration],
  ``Observation of a new boson at a mass of 125 GeV with the CMS experiment at the LHC,''
  Phys.\ Lett.\ B {\bf 716} (2012) 30
  [arXiv:1207.7235 [hep-ex]].



\bibitem{Kim:1983dt}
  J.~E.~Kim and H.~P.~Nilles,
  ``The mu Problem and the Strong CP Problem,''
  Phys.\ Lett.\  {\bf 138B} (1984) 150.



\bibitem{Ellwanger:2009dp}
  U.~Ellwanger, C.~Hugonie and A.~M.~Teixeira,
  ``The Next-to-Minimal Supersymmetric Standard Model,''
  Phys.\ Rept.\  {\bf 496} (2010) 1
  [arXiv:0910.1785 [hep-ph]].


\bibitem{Beuria:2015mta}
  J.~Beuria, A.~Chatterjee, A.~Datta and S.~K.~Rai,
  ``Two Light Stops in the NMSSM and the LHC,''
  JHEP {\bf 1509} (2015) 073
  [arXiv:1505.00604 [hep-ph]].


\bibitem{Beuria:2016mur}
  J.~Beuria, A.~Chatterjee and A.~Datta,
  ``Sbottoms of Natural NMSSM at the LHC,''
  JHEP {\bf 1608} (2016) 004
  [arXiv:1603.08463 [hep-ph]].

\bibitem{Maniatis:2009re}
  M.~Maniatis,
  ``The Next-to-Minimal Supersymmetric extension of the Standard Model reviewed,''
  Int.\ J.\ Mod.\ Phys.\ A {\bf 25} (2010) 3505
  [arXiv:0906.0777 [hep-ph]].



\bibitem{Sher:1988mj}
  M.~Sher,
  ``Electroweak Higgs Potentials and Vacuum Stability,''
  Phys.\ Rept.\  {\bf 179} (1989) 273.



\bibitem{AlvarezGaume:1983gj}
  L.~Alvarez-Gaume, J.~Polchinski and M.~B.~Wise,
  ``Minimal Low-Energy Supergravity,''
  Nucl.\ Phys.\ B {\bf 221} (1983) 495.



\bibitem{Gunion:1987qv}
  J.~F.~Gunion, H.~E.~Haber and M.~Sher,
  ``Charge / Color Breaking Minima and a-Parameter Bounds in Supersymmetric Models,''
  Nucl.\ Phys.\ B {\bf 306} (1988) 1.



\bibitem{Komatsu:1988mt}
  H.~Komatsu,
  ``New Constraints on Parameters in the Minimal Supersymmetric Model,''
  Phys.\ Lett.\ B {\bf 215} (1988) 323.



\bibitem{Casas:1995pd}
  J.~A.~Casas, A.~Lleyda and C.~Munoz,
  ``Strong constraints on the parameter space of the MSSM from charge and color breaking minima,''
  Nucl.\ Phys.\ B {\bf 471} (1996) 3
  [hep-ph/9507294].



\bibitem{Bordner:1995fh}
  A.~J.~Bordner,
  ``Parameter bounds in the supersymmetric standard model from charge / color breaking vacua,''
  hep-ph/9506409.



\bibitem{Strumia:1996pr}
  A.~Strumia,
  ``Charge and color breaking minima and constraints on the MSSM parameters,''
  Nucl.\ Phys.\ B {\bf 482} (1996) 24
  [hep-ph/9604417].



\bibitem{Baer:1996jn}
  H.~Baer, M.~Brhlik and D.~Castano,
  ``Constraints on the minimal supergravity model from nonstandard vacua,''
  Phys.\ Rev.\ D {\bf 54} (1996) 6944
  [hep-ph/9607465].



\bibitem{Abel:1998cc}
  S.~A.~Abel and C.~A.~Savoy,
  ``Charge and color breaking constraints in the MSSM with nonuniversal SUSY breaking,''
  Phys.\ Lett.\ B {\bf 444} (1998) 119
  [hep-ph/9809498].



\bibitem{Abel:1998wr}
  S.~Abel and T.~Falk,
  ``Charge and color breaking in the constrained MSSM,''
  Phys.\ Lett.\ B {\bf 444} (1998) 427
  [hep-ph/9810297].



\bibitem{Ferreira:2000hg}
  P.~M.~Ferreira,
  ``A Full one loop charge and color breaking effective potential,''
  Phys.\ Lett.\ B {\bf 509} (2001) 120
   Erratum: [Phys.\ Lett.\ B {\bf 518} (2001) 333]
  [hep-ph/0008115].



\bibitem{LeMouel:2001ym}
  C.~Le Mouel,
  ``Charge and color breaking conditions associated to the top quark Yukawa coupling,''
  Phys.\ Rev.\ D {\bf 64} (2001) 075009
  [hep-ph/0103341].



\bibitem{LeMouel:2001sf}
  C.~Le Mouel,
  ``Optimal charge and color breaking conditions in the MSSM,''
  Nucl.\ Phys.\ B {\bf 607} (2001) 38
  [hep-ph/0101351].



\bibitem{Brhlik:2001ni}
  M.~Brhlik,
  ``Charge and color breaking minima in supersymmetric models,''
  Nucl.\ Phys.\ Proc.\ Suppl.\  {\bf 101} (2001) 395.



\bibitem{Cerdeno:2003yt}
  D.~G.~Cerdeno, E.~Gabrielli, M.~E.~Gomez and C.~Munoz,
  ``Neutralino nucleon cross-section and charge and color breaking constraints,''
  JHEP {\bf 0306} (2003) 030
  [hep-ph/0304115].



\bibitem{Ferreira:2004yg}
  P.~M.~Ferreira,
  ``One-loop charge and colour breaking associated with the top Yukawa coupling,''
  hep-ph/0406234.



\bibitem{Brandenberger:1984cz}
  R.~H.~Brandenberger,
  ``Quantum Field Theory Methods and Inflationary Universe Models,''
  Rev.\ Mod.\ Phys.\  {\bf 57} (1985) 1.



\bibitem{Kusenko:1995jv}
  A.~Kusenko,
  ``Improved action method for analyzing tunneling in quantum field theory,''
  Phys.\ Lett.\ B {\bf 358} (1995) 51
  [hep-ph/9504418].



\bibitem{Riotto:1995am}
  A.~Riotto and E.~Roulet,
  ``Vacuum decay along supersymmetric flat directions,''
  Phys.\ Lett.\ B {\bf 377} (1996) 60
  [hep-ph/9512401].



\bibitem{Falk:1996zt}
  T.~Falk, K.~A.~Olive, L.~Roszkowski, A.~Singh and M.~Srednicki,
  ``Constraints from inflation and reheating on superpartner masses,''
  Phys.\ Lett.\ B {\bf 396} (1997) 50
  [hep-ph/9611325].



\bibitem{Kusenko:1996jn}
  A.~Kusenko, P.~Langacker and G.~Segre,
  ``Phase transitions and vacuum tunneling into charge and color breaking minima in the MSSM,''
  Phys.\ Rev.\ D {\bf 54} (1996) 5824
  [hep-ph/9602414].



\bibitem{Kusenko:1996xt}
  A.~Kusenko and P.~Langacker,
  ``Is the vacuum stable?,''
  Phys.\ Lett.\ B {\bf 391} (1997) 29
  [hep-ph/9608340].



\bibitem{Kusenko:1996vp}
  A.~Kusenko,
  ``Color and charge breaking minima in the MSSM,''
  Nucl.\ Phys.\ Proc.\ Suppl.\  {\bf 52A} (1997) 67
  [hep-ph/9607287].



\bibitem{Cohen:2013kna}
  T.~Cohen and J.~G.~Wacker,
  ``Here be Dragons: The Unexplored Continents of the CMSSM,''
  JHEP {\bf 1309} (2013) 061
  [arXiv:1305.2914 [hep-ph]].



\bibitem{Datta:2000xy}
  A.~Datta, A.~Kundu and A.~Samanta,
  ``Yukawa unification and unstable minima of the supersymmetric scalar potential,''
  Phys.\ Rev.\ D {\bf 63} (2001) 015008
  [hep-ph/0007148].



\bibitem{Datta:2001dc}
  A.~Datta and A.~Samanta,
  ``Electroweak precision data, light sleptons and stability of the SUSY scalar potential,''
  Phys.\ Lett.\ B {\bf 526} (2002) 111
  [hep-ph/0111222].



\bibitem{Datta:2001qa}
  A.~Datta and A.~Samanta,
  ``Effects of SO(10) D term on Yukawa unification and unstable minima of the supersymmetric scalar potential,''
  J.\ Phys.\ G {\bf 29} (2003) 2721
  [hep-ph/0108056].



\bibitem{Gabrielli:2001py}
  E.~Gabrielli, K.~Huitu and S.~Roy,
  ``Vacuum stability bounds in anomaly and gaugino mediated SUSY breaking models,''
  Phys.\ Rev.\ D {\bf 65} (2002) 075005
  [hep-ph/0108246].



\bibitem{Datta:2004hr}
  A.~Datta and A.~Samanta,
  ``LEP data and the stability of the potential confronts the mSUGRA model,''
  Phys.\ Lett.\ B {\bf 607} (2005) 144
  [hep-ph/0406129].



\bibitem{Kobayashi:2010zx}
  T.~Kobayashi and T.~Shimomura,
  ``Constraints from Color and/or Charge Breaking Minima in the $\nu$SSM,''
  Phys.\ Rev.\ D {\bf 82} (2010) 035008
  [arXiv:1006.0062 [hep-ph]].



\bibitem{Hisano:2010re}
  J.~Hisano and S.~Sugiyama,
  ``Charge-breaking constraints on left-right mixing of stau's,''
  Phys.\ Lett.\ B {\bf 696} (2011) 92
   Erratum: [Phys.\ Lett.\ B {\bf 719} (2013) 472]
  [arXiv:1011.0260 [hep-ph]].



\bibitem{Carena:2012mw}
  M.~Carena, S.~Gori, I.~Low, N.~R.~Shah and C.~E.~M.~Wagner,
  ``Vacuum Stability and Higgs Diphoton Decays in the MSSM,''
  JHEP {\bf 1302} (2013) 114
  [arXiv:1211.6136 [hep-ph]].



\bibitem{Camargo-Molina:2013sta}
  J.~E.~Camargo-Molina, B.~O'Leary, W.~Porod and F.~Staub,
  ``Stability of the CMSSM against sfermion VEVs,''
  JHEP {\bf 1312} (2013) 103
  [arXiv:1309.7212 [hep-ph]].



\bibitem{Chowdhury:2013dka}
  D.~Chowdhury, R.~M.~Godbole, K.~A.~Mohan and S.~K.~Vempati,
  ``Charge and Color Breaking Constraints in MSSM after the Higgs Discovery at LHC,''
  JHEP {\bf 1402} (2014) 110
  [arXiv:1310.1932 [hep-ph]].



\bibitem{Blinov:2013uda}
  N.~Blinov and D.~E.~Morrissey,
  ``Charge and Color Breaking Constraints in the Minimal Supersymmetric Standard Model,''
  arXiv:1309.7397 [hep-ph].



\bibitem{Blinov:2013fta}
  N.~Blinov and D.~E.~Morrissey,
  ``Vacuum Stability and the MSSM Higgs Mass,''
  JHEP {\bf 1403} (2014) 106
  [arXiv:1310.4174 [hep-ph]].



\bibitem{Camargo-Molina:2014pwa}
  J.~E.~Camargo-Molina, B.~Garbrecht, B.~O'Leary, W.~Porod and F.~Staub,
  ``Constraining the Natural MSSM through tunneling to color-breaking vacua at zero and non-zero temperature,''
  Phys.\ Lett.\ B {\bf 737} (2014) 156
  [arXiv:1405.7376 [hep-ph]].



\bibitem{Chattopadhyay:2014gfa}
  U.~Chattopadhyay and A.~Dey,
  ``Exploring MSSM for Charge and Color Breaking and Other Constraints in the Context of Higgs@125 GeV,''
  JHEP {\bf 1411} (2014) 161
  [arXiv:1409.0611 [hep-ph]].



\bibitem{Hollik:2015pra}
  W.~G.~Hollik,
  ``Charge and color breaking constraints in the Minimal Supersymmetric Standard Model associated with the bottom Yukawa coupling,''
  Phys.\ Lett.\ B {\bf 752} (2016) 7
  [arXiv:1508.07201 [hep-ph]].



\bibitem{Hollik:2016dcm}
  W.~G.~Hollik,
  ``A new view on vacuum stability in the MSSM,''
  JHEP {\bf 1608} (2016) 126
  [arXiv:1606.08356 [hep-ph]].



\bibitem{Ellwanger:1996gw}
  U.~Ellwanger, M.~Rausch de Traubenberg and C.~A.~Savoy,
  ``Phenomenology of supersymmetric models with a singlet,''
  Nucl.\ Phys.\ B {\bf 492} (1997) 21
  [hep-ph/9611251].



\bibitem{Ellwanger:1999bv}
  U.~Ellwanger and C.~Hugonie,
  ``Constraints from charge and color breaking minima in the (M+1)SSM,''
  Phys.\ Lett.\ B {\bf 457} (1999) 299
  [hep-ph/9902401].



\bibitem{Kanehata:2011ei}
  Y.~Kanehata, T.~Kobayashi, Y.~Konishi, O.~Seto and T.~Shimomura,
  ``Constraints from Unrealistic Vacua in the Next-to-Minimal Supersymmetric Standard Model,''
  Prog.\ Theor.\ Phys.\  {\bf 126} (2011) 1051
  [arXiv:1103.5109 [hep-ph]].



\bibitem{Kobayashi:2012xv}
  T.~Kobayashi, T.~Shimomura and T.~Takahashi,
  ``Constraining the Higgs sector from False Vacua in the Next-to-Minimal Supersymmetric Standard Model,''
  Phys.\ Rev.\ D {\bf 86} (2012) 015029
  [arXiv:1203.4328 [hep-ph]].


\bibitem{Vilenkin:1984ib}
A.~Vilenkin,
Phys.\ Rept.\  {\bf 121} (1985) 263.
doi:10.1016/0370-1573(85)90033-X

\bibitem{Ellwanger:1983mg}
U.~Ellwanger,
Phys.\ Lett.\  {\bf 133B} (1983) 187.
doi:10.1016/0370-2693(83)90557-9


\bibitem{Bagger:1993ji}
J.~Bagger and E.~Poppitz,
Phys.\ Rev.\ Lett.\  {\bf 71} (1993) 2380
doi:10.1103/PhysRevLett.71.2380
[hep-ph/9307317].

\bibitem{Jain:1994tk}
V.~Jain,
Phys.\ Lett.\ B {\bf 351} (1995) 481
doi:10.1016/0370-2693(95)00373-S
[hep-ph/9407382].

\bibitem{Bagger:1995ay}
J.~Bagger, E.~Poppitz and L.~Randall,
Nucl.\ Phys.\ B {\bf 455} (1995) 59
doi:10.1016/0550-3213(95)00463-3
[hep-ph/9505244].

\bibitem{Abel:1995wk}
S.~A.~Abel, S.~Sarkar and P.~L.~White,
Nucl.\ Phys.\ B {\bf 454} (1995) 663
doi:10.1016/0550-3213(95)00483-9
[hep-ph/9506359].


\bibitem{Abel:1996cr}
  S.~A.~Abel,
  Nucl.\ Phys.\ B {\bf 480} (1996) 55
  doi:10.1016/S0550-3213(96)00470-1
  [hep-ph/9609323].


\bibitem{Kolda:1998rm}
  C.~F.~Kolda, S.~Pokorski and N.~Polonsky,
  Phys.\ Rev.\ Lett.\  {\bf 80} (1998) 5263
  doi:10.1103/PhysRevLett.80.5263
  [hep-ph/9803310].

\bibitem{Panagiotakopoulos:1998yw} 
  C.~Panagiotakopoulos and K.~Tamvakis,
  Phys.\ Lett.\ B {\bf 446}, 224 (1999)
  doi:10.1016/S0370-2693(98)01493-2
  [hep-ph/9809475].


\bibitem{Affleck:1980ac}
  I.~Affleck,
  ``Quantum Statistical Metastability,''
  Phys.\ Rev.\ Lett.\  {\bf 46} (1981) 388.


\bibitem{Linde:1980tt}
  A.~D.~Linde,
  ``Fate of the False Vacuum at Finite Temperature: Theory and Applications,''
  Phys.\ Lett.\  {\bf 100B} (1981) 37.


\bibitem{Linde:1981zj}
  A.~D.~Linde,
  ``Decay of the False Vacuum at Finite Temperature,''
  Nucl.\ Phys.\ B {\bf 216} (1983) 421
   Erratum: [Nucl.\ Phys.\ B {\bf 223} (1983) 544].

\bibitem{Masoumi:2015psa}
  A.~Masoumi,
  ``Topics in vacuum decay (Ph.D Thesis),''
  arXiv:1505.06397 [hep-th].


\bibitem{Brignole:1993wv}
  A.~Brignole, J.~R.~Espinosa, M.~Quiros and F.~Zwirner,
  ``Aspects of the electroweak phase transition in the minimal supersymmetric standard model,''
  Phys.\ Lett.\ B {\bf 324} (1994) 181
  [hep-ph/9312296].


\bibitem{Baer:2012up}
  H.~Baer, V.~Barger, P.~Huang, A.~Mustafayev and X.~Tata,
  ``Radiative natural SUSY with a 125 GeV Higgs boson,''
  Phys.\ Rev.\ Lett.\  {\bf 109} (2012) 161802
  [arXiv:1207.3343 [hep-ph]].


\bibitem{Baer:2013ava}
  H.~Baer, V.~Barger, P.~Huang, D.~Mickelson, A.~Mustafayev and X.~Tata,
  ``Naturalness, Supersymmetry and Light Higgsinos: A Snowmass Whitepaper,''
  arXiv:1306.2926 [hep-ph].


\bibitem{Mustafayev:2014lqa}
  A.~Mustafayev and X.~Tata,
  ``Supersymmetry, Naturalness, and Light Higgsinos,''
  Indian J.\ Phys.\  {\bf 88} (2014) 991
  [arXiv:1404.1386 [hep-ph]].


\bibitem{Baer:2015rja}
  H.~Baer, V.~Barger and M.~Savoy,
  ``Upper bounds on sparticle masses from naturalness or how to disprove weak scale supersymmetry,''
  Phys.\ Rev.\ D {\bf 93} (2016) no.3,  035016
  [arXiv:1509.02929 [hep-ph]].


\bibitem{Camargo-Molina:2013qva}
  J.~E.~Camargo-Molina, B.~O'Leary, W.~Porod and F.~Staub,
  ``$\mathbf{Vevacious}$: A Tool For Finding The Global Minima Of One-Loop Effective Potentials With Many Scalars,''
  Eur.\ Phys.\ J.\ C {\bf 73} (2013) no.10,  2588
  [arXiv:1307.1477 [hep-ph]].


\bibitem{Cerdeno:2004xw}
  D.~G.~Cerdeno, C.~Hugonie, D.~E.~Lopez-Fogliani, C.~Munoz and A.~M.~Teixeira,
  ``Theoretical predictions for the direct detection of neutralino dark matter in the NMSSM,''
  JHEP {\bf 0412} (2004) 048
  [hep-ph/0408102].


\bibitem{Quiros:1999jp}
  M.~Quiros,
  ``Finite temperature field theory and phase transitions,''
  hep-ph/9901312.



\bibitem{Martin:2014bca}
  S.~P.~Martin,
  ``Taming the Goldstone contributions to the effective potential,''
  Phys.\ Rev.\ D {\bf 90} (2014) no.1,  016013
  [arXiv:1406.2355 [hep-ph]].



\bibitem{Belanger:2012tt}
  G.~Belanger, U.~Ellwanger, J.~F.~Gunion, Y.~Jiang, S.~Kraml and J.~H.~Schwarz,
  ``Higgs Bosons at 98 and 125 GeV at LEP and the LHC,''
  JHEP {\bf 1301} (2013) 069
  [arXiv:1210.1976 [hep-ph]].



\bibitem{Bhattacherjee:2013vga}
  B.~Bhattacherjee, M.~Chakraborti, A.~Chakraborty, U.~Chattopadhyay, D.~Das and D.~K.~Ghosh,
  ``Implications of the 98 GeV and 125 GeV Higgs scenarios in nondecoupling supersymmetry with updated ATLAS, CMS, and PLANCK data,''
  Phys.\ Rev.\ D {\bf 88} (2013) no.3,  035011
  [arXiv:1305.4020 [hep-ph]].



\bibitem{Bhattacherjee:2015qct}
  B.~Bhattacherjee, M.~Chakraborti, A.~Chakraborty, U.~Chattopadhyay and D.~K.~Ghosh,
  ``Status of the 98–125 GeV Higgs bosons scenario with updated LHC-8 data,''
  Phys.\ Rev.\ D {\bf 93} (2016) no.7,  075004
  [arXiv:1511.08461 [hep-ph]].



\bibitem{Ellwanger:2011sk}
  U.~Ellwanger,
  ``Higgs Bosons in the Next-to-Minimal Supersymmetric Standard Model at the LHC,''
  Eur.\ Phys.\ J.\ C {\bf 71} (2011) 1782
  [arXiv:1108.0157 [hep-ph]].



\bibitem{Huang:2014ifa}
  W.~Huang, Z.~Kang, J.~Shu, P.~Wu and J.~M.~Yang,
  ``New insights in the electroweak phase transition in the NMSSM,''
  Phys.\ Rev.\ D {\bf 91} (2015) no.2,  025006
  [arXiv:1405.1152 [hep-ph]].



\bibitem{Carena:1995bx}
  M.~Carena, J.~R.~Espinosa, M.~Quiros and C.~E.~M.~Wagner,
  ``Analytical expressions for radiatively corrected Higgs masses and couplings in the MSSM,''
  Phys.\ Lett.\ B {\bf 355} (1995) 209
  [hep-ph/9504316].



\bibitem{Carena:1995wu}
  M.~Carena, M.~Quiros and C.~E.~M.~Wagner,
  ``Effective potential methods and the Higgs mass spectrum in the MSSM,''
  Nucl.\ Phys.\ B {\bf 461} (1996) 407
  [hep-ph/9508343].



\bibitem{Haber:1996fp}
  H.~E.~Haber, R.~Hempfling and A.~H.~Hoang,
  ``Approximating the radiatively corrected Higgs mass in the minimal supersymmetric model,''
  Z.\ Phys.\ C {\bf 75} (1997) 539
  [hep-ph/9609331].



\bibitem{Djouadi:2005gj}
  A.~Djouadi,
  ``The Anatomy of electro-weak symmetry breaking. II. The Higgs bosons in the minimal supersymmetric model,''
  Phys.\ Rept.\  {\bf 459} (2008) 1
  [hep-ph/0503173].



\bibitem{Camargo-Molina:2015qpa}
  J.~E.~Camargo-Molina,
  ``Vacuum stability of models with many scalars,''



\bibitem{hom4ps2}
T. Lee, T. Li and C. Tsai,
 ``HOM4PS-2.0: a software package for solving polynomial systems by the polyhedral
homotopy continuation method'',
Computing 83(2), 109 (2008).

\bibitem{pyminuit}
J. Pivarski, {\tt https://code.google.com/p/pyminuit/}.

\bibitem{minuit}
F. James and M. Roos,
 ``Minuit - a system for function minimization and analysis of the parameter
 errors and correlations'',
Comput. Phys. Commune. 10, 343 (1975).

\bibitem{vevacious-code}
See into the \veva ~code ({\tt https://vevacious.hepforge.org/}).

\bibitem{Coleman:1977py}
  S.~R.~Coleman,
  ``The Fate of the False Vacuum. 1. Semiclassical Theory,''
  Phys.\ Rev.\ D {\bf 15} (1977) 2929
   Erratum: [Phys.\ Rev.\ D {\bf 16} (1977) 1248].



\bibitem{Callan:1977pt}
  C.~G.~Callan, Jr. and S.~R.~Coleman,
  ``The Fate of the False Vacuum. 2. First Quantum Corrections,''
  Phys.\ Rev.\ D {\bf 16} (1977) 1762.






\bibitem{Wainwright:2011kj}
  C.~L.~Wainwright,
  ``CosmoTransitions: Computing Cosmological Phase Transition Temperatures and Bubble Profiles with Multiple Fields,''
  Comput.\ Phys.\ Commun.\  {\bf 183} (2012) 2006
  [arXiv:1109.4189 [hep-ph]].



\bibitem{Djouadi:2002ze}
  A.~Djouadi, J.~L.~Kneur and G.~Moultaka,
  ``SuSpect: A Fortran code for the supersymmetric and Higgs particle spectrum in the MSSM,''
  Comput.\ Phys.\ Commun.\  {\bf 176} (2007) 426
  [hep-ph/0211331].



\bibitem{Ellwanger:2004xm}
  U.~Ellwanger, J.~F.~Gunion and C.~Hugonie,
  ``NMHDECAY: A Fortran code for the Higgs masses, couplings and decay widths in the NMSSM,''
  JHEP {\bf 0502} (2005) 066
  [hep-ph/0406215].



\bibitem{Ellwanger:2005dv}
  U.~Ellwanger and C.~Hugonie,
  ``NMHDECAY 2.0: An Updated program for sparticle masses, Higgs masses, couplings and decay widths in the NMSSM,''
  Comput.\ Phys.\ Commun.\  {\bf 175} (2006) 290
  [hep-ph/0508022].



\bibitem{Ellwanger:2006rn}
  U.~Ellwanger and C.~Hugonie,
  ``NMSPEC: A Fortran code for the sparticle and Higgs masses in the NMSSM with GUT scale boundary conditions,''
  Comput.\ Phys.\ Commun.\  {\bf 177} (2007) 399
  [hep-ph/0612134].



\bibitem{mathematica}
Wolfram Research, Inc., Mathematica, Version 10.0, Champaign, IL (2014).


\bibitem{Djouadi:2008uj}
  A.~Djouadi, U.~Ellwanger and A.~M.~Teixeira,
  ``Phenomenology of the constrained NMSSM,''
  JHEP {\bf 0904} (2009) 031
  [arXiv:0811.2699 [hep-ph]].



\bibitem{Barroso:2013awa}
  A.~Barroso, P.~M.~Ferreira, I.~P.~Ivanov and R.~Santos,
  ``Metastability bounds on the two Higgs doublet model,''
  JHEP {\bf 1306} (2013) 045
  [arXiv:1303.5098 [hep-ph]].



\bibitem{Gamberini:1989jw}
  G.~Gamberini, G.~Ridolfi and F.~Zwirner,
  ``On Radiative Gauge Symmetry Breaking in the Minimal Supersymmetric Model,''
  Nucl.\ Phys.\ B {\bf 331} (1990) 331.



\bibitem{Einhorn:2007rv}
  M.~B.~Einhorn and D.~R.~T.~Jones,
  ``The Effective potential, the renormalisation group and vacuum stability,''
  JHEP {\bf 0704} (2007) 051
  [hep-ph/0702295 [HEP-PH]].



\bibitem{Andreassen:2014eha}
  A.~Andreassen, W.~Frost and M.~D.~Schwartz,
  ``Consistent Use of Effective Potentials,''
  Phys.\ Rev.\ D {\bf 91} (2015) no.1,  016009
  [arXiv:1408.0287 [hep-ph]].



\bibitem{Abel:1998ie}
  S.~A.~Abel and C.~A.~Savoy,
  ``On metastability in supersymmetric models,''
  Nucl.\ Phys.\ B {\bf 532} (1998) 3
  [hep-ph/9803218].



\bibitem{Allanach:2008qq}
  B.~C.~Allanach {\it et al.},
  ``SUSY Les Houches Accord 2,''
  Comput.\ Phys.\ Commun.\  {\bf 180} (2009) 8
  [arXiv:0801.0045 [hep-ph]].



\bibitem{Staub:2013tta}
  F.~Staub,
  ``SARAH 4 : A tool for (not only SUSY) model builders,''
  Comput.\ Phys.\ Commun.\  {\bf 185} (2014) 1773
  [arXiv:1309.7223 [hep-ph]].



\bibitem{Staub:2015kfa}
  F.~Staub,
  ``Exploring new models in all detail with SARAH,''
  Adv.\ High Energy Phys.\  {\bf 2015} (2015) 840780
  [arXiv:1503.04200 [hep-ph]].



\bibitem{Porod:2003um}
  W.~Porod,
  ``SPheno, a program for calculating supersymmetric spectra, SUSY particle decays and SUSY particle production at e+ e- colliders,''
  Comput.\ Phys.\ Commun.\  {\bf 153} (2003) 275
  [hep-ph/0301101].



\bibitem{Porod:2011nf}
  W.~Porod and F.~Staub,
  ``SPheno 3.1: Extensions including flavour, CP-phases and models beyond the MSSM,''
  Comput.\ Phys.\ Commun.\  {\bf 183} (2012) 2458
  [arXiv:1104.1573 [hep-ph]].



\bibitem{Bechtle:2013wla}
  P.~Bechtle, O.~Brein, S.~Heinemeyer, O.~Stål, T.~Stefaniak, G.~Weiglein and K.~E.~Williams,
  ``$\mathsf{HiggsBounds}-4$: Improved Tests of Extended Higgs Sectors against Exclusion Bounds from LEP, the Tevatron and the LHC,''
  Eur.\ Phys.\ J.\ C {\bf 74} (2014) no.3,  2693
  [arXiv:1311.0055 [hep-ph]].



\bibitem{Bechtle:2013xfa}
  P.~Bechtle, S.~Heinemeyer, O.~Stål, T.~Stefaniak and G.~Weiglein,
  ``$HiggsSignals$: Confronting arbitrary Higgs sectors with measurements at the Tevatron and the LHC,''
  Eur.\ Phys.\ J.\ C {\bf 74} (2014) no.2,  2711
  [arXiv:1305.1933 [hep-ph]].



\bibitem{Amhis:2012bh}
  Y.~Amhis {\it et al.} [Heavy Flavor Averaging Group Collaboration],
  ``Averages of B-Hadron, C-Hadron, and tau-lepton properties as of early 2012,''
  arXiv:1207.1158 [hep-ex].



\bibitem{Ade:2015xua}
  P.~A.~R.~Ade {\it et al.} [Planck Collaboration],
  ``Planck 2015 results. XIII. Cosmological parameters,''
  Astron.\ Astrophys.\  {\bf 594} (2016) A13
  [arXiv:1502.01589 [astro-ph.CO]].



\bibitem{Belanger:2014vza}
  G.~Bélanger, F.~Boudjema, A.~Pukhov and A.~Semenov,
  ``micrOMEGAs4.1: two dark matter candidates,''
  Comput.\ Phys.\ Commun.\  {\bf 192} (2015) 322
  [arXiv:1407.6129 [hep-ph]].



\bibitem{lepsusy}
LEPSUSYWG, ALEPH, DELPHI, L3 and OPAL experiments,
note LEPSUSYWG/01-03.1
{\tt (http://lepsusy.web.cern.ch/lepsusy/Welcome.html)}.

%
\end{thebibliography}
\end{document}